\documentclass[review]{elsarticle}

\usepackage[colorlinks,citecolor=blue,linktoc=all,linkcolor=cyan]{hyperref}
\usepackage{graphicx}

\usepackage[T1]{fontenc}
\usepackage{dsfont}               
\usepackage{mathrsfs}             
\usepackage{slashed}              
\usepackage{amsmath}
\usepackage{amssymb}
\usepackage{amsbsy}
\usepackage{amsfonts}

\numberwithin{equation}{section}
\numberwithin{table}{section}
\numberwithin{figure}{section}

\journal{Progress in Particle and Nuclear Physics}

\usepackage{titlesec}
\usepackage{sectsty}
\titleformat{\section}{\normalfont\Large\bfseries}{\thesection}{1em}{}
\titleformat{\subsection}{\normalfont\large\bfseries}{\thesubsection}{1em}{}
\titleformat{\subsubsection}{\normalfont\normalsize\bfseries}{\thesubsubsection}{1em}{}

\usepackage{multirow}
\usepackage{bm}
\usepackage{tikz}
\usetikzlibrary{trees}
\usetikzlibrary{decorations.pathmorphing}
\usetikzlibrary{decorations.markings}
\usetikzlibrary{shapes.misc}
\usetikzlibrary{matrix}
\usetikzlibrary{decorations.pathreplacing}
\usetikzlibrary{arrows.meta}
\usepackage{pgfplots}
\usepgfplotslibrary{colormaps}
\usepgfplotslibrary{fillbetween}
\pgfplotsset{compat=1.18} 

\makeatletter

\pgfdeclaredecoration{gluon}{coil}
{
  \state{coil}[switch if less than=%
    0.5\pgfdecorationsegmentlength+
    \pgfdecorationsegmentaspect\pgfdecorationsegmentamplitude+%
    \pgfdecorationsegmentaspect\pgfdecorationsegmentamplitude to last,
               width=+\pgfdecorationsegmentlength]
  {
    \pgfpathcurveto
    {\pgfpoint@oncoil{0    }{ 0.555}{1}}
    {\pgfpoint@oncoil{0.445}{ 1    }{2}}
    {\pgfpoint@oncoil{1    }{ 1    }{3}}
    \pgfpathcurveto
    {\pgfpoint@oncoil{1.555}{ 1    }{4}}
    {\pgfpoint@oncoil{2    }{ 0.555}{5}}
    {\pgfpoint@oncoil{2    }{ 0    }{6}}
    \pgfpathcurveto
    {\pgfpoint@oncoil{2    }{-0.555}{7}}
    {\pgfpoint@oncoil{1.555}{-1    }{8}}
    {\pgfpoint@oncoil{1    }{-1    }{9}}
    \pgfpathcurveto
    {\pgfpoint@oncoil{0.445}{-1    }{10}}
    {\pgfpoint@oncoil{0    }{-0.555}{11}}
    {\pgfpoint@oncoil{0    }{ 0    }{12}}
  }
  \state{last}[next state=final]
  {
    \pgfpathcurveto
    {\pgfpoint@oncoil{0    }{ 0.555}{1}}
    {\pgfpoint@oncoil{0.445}{ 1    }{2}}
    {\pgfpoint@oncoil{1    }{ 1    }{3}}
    \pgfpathcurveto
    {\pgfpoint@oncoil{1.555}{ 1    }{4}}
    {\pgfpoint@oncoil{2    }{ 0.555}{5}}
    {\pgfpoint@oncoil{2    }{ 0    }{6}}
  }
  \state{final}{}
}

\def\pgfpoint@oncoil#1#2#3{%
  \pgf@x=#1\pgfdecorationsegmentamplitude%
  \pgf@x=\pgfdecorationsegmentaspect\pgf@x%
  \pgf@y=#2\pgfdecorationsegmentamplitude%
  \pgf@xa=0.083333333333\pgfdecorationsegmentlength%
  \advance\pgf@x by#3\pgf@xa%
}

\makeatother

\newcommand{\be}{\begin{equation}}
\newcommand{\ee}{\end{equation}}
\newcommand{\expv}[1]{\langle #1 \rangle}
\newcommand{\Expv}[1]{\left\langle #1 \right\rangle}
\newcommand{\tr}{\textmd{tr}}
\newcommand{\Tr}{\textmd{Tr}}
\newcommand{\A}{\mathcal{A}}
\newcommand{\U}{\mathcal{U}}
\newcommand{\Z}{\mathcal{Z}}
\newcommand{\D}{\mathcal{D}}
\newcommand{\F}{\mathcal{F}}
\newcommand{\B}{\mathcal{B}}
\newcommand{\I}{\mathcal{I}}
\renewcommand{\I}{\mathcal{I}}
\renewcommand{\L}{\mathcal{L}}
\renewcommand{\S}{\mathcal{S}}
\newcommand{\T}{\mathcal{T}}
\newcommand{\Q}{\mathcal{Q}}
\newcommand{\E}{\mathcal{E}}

\newcommand{\M}{\mathcal{M}}
\renewcommand{\P}{\mathcal{P}}
\renewcommand{\O}{\mathcal{O}}
\newcommand{\dd}{\textmd{d}}
\newcommand{\sgn}{\textmd{sgn}}
\newcommand{\Dsf}{\slashed{D}_f}
\newcommand{\GeVin}{\textmd{ GeV}^{-1}}
\newcommand{\GeVsq}{\textmd{ GeV}^2}
\newcommand{\MeV}{\textmd{ MeV}}
\newcommand{\fm}{\textmd{ fm}}
\newcommand{\GeV}{\textmd{ GeV}}

\begin{document}
	
	\begin{frontmatter}
		
		\title{QCD with background electromagnetic fields on the lattice: a review}

		\author[mymainaddress,myotheraddress]{Gergely Endr\H{o}di\corref{mycorrespondingauthor}}
		\cortext[mycorrespondingauthor]{Corresponding author}
		\ead{gergely.endrodi@ttk.elte.hu}
		
		\address[mymainaddress]{Institute of Physics and Astronomy, ELTE E\"otv\"os Lor\'and University, P\'azm\'any P.\ s\'et\'any 1/A, H-1117 Budapest, Hungary}
		\address[myotheraddress]{Universit{\"a}t Bielefeld, Universit{\"a}tsstra{\ss}e 25, 33615 Bielefeld, Germany}
		
		\begin{abstract}
		This review provides a comprehensive summary of results on the physics of strongly interacting matter in the presence of background electromagnetic fields, obtained via numerical lattice simulations of the underlying theory, Quantum Chromodynamics (QCD). Lattice QCD has guided our understanding of magnetized quarks and gluons via landmark results on the phase diagram, the equation of state, the confinement mechanism, anomalous transport phenomena as well as many more fascinating effects. Some of the lattice results lead to completely new paradigms in the description of hot magnetized quark matter and provided useful insights to a broad high-energy particle physics community. Since the first lattice QCD simulations with background fields, this field has been established as an independent research direction. We present the current status and recent developments of this field, together with an outlook including open questions to be answered in the near future.
		\end{abstract}
		
		
	\end{frontmatter}
	
	\newpage
	\thispagestyle{empty}
	\tableofcontents
	

	\newpage
	\section{Introduction}
	\label{chap:intro}

Quantum Chromodynamics (QCD) is the theory of the strong interactions. Being a strongly coupled non-perturbative quantum field theory, QCD explains the origin of more than 99\% of the mass of the visible universe. Compared to the strong force, electromagnetic interactions are weak and can be neglected for most QCD quantities. Notable exceptions are precision calculations of, for example, the proton-neutron mass splitting or the hadronic contribution to the muon anomalous magnetic moment. However, this hierarchy of forces does not persist if one considers strong background electromagnetic fields in a strongly interacting system. In particular, magnetic fields of the order of the strong scale, $eB\sim \Lambda_{\rm QCD}^2$, impact numerous features of QCD in a significant and non-trivial manner. In this domain, electromagnetic fields cannot be considered perturbatively but rather, as elements of this non-perturbative system.

Strong background electromagnetic fields are understood to be relevant for the physics of strongly interacting matter in three major areas: the physics of compact stars, cosmology and heavy-ion collision physics.
Strongly magnetized neutron stars, so-called magnetars, have been known to exist since the 1990s~\cite{Duncan:1992hi}. Strong magnetic fields were conjectured to be generated in various models of cosmological phase transitions, relevant for the evolution of the early Universe, see e.g.~\cite{Vachaspati:1991nm,Enqvist:1993np,Baym:1995fk,Grasso:2000wj}.
In the heavy-ion collision context, electromagnetic fields are created by the spectator particles in off-central events and are expected to affect the expanding quark-gluon plasma~\cite{Kharzeev:2007jp,Skokov:2009qp,Bzdak:2011yy}. 
The year 2008 marked a major advancement in this field: the chiral magnetic effect, and with it the impact of magnetic fields in off-central collisions gained broad visibility~\cite{Fukushima:2008xe}. From this point on, background magnetic fields were widely recognized to be relevant for the physics of strongly interacting matter and the community devoted substantial efforts to the calculation the impact of magnetic fields on QCD observables.

While neutron star matter and the early Universe is believed to be well described by equilibrium QCD, the heavy-ion collision setup is a system exhibiting initial stages far from equilibrium. Still, fast equilibration is observed, allowing hydrodynamic descriptions of the expanding fireball, and the experimental freeze-out conditions can be well explained by hadronic abundances corresponding to the equilibrium distribution, see e.g.\ the review~\cite{Busza:2018rrf}. To what extent the magnetic field is to be viewed as part of the equilibrated system, i.e.\ whether the magnetic field is sufficiently long-lived, is as of today, still one of the most important open questions of the field and the subject of ongoing debates~\cite{Tuchin:2013apa,McLerran:2013hla,Huang:2015oca}. Besides being strongly time-dependent, the magnetic fields present in the initial stages are expected to be strongly localized in space as well~\cite{Deng:2012pc}.

QCD in the presence of a static (i.e.\ time-independent) background electromagnetic field can nonetheless be treated as an equilibrium system and studied via established methods, including effective theories and models of QCD, functional approaches as well as lattice QCD simulations. The primary objective of such studies is to deliver predictions for the above mentioned physical applications. Important examples include the hadron spectrum, the equation of state, thermodynamical properties as well as transport coefficients.
In addition, lattice QCD simulations with background electromagnetic fields turned out to provide important benchmarks for low-energy QCD models, allowing for their systematic improvement. A prominent example for results of this type is the QCD phase diagram in the temperature-magnetic field plane and the associated inverse magnetic catalysis phenomenon discovered in 2011~\cite{Bali:2011qj}. Initially missed in most QCD models, this effect is by now broadly understood as one of the most important features that effective descriptions of QCD are expected to reproduce.

Various general aspects of the response of strongly interacting matter to background electromagnetic fields have been reviewed in the recent years~\cite{Kharzeev:2013jha,Miransky:2015ava}. The discussion of background fields in the quantum field theoretical context allows for the analytic treatment of several subjects, which have been reviewed lately: magnetic catalysis~\cite{Shovkovy:2012zn}, the strong-field behavior in QCD and QED~\cite{Hattori:2023egw}, or anomalous transport and Schwinger pair creation~\cite{Fukushima:2018grm}.
Most of the interesting non-perturbative effects, however, may only be assessed numerically. Indeed, a large portion of our current knowledge on the thermodynamics of strongly interacting matter in general comes from lattice QCD simulations. The most important features in this context are the impact of nonzero densities~\cite{Nagata:2021ugx} and the structure of the QCD phase diagram~\cite{Aarts:2023vsf}. Some of the developments regarding the response of QCD matter to background electromagnetic fields have also been reviewed in the last decade~\cite{DElia:2012ems,Szabo:2014iqa,Endrodi:2014vza,DElia:2015kpi,Buividovich:2016saz,Yamamoto:2021oys}. However, an up-to-date overview of lattice QCD results on background electromagnetic fields has not been presented.

We intend to fill this gap with this review.
Therefore, it is not our intention to discuss the phenomenological applications of the lattice QCD results, but instead to provide a comprehensive summary of lattice simulations 
carried out within this context. Our target audience is the broad community interested in strongly interacting matter under extreme conditions. Thus, this review is structured to be as self-contained as possible, and for completeness we explain several concepts that are standard for lattice experts, but are deemed helpful for readers with different backgrounds.

We will begin with a general introduction to QCD in background electromagnetic fields in Chap.~\ref{chap:lat}. Starting with continuum QCD in infinite volume, we will formulate the theory on a finite torus and discuss the impact of periodic boundary conditions -- these will turn out to play a crucial role throughout this review. Next, some details of the lattice discretizations are provided and a special emphasis is put on the renormalization of various observables. Chap.~\ref{chap:ev} is devoted to the discussion of the eigenvalue spectrum of the so constructed lattice Dirac operators. The magnetic field-dependence of the lattice eigenvalues leads us to Hofstadter's butterfly, a well-known concept in solid state physics. The impact of non-trivial Polyakov loop backgrounds on the eigenvalues is also discussed, laying the ground for a qualitative understanding of the inverse magnetic catalysis phenomenon.

Armed with these concepts, in Chap.~\ref{chap:hadron} we turn to the review of lattice results concerning the low-temperature, confined regime of QCD. Here we will present an overview of lattice calculations of hadron properties, in particular the response to the background fields in terms of magnetic and electric polarizabilities and magnetic moments. The impact of magnetic fields on further gluonic observables relevant for the confining QCD vacuum is also reviewed. In turn, Chap.~\ref{chap:pd} is devoted to nonzero temperatures and the deconfinement/chiral symmetry restoration transition of QCD. This chapter revolves around the magnetic catalysis and inverse magnetic catalysis phenomena. The developments regarding the QCD phase diagram are presented from a historical perspective, and recent results on the impact of quark densities and background electric fields are also discussed.

The next overarching topic in our discussion is the QCD equation of state, discussed in Chap.~\ref{chap:eos}. After summarizing the main effects of the background field on the thermodynamic relations, the main emphasis is put on the determination of the weak-field behavior of observables, which is complicated on the lattice by magnetic flux quantization. We review the latest developments on the equation of state in background electric fields as well as for dense and magnetized QCD. Finally, in Chap.~\ref{chap:anom} we consider observables related to chirality, spin and topology and their response to electromagnetic fields. The central focus is on anomalous transport phenomena: the chiral separation effect and the chiral magnetic effect. The discussion of the latter, as well as that of the electric conductivity of QCD matter, also necessitates an out-of-equilibrium approach. The corresponding lattice results are reviewed in detail.

Each of these chapters ends with a section that contains the key lessons about the corresponding subject. These are the take-home messages that we consider to be the most important aspects of lattice QCD with background magnetic fields.
The review is concluded with a summary in Chap.~\ref{chap:sum}. This chapter also contains an outlook with a list of open questions regarding the subjects of each of the preceding chapters.

\subsection{Glossary}

In this review it has been our aim to cover a broad portion of the literature. Often different conventions and different notations are used, and we attempted to reconcile these as much as possible. Our notation is summarized in Tab.~\ref{tab:glossary}.

\begin{table}[ht!]
\renewcommand*{\arraystretch}{.98}
\centering
\begin{tabular}{c|c|c}
 symbol &  definition & remark  \\ \hline \hline
 $A_\nu$, $F_{\nu\rho}$ & electromagnetic gauge field and field strength & \\
 $\A_\nu$, $\F_{\nu\rho}$ & color gauge field and field strength & \\
 $B$ & background magnetic field & \\
 $E$ & real background electric field & discussion on imaginary (Euclidean) fields in Sec.~\ref{sec:lat_Euclspacetime}\\
 $e$ & elementary electric charge & $e>0$ \\
 $q_f$ & quark electric charge for flavor $f$ & $q_u/e=2/3$, $q_d/e=-1/3$, etc.\ \\
 $N_b$ & magnetic flux quantum & for the down quark, see below~\eqref{eq:lat_flux3} \\
 $N_e$ & Euclidean electric flux quantum & for the down quark, see below~\eqref{eq:lat_flux4} \\
 $m_\ell$ & $m_\ell=m_u=m_d$ & degenerate light quark mass \\
 $F_\pi$ & pion decay constant & $F_\pi\approx93\textmd{ MeV}$ in the vacuum for physical quark masses \\
 $F_\pi^{(0)}$ & chiral limit of $F_\pi$ & $F_\pi^{(0)}\approx86\textmd{ MeV}$~\cite{Colangelo:2003hf} \\
 $L$ & untraced Polyakov loop & \multirow{2}{*}{defined in~\eqref{eq:lat_ploopdef1}} \\
 $P$ & (traced) Polyakov loop & \\
 $\expv{\bar\psi_f\psi_f}$ & quark condensate for flavor $f$ & \\
 $\expv{\bar\psi\psi}$ & $[\expv{\bar\psi_u\psi_u}+\expv{\bar\psi_d\psi_d}]/2$ & its chiral limit appears in the GMOR relation~\eqref{eq:pd_GMOR}\\
 $\Sigma_f$ & (re)normalized quark condensate for flavor $f$ & normalization according to~\eqref{eq:lat_pbpren} \\
 $\Sigma$ & $[\Sigma_u+\Sigma_d]/2$ & \\
 $A^r$ & $A-A(E=B=T=0)$ & renormalized, vacuum-subtracted observable  \\
 $\Delta A$ & $A-A(E=B=0)$ & zero-field subtracted observable or excess observable \\
 $f^{\rm tot}$ & total free energy density & includes background field energy density \\
 $f$ & matter free energy density & excludes background field energy density, cf.~\eqref{eq:lat_ftot_f} \\
 $\chi$ & magnetic susceptibility of medium & \multirow{2}{*}{defined in~\eqref{eq:lat_susc_def}} \\
 $\xi$ & electric susceptibility of medium & \\
 $\pi_m$ & magnetic permeability & \multirow{2}{*}{relation to susceptibilities in~\eqref{eq:lat_perme_permi} and footnote~\ref{fn:lat_perme}}\\
 $\pi_e$ & electric permittivity & \\
 $\tau_f$ & tensor coefficient for flavor $f$ & defined in~\eqref{eq:lat_def_tensorcoeff}, related to photon distribution amplitude\\
 $M_h$ & mass of hadron $h$ & $h=p,n,\pi^+,\ldots$\\
 $\E_h$ & lowest energy of hadron $h$ & coincides with mass in the vacuum, $\E_h(E=B=0)=M_h$\\
 $\hat\mu_h$ & magnetic moment of hadron $h$ & \\
 $\alpha_{h}$ & magnetic polarizability of hadron $h$ & \\
 $\beta_{h}$ & electric polarizability of hadron $h$ & not to be confused with inverse gauge coupling $\beta=6/g^2$\\
\end{tabular}
\caption{\label{tab:glossary} Glossary.}
\end{table}

Concerning our notation, in general $\Tr(A)$ denotes the trace over all open indices of a matrix $A$ including color, spin, coordinates, etc. In turn, we use $\tr(A)$ to indicate the trace in color space only. When discussing local observables, $x_\nu=(\bm x,x_4)$ stands for the coordinates in physical units, while $n_\nu=(\bm n,n_4)$ labels the lattice sites and is understood in lattice units, i.e.\ $x_\nu=a n_\nu$.
Throughout this review, we are working in natural units, $\hbar=c=k_B=1$, as well as unit values for the electric permittivity and magnetic permeability of the classical vacuum.

	\clearpage
	\section{QCD and background fields}
	\label{chap:lat}

Lattice QCD is a first-principles approach to investigate strongly interacting matter.
To construct the lattice formulation of QCD in the presence of background electromagnetic fields, we will start this chapter with the continuum theory in the infinite volume. We will then write down the fields in Euclidean space-time, consider the system in a finite volume with prescribed boundary conditions, and formulate the discretized theory. Finally, we discuss how the renormalization of various observables is affected by the presence of background fields.

\subsection{Continuum QCD}
In continuum QCD, the electromagnetic vector potential $A_\nu$ enters the Dirac operator via minimal coupling, in the same form as the gluon field $\A_\nu$. The latter couples to all quark flavors with the same strength, allowing the strong coupling $g$ to be incorporated into the gluon field by a simple rescaling.
In contrast, the electromagnetic charges $q_f$ of the quark flavors differ, rendering the Dirac-operator different for each flavor $f$,
\be
\Dsf = \gamma_\nu D_{f\nu}, \qquad D_{f\nu} = \partial_\nu + iq_f A_\nu + i \A_\nu\,,
\label{eq:lat_Dslash_cont}
\ee
where $\gamma_\nu$ denote the (Hermitian) Euclidean Dirac-matrices. 
The gluon field is an element of the Lie algebra of $\mathrm{SU}(N_c)$ and unless specified otherwise, in this review we consider $N_c=3$, i.e.\ real QCD. In contrast, the photon field is a real variable. We also introduce the massive Dirac operator,
\be
M_f=\Dsf + m_f\,.
\label{eq:lat_fermionmatrix}
\ee

The partition function of the system is written down using the path integral formulation in Euclidean space-time,
\be
\Z^{\rm tot} = \int \D \A \,\D\bar\psi \,\D\psi \,\exp\left[ -S_F - S_G - S_\gamma \right] \,.
\label{eq:lat_partfunc_cont}
\ee
The superscript $\rm tot$ signals that the system encompasses both the QCD medium as well as the background fields.
The temperature $T$ of the system is set by the inverse of the extent in the Euclidean time direction, $L_4=1/T$.
In this direction, the gluon fields satisfy periodic, while the fermion fields antiperiodic boundary conditions.

The Euclidean action in~\eqref{eq:lat_partfunc_cont} consists of fermion and gluon contributions,
\be
S_F = \sum_f \int \dd^3 x \int_0^{1/T} \!\!\dd x_4 \;\bar\psi_f(x) M_f\psi_f(x), \qquad
S_G = \frac{1}{2g^2} \int \dd^3x \int_0^{1/T} \!\!\dd x_4 \;\tr \,\F_{\nu\rho}(x) \F_{\nu\rho}(x)\,,
\label{eq:lat_action_cont}
\ee
as well as a photon term,
\be
S_\gamma = \frac{1}{4} \int \dd^3 x \int_0^{1/T} \!\!\dd x_4 \; F_{\nu\rho}(x)F_{\nu\rho}(x)\,.
\label{eq:lat_photon_act}
\ee
Above, the gluon and photon field strengths read
\be
\F_{\nu\rho}(x) =\partial_\nu \A_\rho(x) - \partial_\rho \A_\nu(x)  + i[\A_\nu(x),\A_\rho(x)]\,,\qquad
F_{\nu\rho}(x) = \partial_\nu A_\rho(x) - \partial_\rho A_\nu(x)\,.
\label{eq:lat_fieldstr_cont}
\ee

Integrating over the quark fields in~\eqref{eq:lat_partfunc_cont}, we obtain the fermion determinants,
\be
\Z^{\rm tot} =
\int \D\A\, \exp\left[-S_G-S_\gamma\right] \, \prod_f \det M_f\,.
\label{eq:lat_partfunc_cont2}
\ee
Since the electromagnetic field enters as a classical background field that is not integrated over in~\eqref{eq:lat_partfunc_cont2}, $S_\gamma$ can be taken out of the path integral. The remaining path integral corresponds to the partition function of the QCD medium, which still depends on the background field via the Dirac operators.
We denote this partition function by $\Z$,
\be
\Z^{\rm tot} = e^{-S_\gamma}\, \Z, \qquad \Z=
\int \D\A\, \exp\left[-S_G\right] \, \prod_f \det M_f\,.
\label{eq:lat_partfunc_cont2b}
\ee
While the photon action merely sets an overall normalization in $\Z^{\rm tot}$, the relative of weights of $\Z$ and $e^{-S_\gamma}$ will be important when we consider the renormalization of the background gauge field below in Sec.~\ref{sec:lat_renorm}.

\subsubsection{Euclidean space-time}
\label{sec:lat_Euclspacetime}

The Euclidean formulation is related to real, Minkowskian space-time via a Wick rotation of the time coordinate $x_4=-ix_0$. For time-dependent processes and certain types of linear response coefficients, this Wick rotation entails a highly non-trivial analytic continuation of specific expectation values. For equilibrium observables -- which characterize the state of the system after an infinitely long equilibration and are therefore by definition time-independent -- no analytic continuation in the time variable is necessary.
However, the gauge fields also undergo an analogous Wick rotation $A_4=iA_0$, so that altogether $D_4=iD_0$, just like for the partial derivatives $\partial_4=i\partial_0$. While this does not affect the magnetic components of the field strength tensor, it renders the Euclidean time components $F_{4j}$ of the field strength tensor -- i.e.\ the Euclidean electric field -- an imaginary-valued electric field,
\be
B_j(x)=\frac{1}{2}\epsilon_{jkl}F_{kl}(x), \qquad iE_j(x)=F_{4j}(x)\,,
\ee
or, in vector notation in terms of the vector potential,
\be
\bm B(x) = \bm \nabla \times \bm A(x), \qquad
i \bm E(x) = \partial_4 \bm A(x) - \bm \nabla A_4(x)\,.
\label{eq:lat_BfromA}
\ee

Throughout this review, $E_j$ denotes the real, Minkowski electric field.
In terms of the electric and magnetic field components,
the well-known form of the photon action~\eqref{eq:lat_photon_act} is recovered,
\be
S_\gamma=\frac{1}{2}\int\dd^3x \int_0^{1/T}\!\!\dd x_4\left[ \bm B^2(x) - \bm E^2(x) \right]\,.
\label{eq:lat_SgammaBE}
\ee
Notice that the same discussion applies to the notion of a chemical potential, represented by a constant time-like gauge field component $\mu=A_0$. Thus, a real Euclidean $A_4$ corresponds imaginary values of the chemical potential, $\mu=-iA_4$.

Similarly to the photon action, we can also decompose the gluon action~\eqref{eq:lat_action_cont} into individual components. These are the chromomagnetic and chromoelectric gluon fields,
\be
S_G = \frac{1}{2g^2} \int \dd^3x \int_0^{1/T} \!\!\dd x_4 \;\tr \left[\bm\E^2(x) + \bm \B^2(x)\right], \qquad \E_i(x)=\F_{4i}(x), \qquad
\B_i(x)=\frac{1}{2}\epsilon_{ijk}\F_{jk}(x)\,,
\label{eq:lat_action_cont_components}
\ee
which are the Euclidean field components integrated over in the Euclidean path integral.

\subsubsection{External and matter contributions to observables}
\label{sec:lat_extmat}

The partition function encodes all information about the equilibrium thermodynamics of the system. Most importantly, the free energy of the total system -- consisting of both the medium and the background field -- is obtained from~\eqref{eq:lat_partfunc_cont2} as
$
F^{\rm tot} = -T \log\Z^{\rm tot}
$.
The associated intensive quantity is the total free energy density,
\be
f^{\rm tot} \equiv \frac{F^{\rm tot}}{V}= -\frac{T}{V}\log\Z^{\rm tot}\,.
\label{eq:lat_Ftotdef}
\ee
Often, one is only interested in the contribution of the medium to a specific observable. To that end, we use~\eqref{eq:lat_partfunc_cont2b} to separate the energy of the background field,
\be
F^{\rm tot} = TS_\gamma + F\,,
\ee
where $F$ is the matter free energy,
\be
F=-T\log\Z\,.
\label{eq:lat_Fdef}
\ee

All observables relevant for the thermodynamics of the medium can be obtained from $F$ via differentiation with respect to the appropriate variables. We will construct the equation of state based on~\eqref{eq:lat_Fdef} in Chap.~\ref{chap:eos}.
The intensive observable corresponding to $F$ is the matter free energy density,
\be
f\equiv \frac{F}{V}=-\frac{T}{V}\log\Z\,,
\label{eq:lat_fdef}
\ee
which is related to the total and external terms as
\be
f^{\rm tot}  = f^{\rm ext} + f\,,\qquad f^{\rm ext}=\frac{T}{V}\,S_\gamma\,.
\label{eq:lat_ftot_f}
\ee

The background electromagnetic fields are generated by external currents $j_{\nu}^{\rm ext}$ that are kept fixed by some external device. For homogeneous magnetic fields, these are currents $\bm j^{\rm ext}$ encircling the medium, while for homogeneous electric fields, charges $j_0^{\rm ext}$ placed at its boundaries. To realize inhomogeneous background fields, external currents or charges also need to be placed inside the volume. The medium responds to these currents by developing matter contributions $\expv{j_\nu}$, so that the total current is
\be
j_\nu^{\rm tot} = j_\nu^{\rm ext} + \expv{j_\nu}\,.
\label{eq:lat_inducedcurrent_totalcurrent}
\ee

For homogeneous background fields with magnitudes $B=|\bm B|$ and $E=|\bm E|$,~\eqref{eq:lat_ftot_f} takes the form
\be
f^{\rm tot} = \frac{1}{2}\left( B^2 - E^2 \right) + f(eB,eE)\,.
\label{eq:lat_ftot_f_hom}
\ee
Here we made use of the fact that the matter free energy only depends on the electromagnetic fields through the combinations $q_fB$ and $q_fE$. In turn, all quark charges may be expressed as multiples of the elementary electric charge $e>0$. Accordingly, from now on we will always consider the matter free energy density as a function of $eB$ and $eE$ -- which we already indicated in~\eqref{eq:lat_ftot_f_hom}. These combinations will also be useful in the context of renormalization, see Sec.~\ref{sec:lat_renorm}. 

Throughout this review, we will work with the matter contributions as functions of $B$ and $E$ and neglect dynamical QED effects, i.e.\ the back-reaction of the medium on the background fields.
The standard description of magnetization and polarization phenomena~\cite{landau1995electrodynamics} considers this back-reaction, i.e.\ the difference of external and in-medium fields in order to define the magnetic permeability and the electric permittivity of the medium\footnote{
\label{fn:lat_perme}
For magnetization phenomena, one considers the external magnetic field $H$ and the magnetic induction $B$ separately. The two differ by the magnetization $e\M=B-H$ induced by the medium. For weak magnetic fields, $B=\pi_m H$ defines the magnetic permeability~\cite{landau1995electrodynamics} (using natural units for the vacuum permeability). Neglecting dynamical QED effects implies that we do not keep track of $H$ but only of the magnetic field $B$ that charged fermions interact with in the medium. This is the field that enters the Dirac operator~\eqref{eq:lat_Dslash_cont}. The susceptibility defined below in~\eqref{eq:lat_susc_def} corresponds to $\M=\chi \,eB$ so that $\pi_m=B/H=(1-e^2\chi)^{-1}$, in agreement with~\eqref{eq:lat_perme_permi}. In order to find the external field that would be present without the medium, one may reconstruct $H=(1-e^2\chi)B$.
For background electric fields we have a similar situation: we only keep track of the electric field $E$ and not the electric displacement $D$. The two are related by the polarization $e\P= D-E$, which, for weak fields, defines the electric permittivity $\pi_e=\P/E$~\cite{landau1995electrodynamics}. The electric susceptibility~\eqref{eq:lat_susc_def} satisfies $\P=\xi\,e E$, giving $\pi_e=D/E=1+e^2\xi$, compatible with~\eqref{eq:lat_perme_permi}. The electric displacement can be reconstructed as $D=(1+e^2\xi) E$.
}. However, these can be constructed equivalently from the $f(eB,eE)$ dependence, as first discussed in detail in~\cite{Bonati:2013lca}.

To do so, let us consider the weak-field expansion of~\eqref{eq:lat_ftot_f_hom}, which, owing to parity symmetry, takes the form,
\be
f^{\rm tot} = f(B= E=0) + \frac{1}{2}\left(\frac{B^2}{\pi_m} -\pi_e \,E^2 \right) + \mathcal{O}\left[(eB)^4, (eE)^4, (eB)^2(eE)^2\right]\,.
\label{eq:lat_ftot_f_hom_expanded}
\ee
Here we defined the magnetic permeability $\pi_m$ and the electric permittivity $\pi_e$, which encode the linear response of the medium to the background fields and can be written as
\be
\pi_m=\frac{1}{1-e^2\chi}, \qquad \pi_e = 1+e^2\xi\,,
\label{eq:lat_perme_permi}
\ee
in terms of the magnetic susceptibility $\chi$ and the electric susceptibility $\xi$,
\be
\chi = -\left.\frac{\partial^2 f}{\partial (eB)^2}\right|_{B=0}, \qquad
\xi = \left.\frac{\partial^2 f}{\partial (eE)^2}\right|_{E=0}\,.
\label{eq:lat_susc_def}
\ee
As indicated above, here we explicitly included the factor $e$ in the definition of the susceptibilities.

From now on, we will focus on the matter contributions $f$ and $\expv{j_\nu}$, as we are not interested in the energy density of the background field, being completely independent of the properties of QCD matter. Equivalently, we will work with the susceptibilities $\chi$ and $\xi$, which play an important role in the discussion of the equation of state, see Chap.~\ref{chap:eos}. 

\subsection{Flux quantization and twisted boundary conditions}
\label{sec:lat_flux}

Next, we place the system in a finite volume $V$.
For the case of homogeneous electromagnetic fields, this is clearly a physical requirement, otherwise the total photon energy (or the action) would be unbounded and the external currents $j_\nu^{\rm ext}$ need to be pushed to infinity. In fact, for the case of homogeneous electric fields at nonzero temperature, the finite volume is also needed as an infrared regulator, see Sec.~\ref{sec:eos_elsusc}.
For intensive observables, the thermodynamic limit can be approached by carrying out the limit $V\to\infty$.
To be specific, we consider a rectangular box with side lengths $L_j$, so that $V=L_1L_2L_3$. The coordinate system is set such that $0\le x_j\le L_j$. To maintain translational invariance, it is desirable to use periodic boundary conditions, i.e.\ a three-dimensional torus. This choice is also preferred by standard lattice implementations.

Let us consider a background magnetic field first. We will be able to generalize our findings for the case of Euclidean electric fields later. The following discussion bases on the works~\cite{Al-Hashimi:2008quu,Bali:2011qj,DElia:2011koc}.
In contrast to the infinite volume setup, the flux of the magnetic field through the torus becomes a discrete variable. To see why, let us consider an arbitrary two-dimensional slice of our periodic volume -- say, the $x_1-x_2$ two-dimensional plane at $x_3=0$ (we suppress the dependence on $x_3$ below). The flux of the magnetic field through this plane can be calculated using Stokes' theorem involving the line integral of $A_j(x_1,x_2)$ along a contour wrapping around the plane,
\be
\Phi_{12} \equiv
\int \dd x_1 \,\dd x_2 \,B_3(x_1,x_2) = \int_0^{L_1} \dd x_1 \left[ A_1(x_1,0) - A_1(x_1,L_2) \right] +
\int_0^{L_2} \dd x_2 \left[ A_2(L_1,x_2) - A_2(0,x_2) \right]\,,
\label{eq:lat_flux1}
\ee
where we used~\eqref{eq:lat_BfromA}. If the vector potential is exactly periodic, we conclude that $\Phi_{12}=0$. However, $A_j$ is not a physical quantity and it is sufficient that it is periodic up to gauge transformations (twists) $\varphi_{1,2}$,
\be
A_j(x_1,x_2+L_2)=A_j(x_1,x_2)+\partial_j \varphi_2(x_1), \qquad
A_j(x_1+L_1,x_2)=A_j(x_1,x_2)+\partial_j \varphi_1(x_2)\,.
\label{eq:lat_bc_photon}
\ee
In that case,~\eqref{eq:lat_flux1} simplifies to
\be
\Phi_{12} = -\int_0^{L_1}\dd x_1 \,\partial_1\varphi_2(x_1) +
\int_0^{L_2}\dd x_2\, \partial_2\varphi_1(x_2)
= \varphi_1(L_2)-\varphi_1(0)-\varphi_2(L_1)+\varphi_2(0)\,.
\label{eq:lat_flux2}
\ee

\begin{figure}
 \centering
 \begin{tikzpicture}[scale=3]
  \draw[black,-{Latex[length=3mm, width=2mm]}] (0,0) -- (1.5,0);
  \draw[black,-{Latex[length=3mm, width=2mm]}] (1.5,0) -- (1.5,1);
  \draw[black,-{Latex[length=3mm, width=2mm]}] (0,1) -- (1.5,1);
  \draw[black,-{Latex[length=3mm, width=2mm]}] (0,0) -- (0,1);
  \node[left] at (0,0) {$(0,0)$};
  \node[right] at (1.5,0) {$(L_1,0)$};
  \node[left] at (0,1) {$(0,L_2)$};
  \node[right] at (1.5,1) {$(L_1,L_2)$};
 \end{tikzpicture}
 \caption{
 \label{fig:lat_flux}
 Applying the boundary conditions for a quark field in different orders leads to the consistency relation~\protect\eqref{eq:lat_flux2b}.}
\end{figure}
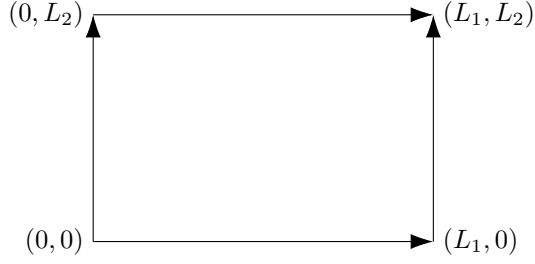

Gauge invariance dictates that the boundary conditions for fermion fields should also involve the same gauge transformation. For the quark field $\psi_f$ with electromagnetic charge $q_f$ this implies,
\be
\psi_f(x_1,x_2+L_2)=e^{-iq_f\varphi_2(x_1)} \,\psi_f(x_1,x_2), \qquad
\psi_f(x_1+L_1,x_2)=e^{-iq_f\varphi_1(x_2)} \,\psi_f(x_1,x_2)\,.
\label{eq:lat_bc_fermion}
\ee
Employing the two boundary conditions to express $\psi_f(L_1,L_2)$ via $\psi_f(0,0)$ in opposite orders -- as depicted in Fig.~\ref{fig:lat_flux} -- we obtain the consistency relation,
\be
\psi_f(L_1,L_2)=e^{-iq_f\varphi_1(L_2)}\,e^{-iq_f\varphi_2(0)}\, \psi_f(0,0) \overset{!}{=}
e^{-iq_f\varphi_2(L_1)}\,e^{-iq_f\varphi_1(0)} \,\psi_f(0,0)\,.
\label{eq:lat_flux2b}
\ee
Comparing to~\eqref{eq:lat_flux2}, we arrive at the well-known condition of magnetic flux quantization~\cite{tHooft:1979rtg},
\be
q_f\Phi_{12} = 2\pi N_b^f , \qquad N_b^f\in\mathds{Z}\,.
\label{eq:lat_flux3}
\ee
Notice that if there are quarks with different electric charges $q_f$ in the system,~\eqref{eq:lat_flux3} must hold for each of them, with integers $N_b^{f}$. This also implies that charged particles can only exist on a torus in a quantum mechanically consistent way, if their charge ratios $q_f/q_{f'}$ are all rational numbers. Luckily\footnote{While QCD with background electromagnetic fields -- in the infinite volume -- would be a perfectly consistent theory for any quark charges, within the Standard Model the quark electric charges are fixed by anomaly cancellation.}, in nature this is the case as the quark electric charges satisfy $q_u/q_d=-2$, $q_d/q_s=1$.
In practice, the down quark has the smallest flux, which we will denote simply by $N_b\equiv N_b^d$. This also fixes the flux for the remaining flavors, e.g.\ $N_b^u=-2N_b$ and $N_b^s=N_b$.

On the torus, we can therefore employ the above configuration, with the photon field $A_j$ and the fermion fields $\psi_f$, which satisfy the twisted boundary conditions~\eqref{eq:lat_bc_fermion}.
Alternatively, we may perform a gauge transformation $\Lambda(x_1,x_2)$ that renders the fermion fields exactly periodic~\cite{Bali:2011qj}. It is given by
\be
\psi_f'(x_1,x_2)=e^{iq_f\Lambda(x_1,x_2)} \,\psi_f(x_1,x_2), \qquad
A_j'(x_1,x_2)=A_j(x_1,x_2)-\partial_j \Lambda(x_1,x_2)\,,
\ee
with
\be
\Lambda(x_1,x_2) = \varphi_1(x_2)\,\Theta(x_1-L_1) + \varphi_2(x_1)\,\Theta(x_2-L_2)\,,
\ee
as can be checked simply using~\eqref{eq:lat_bc_photon}.
This results in new terms in the photon field (using $0\le x_1\le L_1$ and $0\le x_2\le L_2$),
\be
A_1'(x_1,x_2)=A_1(x_1,x_2)-\varphi_1(x_2)\,\delta(x_1-L_1), \qquad
A_2'(x_1,x_2)=A_2(x_1,x_2)-\varphi_2(x_1)\,\delta(x_2-L_2)\,.
\ee
Both for the $\psi_f$ fields and the $\psi_f'$ fields, in the path integral~\eqref{eq:lat_partfunc_cont2} after integrating out fermions, the Dirac operator will contain the same twists. In the former case these originate from the fermion boundary conditions, while in the latter the new terms in the gauge transformed photon fields. This equivalence will become transparent when we consider a specific background magnetic field $B_j(x)$.

In the same manner as above, one finds that the magnetic fluxes $\Phi_{ij}$ through any $x_i-x_j$ plane are also quantized. 
Furthermore, in Euclidean space-time, the imaginary electric field merely differs from the magnetic field by a Euclidean $\mathrm{O}(4)$ rotation, exchanging a spatial with a temporal axis. The above discussion therefore trivially carries over to imaginary background electric fields as well: the `flux' of the electric field (for example in the $x_3-x_4$ plane) is also quantized,
\be
q_f\Phi_{34} \equiv q_f\int \dd x_3 \,\dd x_4 \,iE_3(x_3,x_4) = 2\pi N_e^f, \qquad N_e^f\in\mathds{Z}\,,
\label{eq:lat_flux4}
\ee
and its implementation requires twisted boundary conditions for the fermions -- or an additional gauge transformation of the photon field to make the periodic boundary conditions exact -- just as above for the magnetic field. One difference is that in the temporal direction the fermion fields already satisfy antiperiodic boundary conditions in the electric case.
Just as for the magnetic case, the flux for the down quark, $N_e\equiv N_e^d$ fixes the flux for all other flavors, e.g.\  $N_e^u=-2N_e$ and $N_e^s=N_e$.

Finally, we stress that 
on the torus, the electromagnetic fields $B_j(x)$ and $E_j(x)$ are not sufficient to completely fix the quantum mechanical problem, as winding gauge loops, i.e.\ the $\mathrm{U}(1)$ Polyakov loops also constitute gauge invariant combinations that need to be specified in order for a completely unambiguous setup. These read (with $L_4=1/T$),
\be
P^{\mathrm{U}(1)}_{\nu f}(x)=\exp \left[ iq_f\int_0^{L_\nu} \dd x_\nu A_\nu(x) \right]\,.
\label{eq:lat_U1ploops}
\ee

\subsubsection{Homogeneous magnetic field}

Next we specialize to the case of a homogeneous background magnetic field $\bm B$. Without loss of generality, we may choose it to point along the positive $x_3$ axis. A possible choice for the gauge field is $A_2=B x_1$. The boundary twists~\eqref{eq:lat_bc_photon} are given by $\varphi_1=BL_1x_2$ and $\varphi_2=0$, so that the magnetic flux is
\be
q_f \Phi_{12} = q_fB L_1L_2 = 2\pi N_b^f, \qquad N_b^f\in\mathds{Z}\,,
\label{eq:lat_flux_quant_hom}
\ee
showing explicitly that the amplitude of the magnetic field $B$ is a discrete variable.
After the gauge transformation that makes the fermion fields exactly periodic, the photon field becomes (dropping the prime on $A_\nu$ for brevity),
\be
A_2(x_1)=Bx_1, \qquad A_1(x_1,x_2)=-BL_1x_2\,\delta(x_1-L_1)\,.
\label{eq:lat_hom_Bfield}
\ee

It is instructive to express the gauge field in terms of the flux variable,
\be
q_fA_2(x_1)=\frac{2\pi N_b^f}{L_1L_2}\,x_1, \qquad q_fA_1(x_1,x_2)=-\frac{2\pi N_b^f}{L_2}\,x_2\,\delta(x_1-L_1)\,.
\label{eq:lat_hom_Afield_Nb}
\ee
This last formula shows that the electromagnetic Polyakov loops~\eqref{eq:lat_U1ploops}
encircle the complex unit circle $N_b^f$ times,
\be
P_{1f}^{\mathrm{U}(1)}(x_2) = 
\exp\left[-i\,2\pi N_b^f \frac{x_2}{L_2} \right], \qquad
P_{2f}^{\mathrm{U}(1)}(x_1) = 
\exp\left[i\,2\pi N_b^f \frac{x_1}{L_1} \right]\,.
\label{eq:lat_B_ploops}
\ee
This topological behavior is visualized below in the right panel of Fig.~\ref{fig:lat_Bprofiles} for $P_{2f}^{\mathrm{U}(1)}$.

\subsubsection{Oscillatory magnetic fields}
\label{sec:lat_oscillatory_fields}

Magnetic fields with nonzero flux are topologically distinct from the setup where the photon fields vanishes. This is sometimes inconvenient, for example when a linear response coefficient, encoded in a Taylor-expansion in weak magnetic fields, is sought for, see Sec.~\ref{sec:eos_currentcurrent}. There are two types of alternatives to the homogeneous field that have been discussed in the literature. Still considering a magnetic field that points in the $x_3$ direction everywhere, the harmonic magnetic field profile is given by,
\be
B(x_1)=B\cos(p_1x_1), \qquad
A_2(x_1)=B \,\frac{\sin(p_1x_1)}{p_1}, \qquad
p_1=\frac{2\pi k_1}{L_1}, \qquad k_1\in\mathds{Z}\,,
\label{eq:lat_osc_Bfield}
\ee
with the discrete momentum $p_1$ set such that periodic boundary conditions are satisfied and the total flux is indeed zero. Thus, the amplitude of the magnetic field is now a continuous parameter and no twists are necessary. Now the electromagnetic Polyakov loop $P^{\mathrm{U}(1)}_{2f}(x_1)$ does not wind but oscillates around the real direction on the complex circle in a harmonic fashion, see the right panel of Fig.~\ref{fig:lat_Bprofiles}. In turn, $P^{\mathrm{U}(1)}_{1f}$ is trivial.

Yet another alternative is the so-called half-half field
\be
B(x_1)=B\, \sgn(x_1-L_1/2), \qquad
A_2(x_1)=B\, |x_1-L_1/2|\,.
\label{eq:lat_half_Bfield}
\ee
for which the amplitude $B$ is again a continuous parameter. The $\mathrm{U}(1)$ Polyakov loop in the $x_2$ direction oscillates around the real direction in a piecewise linear fashion.

\subsubsection{Localized magnetic field}

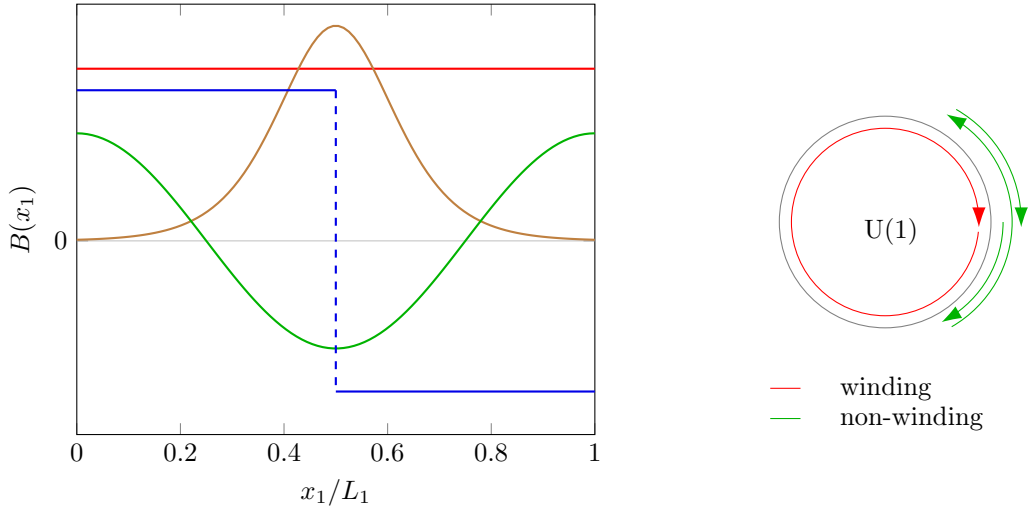
\begin{figure}
 \centering
 \begin{tikzpicture}[scale=1]
 \begin{axis}
    [xmin=0,xmax=1,ymin=-.9,ymax=1.1,xlabel=$x_1/L_1$,ylabel=$B(x_1)$,
    ytick={0},yticklabels={0}]
    \addplot[gray!50,very thin,domain=0:1,samples=100]{0};
    \addplot[red,thick,domain=0:1,samples=100]
     {.8};
    \addplot[brown,thick,domain=0:1,samples=100]
     {1/(cosh((x-0.5)/0.15))^2};
    \addplot[green!70!black,thick,domain=0:1,samples=100]
     {cos(2*3.1415*deg(x))/2};
    \addplot[blue!90!black,thick,domain=0:0.5,samples=100]
     {.7};
    \addplot[blue!90!black,thick,domain=0.5:1,samples=100]
     {-.7};
     \draw [blue!90!black,thick,dashed] (0.5,.7) -- (0.5,-.7);
\end{axis}
 \end{tikzpicture}\qquad\qquad\qquad
 \raisebox{1cm}{\begin{tikzpicture}[scale=4]
	\draw[gray] (3.38,1) circle (.35cm);
	\draw[red] (3.69,1) arc (0:354:.31cm);
	\draw[red, -{Latex[length=3mm,width=2mm]}, draw opacity=0](3,1) to [out=-40,in=90] (3.69,.975);

	\draw[green!70!black] (3.77,1) arc (0:-60:.37cm);
	\draw[green!70!black] (3.80,1) arc (0:-60:.4cm);
	\draw[green!70!black] (3.80,1) arc (0:60:.4cm);
	\draw[green!70!black] (3.83,1) arc (0:60:.43cm);

	\draw[green!70!black, -{Latex[length=3mm,width=2mm]}, draw opacity=0](3,1) to [out=-40,in=30] (3.56,.665);
	\draw[green!70!black, -{Latex[length=3mm,width=2mm]}, draw opacity=0](3,1) to [out=-40,in=-30] (3.576,1.36);
	\draw[green!70!black, -{Latex[length=3mm,width=2mm]}, draw opacity=0](3,1) to [out=-40,in=90] (3.83,.975);

	\node[] at (3.4,.97) {$\mathrm{U}(1)$};

	\draw[red]  (3.,0.45) to (3.1,0.45);
	\draw[green!70!black](3.,.35) to (3.1,.35);
	\node[anchor=west] at (3.2,0.45) {winding};
	\node[anchor=west] at (3.2,.35) {non-winding};
\end{tikzpicture}}
 \caption{\label{fig:lat_Bprofiles}
 Left panel: magnetic field profiles considered in this review: homogeneous (red), harmonic (green), half-half (blue) and localized (brown). Right panel: electromagnetic Polyakov loops in the $x_2$ direction corresponding to the different types of fields. The homogeneous and localized fields wind around the complex unit circle (red), while the harmonic and half-half fields do not (green). Homogeneous electric fields can be represented analogously to the red curves, but in that case the left panel depicts $E(x_3)$ and the right panel the electromagnetic Polyakov loop $P_{4f}^{\mathrm{U}(1)}(x_3)$.}
\end{figure}

In the context of off-central heavy-ion collisions, localized magnetic field profiles are relevant and have been considered recently.
A profile that is expected to capture the corresponding spatial inhomogeneity in the $x_1$ direction, but is still homogeneous in the remaining spatial directions, is given by
\be
B(x_1)=B\, \cosh^{-2}\left(\frac{x_1-L_1/2}{\epsilon}\right), \qquad
A_2(x_1)= B\epsilon\, \tanh\left(\frac{x_1-L_1/2}{\epsilon}\right)\,,
\label{eq:lat_loc_Bfield}
\ee
where $\epsilon$ is the characteristic width of the localized region. This field carries the twists  $\varphi_1=2B\epsilon\tanh[L_1/(2\epsilon)] \,x_2$ and $\varphi_2=0$, and, according to~\eqref{eq:lat_flux2}, has the flux
\be
q_f\Phi_{12}=2q_fB\,\epsilon L_2  \tanh\left(\frac{L_1}{2\epsilon}\right) = 2\pi N_b^f\,, \qquad N_b^f\in\mathds{Z}\,.
\ee
This is the same result as obtained by integrating~\eqref{eq:lat_loc_Bfield} over the $x_1-x_2$ plane, as it should be. Performing the gauge transformation to make the fermion fields exactly periodic, we obtain the photon field (again dropping the prime for brevity),
\be
A_2(x_1)=B\epsilon\, \tanh\left(\frac{x_1-L_1/2}{\epsilon}\right), \qquad
A_1(x_1,x_2)=-2B\,\epsilon\tanh\left(\frac{L_1}{2\epsilon}\right) x_2\, \delta(x_1-L_1)\,.
\ee

Finally, we express the gauge field with the flux variable,
\be
q_fA_2(x_1)=\frac{\pi N_b^f}{L_2 \tanh\left(\frac{L_1}{2\epsilon}\right)}\,\tanh\left(\frac{x_1-L_1/2}{\epsilon}\right), \qquad
q_fA_1(x_1,x_2)= -2\pi N_b^f\,\frac{x_2}{L_2}\, \delta(x_1-L_1)\,.
\ee
Notice that the $A_1$ component coincides with the one in the homogeneous magnetic field case~\eqref{eq:lat_hom_Afield_Nb}. This 
is because the magnetic field is constant in the $x_2$ direction and therefore the twist $\varphi_1$ is linear in $x_2$. In other words, the twists are only sensitive to the total flux of the field and are distributed evenly in the $x_2$ direction.
The electromagnetic Polyakov loops~\eqref{eq:lat_U1ploops} again wind around the complex circle in a topological manner. In comparison to the homogeneous case, the winding of $P^{\mathrm{U}(1)}_{2f}(x_1)$ proceeds faster near $x_1=L_1/2$ and slower towards $x_1=0,L_1$.
The field profiles considered above are illustrated in the left panel of Fig.~\ref{fig:lat_Bprofiles}.

\subsubsection{Imaginary electric fields}

Based on the above results for background magnetic fields, the analogous formulas for imaginary (Euclidean) electric fields also follow. For an imaginary electric field pointing in the $x_3$ direction, the analogue of the boundary conditions~\eqref{eq:lat_bc_photon} are the twists $\varphi_4(x_3)$ and $\varphi_3(x_4)$. Regarding the former, $\varphi_4=0$ must hold in order not to affect the boundary conditions for fermions in the imaginary time direction, necessary to maintain their anticommuting nature. 

For a homogeneous imaginary electric field, represented by $A_4=iEx_3$, we arrive at the twist $\varphi_3(x_4)=iEL_3x_4$, so that the gauge field on the torus becomes
\be
A_4(x_3)=iEx_3, \qquad A_3(x_3,x_4)=-iEL_3x_4\,\delta(x_3-L_3)\,,
\label{eq:lat_hom_Efield}
\ee
with the quantization condition
\be
q_f \Phi_{34}=iq_fEL_3/T=2\pi N_e^f, \qquad N_e^f\in\mathds{Z}\,.
\label{eq:lat_elquant}
\ee
The electromagnetic Polyakov loops are winding around the complex circle $N_e^f$ times as we go around the torus, just as in the homogeneous magnetic field case. The equivalent of~\eqref{eq:lat_B_ploops} reads
\be
P_{3f}^{\mathrm{U}(1)}(x_4) = 
\exp\left[-i\,2\pi N_e^f x_4 T \right], \qquad
P_{4f}^{\mathrm{U}(1)}(x_3) = 
\exp\left[i\,2\pi N_e^f \frac{x_3}{L_3} \right]\,.
\label{eq:lat_E_ploops}
\ee

In turn, harmonic imaginary electric fields have zero flux and therefore no boundary twists are needed. The gauge field in this case is given by
\be
iE(x_3)=iE\cos(p_3x_3), \qquad
A_4(x_3)=iE \,\frac{\sin(p_3x_3)}{p_3}, \qquad
p_3=\frac{2\pi k_3}{L_3}, \qquad k_3\in\mathds{Z}\,.\,
\label{eq:lat_osc_Efield}
\ee
and the Polyakov loops are non-topological.
The formulas~\eqref{eq:lat_hom_Efield} and~\eqref{eq:lat_osc_Efield} reveal the similarity between imaginary electric fields and inhomogeneous imaginary chemical potentials $A_4(x_3)=i\mu(x_3)$.

To obtain the response of the medium to real (Minkowskian) electric fields, an analytic continuation in the electric field variable is necessary -- just like for imaginary chemical potentials. Due to the quantization condition~\eqref{eq:lat_elquant}, for homogeneous fields this is an ill-posed problem at nonzero temperature (and in a finite volume). This problem is avoided by the harmonic fields, for which the amplitude can be taken continuously to zero. This latter setup also allows for a Taylor-expansion in the electric field amplitude. We get back to this point in Sec.~\ref{sec:eos_elsusc}.

\subsection{Lattice discretization}
\label{sec:latdisc}

Next, we need to provide a discretization of the action and the fields in it. We consider an isotropic lattice with spacing $a$, so that the points of Euclidean space-time become discrete, $x_\nu=an_\nu$, with the integer $n_\nu$ labeling the lattice sites. To be specific, the range of lattice coordinates is set to $0\le n_{1,2,3}<N_s$ and $0\le n_4< N_t$. The choice of this lattice geometry, $N_s^3\times N_t$, implies that the temperature is given by $T=1/(N_ta)$ and the spatial extents by $L_1=L_2=L_3=N_sa$.
The lattice spacing $a$ is a function of the bare coupling $g$, which is traditionally parameterized by $\beta=6/g^2$. 
The continuum limit $a\to0$ corresponds to the weak bare coupling limit $\beta\to\infty$, to which we get back to below in Sec.~\ref{sec:lat_contlimit}. 

There are two main alternatives to vary the temperature on the lattice. 
The standard procedure is the so-called fixed-$N_t$ approach, where one changes $T$ by changing $\beta$ (and, via that, the lattice spacing). While this allows for a continuous tuning of $T$, it requires the knowledge of the $a(\beta)$ scale function for a range of inverse gauge couplings.
Alternatively, in the fixed-$\beta$ approach, the temperature is changed by changing $N_t$. In this case, the scale is only needed at one $\beta$, but in turn, the temperature takes on discrete values, which is suboptimal for a studies, where one needs to resolve a potentially narrow transition region.

There are several standard discretizations of the gluon and fermion actions $S_G$ and $S_F$ of~\eqref{eq:lat_action_cont} in the literature. We will not review these here, but instead focus on how the background gauge field appears in $S_F$.
For the basic methods of lattice QCD, we refer the reader to the textbook~\cite{Gattringer:2010zz}. 
Here it suffices to note that the inverse gauge coupling is traditionally scaled out of the gluon action as a prefactor, so that the continuum action is written as $S_G=\beta S_g$. The lattice gauge action $S_g$ can be constructed from products of link variables over small closed loops e.g.\ plaquettes.
In the fermion sector, whichever discretization is used, the derivative operator in $\Dsf$ is in general replaced by a finite difference involving two (or more) neighboring points of the lattice. To maintain gauge invariance, the natural way to represent gauge fields is by means of parallel transporters between these neighboring lattice sites. The gluon field therefore becomes an $\mathrm{SU}(3)$ group element, while the photon field a $\mathrm{U}(1)$ element,
\be
\U_\nu(n)=e^{ia \A_\nu(n)}, \qquad 
u_{\nu f}(n)=e^{ia q_f A_\nu(n)}\,.
\ee
These will also be referred to as link variables. 
The Dirac operator $\Dsf$ contains these link variables in
the form $\U_\nu(n)u_{\nu f}(n)$ in 
its matrix elements that connect the sites $n$ and $n+\hat\nu$, where $\hat\nu$ is the unit vector in the $x_\nu$ direction. In turn, the backward links involve $-\,\U^\dagger_\nu(n)u^*_{\nu f}(n)$ between the sites $n$ and $n-\hat\nu$. The negative sign is consistent with the anti-Hermiticity of $\Dsf$, cf.\ the Hermiticity of the Euclidean Dirac matrices $\gamma_\nu$ in~\eqref{eq:lat_Dslash_cont}.

The gauge invariant integration measure over the $\U_\nu$ links is the so-called Haar measure, which now appears in the path integral,
\be
\Z = 
\int \D\U\, \exp\left[-\beta S_g\right] \, \prod_f \det M_f\,,
\label{eq:lat_partfunc_cont3}
\ee
where $S_g$ is the lattice gauge action and, with a slight abuse of notation, $\Dsf$ the lattice discretization of the Dirac operator. As noted already below~\eqref{eq:lat_partfunc_cont2}, the photon action merely constitutes an overall multiplicative factor in $\Z^{\rm tot}$ and here we only need to work with $\Z$. Note also that such overall factors cancel in expectation values,
\be
\expv{A} = \frac{1}{\Z}
\int \D\U\, A\,\exp\left[-\beta S_g\right] \, \prod_f \det M_f\,.
\label{eq:lat_expvO}
\ee

\subsubsection{Lattice observables}
\label{sec:lat_observables}

Physical quantities correspond to expectation values of gauge invariant observables that can be calculated using the path integral as in~\eqref{eq:lat_expvO}. Purely gluonic observables take the form of traces of products of gluon links along closed loops. The gluon action $S_g$ itself can be written in terms of the smallest possible closed loops on the lattice, like elementary plaquettes $U_{\nu\rho}^{1\times1}$ or rectangles $U_{\nu\rho}^{2\times1}$.
Comparing to~\eqref{eq:lat_partfunc_cont3}, one sees that the expectation value of the gauge action can be obtained via the derivative of $\Z$ with respect to $\beta$,
\be
\frac{\partial \log\Z}{\partial \beta} = -\expv{S_g}\,.
\label{eq:lat_sgdef}
\ee
Using the plaquettes and other small gluonic loops, a lattice version $\F^{\rm lat}_{\nu\rho}$ of the field strength tensor itself can be constructed. From that, we can also define the topological charge and its density,
\be
Q_{\rm top}= \sum_n q_{\rm top}(n), \qquad
q_{\rm top}(n) = \frac{1}{32\pi^2} \,\epsilon_{\nu\rho\alpha\beta} \,\tr \,\F^{\rm lat}_{\nu\rho}(n) \F^{\rm lat}_{\alpha\beta}(n)\,,
\label{eq:lat_topcharge_def}
\ee
which will play an important role in Chaps.~\ref{chap:hadron} and~\ref{chap:anom}.

Larger closed loops also constitute important physical observables. Rectangular loops lying in a spatial-temporal plane (for example the $x_1-x_4$ plane) of the lattice are the so-called Wilson loops,
\be
W(n_1,n_4)=\sum_k U_{14}^{n_1\times n_4}(k)\,.
\label{eq:lat_Wilsonloops}
\ee
Using off-plane paths, Wilson loops $W(\bm n,n_4)$ of arbitrary orientations may also be discretized. These can be used to calculate the potential between static quark and antiquark sources and to discuss confinement in the QCD vacuum.

On a lattice with periodic boundary conditions, one may also consider closed winding gluonic loops. The most important one is the $\mathrm{SU}(3)$ Polyakov loop -- the straight loop winding around the temporal direction,
\be
P(\bm n)=\tr L(\bm n), \qquad L(\bm n) =\prod_{n_4} U_4(\bm n,n_4) \,,
\label{eq:lat_ploopdef1}
\ee
starting at the spatial lattice site $\bm n$. 
Here, we defined the traced Polyakov loop $P$ (which will simply be referred to as Polyakov loop) and the untraced observable $L$ as well.
The spatially averaged Polyakov loop is also of interest,
\be
P=\frac{1}{V}\sum_{\bm n} P(\bm n)\,.
\label{eq:lat_ploopavg}
\ee
Its expectation value is related to the free energy density of a static color charge~\cite{Fukushima:2017csk} and will thus be relevant for the study of deconfinement and the phase diagram, see Sec.~\ref{sec:pd_ploop}. In turn, the untraced observable $L$ will be important when we consider gluonic screening lengths in Sec.~\ref{sec:had_conf_ploop}.

Turning to fermionic observables, we will mostly be concerned with fermion bilinears, that is to say, operators of the form $A=\bar\psi_f \Gamma \psi_f$ in~\eqref{eq:lat_expvO}, where $\Gamma$ is a matrix in spin or flavor space. For the spatially averaged observable, the fermion path integral in this case gives
\be
\expv{\bar\psi_f\Gamma\psi_f} = \frac{T}{V}\frac{1}{\Z}  
\int \D\U\, \exp\left[-\beta S_g\right] \, \prod_{f'} \det M_{f'} \,\Tr \left(\Gamma M_f^{-1}\right) \equiv
\frac{T}{V}\Expv{\Tr
 \left(\Gamma M_f^{-1}\right) }\,.
\label{eq:lat_expvO_fermion}
\ee
In the last equality we also introduced a short notation to indicate expectation values, where the fermion path integral has already been carried out. 
For brevity, it is denoted by the same symbol $\expv{.}$ as the one before the fermionic path integral.
Occasionally, we will also need local observables,
\be
\expv{\bar\psi_f(n)\Gamma\psi_f(n)} = \Expv{\Tr
 \left(\Gamma M_f^{-1} \P_n\right) }\,,
\label{eq:lat_expvO_fermion_local}
\ee
where $\P_n$ denotes a projector localized to the lattice site $n$, and we employed the same notation as in~\eqref{eq:lat_expvO_fermion}.
We will encounter non-trivial local currents when we consider inhomogeneous background magnetic fields in Sec.~\ref{sec:eos_ampere} as well as in Chap.~\ref{chap:anom} for the discussion of anomalous transport phenomena.

Observables of the type~\eqref{eq:lat_expvO_fermion} include the quark condensate, which can also be obtained by differentiating $\Z$ with respect to the quark mass,
\be
\expv{\bar\psi_f\psi_f} 
=
\frac{T}{V}\frac{\partial \log\Z}{\partial m_f} 
=
\frac{T}{V}\Expv{\Tr
 \left(M_f^{-1}\right) }\,,
\label{eq:lat_pbpdef_0}
\ee
which is the primary quantity to describe chiral symmetry breaking in QCD and will be our most important observable for discussing the phase diagram in Chap.~\ref{chap:pd}.
Background electromagnetic fields induce further nonzero expectation values, like that of the tensor bilinear,
\be
\expv{\bar\psi_f\sigma_{\nu\rho}\psi_f} 
=
\tau_f \cdot q_fF_{\nu\rho} + \O(F^3), \qquad
\sigma_{\nu\rho}=\frac{i}{2}[\gamma_\nu,\gamma_\rho]\,,
\label{eq:lat_def_tensorcoeff}
\ee
where we also included the leading weak-field behavior dictated by Lorentz invariance~\cite{Ioffe:1983ju,Balitsky:1983xk}. The coefficient $\tau_f$ of this term will be referred to as the tensor coefficient and will appear in the discussion of the equation of state in Sec.~\ref{sec:eos_tensor}.

Finally, we introduce a notation for the general vector, axial vector, pseudoscalar and tensor bilinear operators, respectively,
\be
V_\nu^f \equiv \bar\psi_f \gamma_\nu \psi_f, \qquad
V_{\nu5}^f \equiv \bar\psi_f \gamma_\nu \gamma_5\psi_f, \qquad
S_5^f \equiv \bar\psi_f \gamma_5\psi_f\\, \qquad
T^f_{\nu\rho} \equiv \bar\psi_f \sigma_{\nu\rho}\psi_f\,.
\label{eq:lat_VV5S5T}
\ee
The induced electromagnetic current appearing in~\eqref{eq:lat_inducedcurrent_totalcurrent} is the linear combination of vector current expectation values,
\be
\expv{j_\nu} = \sum_f\frac{q_f}{e} \expv{V^f_\nu(x)}\,.
\label{eq:lat_jnucurrentdef}
\ee
Note that according to this definition, the current is measured in units of the elementary charge $e$. In addition, we also define the flavor-sum of axial vector currents,
\be
\expv{j_{\nu5}} = \sum_f \expv{V^f_{\nu5}(x)}\,,
\label{eq:lat_jnu5currentdef}
\ee
which will enter in our discussion of the chiral separation effect in Sec.~\ref{sec:anom_CSE}.

\subsubsection{Valence and sea contributions and quenched approximations}
\label{sec:lat_valence_sea}

In the context of fermionic expectation values, one often speaks about valence and sea quarks. Sea quarks are described by the Dirac operator under the determinant, while valence quarks by the Dirac operator under the trace in~\eqref{eq:lat_expvO_fermion}. While both of these Dirac operators are supposed to be the same, in a lattice simulation one may still study them separately, for example, by employing different discretizations of the two Dirac operators or by having different quark masses next to the two operators, $m_f^{\rm val}$ in the valence and $m_f^{\rm sea}$ in the sea term.
An extreme case of the latter is the so-called quenched approximation. Within this approximation, sea quarks are considered to be infinitely heavy ($m_{f'}^{\rm sea}\to\infty$ in the $M_{f'}$ quark matrices in~\eqref{eq:lat_expvO_fermion}), implying that the Dirac determinant becomes an overall constant that cancels from expectation values. The quenched approximation of~\eqref{eq:lat_expvO_fermion} therefore reads
\be
\expv{\bar\psi_f\Gamma\psi_f}^{\rm quenched} = \frac{T}{V}\frac{1}{\Z_G}
\int \D\U\, \exp\left[-\beta S_g\right] \, \Tr \left(\Gamma M_f^{-1}\right)
\equiv\frac{T}{V}
\Expv{\Tr
 \left(\Gamma M_f^{-1}\right) }^{\rm quenched}\,,
\label{eq:lat_expvO_quenched}
\ee
where $\Z_G$ is the partition function of the purely gluonic theory and we again included the short notation for the expectation value.
The relation~\eqref{eq:lat_expvO_quenched} is an often employed -- albeit uncontrolled -- approximation in lattice simulations, allowing one to generate gluon configurations in pure gauge theory, thereby sparing the most expensive part of the calculation. In a perturbative approach, one may think about this approximation as neglecting virtual (sea) quark loops in a diagram containing a (valence) quark propagator.

For the discussion of the impact of background electromagnetic fields on fermionic observables, there is another, similar approximation, often employed on the lattice. In this case one introduces different electric charges for valence and sea quarks -- specifically, by retaining the charge in the valence sector but neglecting it in the sea: $q_f^{\rm val}=q_f$ but $q_f^{\rm sea}=0$. This implies that the background field only affects the measurement of operators but not the gluonic field configurations, which can therefore be generated in the absence of the field. In contrast to the quenched approximation above, this means that gluon fields do feel fermions,  just not the background field. For a fermionic expectation value, it reads
\be
\expv{\bar\psi_f\Gamma\psi_f}^{\rm val}\equiv \expv{\bar\psi_f\Gamma\psi_f}^{\rm electroquenched} \equiv \frac{T}{V}\frac{1}{\Z(q_{f'}=0)}
\int \D\U\, \exp\left[-\beta S_g\right] \, \prod_{f'} \det M_{f'}(q_{f'}=0)\,\Tr \left(\Gamma M_f^{-1}\right) \,.
\label{eq:lat_expvO_electroquenched}
\ee
One often refers to this approximation as the electroquenched or valence approximation.
Perturbatively, this implies setting the quark electric charges in virtual (sea) quark loops to zero, while keeping them in valence propagators. This technique will be important for two discussions in this review: for the understanding of the inverse magnetic catalysis phenomenon for the phase diagram in Chap.~\ref{chap:pd}, as well as for widely used approximations to determine the hadron spectrum in background fields in Chap.~\ref{chap:hadron}.

\subsubsection{Lattice parameters and continuum limit}
\label{sec:lat_contlimit}

The input parameters of a lattice simulation are -- besides the lattice extents $N_s$ and $N_t$ -- the parameters of the QCD action~\eqref{eq:lat_action_cont}: the inverse gauge coupling $\beta$ and the quark masses $m_fa$ in lattice units. As already alluded to above, the inverse gauge coupling sets the lattice scale $a(\beta)$, i.e.\ one can vary the temperature and approach the continuum limit by changing $\beta$.
In turn, the lattice quark masses require tuning in order to describe physics correctly.
The lower the (light) quark masses, the more expensive the inversion of $M_f$ -- required e.g.\ in~\eqref{eq:lat_expvO_fermion} -- is computationally. Often simulations are performed with heavier-than-physical quarks in order to reduce the overall computational complexity. In order to simulate real QCD, it is however important to use physical quark masses. Some of the physical effects that we will consider in this review -- in particular the inverse magnetic catalysis phenomenon at nonzero temperatures -- even disappear if the quarks are not as light as in nature, see Sec.~\ref{sec:pd_phasediagBT}. A determination of the quark masses is often traded for a determination of the mass of the lightest hadron, i.e.\ the pion\footnote{For light quarks, the two are related to each other via the Gell-Mann-Oakes-Renner relation~\eqref{eq:pd_GMOR}.}. A good measure of how close a simulation is to the physical point can therefore be given in terms of the pion mass $M_\pi$ and one often speaks of either physical or heavier-than-physical pion masses.

Physical results are obtained on the lattice once the regularization is gradually removed i.e.\ the continuum limit $a\to0$ is taken. 
To unambiguously define this process, renormalization has to be carried out, see Sec.~\ref{sec:lat_renorm}.
In the fixed-$N_t$ approach described above, one approaches the continuum limit at a given $T$, by performing simulations on a series of temporal extents $N_t$, at values of $\beta$ tuned so that $N_t a(\beta)=1/T$. The continuum extrapolation therefore becomes the limit $N_t\to\infty$. 
Most of the finite temperature results we will discuss in this review have been obtained with this approach. To indicate the range of lattice spacings used in a study, one often quotes the values of $N_t$, e.g.\ $6\le N_t\le 16$ are typical values.

The approach towards the continuum limit depends on the scaling properties of the lattice action and the corresponding lattice artefacts. In Sec.~\ref{sec:lat_latDiracops} we briefly mention the scaling of various lattice discretizations of the fermion action.
In all cases, the continuum limit is a numerical extrapolation based on results obtained at different values of $a$. Lattice spacings of the order of $a=0.1\fm$ are typically necessary for a reliable extrapolation, but this may depend strongly on the observable in question. It is also important to stress that all dimensionful scales should be well resolved by the lattice -- in other words $m_fa\ll1$, $a^2q_fB\ll 1$ and so on should hold. For example for magnetic fields reaching $1/a^2$, severe lattice artefacts appear that are not related to the continuum behavior anymore. We will encounter such lattice artefacts below in Chap.~\ref{chap:ev} when we discuss the recursive patterns in Hofstadter's butterfly.

\subsubsection{Lattice Dirac operators}
\label{sec:lat_latDiracops}

In the literature, three major fermionic lattice discretizations have been used to simulate QCD with background electromagnetic fields: staggered fermions, Wilson fermions as well as overlap fermions.
For completeness, here we explicitly write down the Dirac operators for all three cases. More details, in particular about fermion doubling, can be found in the textbook~\cite{Gattringer:2010zz}. In the staggered discretization, the Dirac operator is diagonalized in spinor space via a local transformation of the fermion field. As a result, the Dirac matrices are replaced by space-dependent phases $\eta_\nu(n)$,
\be
\left(\Dsf\right)_{nm} = \frac{1}{2a} \sum_{\nu} \eta_\nu(n) \left[ \U_\nu(n) u_{\nu f}(n) \,\delta_{m,n+\hat\nu}
- \U^\dagger_\nu(n-\hat\nu)\,u_{\nu f}^*(n-\hat\nu)\, \delta_{m,n-\hat\nu} \right], 
\quad\quad\quad
\eta_\nu(n) = (-1)^{\sum_{\rho=1}^{\nu-1} n_\rho}.
\label{eq:lat_staggered_Dslash}
\ee
Here, $\hat\nu$ again denotes the unit vector in the $\nu$ direction. In the staggered formulation, an important role is played by the diagonal matrix,
\be
\eta_5(n)=(-1)^{n_1+n_2+n_3+n_4}\,.
\label{eq:lat_eta5_def}
\ee
The lattice artefacts of the staggered Dirac operator scale as $\mathcal{O}(a^2)$.
Since the above operator will be used most frequently in this review, we simply denote it as $\Dsf$.

The staggered Dirac operator is expected to describe four fermion flavors (tastes) in the continuum limit. For this reason, often the fourth root of the Dirac determinant is taken under the path integral. This `rooting trick'~\cite{Durr:2005ax,Creutz:2007rk} has been the subject of ongoing research. While rooting has no strict mathematical foundations, empirical evidence suggests that for most thermodynamical observables rooted staggered quarks give results compatible with other lattice formulations (see e.g.~\cite{Borsanyi:2015zva,Borsanyi:2015waa}). Taking the fourth root of the determinant in~\eqref{eq:lat_partfunc_cont3} implies that an overall factor $1/4$ is to be included in the expectation values~\eqref{eq:lat_expvO_fermion} of fermionic bilinears.

While in the continuum limit, the four tastes of the staggered formulation are expected to become degenerate, at nonzero lattice spacings, they have different masses. This taste breaking is the source of potentially large lattice artefacts, which may be reduced by a smearing of the gluon fields in the Dirac operator, in effect removing ultraviolet fluctuations. Two options, which will appear repeatedly in this review, are the stout-smeared staggered action~\cite{Morningstar:2003gk,Aoki:2005vt} and the highly improved staggered quark (HISQ) action~\cite{Follana:2006rc}. We note that in this procedure only the gluon fields $\U_\nu$ are smeared but not the photon links $u_{\nu f}$, cf.\ the discussion in~\cite{DElia:2012ems}.

For Wilson fermions, doublers are separated from the physical fermion by lattice-spacing dependent mass terms at the cost of violating chiral symmetry. The Wilson Dirac operator, denoted by $\Dsf^W$, reads
\be
\big(\Dsf^W\big)_{nm} = \frac{4}{a} \, \mathds{1}\delta_{nm} +\frac{1}{2a} \sum_{\nu} \left[ (\gamma_\nu-\mathds{1}) \,\U_\nu(n) u_{\nu f}(n) \,\delta_{m,n+\hat\nu} -(\gamma_\nu+\mathds{1})
 \,\U^\dagger_\nu(n-\hat\nu)\,u_{\nu f}^*(n-\hat\nu)\, \delta_{m,n-\hat\nu} \right]\,.
\label{eq:lat_Wilson_Dslash}
\ee
The term proportional to $\mathds{1}$ in $\Dsf^W$ -- the so-called the Wilson term -- acts like a mass term that breaks chiral symmetry explicitly. As a consequence, the Wilson formulation implies that the quark mass needs to undergo additive renormalization~\cite{Gattringer:2010zz}. As a further consequence, the discretization errors scale as $\mathcal{O}(a)$ here. This can be improved via a standard improvement scheme~\cite{Sheikholeslami:1985ij}.

Finally, the overlap Dirac operator $\Dsf^{\rm ov}$ can be written as
\be
\Dsf^{\rm ov} = \frac{\mathds{1}}{a}+\gamma_5 \,\sgn\left( \gamma_5 A \right)  = \frac{\mathds{1}}{a}+A \left(\gamma_5A \gamma_5A\right)^{-1/2}\,,
\qquad
A=\Dsf^W - m_0 \mathds{1} \,.
\label{eq:lat_overlap_Dslash}
\ee
Here, $A$ is a kernel operator that satisfies $\gamma_5 A \gamma_5=A^\dagger$. In~\eqref{eq:lat_overlap_Dslash}, it is chosen as the Wilson Dirac operator shifted by a negative mass term with $0<m_0<2/a$. The overlap construction may be thought of as a projection of the eigenvalues of the kernel on the unit circle centered at the projection point $(m_0,0)$ in the complex plane. The operator $\Dsf^{\rm ov}$ exhibits an exact chiral symmetry on the lattice in the sense of the Ginsparg-Wilson relation~\cite{Gattringer:2010zz}. Furthermore, the locality of the operator is expected to hold for gauge fields that are sufficiently smooth on the scale of the lattice spacing~\cite{Hernandez:1998et}.
Finally we note that in two dimensions -- which we will discuss later in Chap.~\ref{chap:ev} -- the Dirac matrices in $\Dsf^W$ and $\Dsf^{\rm ov}$ are replaced by the Pauli matrices, $\gamma_1\to\sigma_1$, $\gamma_2\to \sigma_2$, and the role of $\gamma_5$ is taken over by the spin matrix $\sigma_{3}$.

\subsubsection{Photon fields on the lattice}
\label{sec:lat_photonfieldslat}

Now we only need to provide the $u_{\nu f}$ link variables to completely specify the action. We use the field configurations derived in Sec.~\ref{sec:lat_flux} for the continuum fields on the torus and express everything in dimensionless variables, i.e.\ $n_j=x_j/a$, $N_s=L/a$, $N_t=1/(aT)$ and the flux quanta $N_b^f$ and $N_e^f$. For the homogeneous magnetic case, the continuum photon field~\eqref{eq:lat_hom_Bfield} translates to
\be
u_{2f}(n)=\exp\left( i \frac{2\pi N_b^f n_1}{N_s^2} \right), \qquad
u_{1f}(n)=\exp\left( -i \frac{2\pi N_b^f n_2}{N_s} \,\delta_{n_1,N_s-1}\right), \qquad
u_{3f}(n)=u_{4f}(n)=1\,.
\label{eq:lat_hom_links}
\ee
Here the discretization $\delta(x_1-L)=\delta_{n_1,N_s-1}/a$ was used. The localized magnetic field is implemented similarly, except that $u_{2f}(n)$ in~\eqref{eq:lat_hom_links} needs to be replaced by
\be
u_{2f}(n)=\exp\left[ i\frac{\pi N_b^f}{N_s}\, \tanh\left(\frac{2n_1-N_s}{2\hat\epsilon}\right)\bigg/ \tanh\left(\frac{N_s}{2\hat\epsilon}\right)  \right]\,,
\ee
where $\hat\epsilon=\epsilon/a$.
The flux $N_b^f$ of the magnetic field through the $x_1-x_2$ plane is discrete, just like in the continuum. 
In addition, due to the periodicity of the  links~(\ref{eq:lat_hom_links}) in $N_b^f$, 
there is also a maximal possible magnetic field, 
\be
0\le N_b^f < N_s^2, \qquad N_b^f\in\mathds{Z}\,.
\label{eq:lat_Nbquant}
\ee
Notice that for the localized magnetic field, the link variables are not periodic in $N_b^f$ with this period, except for certain specific choices of the width $\hat\epsilon$~\cite{Brandt:2023dir}.
Nevertheless, in both cases the largest magnetic field is set by the ultraviolet cutoff i.e.\ the inverse lattice spacing: $q_fB\le 1/a^2$. In practice, the magnetic field needs to be much smaller than this upper limit in order to avoid substantial lattice discretization errors, and a reasonable bound is $N_b^f\le N_s^2/16$~\cite{Bali:2012zg}.

The oscillatory magnetic fields have vanishing total flux and can therefore be implemented trivially using $u_{2f}(n)=\exp[iaq_fA_2(n)]$ and the prescriptions~\eqref{eq:lat_osc_Bfield} and~\eqref{eq:lat_half_Bfield} for $A_2(n)$.
The link configurations for imaginary electric fields can be found similarly. The homogeneous setup~\eqref{eq:lat_hom_Efield} implies
\be
u_{4f}(n)=\exp\left( i \frac{2\pi N_e^f n_3}{N_sN_t} \right), \qquad
u_{3f}(n)=\exp\left( -i \frac{2\pi N_e^f n_4}{N_t} \,\delta_{n_3,N_s-1}\right), \qquad
u_{1f}(n)=u_{2f}(n)=1\,.
\label{eq:lat_hom_Elinks}
\ee
The implementation of harmonic imaginary electric fields is again trivial, and follows from $u_{4f}(n)=\exp[iaq_f A_4(n)]$ and the gauge field~\eqref{eq:lat_osc_Efield} for $A_4(n)$.

Beyond the above considered special cases, the implementation of general background fields on the lattice with periodic boundary conditions has also been discussed in~\cite{Davoudi:2015cba}, with a special focus on the impact of the quantization condition on hadronic correlation functions.
We mention moreover that an alternative procedure to implement background fields on the lattice has also been worked out in terms of the
Schr\"odinger functional. It is based on fixing the background field on the boundaries and has been used to simulate the theory with background chromomagnetic fields~\cite{Cea:2002wx,Cea:2005td,Cea:2007yv,Cea:2024efe}.

\subsubsection{Sign problem}
\label{sec:lat_signproblem}

We have left one important aspect of the lattice simulations to the end of this section. 
In order to be able to carry out importance sampling-based lattice Monte-Carlo simulations for expectation values~\eqref{eq:lat_expvO}, it is necessary that the total action is real and positive and thus suitable as a probability weight. 
One therefore needs $S_g$ and $\prod_f\det M_f$ to be real and positive. While the former is always satisfied, the latter only holds under certain circumstances. In general, to show the reality of $\det M_f$, one needs the so-called $\gamma_5$-Hermiticity relation,
\be
\gamma_5 \Dsf \gamma_5=\Dsf^\dagger\,.
\label{eq:lat_gamma5hermiticity}
\ee
This condition, together with $\gamma_5^2=\mathds{1}$ and the cyclicity of the determinant implies that
\be
\det M_f = \det \left[ \gamma_5^2 M_f \right]
= \det \left[\gamma_5 M_f \gamma_5\right]
= \det \big[ M_f^\dagger \big] = \left(\det M_f\right)^*\,,
\label{eq:lat_detM_real}
\ee
i.e.\ that $\det M_f\in\mathds{R}$. The $\gamma_5$-Hermiticity relation~\eqref{eq:lat_gamma5hermiticity} holds for all three Dirac operators\footnote{\label{fn:lat_stagg}For staggered quarks, the relations~\eqref{eq:lat_gamma5hermiticity} and~\eqref{eq:lat_chiralsymm} are understood to hold with $\gamma_5$ replaced by $\eta_5$ from~\eqref{eq:lat_eta5_def}.} discussed in Sec.~\ref{sec:lat_latDiracops}, as long as the hoppings, in particular the photon links $u_\nu$, are unitary. This is the case for background magnetic fields $B$ and imaginary electric fields $iE\in\mathds{R}$, as shown in Sec.~\ref{sec:lat_photonfieldslat}. In contrast, for real electric fields the photon links are not $\mathrm{U}(1)$ phases and the $\gamma_5$-Hermiticity relation becomes,
\be
E\in\mathds{R}:\qquad
\gamma_5\Dsf(E)\gamma_5=\Dsf^\dagger(-E)\,.
\ee
In this case, the action is therefore in general complex, and cannot be interpreted as a probability measure. This is the so-called complex action or sign problem.
The situation with electric fields is completely analogous to the case with chemical potentials: real values of $\mu$ lead to the sign problem whereas imaginary chemical potentials can be simulated directly~\cite{Gattringer:2010zz,Nagata:2021ugx}.

Above we showed that the determinant is real for certain choices of the photon fields. 
In order to show positivity as well, one either needs two degenerate flavors $f$ and $f'$, so that $\det M_f\cdot \det M_{f'}=(\det M_f)^2>0$ (this is typically used for Wilson fermions). Alternatively, one can show positivity for a single flavor as well if the lattice formulation satisfies the chirality relation (as for staggered fermions, see footnote~\ref{fn:lat_stagg}),
\be
\{\Dsf,\gamma_5\} = 0\,.
\label{eq:lat_chiralsymm}
\ee
This, together with~\eqref{eq:lat_gamma5hermiticity} implies that the Dirac operator is anti-Hermitian, its
eigenvalues are purely imaginary and occur in complex conjugate pairs, as we will show below in~\eqref{eq:ev_chiralsymm}. From this the positivity of the determinant, being the product of all eigenvalues, follows. 

We note that the product of Dirac determinants can also be shown to be real and positive for a special (though unphysical) choice of real electric fields -- namely for the setup where up and down quarks have opposite electric charges, $q_u=-q_d$ but are otherwise degenerate. This setting is tantamount to the case of real isospin chemical potentials $\mu_u=-\mu_d$, which are also free of the complex action problem~\cite{Son:2000xc}. We will encounter such isospin electric fields in Sec.~\ref{sec:eos_elsusc}.

\subsection{Renormalization}
\label{sec:lat_renorm}

A consistent definition of the path integral~\eqref{eq:lat_partfunc_cont2} requires a regularization scheme in the ultraviolet.
The bare parameters $m_f$ and $g$ of the action~\eqref{eq:lat_action_cont} need to be tuned as functions of the regulator and are related to renormalized couplings via multiplicative renormalization constants. In the lattice discretization, the regulator is the lattice spacing $a$ and, correspondingly, the bare parameters become functions of it: $m_f(a)$ and $\beta(a)=6/g^2(a)$. In contrast to perturbative approaches, on the lattice it is convenient to fix these functions not by renormalized couplings and multiplicative renormalization constants but rather via other physical observables e.g.\ vacuum hadron masses. Requiring that the mass of the hadron is the same in physical units for different lattice spacings implicitly defines the above functions, which are then called lines of constant physics. Once this trajectory is followed in the space of bare parameters, the regularization can be removed i.e.\ the limit $a\to0$ can be taken. This is referred to as the continuum limit. We note that this trajectory is conveniently parameterized by $\beta$ itself and often the tuning of the quark mass is called a line of constant physics. For example, we will denote the line of constant physics for physical quark masses as $m_f=m_f^{\rm ph}(\beta)$.

Even with such multiplicative renormalizations (or, lines of constant physics) in place, one encounters further additive divergences in e.g.\ the free energy. In a perturbative approach, one cancels these via normal ordering of operators inside expectation values, rendering the free energy of the vacuum (i.e.\ a zero-point energy) zero. Equivalently, one may subtract the vacuum contribution explicitly, utilizing that such divergent terms are in general independent of background electromagnetic fields (or temperature or chemical potentials).
As an example, the total free energy density contains divergent terms $a^{-4}$, $m_f^2 a^{-2}$ and $m_f^4\log (m_fa)$~\cite{Leutwyler:1992yt}. All of these cancel in the subtracted observable such that
\be
f^{\rm tot,r} = f^{\rm tot} - f^{\rm tot}(T=B=E=0)\,,
\label{eq:lat_ftotrenorm}
\ee
is completely free of ultraviolet divergences.
Above in~\eqref{eq:lat_ftot_f} we have seen that it is helpful to separate $f^{\rm tot}$ into matter and background field contributions. However, we need to be careful about the divergent terms in this separation. For this reason, next we discuss renormalizations specific to background fields.

\subsubsection{Renormalization of the matter free energy}
\label{sec:lat_renormfreeenergy}

In the presence of fluctuating electromagnetic fields, i.e.\ dynamical photons, the electromagnetic charges $q_f$ undergo multiplicative renormalization. Expressing the charges in terms of the elementary electric charge $e$, it suffices to have one multiplicative renormalization, $e_r^2=Z_e e^2$. In the perturbative treatment, $Z_e$ can be calculated from the photon vacuum polarization diagram, see e.g.~\cite{itzykson2006quantum}. In our case, where background electromagnetic fields are not integrated over in the path integral, there are no internal photon lines in such diagrams. That means that the renormalization constant only receives one-loop (in QED) contributions. For this reason, for the renormalization of the electric charge, it is more suitable to follow a perturbative description with multiplicative renormalization constants. This is what we do next.

The one-loop renormalization constant in QED reads,
\be
Z_e=1 + \beta_1 e_r^2 \log \left(a^2\mu^2_{\rm QED}\right), \qquad
\beta_1=N_c \sum_f (q_f/e)^2\beta_1^{\rm free}, \qquad \beta_1^{\rm free}=\frac{1}{12\pi^2}\,,
\label{eq:lat_renorm_Ze}
\ee
where $\beta_1$ is the lowest order QED $\beta$-function coefficient and $\mu_{\rm QED}$ the renormalization scale.
In this one-loop diagram, all fermion flavors contribute proportionally to their squared electric charges, as well as the number of colors $N_c=3$. Note that QCD corrections to $\beta_1$ vanish in the continuum limit~\cite{Bali:2014kia}.
The renormalization of the electric charge is tightly connected to that of the photon field due to the QED Ward-Takahashi identity, $e_r A_{\nu r}=e A_\nu$. This also sets the renormalization of the background magnetic and electric fields, $e_rB_r=eB$ and $e_rE_r=eE$, and implies that the photon action~\eqref{eq:lat_SgammaBE} renormalizes as
\be
S_\gamma = Z_e \,S_{\gamma r}\,.
\label{eq:lat_Sgamma_renorm}
\ee
On the one hand, the total free energy density $f^{\rm tot}$ of the system contains $S_\gamma$ according to~\eqref{eq:lat_ftot_f}. On the other hand, it is related to physical, i.e.\ finite, observables as we discussed in Sec.~\ref{sec:lat_extmat} and, apart from a field-independent overall divergence, cannot contain a divergent renormalization constant.

To find the loophole, let us use the separation~\eqref{eq:lat_ftot_f} of the total free energy density into matter and field contributions. Inserting~\eqref{eq:lat_Sgamma_renorm} and rearranging, we obtain
\be
\label{eq:lat_ftotfinite}
f^{\rm tot} -\frac{T}{V}S_{\gamma r} = (Z_e-1)\frac{T}{V}S_{\gamma r} + f\,.
\ee
Since the left hand side of this equation only contains physical observables, it must be ultraviolet finite (up to a field-independent overall constant, as noted in~\eqref{eq:lat_ftotrenorm}). Thus, the same must hold for the right hand side. Using the specific form of the renormalization constant~\eqref{eq:lat_renorm_Ze} and inserting the quadratic form of the photon action~\eqref{eq:lat_SgammaBE}, we conclude that
\be
f=
f^{\rm tot} -\frac{T}{V}S_{\gamma r} - \frac{\beta_1}{2} \frac{T}{V}\int\dd^3 x \int_0^{1/T} \!\!\dd x_4 \left[ e_r^2 \bm B_r^2(x) - e_r^2\bm E_r^2(x) \right]\, \log\left(a^2\mu^2_{\rm QED}\right)\,.
\label{eq:lat_renorm_f_generalEB}
\ee
Thus, the matter free energy density -- more specifically, the contribution in it proportional to the square of the background electromagnetic field -- is logarithmically divergent in the lattice spacing. 
The total free energy density is, in turn, ultraviolet finite, since the logarithmic divergence of $f$ and of $S_\gamma$ cancel each other in it.

As a specific example, the matter free energy density $f$ for a homogeneous background magnetic field contains the divergent term
\be
f=\{B\textmd{-independent}\} + \{\textmd{finite}\} -  \frac{\beta_1}{2}\, (eB)^2 \,\log\left(a^2\mu^2_{\rm QED}\right)\,.
\label{eq:lat_fr_muQED_def}
\ee
The renormalization scale $\mu_{\rm QED}$ is, up to this point, a free parameter. To fix it, a possible choice is to require that the renormalized matter free energy $f^r$ contains no $\mathcal{O}(B^2)$ contribution at zero temperature. This is equivalent to the physical requirement that in the vacuum, the total free energy density~\eqref{eq:lat_ftot_f_hom} receives no $\mathcal{O}(B^2)$ contribution from the medium, i.e.\ in~\eqref{eq:lat_ftot_f_hom_expanded} the magnetic permeability $\pi_m$ of the vacuum is unity\footnote{
\label{fn:lat_muqed}
In the absence of color interactions, this choice for the renormalization scale coincides with the on-shell renormalization condition $\mu_{\rm QED}=m_f$ and is inherent to Schwinger's proper time regularization~\cite{Schwinger:1951nm}. 
The same conclusion also holds for non-interacting charged hadrons, with $m_f$ replaced by the mass $M_h$ of the hadron -- a setup usually considered in hadron resonance gas (HRG) models~\cite{Endrodi:2013cs}. In the latter case, $\mu_{\rm QED}=M_h$ is the only choice for the renormalization scale that ensures that $f^r$ approaches zero in the $M_h\to\infty$ limit. In other words, for this choice static hadrons are insensitive to the magnetic field, as expected on physical grounds~\cite{Endrodi:2013cs}.}. The renormalized matter free energy density in this scheme therefore reads
\be
f^r(T,B) = f(T,B) - f(T=0,B=0) - \frac{(eB)^2}{2}\left.\frac{\partial^2 f(T=0,B)}{\partial (eB)^2}\right|_{B=0}\,.
\label{eq:lat_fr_def}
\ee
Here, the field-independent divergent terms of the vacuum are canceled by the vacuum subtraction, while the magnetic field-induced divergences by the subtraction of the last term.
For inhomogeneous fields, it can be seen from~\eqref{eq:lat_renorm_f_generalEB} that the background field-dependent subtraction takes a similar form, involving the four-volume integral of the square of the fields.

We mention that different renormalization schemes -- differing from the choice~\eqref{eq:lat_fr_def} by finite $\mathcal{O}(B^2)$ terms -- have also been used in the literature, see e.g.~\cite{Menezes:2008qt,Fraga:2012fs}. For a meaningful comparison, it is important to first convert the results to a scheme with the same renormalization scale.

\subsubsection{Renormalization of the susceptibilities}
\label{sec:lat_renormsusc}

From~\eqref{eq:lat_fr_def} we see that the renormalization of the matter free energy for homogeneous fields involves the subtraction of the quadratic contribution at zero temperature. This term is just given by the  magnetic susceptibility~\eqref{eq:lat_susc_def}. Its additive renormalization therefore takes the form,
\be
\chi^r = \chi(T)-\chi(T=0) 
\label{eq:lat_magsusc_renorm}
\,.
\ee
Very similarly, homogeneous electric fields also induce a logarithmic divergence and the electric susceptibility $\xi$ undergoes additive renormalization,
\be
\xi^r = \xi(T)-\xi(T=0) 
\,.
\ee

For the discussion of the equation of state in Chap.~\ref{chap:eos}, we will consider the separation of the magnetic susceptibility into contributions from quark spins and orbital angular momenta. In that context we will encounter the tensor coefficient $\tau_f$, introduced above in~\eqref{eq:lat_def_tensorcoeff} through the weak-field expansion of the tensor quark bilinear.
Similarly to $\chi$, the tensor coefficient also contains a logarithmic divergence~\cite{Bali:2012jv,Bali:2020bcn} which is eliminated by zero-temperature subtraction,
\be
\tau_f^r = Z_T\left[\tau_f(T) - \tau_f(T=0)\right]\,.
\label{eq:lat_renorm_tensorcoeff}
\ee
In addition to the QED-related additive renormalization, the tensor coefficient is also subject to QCD-related multiplicative renormalization by the tensor renormalization constant $Z_T$, introducing a renormalization scale- and scheme-dependence to it.

\subsubsection{Renormalization of the quark condensate}
\label{sec:lat_renormpbp}

Above in~\eqref{eq:lat_pbpdef_0}, we already defined the quark condensate as the derivative of the free energy density with respect to the quark mass,
\be
\expv{\bar\psi_f\psi_f}
\equiv-\frac{\partial f^{\rm tot}}{\partial m_f}
=-\left.\frac{\partial f}{\partial m_f}\right|_{\mu_{\rm QED}}\,.
\label{eq:lat_pbpdef}
\ee
Notice that only the matter free energy density depends on the quark mass and, therefore, $f^{\rm tot}-f=T S_\gamma/V$ does not contribute to the condensate at fixed renormalization scale\footnote{\label{fn:lat_mfder_muqed}
A note about mass-dependent renormalization scales is in order here. In this case an implicit dependence on $m_f$ arises via $\mu_{\rm QED}$ both in $T S_\gamma/V$ and in $f$ with opposite signs. The total free energy density $f^{\rm tot}$ is independent of $\mu_{\rm QED}$ and this implicit dependence cancels in $\expv{\bar\psi_f\psi_f}$.  
The condensate can therefore be equivalently defined by differentiating just the matter part $f$ with respect to $m_f$ at fixed renormalization scale $\mu_{\rm QED}$. Thus, the logarithmic term in~\eqref{eq:lat_fr_muQED_def} does not contribute to the $m_f$-derivative.}.
Following the remark above~\eqref{eq:lat_ftotrenorm}, this observable also contains ultraviolet divergent terms $m_fa^{-2}$ and $m_f^3\log(m_fa)$. While its chiral limit is finite, for any nonzero $m_f$ we again need to consider a difference such as
\be
\Delta\expv{\bar\psi_f\psi_f} = \expv{\bar\psi_f\psi_f} - \expv{\bar\psi_f\psi_f}_{B=E=0}\,,
\label{eq:lat_deltapbp_1}
\ee
in order to cancel ultraviolet divergences. In~\eqref{eq:lat_deltapbp_1} we chose the subtraction point to be at the same temperature but at zero background electromagnetic fields. This subtraction will in general be denoted by $\Delta$ throughout this review.

Furthermore, since the derivative in~\eqref{eq:lat_pbpdef} is with respect to the bare quark mass, the condensate is also subject to multiplicative renormalization, via the scalar renormalization constant $Z_S$. To cancel the latter, it is convenient to consider the product $m_f\expv{\bar\psi_f\psi_f}$.
In particular, for the discussion of the phase diagram in Chap.~\ref{chap:pd} we will often encounter the fully renormalized, vacuum-subtracted combination,
\be
\Sigma_f(B,T) = \frac{2m_{\ell}}{M_\pi^2 F_\pi^{(0)2}} \left[ \expv{\bar\psi_f\psi_f}_{B,T}- \expv{\bar\psi_f\psi_f}_{0,0} \right] + 1\,, \qquad f=u,d\,,
\label{eq:lat_pbpren}
\ee
where the light quark masses are considered degenerate, $m_u=m_d=m_\ell$.
Here we chose a convenient normalization\footnote{In~\eqref{eq:lat_pbpren}, $M_\pi=135 \MeV$ is the vacuum pion mass and $F_\pi^{(0)}=86 \MeV$ the chiral limit of the pion decay constant in the vacuum.}, so that $\Sigma_f$ is unity in the vacuum and approaches zero for high temperatures owing to the Gell-Mann-Oakes-Renner relation~\cite{Gell-Mann:1968hlm},
\be
\expv{\bar\psi\psi}_{0,0} = \expv{\bar\psi_f\psi_f}_{0,0} = \frac{M_\pi^2 F_\pi^{(0)2}}{2m_\ell}+\cdots, \qquad f=u,d\,,
\label{eq:pd_GMOR}
\ee
which connects the average light quark condensate of the vacuum (where the up and down quarks are identical) to the light quark mass and the properties of the pion.

\subsubsection{Renormalization of the induced current}
\label{sec:lat_renormcurrent}

The electromagnetic current is conserved in QCD and therefore does not require any renormalization (a general result with only a few exceptions~\cite{Collins:2005nj}). However, when separating the total current into external and induced terms, as in~\eqref{eq:lat_inducedcurrent_totalcurrent}, we encounter ultraviolet divergences in both of them with opposite signs~\cite{Brandt:2024blb}, just like in the total free energy density~\eqref{eq:lat_ftotfinite}. The renormalization of the induced current $\expv{\bm j}$ can be found from Amp\'{e}re's law for the total current, $e\bm j^{\rm tot}=\bm \nabla\times\bm B$, the separation of the magnetic field into external and magnetization terms, $\bm B=\bm H + e\bm \M$, and the weak-field behavior of the magnetization involving the magnetic susceptibility, $\bm \M = \chi \,e\bm B$, cf.\ footnote~\ref{fn:lat_perme}. Recall here that the electromagnetic current is measured in units of $e$, see~\eqref{eq:lat_jnucurrentdef}.

Putting these together, and using the renormalization~\eqref{eq:lat_magsusc_renorm} of the magnetic susceptibility, we arrive at~\cite{Brandt:2024blb},
\be
\expv{\bm j^r}=\expv{\bm j} - \chi(T=0) \cdot \bm \nabla \times e\bm B\,,
\label{eq:lat_renorm_induced_current}
\ee
which is understood to hold locally, as a function of the coordinate.

\subsubsection{Renormalization of the Polyakov loop}
\label{sec:lat_renormploop}

The spatially averaged traced Polyakov loop $P$, defined above in~\eqref{eq:lat_ploopavg}, acts as a measure of deconfinement in the system. In particular, $\expv{P}$ encodes information about the free energy density $f_Q$ of a static color charge~\cite{Fukushima:2017csk} and may be interpreted as a ratio of two partition functions, one with and one without the static charge,
\be
\expv{P}_{B,T}=\frac{\Z_{Q}(B,T)}{\Z(B,T)}
= \exp\left\{ -\frac{V}{T} \left[ f_Q(B,T) - f(B,T) \right] \right\}\,,
\label{eq:lat_ploopfQ}
\ee
where we also wrote the partition functions in terms of the corresponding free energy densities. Notice that in the above ratio the pure electromagnetic energy $S_\gamma$ cancels, thus we can work with the matter free energy densities. In~\eqref{eq:lat_ploopfQ}, we specialized to the case of a magnetic field, but the discussion is analogous for an electric field, too.

Just like $f$ discussed above, $f_Q$ contains additive divergences, which can be eliminated through the prescription~\eqref{eq:lat_fr_def}. However, the $\mathcal{O}(B^2)$ divergence is the same in both free energy densities, since there are no further renormalization constants beside $Z_e$ that we may use to cancel $B$-dependent divergences.\footnote{Perturbatively, this may be seen by noticing that the coupling between the external photons and the static color charge must involve a sea quark loop with at least two photon and at least two gluon legs. This diagram is ultraviolet finite.} In turn, the vacuum divergences in $f$ and in $f_Q$ are different\footnote{Again resorting to perturbation theory, the diagram involving the gluon vacuum polarization with gluon propagators attached to the static quark is divergent.}.
Altogether, the expression in the square brackets in the right hand side of~\eqref{eq:lat_ploopfQ} contains additive divergences independent of $B$ and $T$. However, because of the $1/T$ prefactor in the exponent, this will translate to a $T$-dependent multiplicative divergence in $\expv{P}_{B,T}$~\cite{Bruckmann:2013oba}.

Since the Polyakov loop vanishes in the vacuum, we cannot set the subtraction point for the renormalization of $f_Q-f$ to $T=0$. Instead, we may require the exponent to vanish at some reference temperature $T=T_\star$ and $B=0$. This implies that the renormalized Polyakov loop equals unity at the temperature $T=T_\star$ and vanishing electromagnetic fields.
This results in the multiplicative renormalization for the Polyakov loop on a lattice with spacing $a$~\cite{Bruckmann:2013oba},
\be
\expv{P^r}_{B,T,a} = \expv{P}_{B,T,a} \cdot Z_L(a,T), \qquad Z_L(a,T) =  \expv{P}_{B=0,T=T_\star,a} ^{-T_\star/T}\,.
\label{eq:lat_ploop_renorm_Tstar}
\ee
Some of the results we will present below in Sec.~\ref{sec:pd_ploop} for this observable were obtained with $T_\star=162\MeV$. Different values for $T_\star$ correspond to different renormalization schemes to which one can convert by a finite renormalization.

We note that alternative methods to renormalize the Polyakov loop also exist, for example using the zero-temperature static potential~\cite{Aoki:2006br} or the gradient flow~\cite{Petreczky:2015yta}.

\subsubsection[Magnetic catalysis, $\beta$-function and paramagnetism for free quarks]{Magnetic catalysis, \boldmath $\beta$-function and paramagnetism for free quarks}
\label{sec:lat_magncat_paramag}

Finally, we return once more to the magnetic field-dependent logarithmic divergence of the matter free energy density~\eqref{eq:lat_fr_muQED_def} and its dependence on the renormalization scale $\mu_{\rm QED}$. Above, we argued that $\mu_{\rm QED}$ may be fixed by requiring the $\O(B^2)$ contributions in $f$ to vanish at zero-temperature and that in the absence of color interactions, this choice amounts to $\mu_{\rm QED}=m_f$ (see footnote~\ref{fn:lat_muqed}). In other words, if we allow for an arbitrary scale $\mu_{\rm QED}$, then the renormalized free energy for a free quark at zero temperature takes the form,
\be
f^r(T=0,B)=-\frac{\beta_1^{\rm free}}{2}(q_fB)^2 \log(m_f^2/\mu_{\rm QED}^2) + \mathcal{O}(B^4)\,.
\label{eq:lat_deltafr_free}
\ee

Next we will show, following~\cite{Bali:2013txa}, that this dependence leads to two notable physical implications: first, the enhancement of the quark condensate for weak fields (relevant for the so-called magnetic catalysis phenomenon) and second, the $\O(B^4)$ paramagnetism of the medium (relevant for the equation of state). While the discussion is based on free quarks, it generalizes to the case of full QCD, as we will demonstrate below in Sec.~\ref{sec:pbp_QCDT0}.

\begin{figure}
 \centering
 \includegraphics[width=8.4cm]{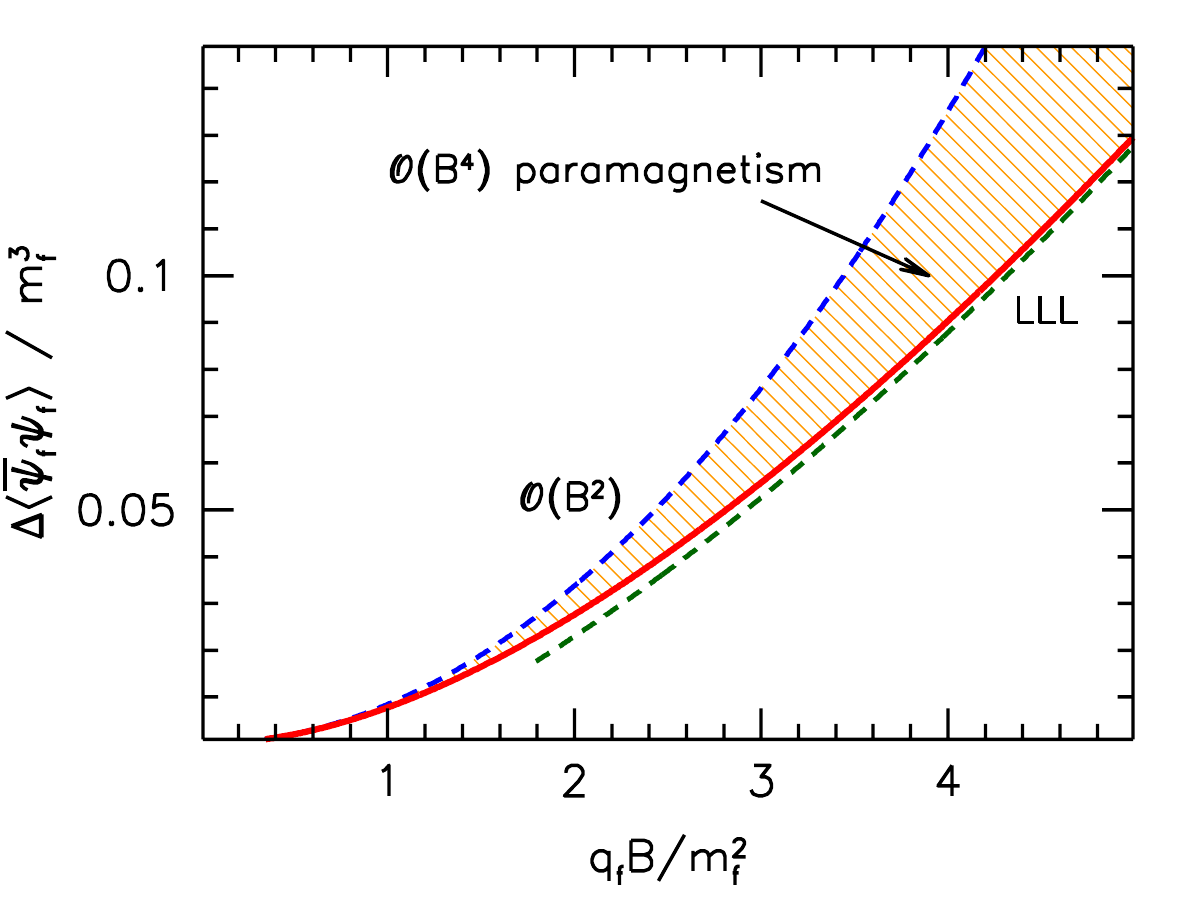} \caption{\label{fig:lat_freecondensate}
 Quark condensate in a homogeneous magnetic field in the free case (solid red line). The leading positive curvature is ensured by the positivity of the first QED $\beta$-function coefficient (dashed blue line), while the strong-field behavior is dictated by the linear degeneracy of the lowest Landau-level (dashed green line). Moreover, the fact that the condensate lies below its quadratic weak-field expansion implies the $\mathcal{O}(B^4)$ paramagnetic response of the medium (yellow dashed area).
 }
\end{figure}

The quark condensate~\eqref{eq:lat_pbpdef_0}, after additive renormalization, is obtained as the derivative of~\eqref{eq:lat_deltafr_free} with respect to $m_f$\footnote{Following footnote~\ref{fn:lat_mfder_muqed}, the renormalization prescription implies that $\mu_{\rm QED}$ is set to the physical quark mass in~\eqref{eq:lat_deltafr_free} {\it after} the derivative with respect to $m_f$ is carried out~\cite{Endrodi:2013cs} (see also footnote~1 in~\cite{Shushpanov:1997sf}).}. Its leading behavior for weak magnetic fields takes the form,
\be
\Delta \expv{\bar\psi_f\psi_f} \equiv \expv{\bar\psi_f\psi_f}_{B}-\expv{\bar\psi_f\psi_f}_{B=0}=
\beta_1^{\rm free} \frac{(q_fB)^2}{m_f}+ \mathcal{O}(B^4)\,,
\label{eq:lat_freecondensate_formula}
\ee
Therefore, in the $\mathcal{O}(B^2)$ term, the first QED $\beta$-function coefficient $\beta_1^{\rm free}>0$ appears. Its positivity ensures the quadratic enhancement of the condensate by the magnetic field, as anticipated above.
In fact,~\eqref{eq:lat_freecondensate_formula} can be calculated for arbitrary magnetic fields analytically (see e.g.~\cite{Bruckmann:2017pft}).
For strong magnetic fields, the leading behavior is $\expv{\bar\psi_f\psi_f}\propto m_fq_fB\log q_fB/m_f^2$. 
This dependence dominantly stems from the contribution of the lowest eigenvalues of $\Dsf$, the so-called lowest Landau-levels, see the discussion in Chap.~\ref{chap:ev}.
The condensate, together with its limiting behaviors, is shown in Fig.~\ref{fig:lat_freecondensate}.

Besides illustrating the weak-field magnetic catalysis of the condensate, this figure also has a non-trivial physical implication regarding the magnetic response of the matter free energy density at zero temperature~\cite{Bali:2013txa}.
To discuss this, we note that the free energy density can be reconstructed by integrating the quark condensate in the quark mass from the physical point $m_f^{\rm ph}$ up to infinity and using that $\Delta f^r$ vanishes in the latter limit.
For our choice of renormalization scale, the $\mathcal{O}(B^2)$ term of the renormalized free energy density vanishes at $T=0$. The higher-order contributions in $B$ can therefore be reconstructed via
\be
\Delta f^r = \left[1-\frac{(eB)^2}{2}\left.\frac{\partial^2}{\partial (eB)^2}\right|_{B=0}\right] \Delta f =  \int_{m_f^{\rm ph}}^\infty \dd m_f \,\left[1-\frac{(eB)^2}{2}\left.\frac{\partial^2}{\partial (eB)^2}\right|_{B=0}\right]\Delta \expv{\bar\psi_f\psi_f}\,.
\label{eq:lat_magncat_paramag}
\ee

Fig.~\ref{fig:lat_freecondensate} shows that the condensate is below its leading, quadratic weak-field expansion (this remains to hold for any quark mass value). This implies that the right hand side of~\eqref{eq:lat_magncat_paramag} is negative, and so must be the left hand side, too. Thus, starting at $\mathcal{O}(B^4)$, magnetic fields reduce the $T=0$ free energy density. In other words, the vacuum is paramagnetic. Note that we will also use the word paramagnetism (and diamagnetism) later in this review to indicate the $\mathcal{O}(B^2)$ response of the medium to the external field (see Chap.~\ref{chap:eos}). In the vacuum the $\mathcal{O}(B^2)$ term is by construction absent, and the magnetic behavior is fixed by the first nonzero coefficient, the one at $\mathcal{O}(B^4)$.

\subsection{Lessons learned}
\label{sec:lat_lessons}

In this chapter we introduced the most important basic elements of lattice QCD and focused on the implementation of background electromagnetic fields in the lattice action. This action differs from the well-known continuum, infinite-volume form in two important aspects. First, the flux of the magnetic field (and of the imaginary electric field) is quantized due to the periodic boundary conditions. Foreseeing the problems related to the discrete nature of the background field amplitudes,
besides homogeneous and localized magnetic fields, we also introduced oscillatory field profiles that have zero flux and are therefore not affected by the quantization condition.

Second, the photon fields appear on the lattice in the form of parallel transporters, which need to be $\mathrm{U}(1)$ phases in order to ensure the reality of the path integral weights, i.e.\ to avoid the complex action problem. While this allows real magnetic fields to be included in the simulations, it forced us to use imaginary electric fields instead of real ones.

Another focus of this chapter was the renormalization of observables in the presence of background magnetic fields. On physical grounds one expects that the background field (being an infrared parameter that one can in principle tune in an experiment) does not introduce new ultraviolet divergences in the theory. We showed that this is indeed the case for the total free energy density $f^{\rm tot}$, but not for the matter term $f$. The additive renormalization of $f$ can be understood in terms of the multiplicative renormalization of the electric charge $e$. Several observables derived from $f$ inherit this kind of divergence, which we will need to take into account for the discussion of the phase diagram and the equation of state.
Finally, the structure of this logarithmic divergence was shown to lead to interesting physical consequences in the free case, which will also persist in full QCD.

	\clearpage
	\section{Dirac eigenvalues}
	\label{chap:ev}

The fermionic weight in the path integral~\eqref{eq:lat_partfunc_cont3} is given by the Dirac determinant. This determinant, together with all fermionic observables derived from it -- like the quark condensate -- can be written using the eigenvalues of the Dirac operator $\Dsf$. Several characteristic features of such observables can be understood already on the level of the Dirac eigenvalues. A prime example for this is the low-temperature magnetic catalysis of the quark condensate by homogeneous background fields, which is related to the proliferation of eigenvalues corresponding to the lowest Landau-level.
The lowest Landau-level has topological properties and plays a crucial role in anomalous transport phenomena like the chiral magnetic effect~\cite{Fukushima:2008xe}.
It has been demonstrated that the concept of Landau-levels -- while naturally defined for weakly interacting fermions -- can also be carried over to strongly interacting quarks. In this chapter we will explore the general behavior of Dirac eigenvalues in the presence of background electromagnetic fields in the continuum and on the lattice, as well as their implications for QCD observables.

\subsection{Continuum Landau-levels}
\label{sec:ev_contLL}

Throughout this chapter, we are working with the (massless) Euclidean Dirac operator $\Dsf$ for one of the quark flavors, say the down quark $f=d$. In the presence of a magnetic field $B$, the operator for the up quark -- and, therefore, the eigenvalues -- follow simply via $\slashed{D}_u(B)=\slashed{D}_d(-2B)$.
The eigensystem of the Dirac operator reads
\be
\Dsf \chi = i \lambda \chi\,,
\label{eq:ev_Diracev}
\ee
where $\chi$ is the eigenvector, corresponding to the eigenvalue $i\lambda$. The anti-Hermiticity of $\Dsf$ ensures that the eigenvalues are purely imaginary, i.e.\ $\lambda\in\mathds{R}$. Moreover, the chiral symmetry of the operator, $\{\gamma_5,\Dsf\}=0$, dictates that nonzero eigenvalues appear in complex conjugate pairs,
\be
\Dsf \left(\gamma_5\chi\right) = -\gamma_5 \Dsf \chi = -i\lambda\left(\gamma_5\chi\right)\,.
\label{eq:ev_chiralsymm}
\ee
These properties hold in the continuum theory as well as for lattice discretizations that respect the anti-Hermiticity property and the chiral symmetry, respectively. The symmetries of the lattice operators were discussed above in Sec.~\ref{sec:lat_signproblem}.

In the continuum theory, the impact of background magnetic fields on charged and otherwise non-interacting particles is well known~\cite{Shovkovy:2012zn}. For a homogeneous magnetic field (oriented along the $x_3$ direction), the free Dirac spectrum consists of so-called Landau levels, so that the squared eigenvalues read
\be
\lambda_{n,p_3,p_4}^2=2n|q_fB|+p_3^2+p_4^2, 
\qquad n\in\mathds{Z}_0^+,
\qquad p_3\in\mathds{R}, \qquad p_4=2\pi(k_4+1/2)T, \qquad k_4\in\mathds{Z} \,,
\label{eq:ev_contLL_evs}
\ee
where the temporal momentum corresponds to the fermionic Matsubara frequencies at nonzero temperature $T$.
All levels with $n>0$ have a two-fold degeneracy, corresponding to the two solutions for the spin quantum number,
\be
\frac{\sigma_3}{2} \,\chi = s_3\chi, \qquad s_3=\pm1/2\,,
\label{eq:ev_sigma3_s3}
\ee
where $s_3$ represents the spin of the particle along the magnetic field. The only exception is the lowest Landau-level $n=0$, which only accommodates one spin direction,\footnote{\label{fn:lat_zeroeigs}Here, $\sigma_3$ is identified with $\sigma_{12}=-i[\gamma_1,\gamma_2]/2$. The squared eigenvalues have a specific spin eigenvalue due to $[\Dsf^2,\sigma_{12}]=0$. Moreover, in the subspace corresponding to the lowest Landau-level, $[\Dsf,\sigma_{12}]=0$.} namely $s_3=\textmd{sgn}(q_fB)/2$. Moreover, there is an additional degeneracy proportional to the flux of the magnetic field. In a finite volume, this degeneracy is given by $\Phi_{12}/(2\pi)$, with the flux defined in~\eqref{eq:lat_flux1} above. 
Altogether, the lowest Landau-level has a degeneracy of $N_b^f$, while higher Landau-levels $2N_b^f$.
The spatial momentum along the magnetic field takes on the values $p_3=2\pi k_3/L$ with $k_3\in\mathds{Z}$.

Most physical observables are dominated by infrared physics i.e.\ the lowest eigenvalues. In this case, one may therefore expect that for strong magnetic fields, the lowest Landau-level is dominant and neglecting $n>0$ levels is reasonable. This is the basis of the so-called lowest Landau-level approximation, which is employed frequently in model calculations, see e.g.\ the review~\cite{Shovkovy:2012zn}.

The Dirac eigenvalues can also be found analytically for specific inhomogeneous magnetic field backgrounds, including the spatially localized profile~\eqref{eq:lat_loc_Bfield} with width $\epsilon$. The solutions are plane waves in the direction orthogonal to the modulation, described by the momentum $p_2$, and localized functions in the $x_1$ direction.
The result for the squared eigenvalues reads~\cite{Cangemi:1995ee},
\be
\lambda_{n,s_3,p_2,p_3,p_4}^2=\frac{\alpha_{p_2}^2+\alpha_{-p_2}^2}{2} + \frac{(n-\gamma)^2}{\epsilon^2}-\frac{\epsilon^2}{16}\frac{(\alpha_{p_2}^2-\alpha_{-p_2}^2)^2}{(n-\gamma)^2}
+p_3^2+p_4^2,
\qquad n\in\mathds{Z}_0^+,
\qquad s_3=\pm\frac{1}{2}\,,
\label{eq:ev_contloc_evs}
\ee
where
\be
\alpha_{p_2}=|p_2-q_fB\,\epsilon|, \qquad
\gamma=2q_fB\,\epsilon s_3, \qquad
\qquad p_2,p_3\in\mathds{R}, \qquad p_4=2\pi(k_4+1/2)T, \qquad k_4\in\mathds{Z}\,,
\ee
and the integer label is constrained as
\be
0\le n \le \gamma-\frac{\epsilon}{2}\sqrt{|\alpha_{p_2}^2-\alpha_{-p_2}^2|}\,.
\ee
In a finite volume, the spatial momenta $p_2$ and $p_3$ become discrete, e.g.\ $p_2=2\pi k_2/L$ with $k_2\in\mathds{Z}$.
While the eigenvalues above correspond to the bound states, this potential also allows for a continuum of scattering states~\cite{Cangemi:1995ee}. 

We mention that there are several further special profiles for which the Dirac equation may be solved analytically~\cite{Dunne:2004nc,Raya:2010id}, which we do not discuss here.

\subsection{Hofstadter's butterfly}

The Dirac eigenvalues in the presence of a homogeneous background magnetic field on the lattice have a long history. Well before the relevance of this setup was recognized in the context of lattice QCD, the analogous eigensystem was discussed in a solid state physics model: the so-called Hofstadter model~\cite{Hofstadter:1976zz}.
The generalization to lattice QCD reveals several similarities as well as novel aspects that are relevant for the quantum field theory context. Some of these have been discussed in the review~\cite{Endrodi:2014vza}, which we also follow here. 

We first consider the free Dirac operator, i.e.\ $\A_\nu=0$. Moreover we discuss the two-dimensional case -- this is the interesting part of the spectrum, obtained from~\eqref{eq:ev_contLL_evs} by setting $p_3=p_4=0$. The impact of gluonic interactions in four dimensions will be considered below in Sec.~\ref{sec:lat_landaulevelQCD}.

\subsubsection{Staggered fermions}

In this section the staggered discretization~\eqref{eq:lat_staggered_Dslash} of the Dirac operator is used. We will comment on the spectrum for Wilson and overlap quarks later. The staggered formulation respects the anti-Hermiticity of $\Dsf$ and also maintains a remnant chiral symmetry (see footnote~\ref{fn:lat_stagg}), therefore the eigenvalues continue to be purely imaginary and occur in complex conjugate pairs.

\begin{figure}
 \centering
 \mbox{
 \includegraphics[width=8.4cm]{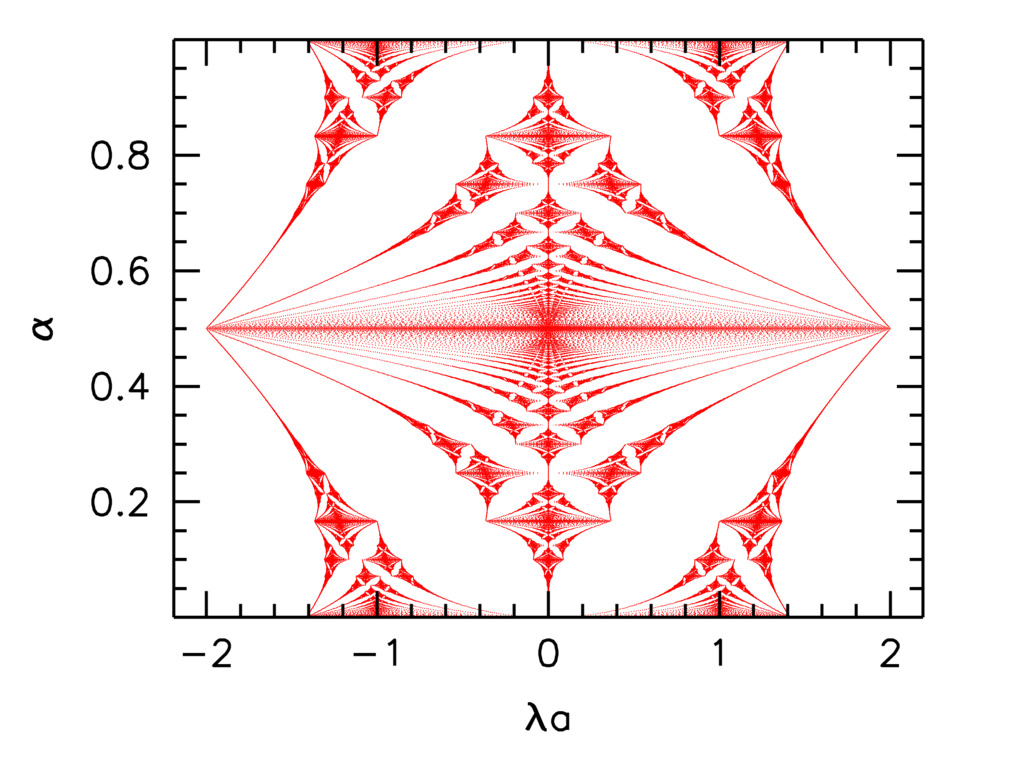}
 \includegraphics[width=8.4cm]{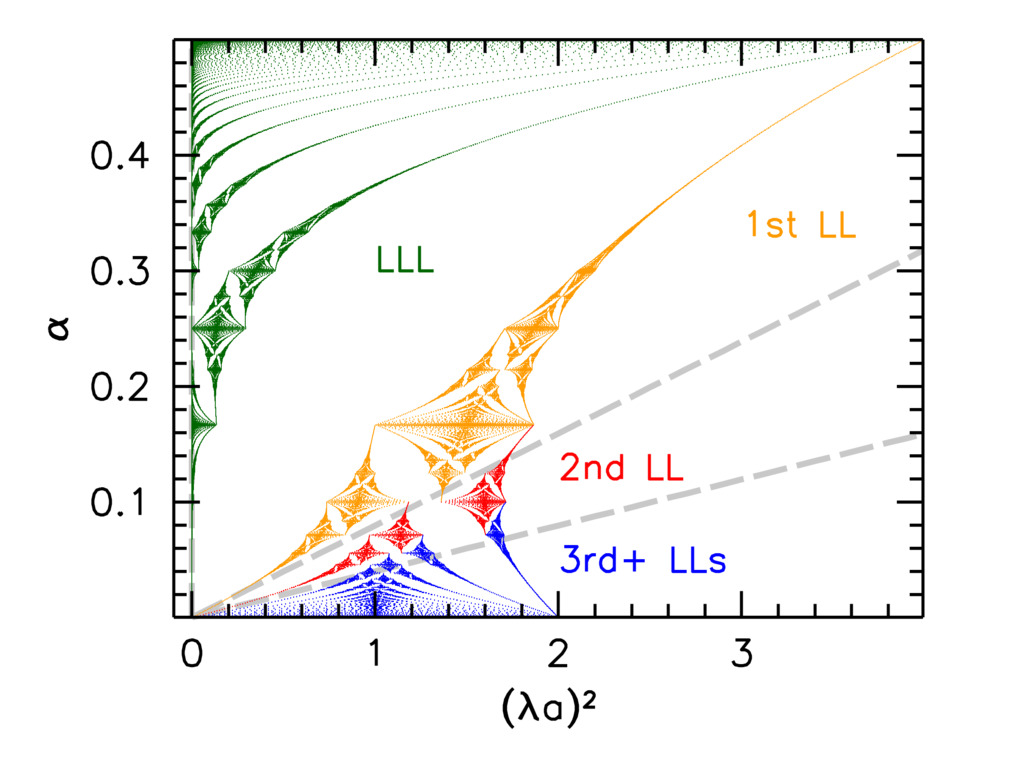}
 }
 \caption{\label{fig:ev_butterflies}
 Left panel: Hofstadter's butterfly on a $40^2$ lattice.
 The eigenvalues in lattice units are plotted against the flux of the 
 magnetic field. Taken from~\protect\cite{Endrodi:2014vza}. Right panel: classification of the branches of Hofstadter's butterfly into Landau-levels (colors) and a comparison to the continuum eigenvalues (gray dashed lines). Figure adapted from~\cite{Bruckmann:2017pft}.}
\end{figure}

The traditional representation of the Dirac spectrum in two dimensions is provided in terms of Hofstadter's butterfly, shown in Fig.~\ref{fig:ev_butterflies} from~\cite{Endrodi:2014vza}. Here, the Dirac eigenvalues $\lambda$ are shown in lattice units (on the horizontal axis) for different values of the magnetic flux (on the vertical axis).
Collected in the $\Phi_{12}-\lambda$ plane in this manner, the spectra reveal a recursive pattern with a non-trivial fractal structure, as demonstrated originally for the non-relativistic, spinless case\footnote{Note that moving from the energy eigenvalues in the spinless setup to the Dirac spectrum merely amounts to a mirror transformation of the spectrum, as demonstrated in~\cite{Endrodi:2014vza}, see also~\cite{Kimura:2012de}.} in~\cite{Hofstadter:1976zz}.
Notice that this graph is not continuous: as shown above in~\eqref{eq:lat_Nbquant}, on the lattice the flux values are discrete, and the eigenvalues themselves are also discrete. Moreover, both the eigenvalues and the flux values are bounded from above by the ultraviolet regulator. In Fig.~\ref{fig:ev_butterflies}, the flux variable is normalized as
\be
\alpha\equiv \frac{a^2q_fB}{2\pi} = \frac{N_b^f}{N_s^2}\,, \qquad 0\le\alpha\le1\,.
\label{eq:ev_alphadef}
\ee
This variable can be written as the ratio of the two characteristic scales appearing in this system -- the variation of the vector potential in the Landau-problem over one lattice cell, $aq_fB$, and the largest Bloch momentum $2\pi/a$. 
In the thermodynamic limit, $N_s\to\infty$, an arbitrary real value of $\alpha$ can be approached. In this limit, $\alpha\in\mathds{Q}$ corresponds to the case, where the two characteristic scales are commensurable, while $\alpha\not\in\mathds{Q}$ implies their incommensurability. It turns out that in the former case the spectrum consists of bands well known from conductors, while in the latter one has a zero-measure nowhere dense spectrum that is isomorphic 
to the Cantor set~\cite{Hofstadter:1976zz,Last:1994}. One can therefore conclude that the Landau- and Bloch-problems are incompatible with each other in this latter case, resulting in a frustration for the charged particle and, eventually, in the fractal structure of the butterfly.

In the solid state physics setup, the structure of the butterfly for $a^2q_fB=\mathcal{O}(1)$ is a physical notion, actively sought for in experiments~\cite{Dean_2013,Ponomarenko_2013,Bodesheim:2022tse}.
In contrast, in lattice QCD the lattice spacing is a mere regulator that vanishes in the continuum limit, $a\to0$ and with it also $\alpha\to0$. Still, the $\alpha\ll 1$, $a\lambda\ll1$ corner of the butterfly encodes physical information, as demonstrated in the right panel of Fig.~\ref{fig:ev_butterflies}.
Here, the squared lattice eigenvalues are shown to follow the continuum Landau-levels (straight lines in the plot) as long as neither $a\lambda$ nor $\alpha$ is too large. Moreover, the Dirac eigenvalues in the plot are colored in terms of their index according to the continuum degeneracies described below~\eqref{eq:ev_sigma3_s3} -- the first $N_b^f\times2$ are associated\footnote{The eigenvalues of the two-dimensional staggered Dirac operator exhibit an additional two-fold degeneracy due to fermion doubling, similarly to the four-fold doubling in four dimensions, as mentioned in Sec.~\ref{sec:lat_latDiracops}.} to the lowest Landau-level, the next $2N_b^f\times2$ to the first Landau-level and so on~\cite{Bruckmann:2017pft}.

Further away from the origin, the lattice eigenvalues tend to deviate from the continuum curves, 
mix with each other and eventually form the peculiar recursive pattern that has fascinated physicists and mathematicians since the discovery of Hofstadter's butterfly. We mention that the Landau-level structure also emerges in the spectrum near the corner $\alpha=0.5$, $\lambda a=\pm 2$ -- here the Landau levels for spinless charged particles can be identified~\cite{Endrodi:2014vza}, see also~\cite{Hatsuda_2016}. The self-similarity of the subsets of the butterfly has been investigated in a renormalization group approach~\cite{Satija_2020}. We also mention that the spectrum can be generalized to hexagonal~\cite{Bermudez:2009fw}, triangular and Kagome lattices as well~\cite{Du_2018}.

The lowest Landau-level is separated from the rest of the spectrum due to its topological nature. Indeed, in the continuum the 
vanishing of the eigenvalues belonging to the lowest Landau-level is guaranteed by the two-dimensional index theorem. More specifically, the index theorem in two dimensions states that the flux~\eqref{eq:lat_flux1} is a topological invariant,
\be
N_b^f=\frac{1}{2\pi}\int \dd x_1 \dd x_2 \,q_fB(x_1,x_2) = N_0^{\sigma_3 +}-N_0^{\sigma_3 -}\,.
\label{eq:ev_indextheorem}
\ee
where $N_0^{\sigma_3\pm}$ are the number of eigenvectors $\chi$ with zero eigenvalue that have eigenvalue $\pm1$ of the spin operator $\sigma_3$ (see footnote~\ref{fn:lat_zeroeigs}). Moreover, the so-called vanishing theorem~\cite{Kiskis:1977vh,Nielsen:1977aw,Ansourian:1977qe} 
ensures that either $N_0^{\sigma_3+}$ or $N_0^{\sigma_3-}$ is zero. Thus, for $q_fB>0$, we only have spin-up states in the lowest Landau-level and, according
to~(\ref{eq:ev_indextheorem}), $N_b=N_0^{\sigma_3+}$. 
This result holds for any magnetic field profile with nonzero flux and also in the presence of QCD interactions\footnote{In this case an additional degeneracy due to the number $N_c=3$ of colors also emerges.}. Moreover it remains valid on the lattice, too, as long as the lattice vector potential is sufficiently smooth. In the overlap
formulation~\cite{Neuberger:1997fp,Neuberger:1998wv}, the topological modes are exact zero modes, just like in the continuum.
In the staggered discretization, the zero eigenvalues become would-be zero modes that are still well separated from the rest of the spectrum.
In this case, the spin matrix $\sigma_3$ is approximately diagonal in the lowest Landau-level subspace and zero on its complement~\cite{Bruckmann:2017pft}.

\begin{figure}
 \centering
 \includegraphics[width=\textwidth]{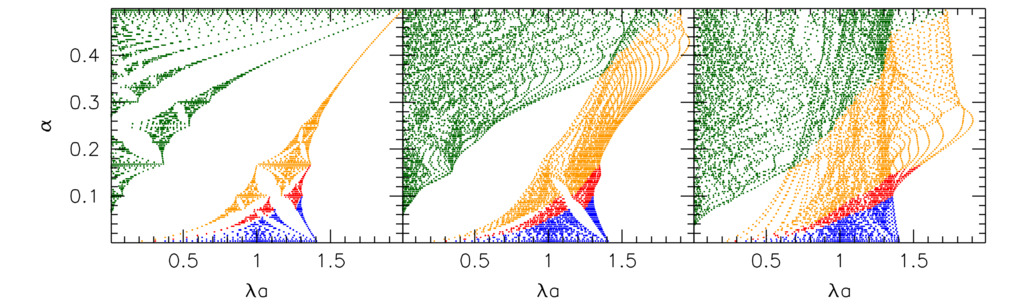} \caption{\label{fig:ev_local_butterfly}
 Deformation of Hofstadter's butterfly due to localized magnetic fields. The three panels correspond to the
homogeneous case ($\epsilon\to\infty$, left), $\epsilon=10a$ (middle) and $\epsilon=4a$ (right). The equivalents of the lowest Landau-level, the first Landau-level and the rest of the spectrum are indicated by blue, red and yellow dots, respectively. Figure adapted from~\cite{Brandt:2023dir}.
 }
\end{figure}

The butterfly has been generalized for the case of the localized magnetic field profile~\eqref{eq:lat_loc_Bfield} in~\cite{Brandt:2023dir}.
The Dirac spectrum on a $16^2$ lattice is shown in Fig.~\ref{fig:ev_local_butterfly} for three values of the profile width $\epsilon$.
The lattice eigenvalues again follow the continuum results~\eqref{eq:ev_contloc_evs} if the local magnetic field is small in lattice units. The lowest edge of the spectrum -- the equivalents of a lowest Landau-level of the homogeneous setup -- are protected from mixing with other states due to the index theorem, as we argued above.

\subsubsection{Wilson and overlap fermions}
\label{sec:ev_Wilsonoverlap}

To conclude the discussion about the free Landau-levels on the lattice, we show results for two more fermion discretizations besides the staggered formulation. Here we only consider the homogeneous magnetic field case.

\begin{figure}
 \centering
 \mbox{
 \includegraphics[width=8.4cm]{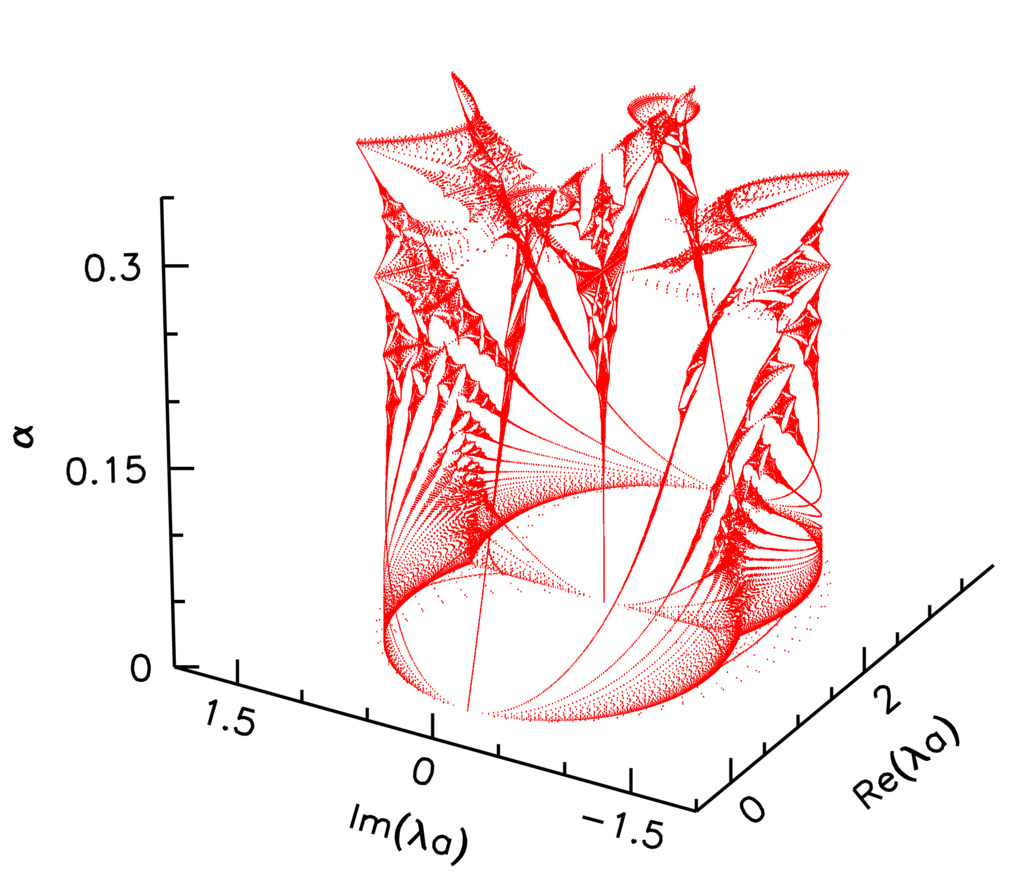}
 \includegraphics[width=8.4cm]{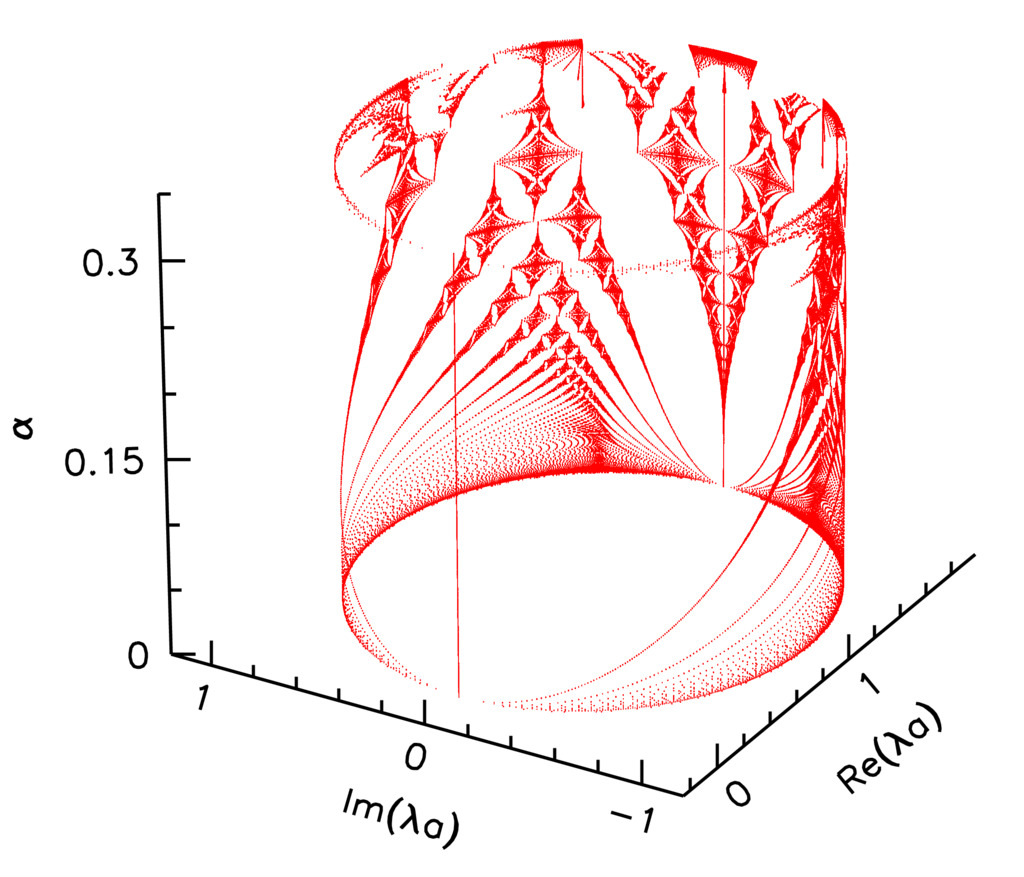}
 }
 \caption{\label{fig:ev_wilsonoverlap}
 Hofstadter's butterfly with Wilson fermions (left panel) and overlap fermions (right panel) on a $32^2$ lattice.}
\end{figure}

The left panel of Fig.~\ref{fig:ev_wilsonoverlap} shows the spectrum with Wilson fermions on a $32^2$ lattice. Unlike the staggered case, the Wilson operator~\eqref{eq:lat_Wilson_Dslash} is not anti-Hermitian, therefore the eigenvalues become complex. The figure contains a set of complex spectra for different values of the flux, labeled by the normalized flux variable $\alpha$ from~\eqref{eq:ev_alphadef}. Similarly to the staggered case in Fig.~\ref{fig:ev_butterflies}, the Landau-levels can be identified for small $\alpha$ and small $\lambda$. The fractal recursive pattern emerges further away from the origin\footnote{Some of these aspects have been discussed in~\cite{Schulze:2022}.}. Notice that the lowest Landau-level has a nonzero real part proportional to $\alpha$. In the massive Dirac operator $\Dsf+m_f$, such a shift in the real part of the spectrum is equivalent to a shift of the quark mass. Therefore, this signals the impact of the magnetic field on additive mass renormalization, even in the full four-dimensional case~\cite{Brandt:2015hnz}. One may take care of this by a careful, $B$-dependent tuning of the quark mass, for example by means of requiring that the renormalized mass extracted from the axial Ward identity remains $B$-independent~\cite{Brandt:2015hnz,Bali:2017ian}. If one only intends to remove the leading, $\mathcal{O}(a)$ additive shift of the quark mass, an alternative is to extend the standard $\mathcal{O}(a)$ improvement of the Wilson operator to include the electromagnetic fields, too~\cite{Bignell:2019vpy}. Such procedures will be important for the studies of hadron masses with Wilson quarks, discussed in Sec.~\ref{sec:had_latdethadprop}.

The right panel of Fig.~\ref{fig:ev_wilsonoverlap} shows the two-dimensional spectrum for the overlap Dirac operator~\eqref{eq:lat_overlap_Dslash}. Since the overlap spectrum always lies on the complex unit circle, here the magnetic field-dependent spectra build up the surface of a cylinder in the figure.
The fractal structure emerges on the far side of the cylinder, where lattice artefacts dominate. In turn, one can again identify the lowest Landau-level near the origin. Due to the exact chiral symmetry of this formulation, these modes are exact zero modes. However, we can see that around $\alpha=1/3$, the separation of the lowest Landau-level from the rest is spoiled.
This is due to the fact that for such strong magnetic fields, the physical modes of the kernel operator become larger than the kernel mass $m_0$, thus even the eigenvalues corresponding to the lowest Landau-level are projected to the far end of the spectrum (cf.\ the remark below~\eqref{eq:lat_overlap_Dslash}). Here we are working with the standard value for the kernel mass, $am_0=1$.

\begin{figure}
 \centering
 \mbox{
 \includegraphics[width=8.4cm]{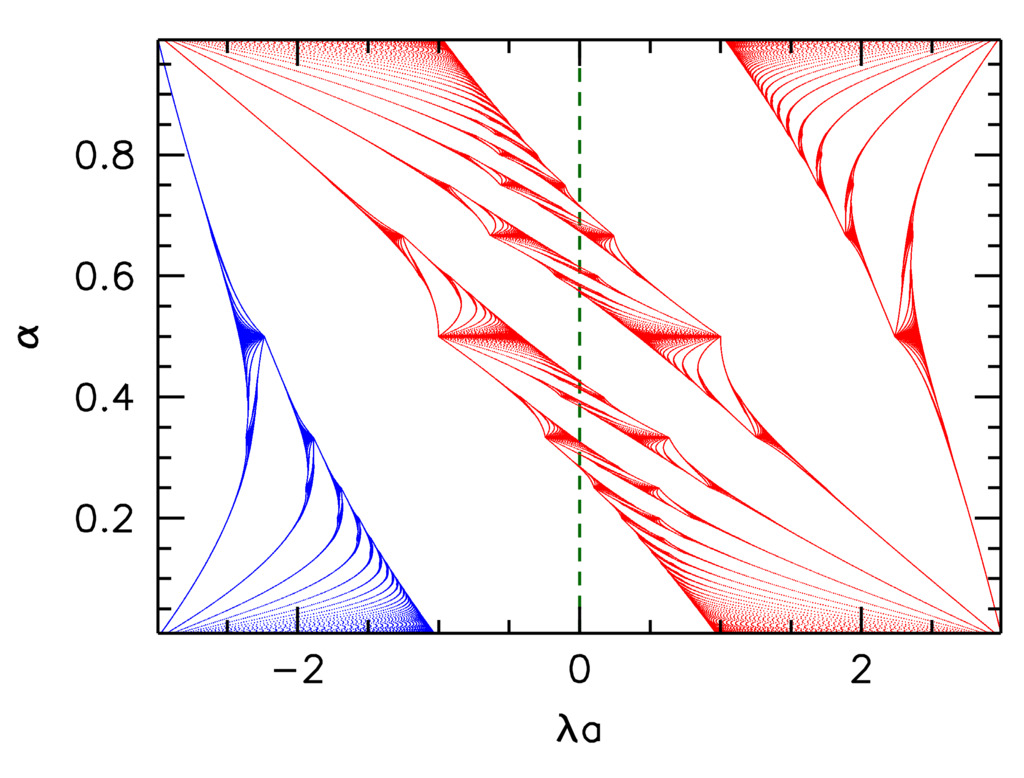}
 }
 \caption{\label{fig:ev_kernel}
 The spectrum of $\sigma_3 A$, where $A=\Dsf^W - m_0 \mathds{1}$ is the kernel of the overlap operator. A Wilson kernel with $am_0=1$ is used. The distinction between positive and negative modes, necessary for the unambiguity of the sign function, is spoiled at $\alpha\approx0.284$. This is the point, where the index theorem breaks down, see the discussion in the text.}
\end{figure}

In fact, already at a slightly lower magnetic field, $\alpha\approx0.284$, the index theorem is not fulfilled by the overlap operator anymore, because the electromagnetic gauge field is not sufficiently smooth anymore for the overlap construction to work, cf.\ the discussion below~\eqref{eq:lat_overlap_Dslash}.
The index $I$ of the two-dimensional overlap operator is written as~\cite{Gattringer:2010zz},
\be
I=\frac{a}{2}\,\Tr \left\{\sigma_{3}\Dsf^{\rm ov}\right\}
= \frac{1}{2}\,\Tr \left\{\sigma_{3}\left[\frac{\mathds{1}}{a}+\sigma_{3}\,\sgn\left(\sigma_{3}A\right)\right]\right\}
=\frac{1}{2}\,\Tr\, \sgn\left(\sigma_{3}A\right) = \frac{N^{\sigma_3A}_{+}-N^{\sigma_3A}_{-}}{2}\,,
\label{eq:ev_overlap_index}
\ee
equal to half\footnote{Note that the index can also be written as the difference of the number of zero eigenvalues of $\Dsf^{\rm ov}$ with positive and negative $\sigma_3$-eigenvalues, as in~\eqref{eq:ev_indextheorem}.} of the difference between the numbers of positive and negative eigenvalues of $\sigma_3A$. The two-dimensional index theorem states that this is equal to the topological invariant, i.e.\ the magnetic flux $N_b^f$.
To confirm this, in Fig.~\ref{fig:ev_kernel} we plot the spectrum of $\sigma_3A=\sigma_3(\slashed{D}_W^f-m_0)$. The eigenvalues are colored so that for each magnetic field, there are $N_s^2+N_b^f$ red dots and $N_s^2-N_b^f$ blue dots. One sees from the plot that for low magnetic fields all red eigenvalues are positive and all blue eigenvalues are negative, therefore~\eqref{eq:ev_overlap_index} gives $I=N_b^f$ and the index theorem is valid. However, as soon as the red eigenvalues hit zero, the index theorem is violated. We conclude that for the overlap construction to work properly, the magnetic flux quantum needs to be sufficiently small and extra care has to be taken in order to ensure that the index theorem holds.

\subsubsection{Quantum phase diagrams}

As argued in~\eqref{eq:ev_indextheorem}, the degeneracy of the lowest Landau-level is fixed by topology in the continuum, and this remains to hold on the lattice, even if the eigenvalues do not vanish exactly, but are rather would-be zero modes. This implies that it is in fact the {\it gap} between the lowest and the first Landau-levels (the green and yellow dots in the left panel of Fig.~\ref{fig:ev_local_butterfly} for the staggered Dirac operator and the steepest branches starting from the origin in Fig.~\ref{fig:ev_wilsonoverlap} for Wilson and overlap quarks), which has a distinct topological nature. Indeed, as the magnetic flux quantum is increased by one unit, always one Dirac eigenvalue jumps from the right hand side of the gap to the left hand side. This largest gap in Hofstadter's butterfly has therefore a unit topological number, or Chern-Simons number $C$.

Such Chern-Simons numbers can also be associated to the other gaps in the spectrum. For the gap between the staggered lattice equivalents of the first and second Landau-levels (yellow and red dots in the left panel of Fig.~\ref{fig:ev_local_butterfly}), this number is for example $C=3$: the degeneracy of the first Landau-level increases by two and that of the lowest Landau-level by one as the flux grows by one unit. For negative eigenvalues (not shown in Fig.~\ref{fig:ev_local_butterfly}), the eigenvalues jump from the left side of the largest gap to its right side instead, i.e.\ the largest gap has $C=-1$. For an arbitrary point in the $N_b^f-\lambda$ plane with $\lambda>0$, it can be shown that the Chern-Simons number is the integer of smallest magnitude that satisfies the implicit Diophantine equation~\cite{10.1063/1.530758,Osadchy_2001,NAUMIS20161772},
\be
N_b^f \cdot C(N_b^f,\lambda) = \sum_{\lambda_j\ge 0} \Theta(\lambda-\lambda_j) \;\;\textmd{mod}\;\; N_\lambda\,,
\label{eq:ev_ChernSimons}
\ee
where $N_\lambda$ is the number of positive eigenvalues.
For $\lambda<0$, the Chern-Simons numbers follow from charge conjugation symmetry, $C(N_b^f,-\lambda)=-C(N_b^f,\lambda)$.
Remarkably, in the solid state physics context, these Chern-Simons numbers can be shown to characterize the integer Hall conductance of the system, providing yet another interpretation of the butterfly as a quantum phase diagram~\cite{Osadchy_2001}.

\begin{figure}
 \centering
 \mbox{
 \raisebox{.5cm}{\includegraphics[width=9cm]{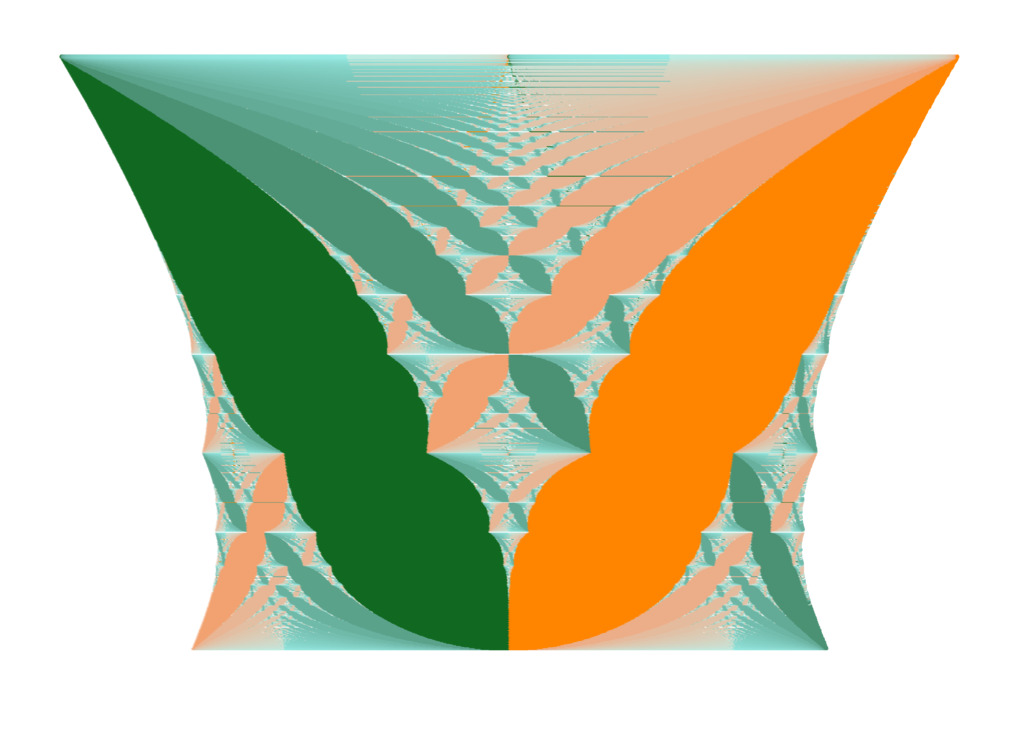}}\qquad
 \includegraphics[width=7cm]{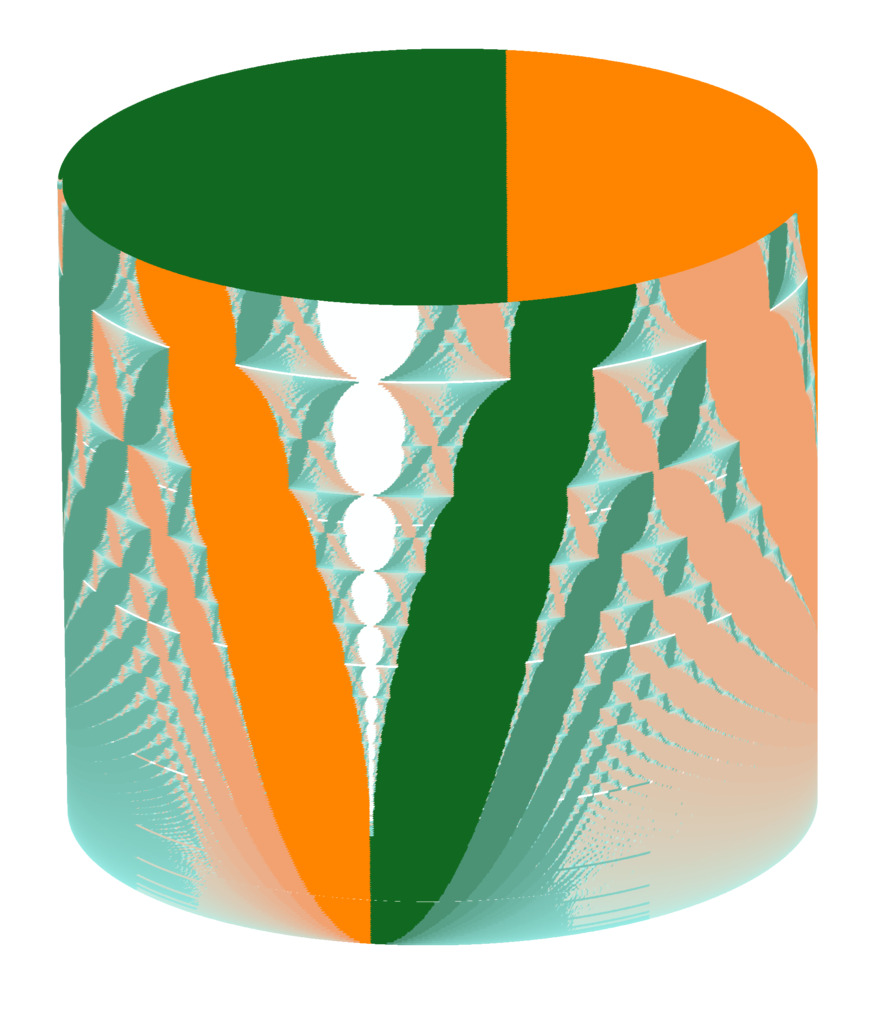}
 }
 \caption{\label{fig:ev_quantumphd}Quantum phase diagram of the two-dimensional staggered (left panel) and the overlap (right panel) Dirac operators on the lattice. The coordinate axes are oriented as in Fig.~\ref{fig:ev_butterflies} and in the right panel of Fig.~\ref{fig:ev_wilsonoverlap}, respectively, except that the $\textmd{Re}(\lambda a)$ axis is flipped in the overlap case. The colors encode the Chern-Simons number and indicate positive (orange) and negative (green) values with small (bright) and large (faint) magnitudes. For further details, see the text.}
\end{figure}

In the left panel of Fig.~\ref{fig:ev_quantumphd}, we plot the so constructed quantum phase diagram of staggered fermions (where $N_\lambda=N_s^2/2$). Here, orange colors denote positive and green colors negative values of $C$. Moreover, bright colors indicate Chern-Simons numbers of small magnitude, while faint colors correspond to large $C$ values. The special cases discussed above may be recognized here: the largest gap between the lowest and the first Landau-levels has $C=1$ and is marked by the brightest orange color, its mirror image at negative eigenvalues has $C=-1$ and is marked by bright green, while the second-largest gap between the first and second Landau-levels has $C=3$ and is marked by the second brightest orange. This construction analogously carries over to the overlap formulation. To apply~\eqref{eq:ev_ChernSimons} in this case, we interpolate the surface of the cylinder (see the right panel of Fig.~\ref{fig:ev_wilsonoverlap}), i.e.\ we replace $\lambda$ by $\arg \lambda$\footnote{The phase of the eigenvalue near $\arg\lambda=\pm\pi$ is ambiguous. To find $C(N_b^f,\lambda>0)$, we require that the number $N_\lambda$ of eigenvalues with $0\le\arg\lambda\le\pi$ equals $N_s^2$.}. The quantum phase diagram for overlap fermions is shown in the right panel of Fig.~\ref{fig:ev_quantumphd}. Near $\lambda=0$, this phase diagram is dominated by the $C=\pm1$ gaps. Therefore, the viewpoint of the figure is chosen in this case so that the richer part of the spectrum, around $\arg\lambda=\pi$, is visible.

Finally, we note that the overlap quantum phase diagram also reveals some information about the behavior of the overlap construction as the projection point is changed. Recall from the definition~\eqref{eq:lat_overlap_Dslash} that the overlap operator involves the projection of the Wilson spectrum on a unit circle centered at $(m_0,0)$. Moving this projection point along the real line to $m_0>2/a$, for example, implies that the branches at the center of the Wilson spectrum (see the left panel of Fig.~\ref{fig:ev_wilsonoverlap}) also end up as exact zero modes, thereby modifying the index theorem satisfied by $\Dsf^{\rm ov}$, and rendering it topologically improper~\cite{Chiu:1998ce,Adams:2000rn}, see also~\cite{Gattringer:1997ci}. The regions in the quantum phase diagram in the right panel of Fig.~\ref{fig:ev_quantumphd} with $|C|\neq1$ correspond to similar improper operators, defined with a projection point that is shifted away from the real line in a magnetic field-dependent way.

\subsection{Landau-levels in QCD}
\label{sec:lat_landaulevelQCD}

Above, we discussed the Dirac eigenvalues in the presence of a homogeneous magnetic field for otherwise non-interacting fermions, in two dimensions. In order to make contact with full QCD, one needs to incorporate gluonic interactions and also extend the discussion to four-dimensional space-time. 

\begin{figure}
 \centering
 \includegraphics[width=8.4cm]{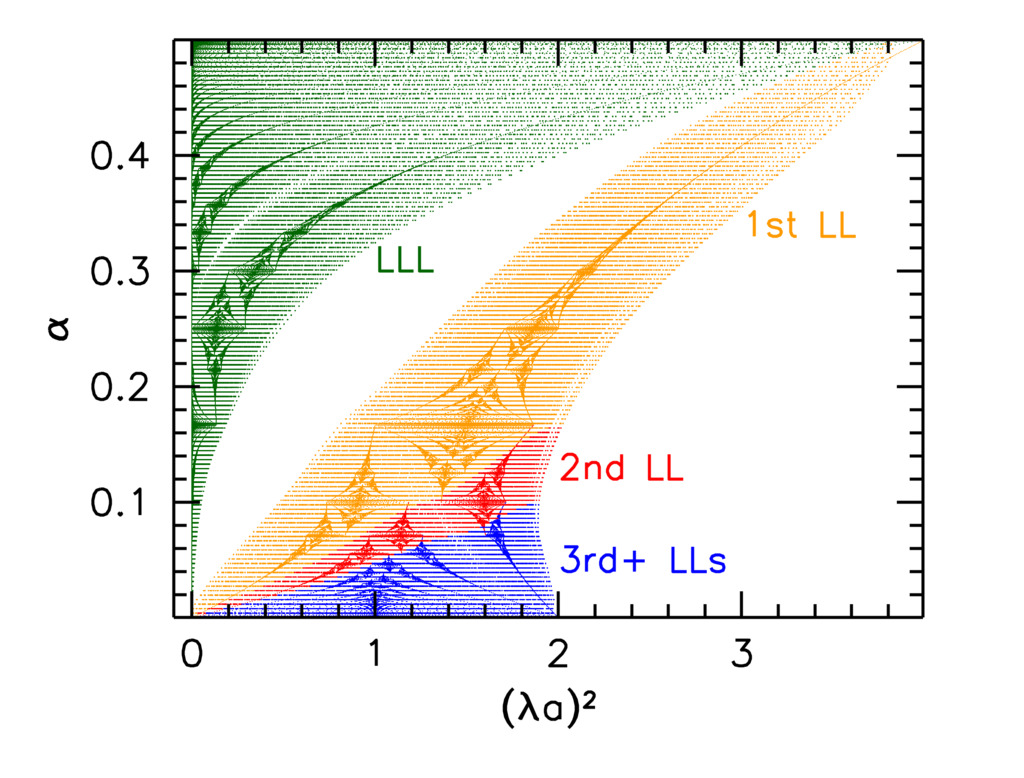}
 \caption{\label{fig:ev_LLLQCD}
 Dirac eigenvalues for staggered quarks, calculated on a two-dimensional slice of a typical four-dimensional gauge configuration~\cite{Bruckmann:2017pft}. The free-case eigenvalues, also shown in the right panel of Fig.~\ref{fig:ev_butterflies} are also included for comparison. Figure adapted from~\cite{Bruckmann:2017pft}.
 }
\end{figure}

Let us first attend to the effect of gluons on the two-dimensional spectrum. While the individual eigenvalues are clearly strongly affected, the index theorem~\eqref{eq:ev_indextheorem} still holds, fixing the number of zero eigenvalues (in the continuum). Thus, while higher Landau-levels will in general mix among each other, the lowest Landau-level is still separated from the rest of the spectrum and has nonzero matrix element for the $\sigma_{3}$ operator. On the lattice, the corresponding eigenmodes will be near-zero modes for the staggered and Wilson discretizations, and exact zero modes for the overlap formulation.
This is demonstrated for staggered quarks in Fig.~\ref{fig:ev_LLLQCD}, where the spectrum is shown for one $x_1-x_2$ slice of a high-temperature $16^3\times4$ lattice~\cite{Bruckmann:2017pft}. The eigenvalues are again colored based on their index, i.e.\ the lowest $N_cN_b^f\times2$ are associated with the would-be lowest Landau level and so on. While QCD interactions clearly smear out the fractal structure, the would-be lowest Landau-level remains well separated.

Next, we proceed to the definition of Landau-levels in the full, four-dimensional setting, following~\cite{Bruckmann:2017pft}. Already in the free case, the actual value of the eigenvalues~\eqref{eq:ev_contLL_evs} is insufficient to define Landau-levels. Instead, one needs to keep track of the behavior of the eigenmode in the $x_1-x_2$ plane, indicated by the Landau-index $n$. In the full QCD case, the analogous procedure is to define a projector $P_{\rm LLL}$ that projects any four-dimensional vector on the lowest Landau-level subspace for each $x_1-x_2$ plane located at any $0\le n_3 < N_s$ and $0\le n_4< N_t$. The details on how to construct this projector were given in~\cite{Bruckmann:2017pft}, together with a numerical proof that $P_{\rm LLL}$ is an operator localized over a range $\ell_B=1/\sqrt{q_fB}$. The study~\cite{Bruckmann:2017pft} also demonstrated that typical low-lying modes of the four-dimensional Dirac operator have large overlap with the so-defined lowest Landau-level subspace, while high-lying modes tend to lie more in the complement subspace.

Altogether, this leads us to an important general observation: 
the Landau-level structure of the Dirac spectrum persists in QCD and, thus,
on practically any gluonic configuration, the magnetic field always leads to a proliferation of low eigenvalues of the Dirac operator. This finding will be important when we consider the magnetic catalysis phenomenon in Sec.~\ref{sec:pd_magncat}.
In addition, using the projector $P_{\rm LLL}$ defined above, the contribution of the lowest Landau-level to fermionic observables can also be calculated on the lattice. This was carried out in~\cite{Bruckmann:2017pft} for the quark condensate~\eqref{eq:lat_pbpdef_0} and the spin polarization~\eqref{eq:lat_def_tensorcoeff} and the lowest Landau-level dominance was demonstrated for strong magnetic fields.

The fact that the lowest Landau-level remains a well-defined concept in full QCD, is also useful for various further observables. In particular, the construction discussed above turns out to be important for hadronic correlation functions. For the evaluation of the latter, it is important to construct hadron interpolating operators with the largest possible overlap with the ground state. It has been demonstrated that applying the projector $P_{\rm LLL}$ (and similar projectors based on the two-dimensional Laplace operator) to quark operators, one can significantly suppress the noise in various hadron correlators~\cite{Bignell:2018acn,Bignell:2020xkf}. This technique has been used to calculate the magnetic polarizability of the neutron, for example. We will review results on the electromagnetic field-dependence of hadron correlators in Chap.~\ref{chap:hadron}.

\subsection{Electric fields}
\label{sec:ev_electricfields}

Based on the above, the generalization of the eigenvalue spectrum to the case of background electric fields is straightforward. In fact, the Dirac eigenvalues for homogeneous imaginary electric fields follow the same pattern as the ones for the Landau problem~\eqref{eq:ev_contLL_evs}.
To see this, consider the gauge field~\eqref{eq:lat_hom_Efield} for the electric field and compare it to the magnetic case~\eqref{eq:lat_hom_Bfield}. Besides a relabeling of the coordinate axes $(x_1,x_2)\to(x_3,x_4)$, the Dirac operators in the two settings merely differ by the antiperiodic boundary conditions in the temporal direction. The difference is tantamount to the introduction of a boundary twist $A_4^{\rm twist}=-i\pi\delta(x_4-1/T)$ or, equivalently, a homogeneous imaginary chemical potential $A_4^{\rm twist}=i\mu=-i\pi T$. This changes the temporal Polyakov loops~\eqref{eq:lat_E_ploops} to
\be
P_{4f}^{\mathrm{U}(1)}(x_3) = 
\exp\left[i\,2\pi N_e^f \frac{x_3}{L_3} + i\frac{\mu}{T} \right]
\,.
\label{eq:ev_electricPloop}
\ee
As we illustrated in Fig.~\ref{fig:lat_Bprofiles}, the $P_{4f}^{\mathrm{U}(1)}$ Polyakov loop winds around the complex circle completely (in fact, $N_e^f$ times). Therefore, a shift of the starting phase by $\mu/T=-\pi$ does not affect the system. We will get back to this argument below in Sec.~\ref{sec:eos_elsusc}.

Altogether, the homogeneous magnetic and imaginary electric field setups therefore merely differ by the geometry of the finite volume, $N_s\times N_s\to N_s\times N_t$. Thus, the eigenvalues are the same as in~\eqref{eq:ev_contLL_evs}, except for a relabeling of the momenta,
\be
\lambda_{n,p_1,p_2}^2=2n|q_fE|+p_1^2+p_2^2, 
\qquad n\in\mathds{Z}_0^+,
\qquad p_1,p_2\in\mathds{R} \,,
\label{eq:ev_contLL_electric_evs}
\ee
and $p_1=2\pi n_1/L_1$, $p_2=2\pi n_2/L_2$ with $n_1,n_2\in\mathds{Z}$. The analogue of the spin operator $\sigma_{12}$ (identified with $\sigma_3$ in Sec.~\ref{sec:ev_contLL}) is in this case $\sigma_{34}$. The lowest `Landau-level' $n=0$ has fixed eigenvalue of $\sigma_{34}$ and a degeneracy $N_e^f$. Higher levels have both eigenvalues $\pm1$ for $\sigma_{34}$ and a degeneracy of $2N_e^f$.

Notice that -- generalizing our argument from above -- the eigenvalues~\eqref{eq:ev_contLL_electric_evs} are not only insensitive to the fermionic boundary twist, but independent of the imaginary chemical potential $i\mu$ altogether. Thus, the Dirac determinant $\det M_f$ is likewise $i\mu$-independent. In fact, its dependence on the temperature is only through the degeneracy factor $N_e^f\propto iq_fE L_3/T$ of the eigenvalues. This is in stark contrast to the case at $E=0$ and $T>0$, where the determinant depends non-trivially on both $i\mu$ and $T$. The independence of $\det M_f$ on $i\mu$ has important consequences for the simulations of QCD with homogeneous imaginary electric fields that we discuss now briefly, following~\cite{Endrodi:2022wym,Endrodi:2023wwf}.

We just showed that at $iE>0$, derivatives of the free energy density of free fermions, $f\propto -\log\det M_f$ with respect to $i\mu$ vanish. In other words, the equilibrium system in a homogeneous imaginary electric field in a finite periodic volume is necessarily globally neutral, i.e.\ $n=-\partial f/\partial \mu=0$. Thus, at any $iE>0$, the system is automatically projected to the canonical sector with zero particle number. This is very different from the $iE=0$ system, which is defined in the grand canonical ensemble with $\mu=0$. Therefore, there is a singular change of relevant thermodynamic ensembles, which leads to a discontinuity of physical observables as a function of $iE$ and was discussed in~\cite{Endrodi:2022wym,Endrodi:2023wwf}.  Notice that at $T=0$, this issue is absent since in the vacuum, fluctuations of the charge vanish anyway for $\mu<m_f$ due to the mass gap.

\begin{figure}
 \centering
 \includegraphics[width=9.5cm]{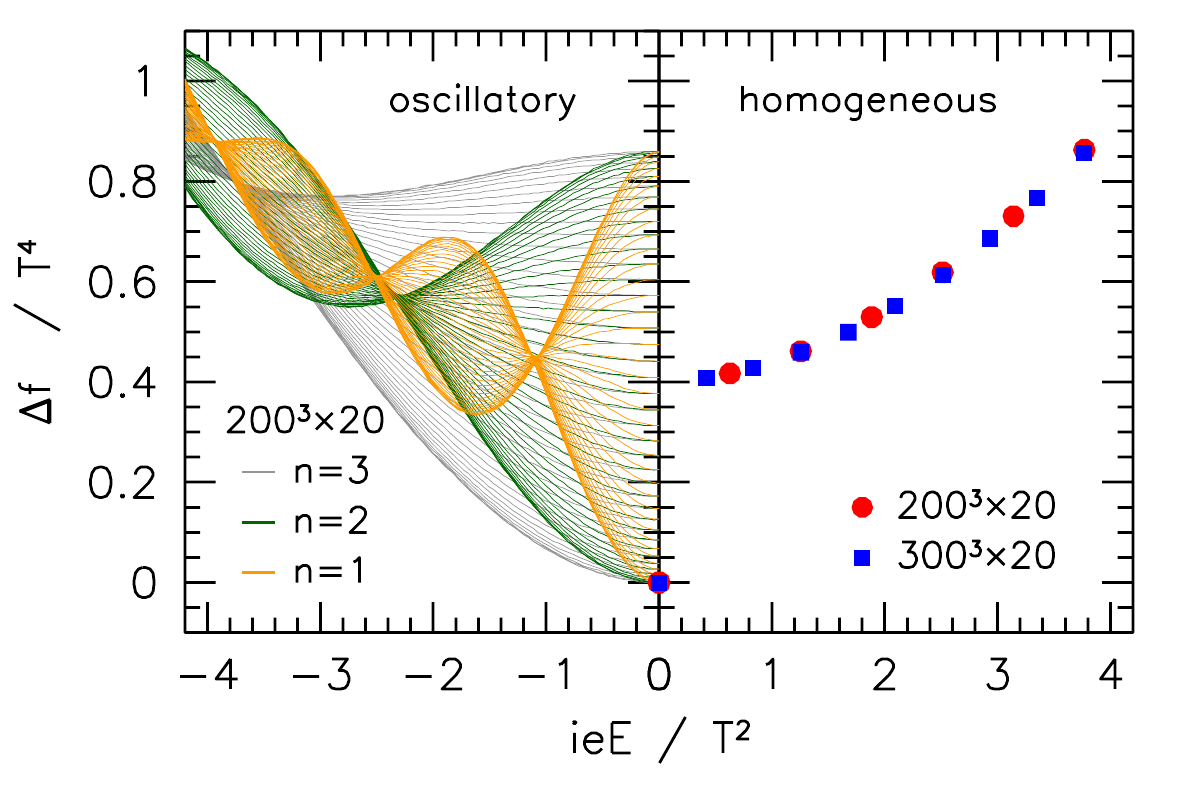}
 \caption{ 
 \label{fig:ev_electric_logdet}
 The excess free energy density $\Delta f$ due to the imaginary electric field for free fermions as a function of $iE$. The two sides show the dependence on harmonic (left side) and homogeneous (right side) imaginary fields. In the left side, the different curves correspond to different momenta $k_3=p_3L_3/(2\pi)$ (denoted here by $n$, indicated by the different colors) and different values of the imaginary chemical potential $0\le i\mu\le \pi T$ (curves from bottom to top). In the right side, $i\mu$ has no impact on $\Delta f$. Figure from~\cite{Endrodi:2023wwf}.
 }
\end{figure}

To demonstrate the singular behavior at $E=0$, we show $\Delta f=f-f(E=0)$ in the right side of Fig.~\ref{fig:ev_electric_logdet} for homogeneous fields~\cite{Endrodi:2023wwf}. As the volume grows, the smallest electric field quantum approaches zero but $\Delta f$ for this field converges to a nonzero value. The jump $\lim_{iE\to0}\Delta f(iE)$ is proportional to $T^4$.
This jump can also be described with harmonic imaginary electric fields~\eqref{eq:lat_osc_Efield}, where the amplitude $iE$ is a continuous variable but the momentum $k_3$ is discrete. In this case, the $\mathrm{U}(1)$ Polyakov loops do not wind completely around the complex circle (see Fig.~\ref{fig:lat_Bprofiles}) and, thus, the chemical potential shifts them non-trivially. Correspondingly, the dependence of $\Delta f$ on $i\mu$ is maintained, see the left side of Fig.~\ref{fig:ev_electric_logdet}. However, in the homogeneous limit $k_3\to0$, the results again approach a singular behavior, where the curves collapse on a set of $i\mu$-independent nodepoints approaching the $iE=0$ axis.
From this figure one can also conclude that for homogeneous fields the discontinuity at $E=0$ coincides with the average of $\Delta f$ over all chemical potentials,
\be
\lim_{iE\to0} \Delta f(iE) = \frac{1}{\pi T}\int_0^{\pi T} \dd (i\mu) \,\Delta f(iE=0,i\mu)\,.
\ee
which is indeed the $E=0$ free energy in the canonical sector with zero charge~\cite{Endrodi:2022wym,Endrodi:2023wwf}.

A final comment about the nature of equilibrium simulations with nonzero background electric fields is in order.
Unlike the situation for $B>0$, electric fields lead a homogeneous medium out of equilibrium by accelerating charge carriers. At $E>0$, the system equilibrates by developing an inhomogeneous profile for the electric charge $\expv{j_0(x)}$. In this equilibrium, the gradient of the charge distribution is such that it exerts a degeneracy force that balances the electric force on any charge. A homogeneous electric field, for example, implies a linearly rising equilibrium charge profile and a weak harmonic electric field induces a harmonic charge distribution.
An analogous behavior occurs for imaginary electric fields, too, for which the imaginary density profile becomes inhomogeneous. Many of these fascinating features can be understood already on a perturbative level~\cite{Endrodi:2022wym}.

\subsection{Polyakov loop backgrounds}
\label{sec:ev_ploop}

We close this chapter by discussing the Dirac spectrum in the presence of an $\mathrm{SU}(3)$ Polyakov loop background and magnetic fields. The results will be relevant for the understanding of the qualitative behavior of the transition temperature as a function of the magnetic field in the QCD phase diagram, to be discussed in Sec.~\ref{sec:pd_ploop}. The discussion below follows~\cite{Bruckmann:2013oba}, see also the review~\cite{Fukushima:2017csk}.

A homogeneous Polyakov loop background appears in the Dirac operator in the form of a constant $\mathrm{SU}(3)$ matrix $i\A_4$. Being independent of the coordinate, this background is covariantly constant, i.e.\ it can be diagonalized to the form $\A_4/T=\textmd{diag}(\varphi_1,\varphi_2,\varphi_3)$~\cite{Dunne:2004nc}. 
In this case, the traced Polyakov loop, defined above in~\eqref{eq:lat_ploopavg}, simply reads
\be
P=\tr \exp \left[i\A_4/T\right] = \sum_{c=1}^{3} \exp \left[i\varphi_c\right]\,.
\ee
The three angles here satisfy $\varphi_1+\varphi_2+\varphi_3=0 \textmd{ mod }2\pi$, since $\exp[i\A_4/T]$ is an element of $\mathrm{SU}(3)$.
Thus, homogeneous Polyakov loop backgrounds may be thought of as imaginary chemical potentials that couple differently to the three quark colors. The deconfined phase ($P=3$) corresponds to $\varphi_1=\varphi_2=\varphi_3=0$. In turn, the confined phase ($P=0$) is approached for nontrivial imaginary chemical potentials $\varphi_1=2\pi /3$, $\varphi_2=-2\pi/3$ and $\varphi_3=0$. An interpolation is realized by a single angle $\varphi$,
\be
\varphi_1=\varphi, \qquad \varphi_2=-\varphi, \qquad \varphi_3=0,\qquad
\begin{cases}
 \varphi=0 & \textmd{deconfinement}\,,\\
 \varphi=2\pi/3 & \textmd{confinement}\,.
\end{cases}
\ee

Here we will discuss the impact of this Polyakov loop background on the continuum Dirac eigenvalues in the free case. The imaginary chemical potential merely shifts the temporal momenta, therefore the eigenvalues read
\be
\lambda^2_{c,p_\nu} = p_1^2+p_2^2+p_3^2+(p_4+\varphi_{c}T)^2, \qquad p_j\in\mathds{R}, \qquad p_4=2\pi (k_4+1/2)T, \quad k_4\in\mathds{Z}\,,
\ee
where $1\le c\le 3$ runs over the colors.
In Sec.~\ref{sec:pd_ploop} we will discuss the dependence of the Dirac determinant on the Polyakov loop background. This dependence is expected to be dominated by the lowest Dirac eigenvalues, because the Polyakov loop background, as a specific colored imaginary chemical potential, is an infrared parameter. Notice that the lowest eigenvalue is $\lambda^2_{c=2,p_j=0,p_4=\pi T}=(\pi - \varphi)^2T^2$, which is maximal for $\varphi=0$ and is reduced as $\varphi$ grows. Thus -- assuming low-mode dominance --  the Dirac determinant is reduced when $\varphi$ grows~\cite{Bruckmann:2013oba}. Since smaller determinants amount to smaller weights in the path integral, this means that fermions favor deconfinement over confinement. Indeed, the deconfinement transition temperature is lower in the presence of dynamical quarks as compared to pure gauge theory~\cite{Gattringer:2010zz}.

We may also combine the effect of the magnetic field and that of the Polyakov loop. The corresponding continuum Dirac eigenvalues in the free case are 
\be
\lambda^2_{c,n,p_3,p_4} = 2n|q_fB|+p_3^2+(p_4+\varphi_{c}T)^2, \qquad n\in\mathds{Z}_0^+,
\qquad p_3\in\mathds{R}, \qquad p_4=2\pi(k_4+1/2)T, \qquad k_4\in\mathds{Z} \,.
\ee
Again we can look at the lowest eigenvalues to learn about the tendency of the Dirac determinant for confining and deconfining backgrounds. Just like above, the lowest eigenvalue is $\lambda_{c=2,n=0,p_3=0,p_4=\pi T}^2=(\pi-\varphi)^2T^2$. But this time, these lowest eigenvalues appear in the spectrum with a degeneracy proportional to $\Phi_{12}/(2\pi)$. Thus, the tendency of the Dirac determinant to favor deconfinement is stronger at $B>0$ than at $B=0$. This suggests that the deconfinement transition temperature is lowered as $B$ grows. We will elaborate more on this point in Sec.~\ref{sec:pd_ploop}.

Notice that in the case of homogeneous imaginary electric fields, the Dirac eigenvalues become insensitive to homogeneous Polyakov loop backgrounds. Indeed, the homogeneous Polyakov loop background corresponds to color-dependent imaginary chemical potentials, and as we showed above in~\eqref{eq:ev_electricPloop}, these cancel from the Dirac operator upon a shift of the coordinate origin. 
This is yet another manifestation of the singular change of relevant thermodynamic ensembles that occurs as a weak imaginary electric field is switched on for free fermions.
For arbitrary color interactions, such a cancellation does not take place in general.

\subsection{Lessons learned}

In this chapter we studied the impact of background magnetic fields on the eigenvalues of the Dirac operator. Through the Dirac determinant, these eigenvalues -- predominantly the lowest ones -- govern the infrared behavior of fermionic observables. In the two-dimensional case and in the absence of color interactions, we found the equivalent of the continuum Landau-levels and related the emerging spectrum to that of the Hofstadter model in solid state physics. This was discussed using three different fermion discretizations: staggered, Wilson and overlap quarks.

Subsequently, we proceeded to the physical setting of full, four-dimensional QCD.
Our most important observation is that due to its topological nature, the separation of the lowest-Landau level from the rest of the spectrum persists even here. This allowed us to conclude that magnetic fields in general lead to a proliferation of low Dirac eigenvalues, independent of the gluonic fields. This will lead directly to the magnetic catalysis phenomenon, to be discussed in Sec.~\ref{sec:pd_magncat}.
In turn, we also showed that -- in the free case with constant Polyakov loop backgrounds -- magnetic fields in general tend to enhance the Polyakov loop, an important guideline for the study of deconfinement in a magnetic background in Sec.~\ref{sec:pd_ploop}.

Finally, the generalization of these concepts to the case of background electric fields revealed a very different picture, namely that equilibrium lattice simulations at $iE\neq0$ and $E=0$ correspond to different thermodynamic ensembles.  This complicates the discussion of the response of the medium to weak electric fields based on such simulations and calls for an alternative approach, which we will discuss in Sec.~\ref{sec:eos_elsusc}.

	\clearpage
	\section{Confinement and hadron properties}
	\label{chap:hadron}

The first lattice QCD simulations involving background electromagnetic fields targeted various, experimentally observable properties of the bound states of the strongly interacting vacuum: hadrons. These seminal lattice studies~\cite{Martinelli:1982cb,Bernard:1982yu,Fiebig:1988en,Martinelli:1988rr,Draper:1989pi,Wilcox:1991cq,Leinweber:1992hy} on magnetic moments and electric and magnetic polarizabilities from the 1980s and the 1990s laid the ground for a wide range of calculations of hadron properties in background fields. This field has undergone significant improvements since then -- not just by enhancing precision, but also in terms of conceptual understanding that lead to a series of methodological advancements. These conceptual aspects include the issue of gauge choices, boundary conditions, quenched and electroquenched approximations, flavor mixing, lattice artefacts, optimized interpolating operators and signal-to-noise improvement. 
In this chapter, we review the current status in this context.

In addition to the hadron properties encoded in the weak-field behavior, research interest has also partially shifted to hadron energies at strong magnetic fields. One motivation to look for the impact of $B$ on vector meson energies comes from the conjectured existence of a superconducting phase at strong fields, originally put forward in~\cite{Chernodub:2010qx}. While according to our current understanding, such a phase does not exist in QCD, this hypothesis gave rise to various theoretical developments and enriched the discussion of the low-temperature region of the QCD phase diagram.

Hadrons exist in QCD due to confinement, encoded in the features of the gluon fields dominating the path integral in the QCD vacuum. The confinement mechanism becomes more intricate in the presence of background magnetic fields: Lorentz-structures constructed from the gluon field strength become anisotropic and spatially extended gluonic objects also behave differently in the directions parallel and perpendicular to the magnetic field.
This chapter is also devoted to summarizing the existing results about these gluonic characteristics.

\subsection{Hadrons in electromagnetic fields}
\label{sec:had_hadelmagfields}

In the vacuum of QCD, the relevant degrees of freedom are hadrons, and the low-energy behavior of the theory is characterized by their spins and masses. The expansion of the QCD Lagrangian in this low-energy limit defines chiral perturbation theory ($\chi$PT), where the lightest hadrons, pions, enter. An alternative approach is the hadron resonance gas (HRG) model, which approximates QCD as a non-interacting gas of all possible hadrons and resonances.
However, hadrons are not point-like particles but bear non-trivial internal structures. This becomes most transparent in the presence of background electromagnetic fields, which deform hadrons and thereby resolve their spatial composition. To make contact to experimental measurements~\cite{Griesshammer:2012we,Moinester:2019sew}, one needs long-wavelength background fields, and here we will concentrate on the response of hadrons and light nuclei to homogeneous fields. 

When exposed to background electromagnetic fields, the lowest possible energy $\E$ that a hadron may acquire is not equal to its mass $M$ anymore. Charged pions, for example, are forced to occupy Landau-levels with nonzero orbital angular momentum, contributing to the total energy. In the case of baryons, the background field also interacts with their spins, again impacting on the lowest possible energy. Loosely, one often speaks of a change of the hadron mass due to the background field, but one should keep in mind that this really corresponds the lowest possible energy of the hadron state in the presence of the field. We will make this fact explicit by using the notation $\E(B)$ and $\E(E)$. This behavior depends on the charge of the hadron, which we will indicate with a superscript, e.g.\ $\pi^+$ or $K^0$.

Let us first concentrate on background magnetic fields. We consider a non-interacting hadron with mass $M$, charge $q$ and spin $s$, placed in a homogeneous magnetic field, oriented in the direction of the $x_3$ axis. Along the magnetic field, the momentum component $p_3$ is conserved, just like the spin component, which takes the values $-s\le s_3\le s$. 
The coupling between the spin and the background field is set by the gyromagnetic ratio $g$ of the hadron. 
The relativistic energy $\E$ of this non-interacting particle is written in terms of the Landau-levels that we already encountered in Chap.~\ref{chap:ev},
\be
\E_{p_3,\ell,s_3}(B)=\sqrt{M^2+p_3^2+(2\ell+1)|qB| - g s_3 qB}\,,
\label{eq:had_generalenergy}
\ee
where $\ell$ labels the Landau-levels.

In relativistic quantum mechanics, the gyromagnetic ratio is fixed to $g=2$. In this case, for spin-$1/2$ particles, the two magnetic field-dependent terms under the square root in~\eqref{eq:had_generalenergy} add up to $2n |qB|$ with $n\in\mathds{Z}_0^+$, as we have already seen for the Dirac eigenvalues~\eqref{eq:ev_contLL_evs}. In turn, in quantum field theory, the elementary particles receive corrections to this tree-level value, giving rise to anomalous magnetic moments, e.g.\ of the muon~\cite{Meyer:2018til}. Hadrons possess, in general, non-trivial values of the gyromagnetic ratio. In order to describe these, it is convenient to consider the weak-field expansion of~\eqref{eq:had_generalenergy}. We are interested in the minimal energy, therefore we set $p_3=\ell=0$ to arrive at
\be
\E_{s_3}(B)=M-\hat\mu \frac{s_3}{s}\,eB + \frac{|qB|}{2M} - 2\pi \beta\, (eB)^2 + \mathcal{O}(B^3)\,,
\label{eq:had_energy_hadron_spin}
\ee
where $\hat\mu$ is the magnetic moment and $\beta$ the magnetic polarizability of the hadron\footnote{We note that the magnetic polarizability can alternatively be defined in terms of the relativistic formula~\eqref{eq:had_generalenergy}, as the coefficient of an $\O(B^2)$ term under the square root, see~\cite{Bignell:2018acn,Bignell:2019vpy}. While for the quadratic expansion around $B=0$, both definitions coincide, the latter may be advantageous when $\beta$ is evaluated via fitting $B>0$ data~\cite{Bignell:2019vpy}.
},
\be
\hat\mu=
-\frac{\partial}{\partial (eB)}\left[ \E_{s_3=s}-\frac{|qB|}{2M}\right]_{B=0}, \qquad
\beta = -\frac{1}{4\pi}\left.\frac{\partial^2 \E_{s_3}}{\partial (eB)^2}\right|_{B=0}\,.
\label{eq:had_mu_alpha_defs}
\ee
Notice that while the magnetic moment $\hat\mu$ couples to the magnetic field in proportion to the spin projection $s_3$, the polarizability $\beta$ is independent of the spin orientation. Above, we defined both parameters in terms of the renormalization group invariant combination $eB$ (see the discussion in Sec.~\ref{sec:lat_renormfreeenergy}). With these definitions, it is clear that the $\mathcal{O}(B)$ dependence of the energy of charged hadrons has two sources: the quantized orbital angular moment of the lowest Landau-level, as well as the magnetic moment. 
For baryons ($s=1/2$), the deviation from the quantum mechanical value, $g=2$, is usually quantified by the anomalous magnetic moment $\delta \hat\mu = \hat\mu - q/(2Me)$.

The magnetic moment therefore describes the tendency of the hadron spin to align with the background magnetic field -- an effect linear in $B$ for the energy. In turn, the magnetic polarizability measures the deformation of the hadron when exposed to the background field, affecting the energy proportionally to $B^2$.
We point out that for charged mesons, the energy is always increased to leading order due to the coupling to orbital angular momentum in~\eqref{eq:had_energy_hadron_spin}, irrespectively of the value of $\beta$. In turn, for charged baryons the leading-order effect depends on the magnetic moment: for $\delta\hat\mu>0$, the energy of the $s_3=s$ component is reduced and for $\delta\hat\mu<0$ it is increased by the magnetic field. We will get back to this point when we discuss the magnetic susceptibility of the QCD medium in Sec.~\ref{sec:eos_magsusc_QCD}.

For homogeneous background electric fields $E$, the polarizability $\alpha$ is defined similarly to the magnetic one in~\eqref{eq:had_mu_alpha_defs},
\be
\alpha = -\frac{1}{4\pi}\left.\frac{\partial^2 \E}{\partial (eE)^2}\right|_{E=0}\,.
\label{eq:had_alphae_def}
\ee
This coefficient is usually referred to as the static electric polarizability.
Unlike for magnetic fields, in the electric case a linear effect in $\E$ is absent due to time reversal symmetry in QCD, except if the latter is broken by another external parameter. We get back to this point in Sec.~\ref{sec:had_eldipmom}. 

The polarizabilities can be used to determine cross sections for Compton scattering reactions, relevant for experiments~\cite{Fiebig:1988en}.
However, for spin-$1/2$ hadrons, not only the electric polarizability affects the energy at $\O(E^2)$ but also the magnetic moment $\hat\mu$. The static polarizability appearing in~\eqref{eq:had_alphae_def} receives contributions from both, so that for example for the neutron, the Compton scattering polarizability $\bar\alpha_n$, relevant for experiments, can be found via~\cite{Lvov:1993fp,Detmold:2010ts,Lujan:2014kia},
\be
\bar\alpha_n=\alpha_n+\frac{\hat\mu_n^2}{M_n}\,.
\label{eq:had_compton_neutron}
\ee
Therefore, one needs separate determinations of $\hat\mu_n$ and $\alpha_n$ to find the Compton scattering polarizability. Alternatively, the magnetic moment can be eliminated directly on the lattice by using boost-projected correlators~\cite{Detmold:2009dx}, also referred to as Born subtraction~\cite{Detmold:2010ts}.

\subsection{Lattice techniques}
\label{sec:had_lattechn}

Hadron energies can be determined directly in lattice simulations in terms of the correlators of interpolating operators with the corresponding quantum numbers~\cite{Gattringer:2010zz}. The correlator $C$ is constructed by
\be
C(n_4)=\expv{A(0) A^\dagger(n_4) } = c \, e^{-n_4 a\E} + \ldots\,,
\label{eq:had_correlator1}
\ee
creating a hadron state at the source by the operator $A$ and destroying it at the sink by $A^\dagger$. 
The hadron energy $\E$ is encoded in the leading exponential decay of the correlator, as we indicated in~\eqref{eq:had_correlator1}. Next, we discuss the key aspects of the lattice calculations of $C(n_4)$, as well as typically employed approximations.

\paragraph{Connected and disconnected diagrams}

The fermionic expectation value in the middle of~\eqref{eq:had_correlator1} can be rewritten as a combination of products of quark propagators. For mesons, the operator is of the form $A=\bar\psi_f \Gamma \psi_{f'}$, so that~\cite{Gattringer:2010zz},
\be
C(n_4)=-\Expv{\bar\psi_f(0) \Gamma \psi_{f'}(0)\bar\psi_{f'}(n_4) \bar\Gamma \psi_{}(n_4)} = 
\Expv{\tr \left [M^{-1}_f \Gamma  \,\P_{0} M^{-1}_{f'}\bar\Gamma \P_{n_4}\right]} - \delta_{ff'} \Expv{\tr\left[M_f^{-1}\Gamma\P_{0}\right]\tr\left[M_f^{-1}\bar\Gamma\P_{n_4}\right]}\,,
\label{eq:had_mesoncorr}
\ee
where $\bar\Gamma=\gamma_4\Gamma^\dagger\gamma_4$, $\P_{n_4}$ denotes a projector on the time-like slice of the lattice at coordinate value $n_4$ and the trace is over internal degrees of freedom (color and spin). On the left hand side, the expectation value is both over quark and gluon degrees of freedom, while on the right hand side, it only denotes the gluonic path integral.

One sees that in general, $C(n_4)$ consists of a connected diagram (the first term in the right hand side of~\eqref{eq:had_mesoncorr}) as well as a disconnected one (the second term), but the latter is only present for flavor-singlet mesons. Disconnected diagrams also cancel for the neutral pion in the vacuum, $\pi^0=(\bar u \gamma_5 u - \bar d \gamma_5 d)/\sqrt{2}$ if isospin symmetry is intact~\cite{Gattringer:2010zz}, but this is not the case at nonzero background fields anymore due to $q_u\neq q_d$.
In turn, disconnected diagrams are always absent for baryons.
While connected diagrams can typically be calculated using a Hermiticity relation via a single inversion for all values of $n_4$, this is not the case for the disconnected term, rendering it substantially more expensive. Often, the disconnected contribution is therefore neglected in the literature.

\paragraph{Flavor mixing}

For flavor-singlet mesons, discarding disconnected diagrams by construction implies neglecting mixing in the meson sector. For the neutral pion at nonzero magnetic field, for example, the physical pion field would be $\pi^0(B)=c_u \,\bar u \gamma_5 u + c_d \,\bar d\gamma_5d$, with the $B$-dependent coefficients $c_u$, $c_d=\sqrt{1-c_u^2}$ encoding the physical flavor content of the neutral pion in this isospin-asymmetric system. To determine these flavor coefficients self-consistently, one considers so-called `connected pions' $\pi^u=\bar u \gamma_5 u$ and $\pi^d=\bar d \gamma_5 d$, and a matrix of their correlators $\expv{\pi^{u,d}(0) \pi^{u,d\dagger}(n_4)}$ in place of~\eqref{eq:had_correlator1}. 
The physical eigenstates can be found from the eigenvectors of this matrix via solving a generalized eigenvalue problem~\cite{Blossier:2009kd}. The off-diagonal components of this matrix contain only disconnected diagrams -- thus, neglecting them immediately results in considering the connected pion states $\pi^u$ and $\pi^d$. This approximation is often made in the literature.
We note moreover that connected pions carry further relevant information, because their mass gives an exact lower bound for the charged vector meson mass through a QCD inequality~\cite{Hidaka:2012mz}.

\paragraph{Electroquenched approximation}

Another simplification, often employed in the literature, is the valence approximation that we introduced already in~\eqref{eq:lat_expvO_electroquenched}. In this context, it is referred to as the electroquenched approximation.
Since in this case sea quarks are electrically neutral, the same gauge field configurations can be used for measurements at all values of the electromagnetic field. Besides being computationally less expensive, this also implies that one can take advantage of correlations between the correlators at different values of the background field. 
In order to go beyond this approximation, one needs to consider different gauge field ensembles at zero and nonzero background field values. This implies that the cross-correlations between measurements with and without background field cannot be used, and one is left with the task of extracting the tiny energy shift as the difference of two noisy energy values.
An alternative approach is provided by reweighting the zero-field configurations to nonzero background fields~\cite{Freeman:2014kka}.
The exact evaluation of the fermion determinant, necessary for this, is in practice not feasible. Instead, the weight factor can be expanded in the background field and computed stochastically using noisy estimators~\cite{Freeman:2014kka}.

We note that to linear order in the background field, charged sea-quark contributions are proportional to $\sum_f q_f/e$. This happens to vanish for the three-flavor symmetric theory ($m_u=m_d=m_s$), therefore for this system, the electroquenched approximation of linear effects in fact becomes exact.

\paragraph{Interpolating operators} One is in principle free to choose any operator $A$ that shares the same quantum numbers as the hadron under study. However, increasing the overlap with the hadron state can significantly reduce noise, which is crucial for measuring the response to weak fields.
A standard strategy is to use spatially extended smeared sources as a generalization of point sources. At nonzero magnetic field however, hadronic states resemble Landau-levels, similarly to the Dirac eigenstates we discussed in Chap.~\ref{chap:ev}. Methods to project the sources to lowest hadronic Landau-levels were developed recently and put into practice in~\cite{Tiburzi:2012ks,Bignell:2020xkf,Bignell:2020dze,Kabelitz:2024aye}.

\paragraph{Lattice artefacts for Wilson fermions}

Most hadron spectrum calculations in the literature have been performed with variants of the Wilson fermion formulation. 
This discretization entails a technical complication that we encountered already in Sec.~\ref{sec:ev_Wilsonoverlap}: magnetic field-dependent quark mass renormalization~\cite{Brandt:2015hnz,Bali:2017ian}. This is a lattice artefact that contributes to quark masses as $m_f\to m_f + \O(aq_fB)$. While it disappears in the continuum limit, for typical values of the magnetic field it causes significant shifts in hadron masses~\cite{Bali:2017ian}. It can be removed either by explicitly tuning the quark mass as a function of $B$~\cite{Bali:2017ian} or by using $\mathcal{O}(a)$ improvement in the electromagnetic fields for the Wilson action, referred to as background field-corrected clover action~\cite{Bignell:2019vpy}.

\paragraph{Resonances}

In nature, some of the hadrons are not stable particles but resonances. For example, the $\rho$ meson decays into pions via the strong interactions. In dynamical lattice QCD simulations, this complicates the determination of the $\rho$ meson mass already at $B=0$ and demands a careful finite size analysis. This can be avoided if one works with pions heavier than in nature, so that the decay is kinematically blocked. 
Moreover, the $\rho$ meson also remains stable if dynamical sea quarks are absent, i.e.\ in the quenched approximation. 
In addition, some of the hadrons to be discussed in this chapter are stable in QCD but decay via the weak interactions. One example is the charged pion. Such weak decays are parameterized by decay constants, to be discussed below in Sec.~\ref{sec:had_decay_constant}.

\paragraph{Gauge choice and boundary conditions}

For the response to electric fields, it turns out to be important to distinguish between different electromagnetic gauges. For a non-static gauge $A_3=-iE x_4$, translational invariance is exact, while for a static one $A_4=iEx_3$ it is lost at the level of the gauge field. 
In particular, these different gauges (and, shifts of the coordinate origin) correspond to different $\mathrm{U}(1)$ Polyakov loops that might influence the correlators.
This effect is also intertwined with the choice for the boundary conditions, to which we get back to below.

\paragraph{Further systematics}

Finally, we note that carrying out simulations at the physical point is in most cases prohibitively expensive. To alleviate this issue, often one simulates with heavier-than-physical sea quarks, and also uses heavier-than-physical valence quarks in the measurements. In order to recover the physical results, an extrapolation to the physical point is required.
Besides this aspect, other important sources of systematics are the impact of finite volume as well as the already mentioned quenched and electroquenched approximations. Many of these aspects may be understood within chiral effective theory, see e.g.~\cite{Detmold:2006vu,Hall:2013dva,He:2021eha}.

\subsubsection{Direct method}
\label{sec:had_directmethod}

There are two alternatives to determine the impact of electromagnetic fields on hadron energies. The first one is the direct method -- often referred to in the literature simply as the background field method -- for which the correlators are measured at nonzero background field values. A subsequent analysis for weak fields gives the magnetic moment and the magnetic or electric  polarizabilities. 
A crucial issue is the choice of field values for the construction of the correlators. For overly strong fields, higher-order effects may become non-negligible, overshadowing the leading-order effects. In turn, for very weak fields the signal-to-noise ratio is too low.

There are various proposals in the literature for how correlators at different values of the background field may be combined in order to improve the signal-to-noise ratio. These typically work only in the electroquenched approximation, where the correlations among different field values can be exploited. For polarizabilities, one can get rid of unwanted linear effects by averaging correlators at positive and negative values of the background field (as done e.g.\ in~\cite{Lujan:2014kia}) or by considering ratios and products of correlators at nonzero and zero background field values to partially cancel correlated fluctuations~\cite{Detmold:2009dx,Detmold:2010ts,Bignell:2020dze}.
For baryons, this may be further improved by combining setups with different magnetic field orientations, while always keeping the baryon spin aligned (or anti-aligned) with it~\cite{Bignell:2018acn}.
For the magnetic moment, in turn, the difference of energies with opposite spin is necessary, which can be found effectively using ratios of correlators with opposite spin projection~\cite{Primer:2013pva}.

For the direct method, one may either use periodic boundary conditions (e.g.~\cite{Detmold:2009dx,Detmold:2010ts}), which maintain translational invariance but the value of the magnetic field $B$ or that of the imaginary electric field $iE$ are quantized according to~\eqref{eq:lat_flux_quant_hom} and~\eqref{eq:lat_elquant}. Alternatively, Dirichlet boundary conditions have also been employed in the valence sector (e.g.~\cite{Lujan:2014kia,Freeman:2014kka}) for nonzero electric fields. In this case, the field value is a continuous variable, allowing for arbitrarily small field values, at the cost of translational invariance. However, the latter implies that hadron states with exactly zero momentum are absent, resulting in an induced momentum, i.e.\ an $\O(1/L)$ shift to the energy, which needs to be accounted for~\cite{Lujan:2014kia}.

Finally, we note that for background electric fields, the spectral representation~\eqref{eq:had_correlator1} of hadron correlators is not a simple sum of exponentials, as usual in the absence of background fields. The appropriate, infinite-volume correlators have been derived in~\cite{Detmold:2009dx}, and the finite volume corrections for Dirichlet boundary conditions in~\cite{Niyazi:2021jrz}. The latter turn out to be substantial and important for typical lattice analyses. The fitting procedure for lattice correlators becomes more involved in this case~\cite{Niyazi:2021jrz}. One particular issue is the need to correct the effective distance between the Dirichlet walls in order to account for the interaction between the hadron and the boundaries~\cite{Niyazi:2021jrz}.

\subsubsection{Weak-field expansion}
\label{sec:had_weakfield}

The second method to determine the dependence $\E(B)$ or $\E(E)$ does not focus on the energy, but is instead based on an expansion of the correlators themselves in the background field, sometimes also called the form factor method. Differentiating the two-point function~\eqref{eq:had_correlator1} with respect to the background field results in observables involving three-point functions for the magnetic moment and four-point functions for the polarizabilities, to be evaluated in the vacuum.
The fundaments of this approach date back to the pioneering works~\cite{Martinelli:1988rr,Draper:1989pi,Wilcox:1991cq,Leinweber:1992hy,Gadiyak:2001fe,Gockeler:2003ay}.

The weak-field expansion was considered in the influential work~\cite{Engelhardt:2007ub}, where the diagrams relevant to the second order expansion were considered specifically.
This approach was later on reconsidered in~\cite{Wilcox:2021rtt}, where the electric and magnetic polarizabilities were formulated in terms of zero-momentum limits of four-point functions. This machinery was employed in~\cite{Lee:2023rmz} to calculate $\alpha_{\pi^\pm}$ and extended to $\beta_{\pi^\pm}$~\cite{Lee:2023lnx} and to $\alpha_{\pi^0}$ and $\beta_{\pi^0}$~\cite{Lee:2024dgf}. A similar approach involving the four-point function was worked out in~\cite{Wang:2023omf} using position-space formulas for the Compton tensor for proton and neutron electric polarizabilities.

We note that the operators discussed within the form factor method also carry important information when evaluated not within hadron states but in the vacuum. These matrix elements are related to the magnetic susceptibility of the QCD medium~\cite{Bali:2015msa}, as we discuss in Sec.~\ref{sec:eos_currentcurrent}. Moreover, these also appear in the hadronic contribution to the anomalous magnetic moment of the muon, in the focus of current precision lattice QCD simulations~\cite{Meyer:2018til}.

\subsection{Hadron properties for weak background fields}
\label{sec:had_latdethadprop}

The above discussed hadron properties -- magnetic moments and magnetic and electric polarizabilities -- have been calculated in numerous lattice studies. In this section we provide a summary of these efforts and illustrate the results in a selected set of figures.
In Tab.~\ref{tab:had_list} we list the lattice results for magnetic moments $\hat\mu$, magnetic polarizabilities $\beta$ and electric polarizabilities $\alpha$ of hadrons. We indicate whether the corresponding study used the direct method or a weak-field expansion; quenched, electroquenched or dynamical ensembles; which hadrons were analyzed; and what was the main observable that was calculated.

\begin{table}[!]
  \renewcommand*{\arraystretch}{.86}
 \centering
 \begin{tabular}{c|c|c|c}
  reference & approach & hadrons & observable \\ \hline\hline
  \cite{Martinelli:1982cb} & direct, quenched, & $n, p$ & $\hat\mu$ \\
  \cite{Bernard:1982yu} & direct, quenched  & $n, p, \Omega^-$ & $\hat\mu$ \\
  \cite{Fiebig:1988en} & direct, quenched  & $\pi^0, n, \rho^0$ & $\alpha$ \\
  \cite{Draper:1989pi} & weak-field expansion, quenched  & $n, p$ & $\hat\mu$ \\
  \cite{Wilcox:1991cq} & weak-field expansion, quenched  & $n, p$ & $\hat\mu$ \\
  \cite{Leinweber:1992hy} & weak-field expansion, quenched  & $n, p, \Lambda, \Delta, \Omega$ & $\hat\mu$ \\
  \cite{Rubinstein:1995hc} & direct, quenched  & $n, p$ & $\hat\mu$ \\
  \cite{Gadiyak:2001fe} & weak-field expansion, quenched  & $n, p$ & $\hat\mu$ \\
  \cite{Gockeler:2003ay} & weak-field expansion, quenched  & $n, p$ & $\alpha$ \\
  \cite{Christensen:2004ca} & direct, quenched  & $\rho^0, K^{*0}, n, \Sigma^0, \Lambda^0, \Xi^0, \Delta^0, \Sigma^{*0}, \Xi^{*0}$ & $\alpha$ \\
  \cite{Lee:2005ds} & direct, quenched  & $n, p, \Sigma, \Xi, \Delta$ & $\hat\mu$  \\
  \cite{Lee:2005dq} & direct, quenched  & $n, p, \Sigma, \Xi, \Delta, \pi, \rho, K$ & $\beta$  \\
  \cite{Boinepalli:2006xd} & direct, quenched  & $n, p$ & $\hat\mu$ \\
  \cite{Shintani:2006xr} & direct, quenched  & $n$ & $\alpha$\\
  \cite{Engelhardt:2007ub} & weak-field expansion, dynamical  & $n$ & $\alpha$ \\
  \cite{Lee:2008qf} & direct, quenched  & $\rho, K^*, a_1, b_1, K_t^*$ & $\hat\mu$ \\
  \cite{Aubin:2008qp} & direct, electroquenched  & $\Delta, \Omega^-$ & $\hat\mu$ \\
  \cite{Detmold:2009dx} & direct, electroquenched  & $\pi^0, \pi^\pm, K^0, K^\pm$ & $\alpha$  \\
  \cite{Detmold:2010ts} & direct, electroquenched  & $n,p$ & $\hat\mu$, $\alpha$  \\
  \cite{Primer:2013pva} & direct, electroquenched  & $n, p$ & $\hat\mu, \beta$  \\
  \cite{Beane:2014ora} & direct, electroquenched  & $n, p, d, {}^3\!He, {}^3\!H$ & $\hat\mu$  \\
  \cite{Luschevskaya:2014lga} & direct, quenched  & $\pi^0, \rho^0$ & $\beta$ \\
  \cite{Freeman:2014kka} & direct, reweighted   & $\pi^0, K^0, n$ & $\alpha$  \\
  \cite{Lujan:2014kia} & direct, electroquenched  & $\pi^0, K^0, n$ & $\alpha$ \\
  \cite{Luschevskaya:2015cko} & direct, quenched  & $\pi$ & $\beta$ \\
  \cite{Chang:2015qxa} & direct, electroquenched  & $n, p, d, nn, pp, {}^3\!He, {}^3\!H, {}^4\!He$ & $\hat\mu, \beta$  \\
  \cite{Luschevskaya:2016epp} & direct, quenched  & $\rho^\pm, K^{\pm*}$ & $\hat\mu, \beta$ \\
  \cite{Parreno:2016fwu} & direct, electroquenched  & $n, p, \Sigma, \Xi, \Lambda$ & $\hat\mu$ \\
  \cite{Lujan:2016ffj} & direct, electroquenched  & $\pi^0, K^0, n$ & $\alpha$  \\
  \cite{Bali:2017ian} & direct, quenched  & $\pi, \rho$ & $\hat\mu, \beta$ \\
  \cite{Bignell:2018acn} & direct, electroquenched  & $n$  & $\beta$  \\
  \cite{Bignell:2019vpy} & direct, electroquenched  & $\pi^0$ & $\beta$  \\
  \cite{Ding:2020hxw} & direct, dynamical  & $\pi, K, \eta_s$ & $\beta$ \\
  \cite{Bignell:2020dze} & direct, electroquenched  & $\pi^0, \pi^\pm$ & $\beta$ \\
  \cite{Bignell:2020xkf} & direct, electroquenched  & $n, p$ & $\beta$ \\
  \cite{Wilcox:2021rtt} & weak-field expansion  & $\pi^\pm, p$ & $\alpha, \beta$ \\
  \cite{Niyazi:2021jrz} & direct, electroquenched  & $\pi^\pm$ & $\alpha$  \\
  \cite{Wang:2023omf} & weak-field expansion, dynamical  & $n, p$ & $\alpha$ \\
  \cite{Lee:2023lnx} & weak-field expansion, quenched  & $\pi^\pm$ & $\beta$ \\
  \cite{Lee:2023rmz} & weak-field expansion, quenched  & $\pi^\pm$ & $\alpha$ \\
  \cite{Kabelitz:2024aye} & direct, electroquenched & $n, p, \Sigma, \Xi$ & $\beta$ \\
  \cite{Lee:2024dgf} & weak-field expansion, quenched & $\pi^0$ & $\alpha, \beta$ \\
 \end{tabular}
 \caption{
 \label{tab:had_list}
 Lattice studies of magnetic moments and magnetic and electric polarizabilities of hadrons.
 }
\end{table}

\subsubsection{Magnetic moment}

Starting with the pioneering works of~\cite{Martinelli:1982cb,Bernard:1982yu,Draper:1989pi,Wilcox:1991cq,Leinweber:1992hy,Rubinstein:1995hc}, there have been a number of lattice QCD calculations of the magnetic moments of octet baryons~\cite{Lee:2005ds,Detmold:2010ts,Primer:2013pva}, decuplet
baryons~\cite{Aubin:2008qp} and of tensor, axial~\cite{Lee:2008qf} and vector mesons~\cite{Luschevskaya:2016epp,Bali:2017ian}, among other studies, see Tab.~\ref{tab:had_list}. 

\begin{figure}
 \centering
 \includegraphics[width=8.cm]{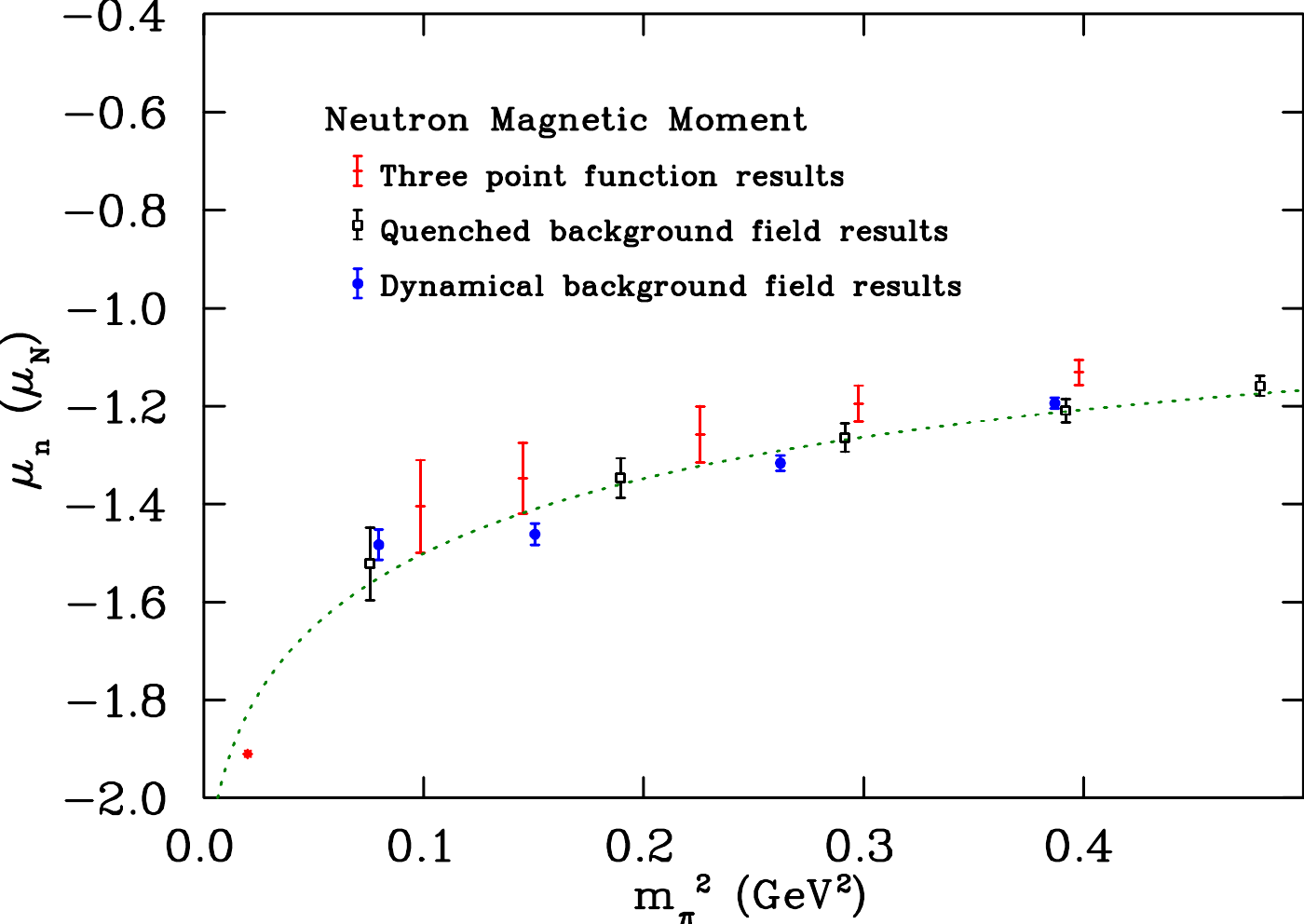}\qquad
 \includegraphics[width=7.9cm]{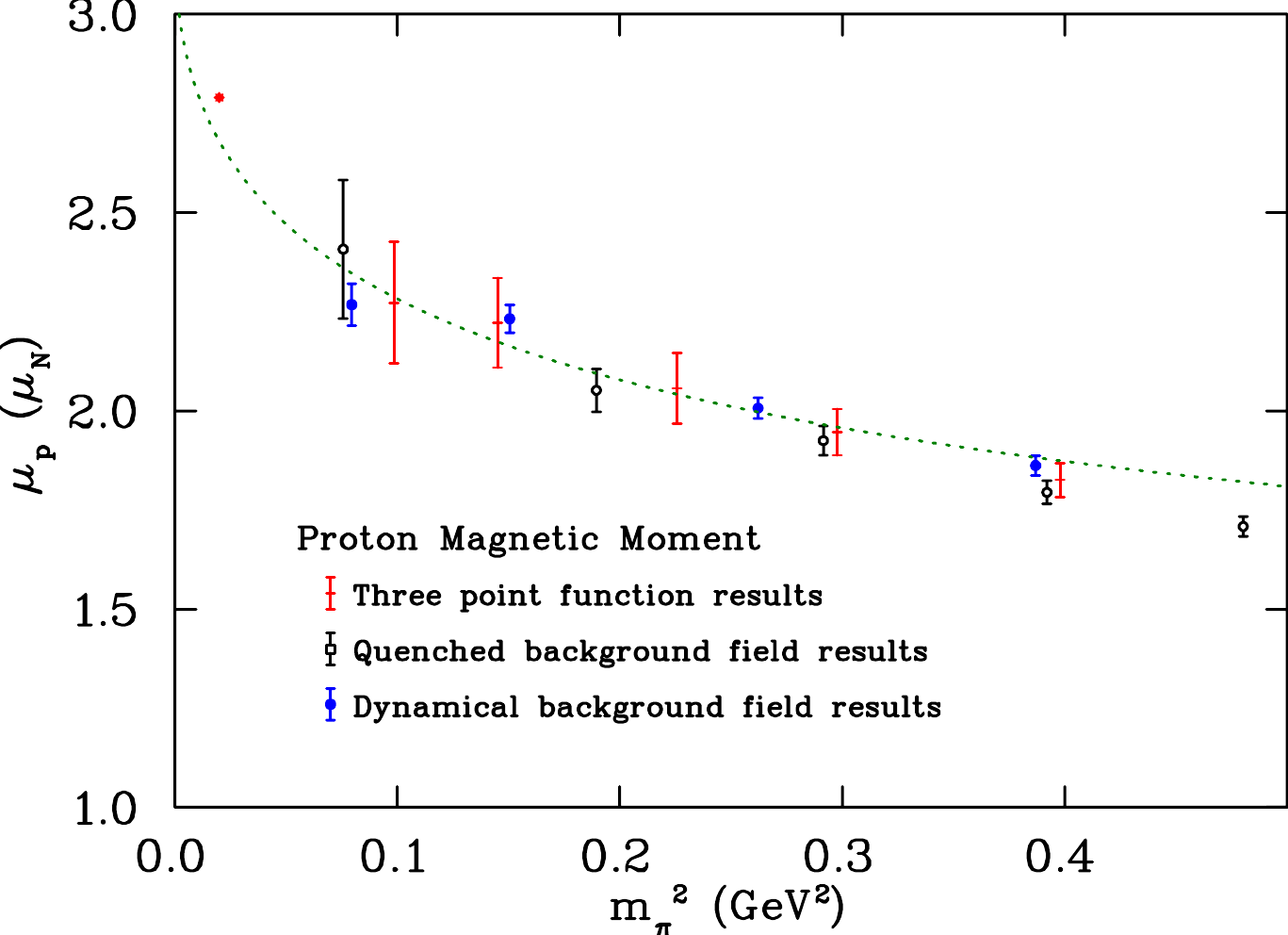}
 \caption{
 \label{fig:had_nucleon_magmom}
 Magnetic moments of the neutron (left panel) and of the proton (right panel) as a function of the valence pion mass~\cite{Primer:2013pva}. The leftmost points represent the experimental values and the dashed lines a chiral fit. Besides the quenched and electroquenched results of~\cite{Primer:2013pva} (the latter labeled as dynamical) based on the direct approach, also included are the weak-field expansion results of~\cite{Boinepalli:2006xd}.
 }
\end{figure}

Most of the results in the literature followed the direct approach and used quenched ensembles with Wilson valence quarks. An example for this setup is~\cite{Lee:2005ds}, where the magnetic moments of octet and decuplet baryons have been measured. We note that this study used a photon field consisting of only the $u_{2f}(n_1)$ links of~\eqref{eq:lat_hom_links}, while the $u_{1f}(n)$ were set to unity. Abandoning the periodic boundary conditions in the $x_1$ direction and using Dirichlet boundary conditions instead, allowed to achieve arbitrary small magnetic field amplitudes. The resulting breaking of translational invariance, on the other hand, was observed to lead to enhanced finite volume effects~\cite{Lee:2005ds}.
A study using the direct approach with periodic boundary conditions and quantized magnetic fields, focusing on the proton and the neutron, was carried out in~\cite{Primer:2013pva}. The results for $\hat\mu_p$ and $\hat\mu_n$ are included in Fig.~\ref{fig:had_nucleon_magmom}, revealing a smooth approach to the physical point and reasonable agreement with the experimental values.

\begin{figure}
 \centering
 \includegraphics[width=7.9cm]{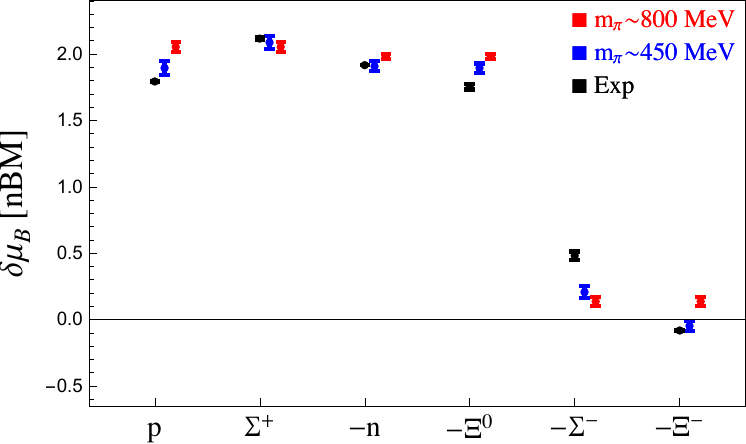}\qquad
 \raisebox{-.4cm}{\includegraphics[width=7.7cm]{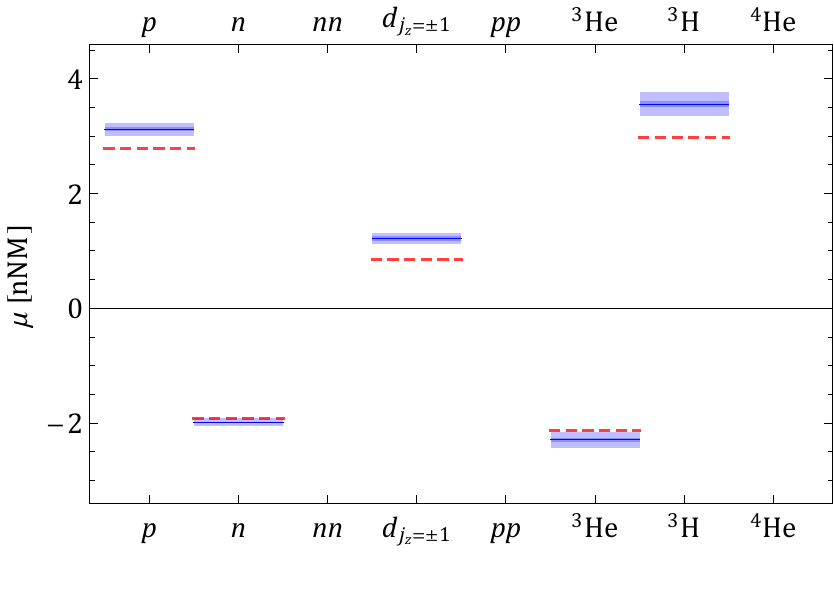}}
\caption{
 \label{fig:had_anom_magmom}
 Left panel: anomalous magnetic moments $\delta\hat\mu=\hat\mu-q/(2Me)$ of the baryon octet for two different vacuum pion masses~\cite{Parreno:2016fwu}.
 Right panel: magnetic moment of hadrons and light nuclei in natural nuclear magnetons~\cite{Chang:2015qxa} in the three-flavor symmetric theory with a vacuum pion mass $M_\pi\approx800\MeV$. The lattice results (blue bands) are compared to the experimental values (red dashed lines).
 }
\end{figure}

The magnetic moments of the baryon octet were calculated using the direct method in~\cite{Parreno:2016fwu}. Here, two vacuum pion masses of $m_\pi\approx800\MeV$ and $m_\pi\approx450\MeV$ were used to perform chiral extrapolations for two lattice spacings. The heavier setup corresponds to the three-flavor symmetric point, where the electroquenched approximation, employed by this study, becomes exact as we argued above. The importance of the choice of magnetic moment units for the comparison to experimental values is discussed in detail. In particular, the best agreement is achieved when the magnetic moments are normalized by natural baryon magnetons (i.e.\ those obtained from the vacuum baryon masses measured on the lattice). The results for the anomalous magnetic moments $\delta\hat\mu$ are shown in the left panel of Fig.~\ref{fig:had_anom_magmom}. While $\Sigma^-$ and $\Xi^-$ baryons behave, in this respect, as point-like particles with $\delta\hat\mu\approx0$, the other baryons show large anomalous contributions.

The direct method has also been generalized to determine the magnetic moments of light nuclei~\cite{Beane:2014ora,Chang:2015qxa} containing up to four nucleons. These studies discussed the three-flavor theory, where 
the electroquenched approximation becomes exact. 
For the pion mass used here, $M_\pi\approx800\MeV$, the dineutron and diproton also form bound states (of the strong interactions) and have also been considered. The results for the magnetic moments are shown in the right panel of Fig.~\ref{fig:had_anom_magmom}.

In this context it is worth mentioning that the shift of such two-nucleon energy levels due to a background magnetic field in a finite volume also contains information about nucleon scattering amplitudes and transition matrix elements. Based on L\"uscher's formalism~\cite{Luscher:1986pf,Luscher:1990ux}, the corresponding methodology was developed in~\cite{Detmold:2004qn,Meyer:2012wk} and employed to calculate the cross sections for the radiative capture process $np\to d\gamma$ and the inverse deuteron photo-disintegration process~\cite{Beane:2015yha}. Moreover, strong magnetic fields were also found to reduce the binding energy of two-hyperon and two-nucleon states, potentially unbinding the deuteron~\cite{Detmold:2015daa}. It will be interesting to see whether these findings, obtained in~\cite{Detmold:2015daa} with larger-than-physical quark masses, persist at the physical point as well.

\subsubsection{Electric dipole moment}
\label{sec:had_eldipmom}

As mentioned below~\eqref{eq:had_alphae_def}, the electric field  does not introduce a linear effect in hadron energies due to time reversal symmetry. However, in a CP-odd environment, like in the presence of a $\theta\cdot Q_{\rm top}$ term in the QCD action, containing the topological charge~\eqref{eq:lat_topcharge_def}, a linear electric field-dependence does appear in the energy. This dependence, which characterizes the electric dipole moment of the hadron, was determined on the lattice for the neutron by considering the shift in the neutron energy in the presence of background electric fields and a nonzero $\theta$ term~\cite{Shintani:2006xr}. Similarly to the analyses of magnetic moments, the electric dipole moment was found by investigating the difference of neutron energies for the spin being aligned and anti-aligned with the electric field. This approach was employed in the quenched~\cite{Shintani:2006xr} and the electroquenched approximation~\cite{Shintani:2008nt} using domain wall and Wilson-clover quarks. In these studies, the electric field is introduced with non-periodic boundary conditions and the $\theta$ term is included via reweighting. The electric dipole moment of the neutron has also been calculated via the form factor method using domain wall fermions~\cite{Berruto:2005hg,Shintani:2015vsx}, Wilson-clover quarks~\cite{Guo:2015tla}, twisted-mass clover fermions~\cite{Alexandrou:2015spa,Alexandrou:2020mds}, HISQ fermions~\cite{Bhattacharya:2021lol} and overlap quarks~\cite{Liang:2023jfj}.
A comparison of the background electric field method and the form factor method was performed in~\cite{Abramczyk:2017oxr}.

\subsubsection{Magnetic polarizability}

\begin{figure}
 \centering
 \includegraphics[width=8.2cm]{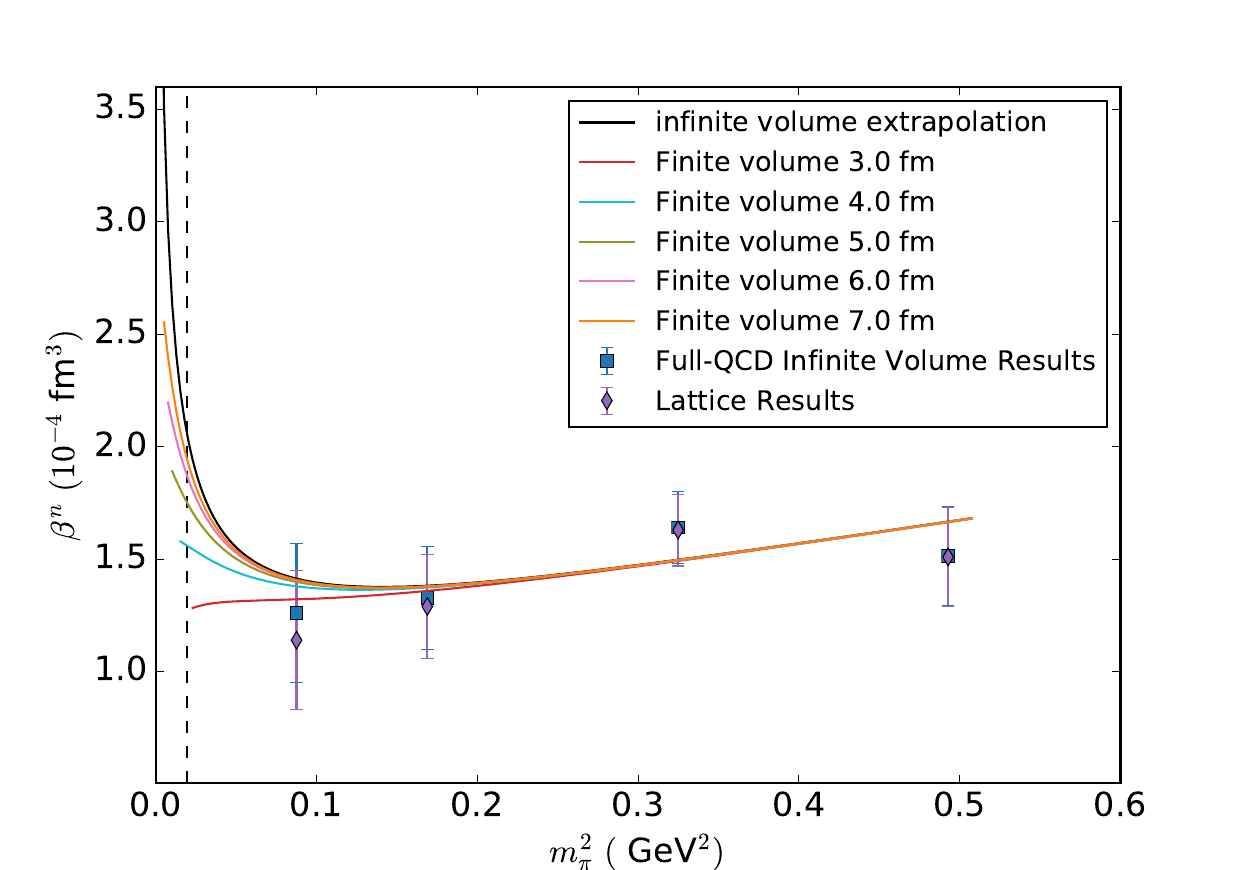}
 \raisebox{-.7cm}{\includegraphics[width=8.3cm]{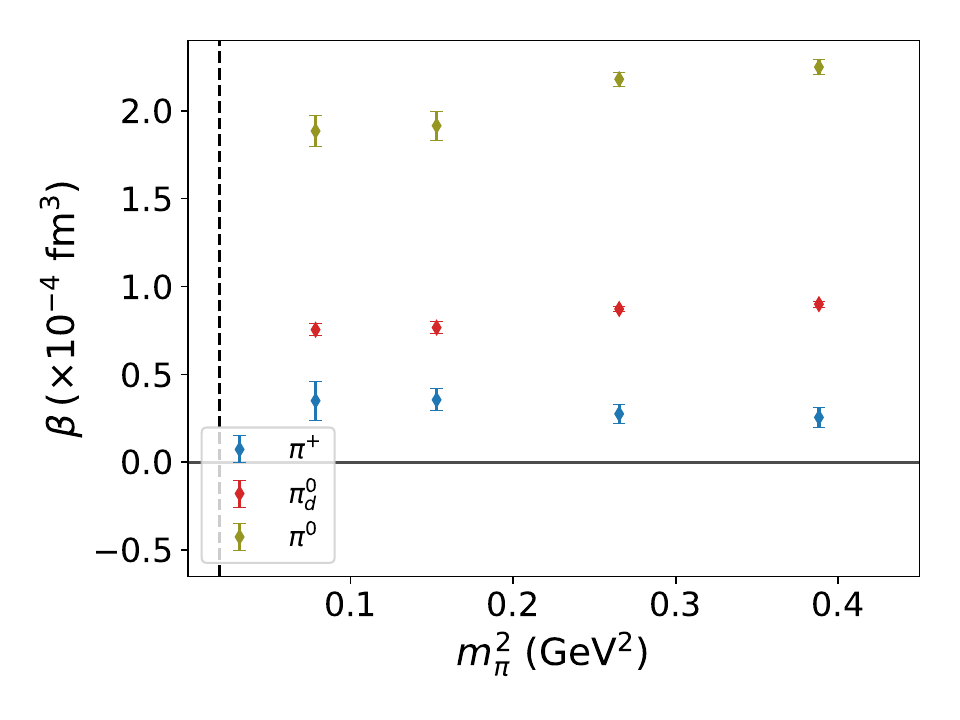}}
 \caption{
 \label{fig:had_pion0_mass1}
  Left panel: magnetic polarizability of the neutron as a function of the vacuum pion mass, together with a chiral extrapolation guided by chiral perturbation theory for different volumes~\cite{Bignell:2018acn}.
  Right panel: magnetic polarizability $\beta_\pi$ of pions (denoted here as $\beta$) as a function of the vacuum pion mass~\cite{Bignell:2020dze}. 
 }
\end{figure}

Many of the above discussed studies of magnetic moments that used the direct approach, also presented results for magnetic polarizabilities of hadrons, see Tab.~\ref{tab:had_list}. 
For example, the same lattice setup as in~\cite{Lee:2005ds} was used in~\cite{Lee:2005dq} to determine the magnetic polarizabilities of octet and decuplet baryons as well as of various mesons. 
A similar analysis, dedicated to calculate $\beta_n$, was also performed in~\cite{Primer:2013pva}.
The direct method has been applied to determine the neutral pion and neutron magnetic polarizabilities with Wilson fermions. Employing sink operators projected to the lowest Landau-level turned out to efficiently increase the signal-to-noise ratio in the correlators~\cite{Bignell:2018acn,Bignell:2019vpy}. The results for the neutron magnetic polarizability are shown in the left panel of Fig.~\ref{fig:had_pion0_mass1}. As visible from the figure, the extrapolation towards the physical point involves large systematic effects due to the finite volume.

\begin{figure}
 \centering
 \includegraphics[width=8.cm]{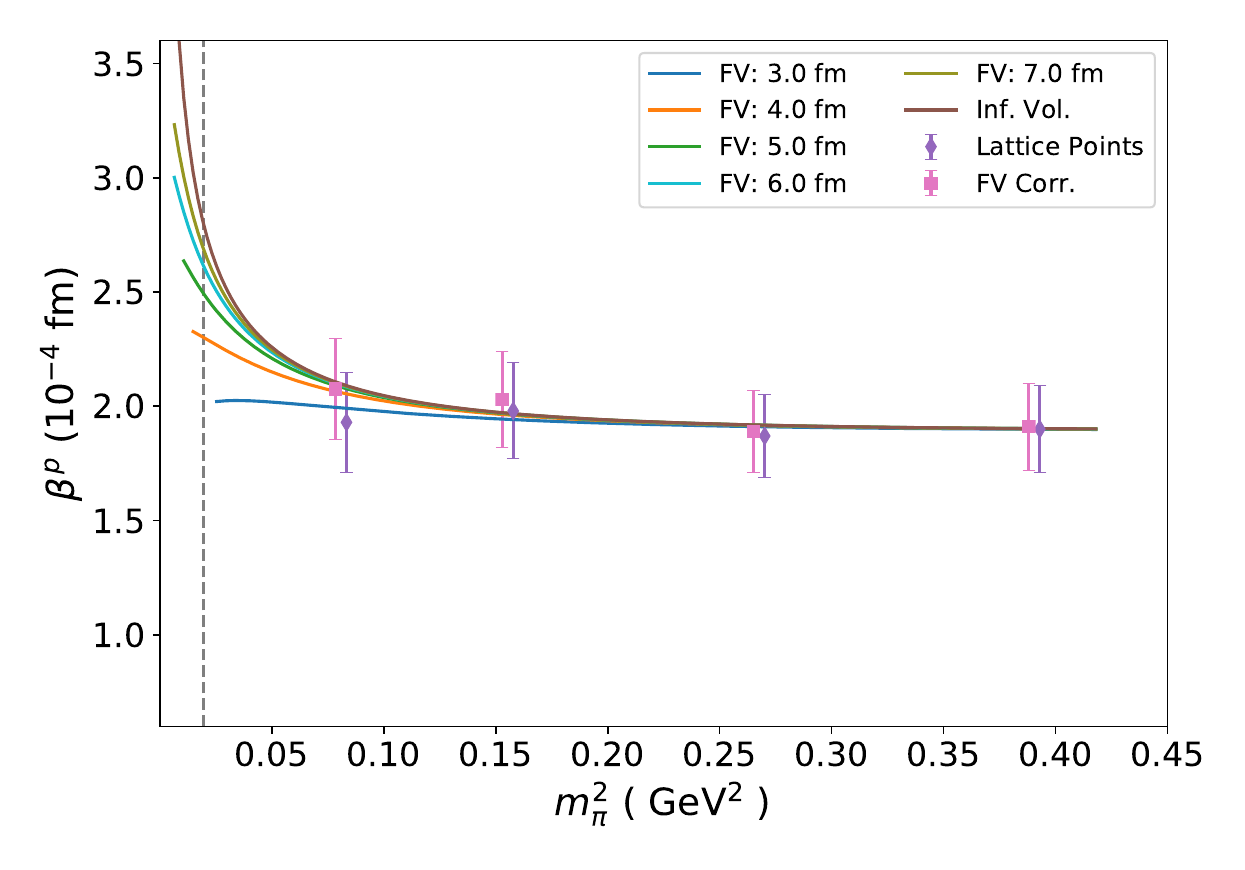}\qquad
 \raisebox{-.1cm}{\includegraphics[width=8.cm]{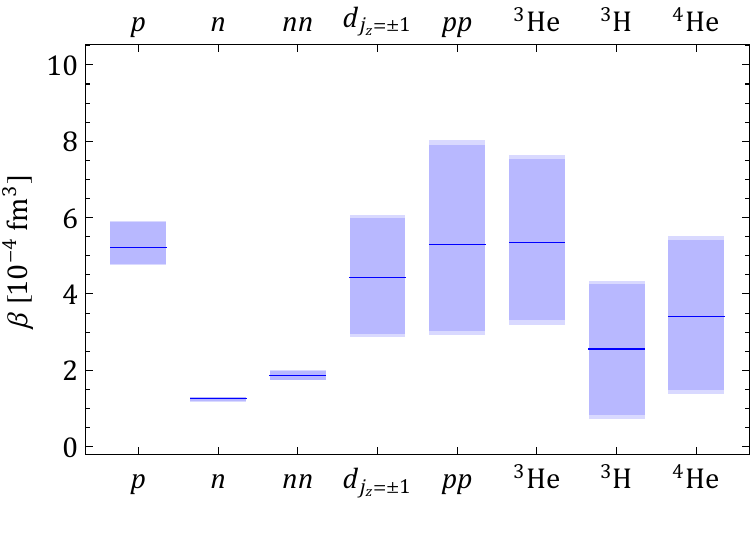}}
 \caption{
 \label{fig:had_proton}
 Left panel: magnetic polarizability of the proton as a function of the vacuum pion mass, together with an extrapolation guided by chiral perturbation theory for different volumes~\cite{Bignell:2020xkf}.
 Right panel: magnetic polarizability of hadrons and light nuclei~\cite{Chang:2015qxa} in the three-flavor symmetric theory with a vacuum pion mass $M_\pi\approx800\MeV$.
 }
\end{figure}

Pion magnetic polarizabilities were calculated using the background field-corrected Wilson clover action in~\cite{Bignell:2020dze}. This study used sources projected to the lowest Landau-level in order to optimally isolate the ground state for the charged pion. The $\beta_\pi$ values obtained for four different vacuum pion masses are shown in the right panel of Fig.~\ref{fig:had_pion0_mass1}. In the neutral sector, the polarizabilities of the connected states $\pi^u$ and $\pi^d$ were calculated and their average was quoted as an estimate for the true neutral pion, also shown in the figure.
The same setup was used in~\cite{Bignell:2020xkf} to calculate $\beta_p$ and $\beta_n$ at different vacuum pion masses and extrapolated to the physical point using $\chi$PT, see the left panel of Fig.~\ref{fig:had_proton}. Very recently, these techniques were further developed to determine the magnetic polarizabilities of other members of the baryon octet~\cite{Kabelitz:2024aye}.

Neutral connected pions were also considered in~\cite{Bali:2017ian} employing quenched Wilson quarks. Here, the removal of magnetic field-dependent lattice artefacts from the quark masses was found to alleviate the continuum limit substantially.
Overlap fermions were also used to determine $\beta$ in the quenched approximation for neutral (connected) pions and neutral $\rho$ mesons in~\cite{Luschevskaya:2014lga}, for charged and neutral pions in~\cite{Luschevskaya:2015cko} and for charged $\rho$ and $K^*$ mesons in~\cite{Luschevskaya:2016epp}. The latter studies also discussed the $\O(B^4)$ contributions to the meson energies, the so-called hyperpolarizability.
Dynamical HISQ fermions were also used to determine $\beta$ for $\pi^{0,\pm}$ and $K^{0,\pm}$ pseudoscalar mesons~\cite{Ding:2020hxw}. This study used a fully dynamical setup, i.e.\ the background field was taken into account also for sea quarks. Distinguishing the polarizability effect from the lowest Landau-level shift, proportional to $B$ in the charged pion energy, becomes a complicated task in this case.

Finally, the study~\cite{Chang:2015qxa} also extended the background field analysis of magnetic polarizabilities to light nuclei. The results, obtained for the three-flavor symmetric theory with a vacuum pion mass $M_\pi\approx 800\MeV$ (where the dineutron and diproton also form bound states) are shown in the right panel of Fig.~\ref{fig:had_proton}.

\subsubsection{Electric polarizability}
\label{sec:had_electr_pol}

The first lattice result for meson and baryon electric polarizabilities were obtained in quenched QCD with staggered valence quarks~\cite{Fiebig:1988en} and in the Wilson formulation~\cite{Christensen:2004ca}. 
More recently, neutral pion, kaon and neutron electric polarizabilities were calculated in~\cite{Lujan:2014kia} using the electroquenched approximation with Wilson clover fermions on dynamical ensembles. This study found several interesting and unexpected results.

\begin{figure}
 \centering
 \includegraphics[width=8.8cm]{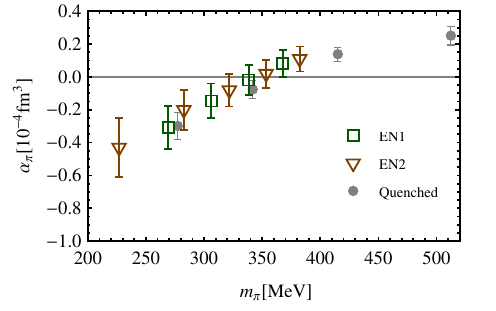}\quad
 \raisebox{.1cm}{\includegraphics[width=8.5cm]{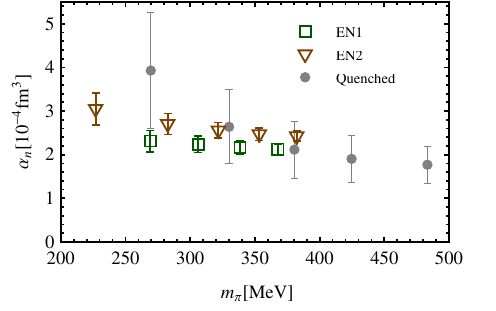}}
 \caption{
 \label{fig:had_neutron_elpol}
 Electric polarizability of the neutral pion (left panel) and of the neutron (right panel) for two different valence quark masses (labeled as EN1 and EN2), as a function of the sea pion mass~\cite{Lujan:2014kia}, and compared to the quenched results (gray dots)~\cite{Alexandru:2010dx}. 
 }
\end{figure}

First, a negative trend for $\alpha_{\pi^0}$ was observed, see the left panel of Fig.~\ref{fig:had_neutron_elpol}. This tendency, found already in previous studies~\cite{Detmold:2009dx}, is in disagreement with (electroquenched) $\chi$PT, which predicts positive values of $\alpha_{\pi^0}$ around the physical point.  
It was speculated that a possible reason for this deviation are finite volume effects for the Dirichlet boundary conditions used in~\cite{Lujan:2014kia}. These were analyzed in detail in~\cite{Lujan:2016ffj}, revealing that finite volume effects in the direction of the electric field only decay as an inverse power of the lattice extent $L$ and not exponentially, as at $E=0$~\cite{Lujan:2016ffj}. 
However, it was found that even after the infinite volume extrapolation guided by $\chi$PT was carried out, the negative trend for $\alpha_{\pi^0}$ persists~\cite{Lujan:2016ffj}.
Another possible explanation for the negative values of the electric polarizability for light pions is the electroquenched approximation used by all of these studies. Going beyond this approximation via reweighting~\cite{Freeman:2014kka} did indicate that the sign of $\alpha_{\pi^0}$ changes, albeit within large statistical errors. This is an important point for future research.

\begin{figure}
 \centering
 \includegraphics[width=8.8cm]{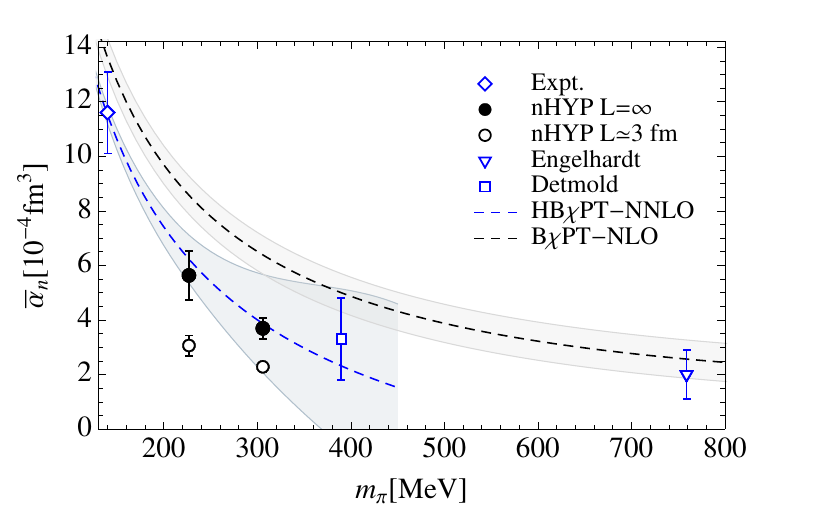}
 \raisebox{.1cm}{\includegraphics[width=8.cm]{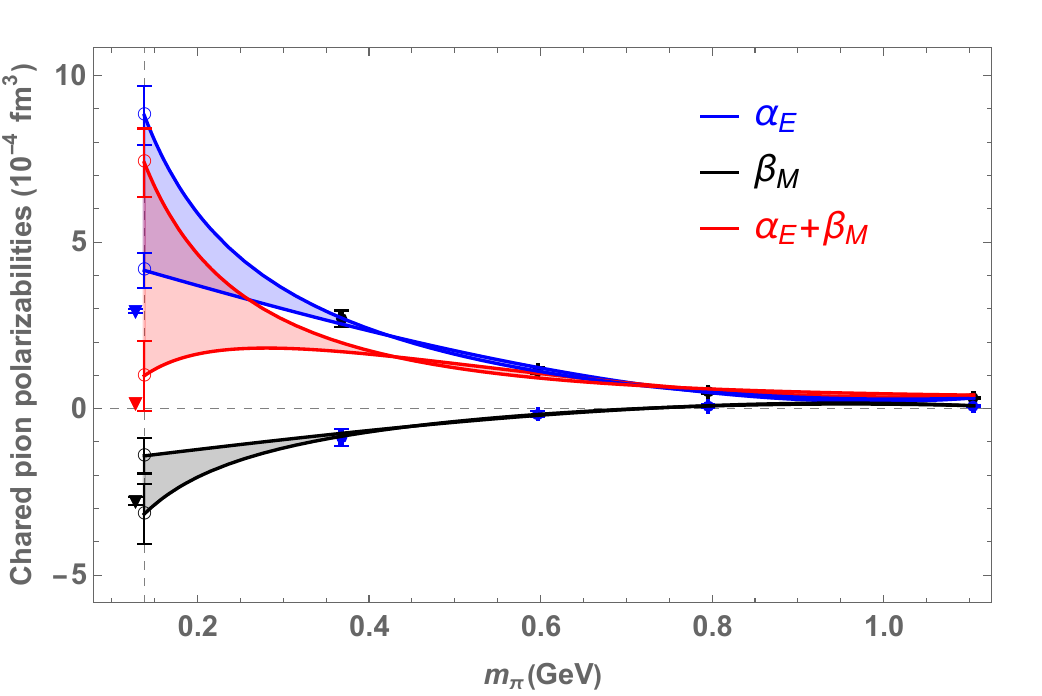}}
 \caption{
 \label{fig:had_neutron_elpol_infvol}
 Left panel: Compton scattering polarizability of the neutron $\bar\alpha_{n}$ as a function of the vacuum pion mass. The results of~\cite{Lujan:2014kia} for light pions (open circles) are extrapolated to the infinite volume limit (filled circles)~\cite{Lujan:2016ffj}, closing the gap to the chiral perturbation theory prediction (shaded bands). Comparisons to the lattice results of~\cite{Engelhardt:2007ub} and~\cite{Detmold:2010ts} are also included, as well as the experimental value.
 Right panel: magnetic and electric polarizabilities of the charged pion (denoted here as $\beta_M$ and $\alpha_E$, respectively) as well as their sum, as functions of the vacuum pion mass, together with a chiral extrapolations inspired by $\chi$PT~\cite{Lee:2023lnx}.
 }
\end{figure}

The study~\cite{Lujan:2014kia} also calculated $\bar\alpha_n$ (based on results for $\hat\mu_n$ from~\cite{Lee:2005ds}, cf.\ the relation~\eqref{eq:had_compton_neutron}). 
The results, shown in the right panel of Fig.~\ref{fig:had_neutron_elpol}, compatible with earlier lattice findings~\cite{Shintani:2006xr}, revealed a disagreement with the $\chi$PT prediction, which suggests considerably larger values for the polarizability at these pion masses.
In order to eliminate finite volume effects, the study~\cite{Lujan:2016ffj} performed an infinite volume extrapolation guided by $\chi$PT. The so obtained corrected results are shown in the left panel of Fig.~\ref{fig:had_neutron_elpol_infvol}, together with the prediction of $\chi$PT in the infinite volume. The lattice and the NNLO $\chi$PT results are now consistent with each other as well as with the experimental value. 
Charged sea quark effects for the neutron were also considered in~\cite{Freeman:2014kka} via perturbative reweighting. 
Here, the reweighting performed to $\O(E)$ and to $\O(E^2)$ mostly only lead to enhanced statistical errors.
We note that the work~\cite{Engelhardt:2007ub} also considered a perturbative expansion of the reweighting factors, including disconnected diagrams. 
Here, $\alpha_n$ was calculated using domain wall valence quarks at a pion mass of about $750\textmd{ MeV}$.

\begin{figure}
 \centering
 \includegraphics[width=8.3cm]{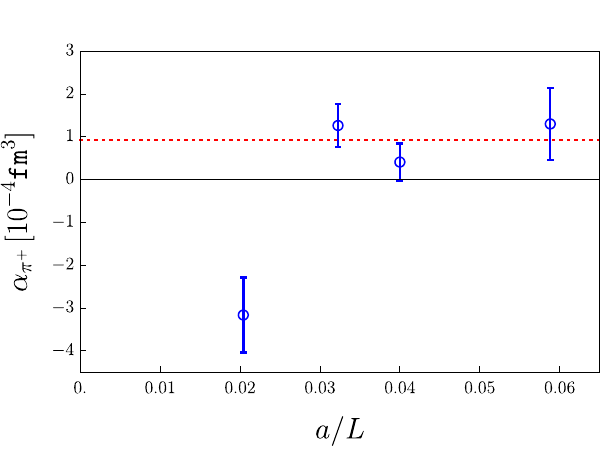}
 \caption{
 \label{fig:had_pionplus_elpol}
 Volume-dependence of the electric polarizability $\alpha_{\pi^+}$ of the charged pion. The lattice spacing is $a=0.1245\fm$ and the red dashed line indicates the prediction of leading-order chiral perturbation theory for this vacuum pion mass, $m_\pi=315\MeV$~\cite{Niyazi:2021jrz}.
 }
\end{figure}

Concerning further hadrons, the neutral kaon electric polarizability was found to depend very mildly on the light and strange quark mass, in contrast to $\alpha_{\pi^0}$ and $\alpha_n$~\cite{Lujan:2014kia,Lujan:2016ffj}.
Moreover, finite volume effects were found to be relatively mild also in this case~\cite{Lujan:2016ffj}. 
Finally, a study based on four-point functions calculated the magnetic and electric polarizabilities of the charged pion, see the right panel of Fig.~\ref{fig:had_neutron_elpol_infvol}.
The latest study of $\alpha_{\pi^\pm}$~\cite{Niyazi:2021jrz} concentrated on the investigation of finite volume corrections, see Fig.~\ref{fig:had_pionplus_elpol}. For the largest volume, the results turn negative -- a surprising finding that is yet to be understood.

\subsection{Hadron properties for strong magnetic fields}
\label{sec:had_massesB}

The results above concerned the weak-field behavior of hadron energies, directly relevant for scattering experiments. Turning to strong fields, an initial motivation to study the dependence of $\E(B)$ in this case was supplied by the conjecture that the QCD vacuum might exhibit a superconducting phase. This phase was claimed to involve vector meson condensation, expected to leave an imprint on charged $\rho$ meson energies~\cite{Chernodub:2010qx}.
This argument sparked several subsequent studies on the strong magnetic field-behavior of meson masses, which we review next.

At nonzero $B$, the hadron states may be characterized by their electric charges and spin projections $s_3$ in the direction of the magnetic field. Retaining the superscripts to denote the charge, we will indicate the spin by subscripts, e.g.\ $\rho^+_0$ is the positively charged $\rho$ meson with $s_3=0$. The hadron spectrum has degenerate states due to the symmetries of the system. Parity and charge conjugation symmetry imply that the energies of the following pairs coincide,
\be
\pi^+\leftrightarrow \pi^-, \quad
\rho^+_0\leftrightarrow \rho^-_0, \quad
\rho^+_+\leftrightarrow \rho^-_-, \quad
\rho^+_-\leftrightarrow \rho^-_+, \quad
\rho^0_+\leftrightarrow \rho^0_-\,,
\ee
and similarly for other (pseudo)scalar and (axial)vector mesons.
An important issue, relevant for the discussion of the results below, is the mixing between mesonic states in various sectors, as we mentioned already in Sec.~\ref{sec:had_lattechn}. 
In particular, the neutral pion and the isosinglet meson $\eta$ mix in a non-trivial manner. The neutral pion also mixes with the neutral $\rho$ meson with spin projection $s_3=0$ (but not with $s_3=\pm1$ due to the conservation of angular momentum). Thus, the correlation functions of both corresponding operators decay exponentially with the same, lower energy. To extract the higher energy, a two-state fit of the correlators must be performed or a generalized eigenvalue problem has to be considered.

\begin{figure}
 \centering
 \includegraphics[width=8.cm]{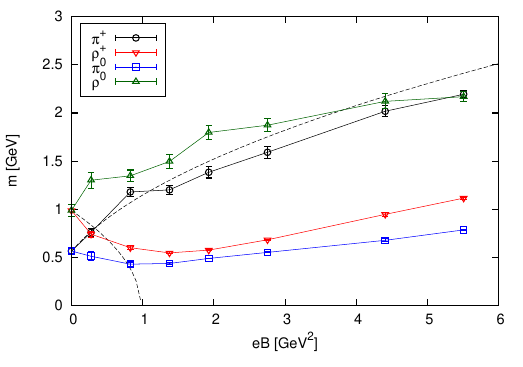}\quad
 \includegraphics[width=8.1cm]{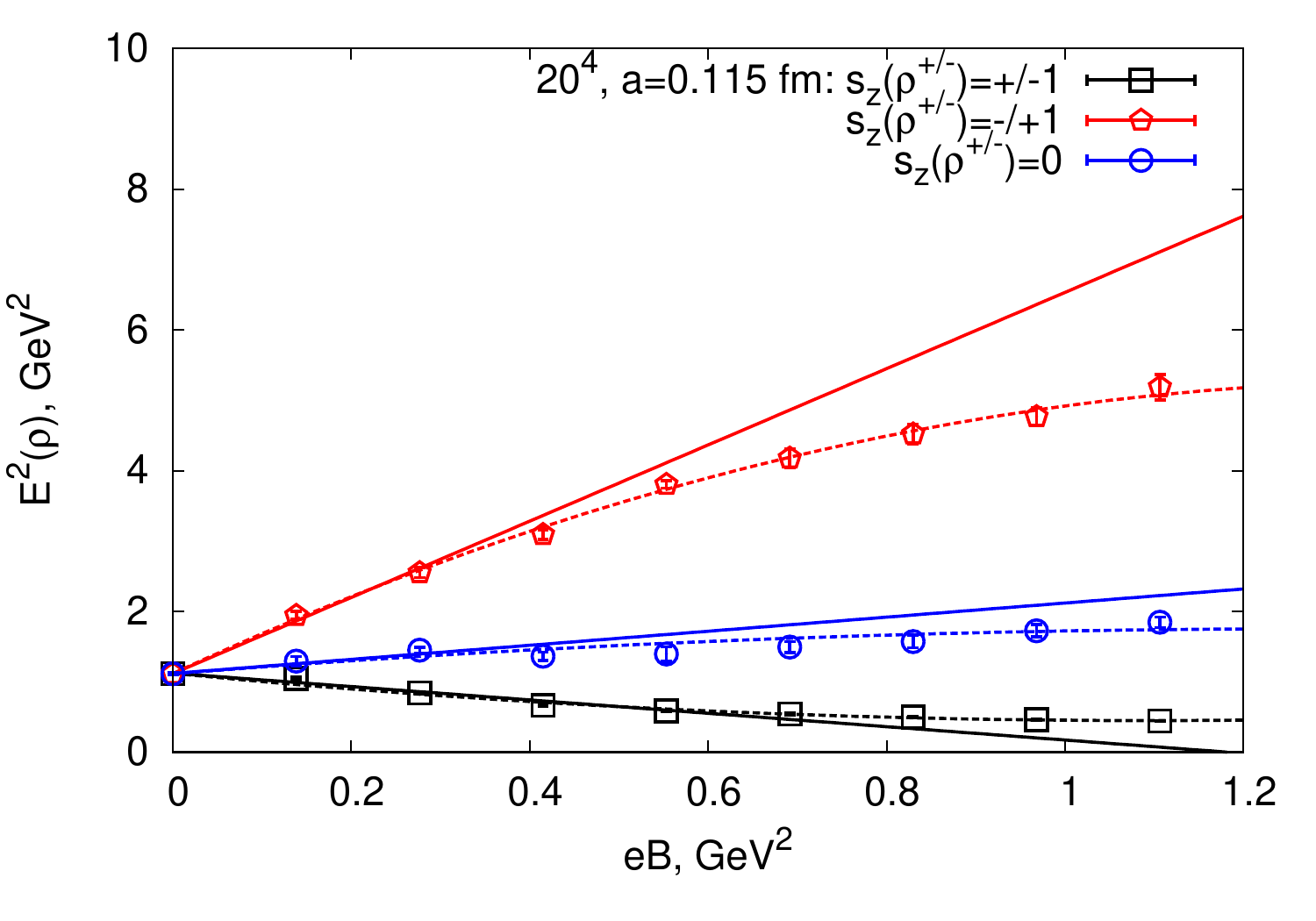}
 \caption{
 \label{fig:had_pi_rho_masses1}
 Left panel: energies of pions and $\rho$ mesons (with nonzero spin), denoted here as $m$, as a function of the magnetic field for a vacuum pion mass of $M_\pi\approx520\MeV$~\cite{Hidaka:2012mz}. The dashed lines represent the prediction for free, point-like particles.
 Right panel: energies of charged $\rho$ mesons with different spin projections~\cite{Luschevskaya:2016epp} with linear fits in $B$ (solid lines) and linear plus quadratic fits (dotted lines).
 }
\end{figure}

The superconducting phase involving $\rho$ meson condensates, predicted in~\cite{Chernodub:2010qx}, was supported initially by first analyses of $\rho$ meson correlators in a magnetic field on moderate lattice sizes~\cite{Braguta:2011hq}, as well as indications for a vortex fluid type behavior~\cite{Braguta:2012fol} in quenched $\mathrm{SU}(2)$ theory.
In turn, this scenario was argued against in the study~\cite{Hidaka:2012mz}, which showed that vector meson condensates are forbidden in QCD by the Vafa-Witten theorem. Moreover, this study demonstrated using quenched Wilson quarks that the energies of charged $\rho$ mesons do not approach zero but always remain positive, see the left panel of Fig.~\ref{fig:had_pi_rho_masses1}. In fact, in this work QCD inequalities were used to derive a lower bound for the charged $\rho$ meson energy in terms of connected neutral pion energies, $E_\rho\ge E_{\pi^{u,d}}$. 
Note that the latter energy was observed to grow with $B$ for strong magnetic fields -- later this finding was shown to be caused by a lattice artefact specific to Wilson fermions, the magnetic field-dependent quark mass renormalization mentioned in Sec.~\ref{sec:had_lattechn}. We get back to this point below.

Neutral mesons in the vector and axial vector channel were also studied in quenched two-color QCD with overlap quarks~\cite{Luschevskaya:2012xd}. A similar analysis for three colors was performed for neutral pions and vector mesons as well in~\cite{Luschevskaya:2014lga}, for charged $\rho$ and $K^*$ vector mesons in~\cite{Luschevskaya:2016epp} and for neutral and charged $K^*$ mesons in~\cite{Luschevskaya:2024iic}. The results of~\cite{Luschevskaya:2016epp} for the charged $\rho$ meson energies are shown in the right panel of Fig.~\ref{fig:had_pi_rho_masses1}. We note that in these studies, the mixing between pseudoscalar mesons and vector mesons with $s_3=0$ was not considered.

The energies for pseudoscalar and vector mesons were calculated for a broad range of magnetic fields in~\cite{Bali:2017ian} using quenched Wilson fermions at three different vacuum pion masses $415\MeV\le M_\pi\le 811\MeV$ and a continuum extrapolation. In this study, the magnetic field-dependent quark mass renormalization, necessary for the unimproved Wilson formulation at $B>0$, was carried out for the first time. This was found to improve the convergence towards the continuum limit significantly, in particular for hadron masses at large magnetic fields.
Another theoretical development in~\cite{Bali:2017ian} was the treatment of the mixing between the charged pion and the charged $\rho$ meson with spin $s_3=0$. 

\begin{figure}
 \centering
 \hspace*{-.6cm}
 \includegraphics[width=9.4cm]{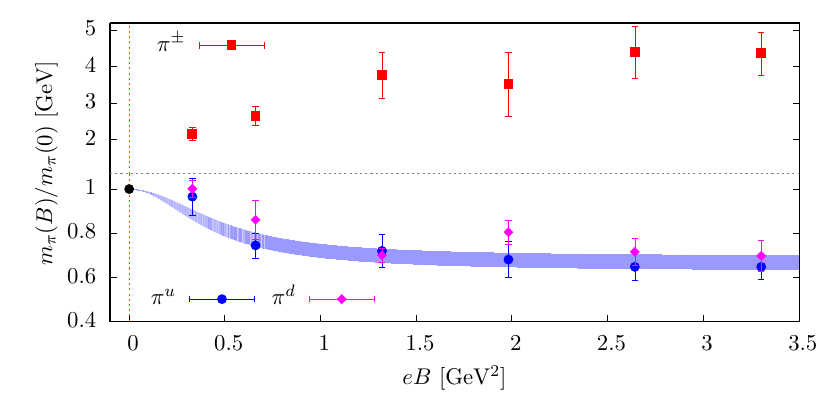}
 \includegraphics[width=9.4cm]{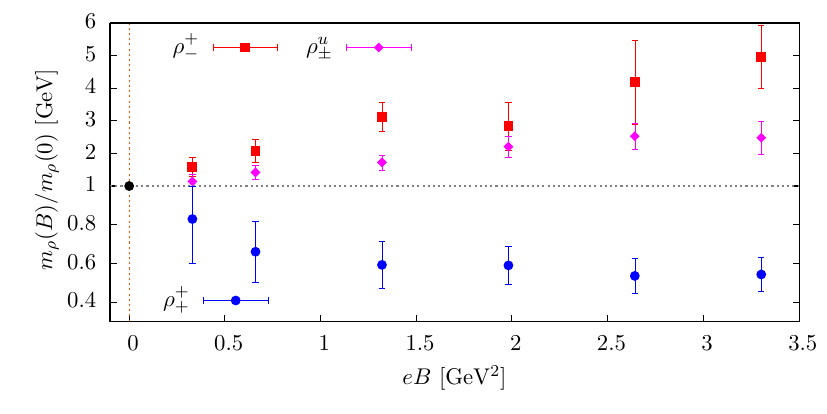}
 \caption{
 \label{fig:had_pi_rho_masses2}
 Energies of pions (left panel) and $\rho$ mesons (right panel), denoted here as $m$, as a function of the magnetic field for a vacuum pion mass of $M_\pi=415\MeV$. The results are continuum extrapolated and shown in units of the corresponding $B=0$ masses~\cite{Bali:2017ian}. The shaded band in the left panel is an interpolation for the connected neutral pion $\pi^u$. The notation in the legends indicates the electric charge as superscripts and the spin projection along the magnetic field as subscripts. 
 }
\end{figure}

The main results of~\cite{Bali:2017ian} are shown in Fig.~\ref{fig:had_pi_rho_masses2} for a vacuum pion mass of $M_\pi=415\MeV$. The charged pion is observed to increase its energy similarly to the point-like approximation~\eqref{eq:had_generalenergy}, giving $\E_{\pi^\pm}=\sqrt{M_\pi^2+eB}$, with a tendency to undershoot this formula for strong fields. This is consistent with the results of~\cite{Bali:2011qj}, obtained using improved staggered quarks with physical masses on four different lattice spacings for magnetic fields $0\le eB\le 0.4 \GeVsq$. 
In turn, the (connected) neutral pion energy was found to be reduced by $B$ and to saturate at about $60\%$ of the vacuum mass. 
Comparing the neutral pion energies in the left panel of Fig.~\ref{fig:had_pi_rho_masses1} and in the left panel of~\ref{fig:had_pi_rho_masses2}, the impact of the removal of magnetic field-dependent lattice artefacts is clearly visible.\footnote{We note that this lattice artefact can be eliminated for weak fields by using the background-field-corrected clover action as well -- this approach was followed in~\cite{Bignell:2019vpy}, giving consistent results.}
Concerning the $\rho$ mesons, the results of~\cite{Bali:2017ian} revealed that the charged $\rho$ meson energy does not approach zero for magnetic fields $eB<3.5\GeVsq$. In fact, the bound due to the QCD inequality~\cite{Hidaka:2012mz} $E_\rho \ge E_{\pi^u}$ was found to be fulfilled. The $\rho$ meson energies are shown in the right panel of Fig.~\ref{fig:had_pi_rho_masses2}.

\begin{figure}
 \centering
 \includegraphics[width=8.6cm]{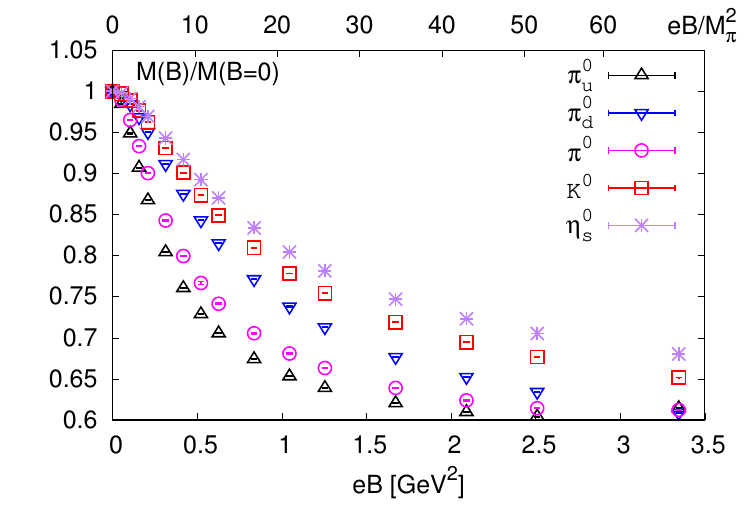}\quad
 \includegraphics[width=8.6cm]{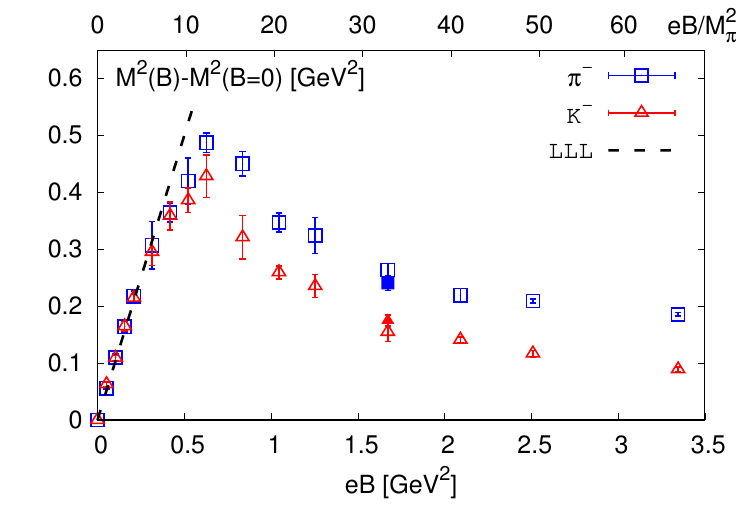}
 \caption{
 \label{fig:had_pi_masses2}
  Left panel: 
  dependence of neutral pseudoscalar meson energies on the magnetic field~\cite{Ding:2020hxw}.
  Right panel: squared charged pion and kaon energy as a function of the magnetic field~\cite{Ding:2020hxw}. The point-like approximation at weak fields (denoted by the dashed line) breaks down at around $eB\approx0.4\GeVsq$. Note that the $B=0$ value is subtracted from the energies.
 }
\end{figure}

Next, we turn to the study of pseudoscalar meson masses with dynamical HISQ fermions at a vacuum pion mass of $M_\pi=220\MeV$~\cite{Ding:2020hxw}.
These results confirm some of the expectations based on earlier Wilson and overlap simulations, namely that the (connected) neutral pion mass reduces monotonously and saturates to around $60\%$ of its vacuum mass as the magnetic field grows. A similar reduction for the energies of $K^0$ and $\eta_s^0=\bar s \gamma_5 s$ meson states was also observed~\cite{Ding:2020hxw}, see the left panel of Fig.~\ref{fig:had_pi_masses2}. Concerning charged pions, the point-like approximation was found to break down at around $eB=0.4\GeVsq$. For $eB>0.6\GeVsq$, the charged pion energy was even observed to decrease, in contrast to the quenched Wilson results~\cite{Bali:2017ian} in the left panel of Fig.~\ref{fig:had_pi_rho_masses2}. This is shown in the right panel of Fig.~\ref{fig:had_pi_masses2}, calling for dedicated future studies and a better understanding of this discrepancy.

We note that besides the impact of the magnetic field on the hadron energy, the effect of $B$ on the spatial deformation of hadrons can also be investigated directly on the lattice. In particular, 
density-density correlators have been evaluated between charged pion and charged and polarized $\rho$ meson states using quenched Wilson quarks in~\cite{Hattori:2019ijy}, revealing an elongation of meson states along $B$.
Finally, we mention that baryon masses were also measured for strong magnetic fields using improved staggered quarks at the physical point~\cite{Endrodi:2019whh}. The dependence of the baryon octet masses on $B$ was used to define a constituent quark mass and employed in matching QCD to low-energy models.

\subsubsection{Neutral pion operators for strong magnetic fields}

We close this section with a discussion on the composition of the neutral pion state $\pi^0$. We already mentioned in Sec.~\ref{sec:had_lattechn} that for $B\neq0$ this state can in general differ from the usual expression $\pi^{0,\rm vac}=(\pi^u-\pi^d)/\sqrt{2}$, valid in the vacuum. A closely related feature of pion correlators is the contribution of disconnected contributions.
According to the arguments of~\cite{Ding:2020hxw} based on Ward identities for integrated correlators, disconnected diagrams give negligibly small contributions, and the physical neutral pion state remains close to the vacuum expression $\pi^{0,\rm vac}$ even for strong magnetic fields. This implies that the relevant correlator is well approximated as $C_{\pi^0}\approx C_{\pi^u}+C_{\pi^d}$.
Disconnected contributions to pion correlators were also found to be suppressed at $B>0$ by the study~\cite{Luschevskaya:2015cko} using overlap fermions and the quenched approximation.

We note that this picture cannot be complete in this form. All lattice results so far have lead to the conclusion that the correlators $C_{\pi^u}$ and $C_{\pi^d}$ give different exponential decays at large distances, see the left panel of Fig.~\ref{fig:had_pi_masses2} for a broad range of magnetic fields. It follows that
it is not possible that the physical pion state at these magnetic fields is $\pi^0=\pi^{0,\rm vac}$ {\it and} that disconnected diagrams are heavily suppressed. Instead, one of the following two scenarios can hold:

\begin{enumerate}
 \item[a)] Disconnected diagrams contributing to $C_{\pi^{u,d}}$ are negligible. The two operators $\pi^{u,d}$ couple to different physical states in Fock space, and the physical light pseudoscalar states are indeed $\pi^u$ and $\pi^d$ and not $\pi^{0,\rm vac}$ as at $B=0$. The correlators $C_{\pi^{u,d}}$ are saturated by their respective connected contributions and they have different exponential decays.
 \item[b)] The physical neutral pion state at nonzero $B$ is still very close to $\pi^{0,\rm vac}$. Both the $\pi^u$ and the $\pi^d$ operators couple to this one state and both correlators $C_{\pi^{u,d}}$ decay with its energy. These correlators must contain connected as well as sizeable disconnected diagrams, and the lattice studies see different exponential decays because they neglected the (important) disconnected contributions.
\end{enumerate}

Given that for asymptotically strong magnetic fields one expects the strong coupling to decrease due to asymptotic freedom and, therefore, disconnected diagrams in general to get suppressed, scenario a) from above seems more plausible. Nevertheless, a dedicated lattice study is needed in order to settle this issue.

\subsubsection{Pseudoscalar meson decay constants}
\label{sec:had_decay_constant}

As mentioned above in Sec.~\ref{sec:had_lattechn}, some of the hadrons decay via the weak interactions, with a decay rate parameterized in terms of the corresponding decay constants $F_h$. For nonzero background magnetic fields these have been discussed for pions and kaons, which we discuss next.

The decay rate is given by the matrix element of the weak current between the vacuum and the pseudoscalar meson state. For $B=0$, only the weak axial-vector current contributes due to parity symmetry and this term is proportional to the decay constant $F_h$. In contrast, for $B\neq0$, the weak vector current also contributes and there are several decay constants that parameterize the amplitude~\cite{Coppola:2018ygv}. 
This was first pointed out in the lattice study~\cite{Bali:2018sey}, where two decay constants entering the decay rate were determined for charged pions,
\be
\langle 0 | \bar d\gamma_0\gamma_5 u | \pi^-\rangle \propto F_{\pi^{\pm}}\, \E_{\pi^{\pm}}, \qquad 
\langle 0 | \bar d\gamma_3 u | \pi^-\rangle \propto F_{\pi^{\pm}}' \,eB \,\E_{\pi^{\pm}}\,.
\ee
We note that these pion decay constants also appear in the radiative decay of the charged pion for specific outgoing photon momenta~\cite{Desiderio:2020oej}.

\begin{figure}
 \centering
 \includegraphics[width=8cm]{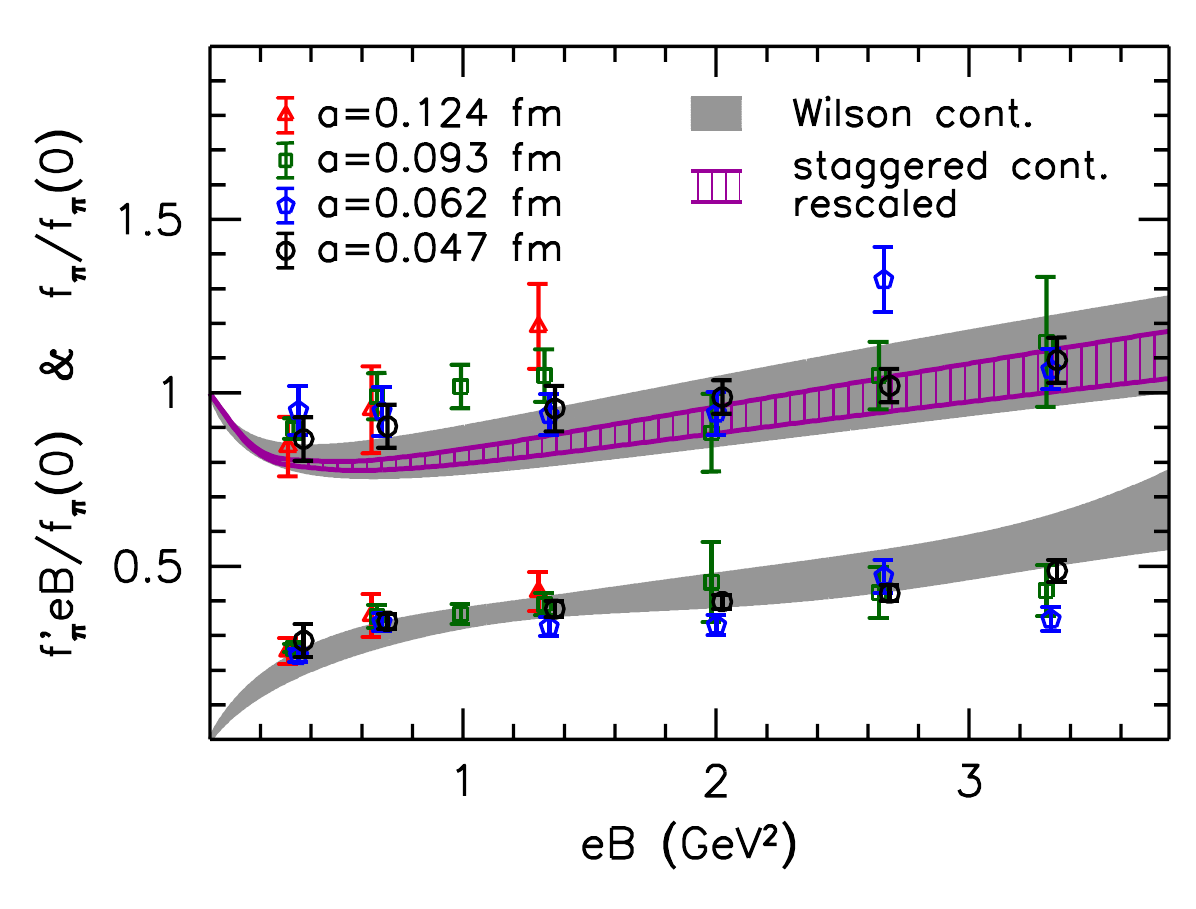}\qquad
 \raisebox{.1cm}{\includegraphics[width=8.5cm]{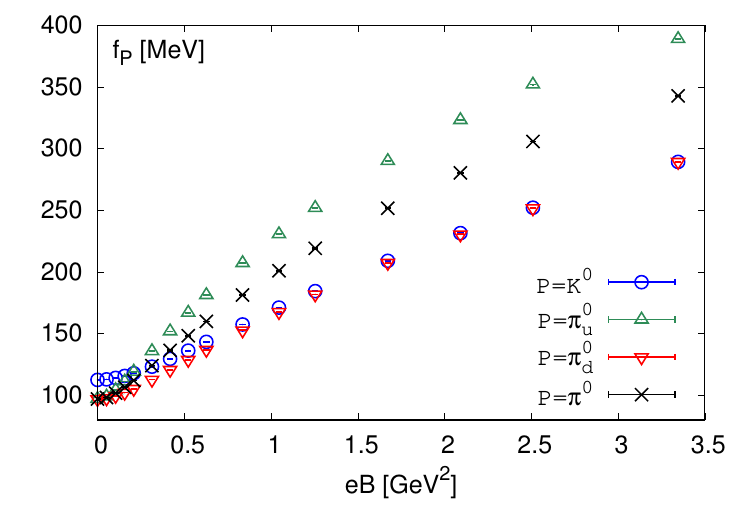}}
 \caption{
 \label{fig:had_decayconstant}
 Left panel: two decay constants $F_\pi$ and $F_\pi'$ (denoted here as $f_\pi$ and $f_\pi'$) entering the decay rate of charged pions as functions of the magnetic field in units of the vacuum value of $F_\pi$~\cite{Bali:2018sey}. The gray band shows the continuum limit for quenched Wilson results with a vacuum pion mass of $M_\pi=415\MeV$ for both decay constants. For $F_\pi$, the continuum limit for dynamical staggered quarks with physical masses is also shown (purple band) after a rescaling of the magnetic field by the squared vacuum pion mass.
 Right panel: decay constant $F$ of neutral pseudoscalar mesons (denoted here as $f_P$) as a function of the magnetic field using dynamical HISQ fermions~\cite{Ding:2020hxw}.
 }
\end{figure}

The decay constants for $\pi^\pm$ were calculated in~\cite{Bali:2018sey} using continuum extrapolated stout-improved staggered quarks with physical masses as well as heavier-than-physical quenched Wilson quarks. The left panel of Fig.~\ref{fig:had_decayconstant} shows the results for both discretizations, revealing significant nonzero values of $F_{\pi^\pm}'$ at $B>0$. This study also observed that the Wilson and staggered results, albeit obtained at different pion masses, are compatible with each other if the magnetic field is rescaled by the respective squared vacuum pion mass. In addition to $\pi^\pm$, the case of the charged kaon was also considered in the staggered analysis of~\cite{Bali:2018sey}. Recently, $F_{\pi^\pm}'$ was calculated for the first time in chiral perturbation theory for weak magnetic fields~\cite{Adhikari:2024vhs}, giving slightly higher values as the lattice results. 
The decay constant $F$ of the neutral pion and the neutral kaon was also calculated on the lattice in~\cite{Ding:2020hxw} using dynamical HISQ quarks. A monotonous increase was observed both for the neutral kaon and the (connected) neutral pion, see the right panel of Fig.~\ref{fig:had_decayconstant}.

\subsection{Gluonic observables in strong magnetic fields}
\label{sec:had_conf_ploop}

In general, the vacuum structure of QCD, in particular confinement, is a notion that gluonic degrees of freedom are responsible for.
Gluons are charge neutral and therefore do not respond directly to the magnetic field. Gluonic observables, however, can be affected by $B$ via sea quark loops. The prime example for this behavior is the dependence of the static quark-antiquark potential on the magnetic field. Next, we focus on such magnetic field-induced effects at zero temperature, and for completeness mention some of the interesting results obtained at nonzero temperatures as well.

\subsubsection{Gluon action and interaction measure}

In this context, one of the first analyses was carried out for the gluon action itself, as defined in the continuum in~\eqref{eq:lat_action_cont}, with a focus on the anisotropies among the individual components contributing to the lattice $S_g$~\cite{Ilgenfritz:2012fw}. This study simulated two-color QCD with four flavors of unimproved staggered quarks. To describe the results, let us use the notation of chromomagnetic $\B_i$ and chromoelectric $\E_i$ components introduced in~\eqref{eq:lat_action_cont_components}.
In terms of these components, the magnetic field was found to induce a splitting between the parallel ($i=3$, denoted $\parallel$) and perpendicular ($i=1,2$, denoted $\perp$) components, while the temperature a splitting between the chromomagnetic and chromoelectric contributions. The observed hierarchy, $\expv{\tr \,\B_\parallel^2}>\expv{\tr \,\B_\perp^2}>\expv{\tr \,\E_\perp^2}>\expv{\tr \,\E_\parallel^2}$, is expected already from a perturbative treatment of this problem via the Euler-Heisenberg effective action~\cite{Bali:2013esa}. This ordering of anisotropies was observed to hold in simulations of dynamical QCD with physical quark masses as well~\cite{Bali:2013esa}. The latter study also demonstrated how these anisotropies can be used to calculate the magnetization and the magnetic susceptibility of the QCD medium -- we get back to this point in Sec.~\ref{sec:eos_aniso}.

The sum of all chromo-field components carries information about the gluon condensate carried by the QCD vacuum. Up to a proportionality factor involving the QCD $\beta$-function, the gluon condensate equals the gluonic contribution to the interaction measure $I$, relevant for the equation of state, to be defined in~\eqref{eq:eos_intmeasure} in Sec.~\ref{sec:eos_thermo_rel}. It is given through the response of the partition function to an overall change of the scale (i.e., the lattice spacing),
\be
I=-\frac{T}{V}\frac{\partial\log\Z}{\partial \log a}\,.
\label{eq:had_intmeasdef1}
\ee
On the lattice, a change in $a$ is manifested via changes in the lattice parameters $\beta$ and $m_f$. Therefore, the interaction measure receives contributions both from gluons and from fermions,
\be
I=I_G+I_F, \qquad I_G=-\frac{\partial \beta}{\partial \log a}\,\frac{T}{V}\frac{\partial \log\Z}{\partial \beta} = \frac{\partial \beta}{\partial \log a}\frac{T}{V}\expv{S_g}, 
\qquad
I_F=-\sum_f\frac{\partial m_f}{\partial \log a} \,\frac{T}{V}\frac{\partial\log\Z}{\partial m_f}
=-\sum_f\frac{\partial m_f}{\partial \log a} \expv{\bar\psi_f\psi_f}\,.
\ee
Thus, we see that the interaction measure is built up from the gluon condensate and the quark condensate.

\begin{figure}
 \centering
 \includegraphics[width=8cm]{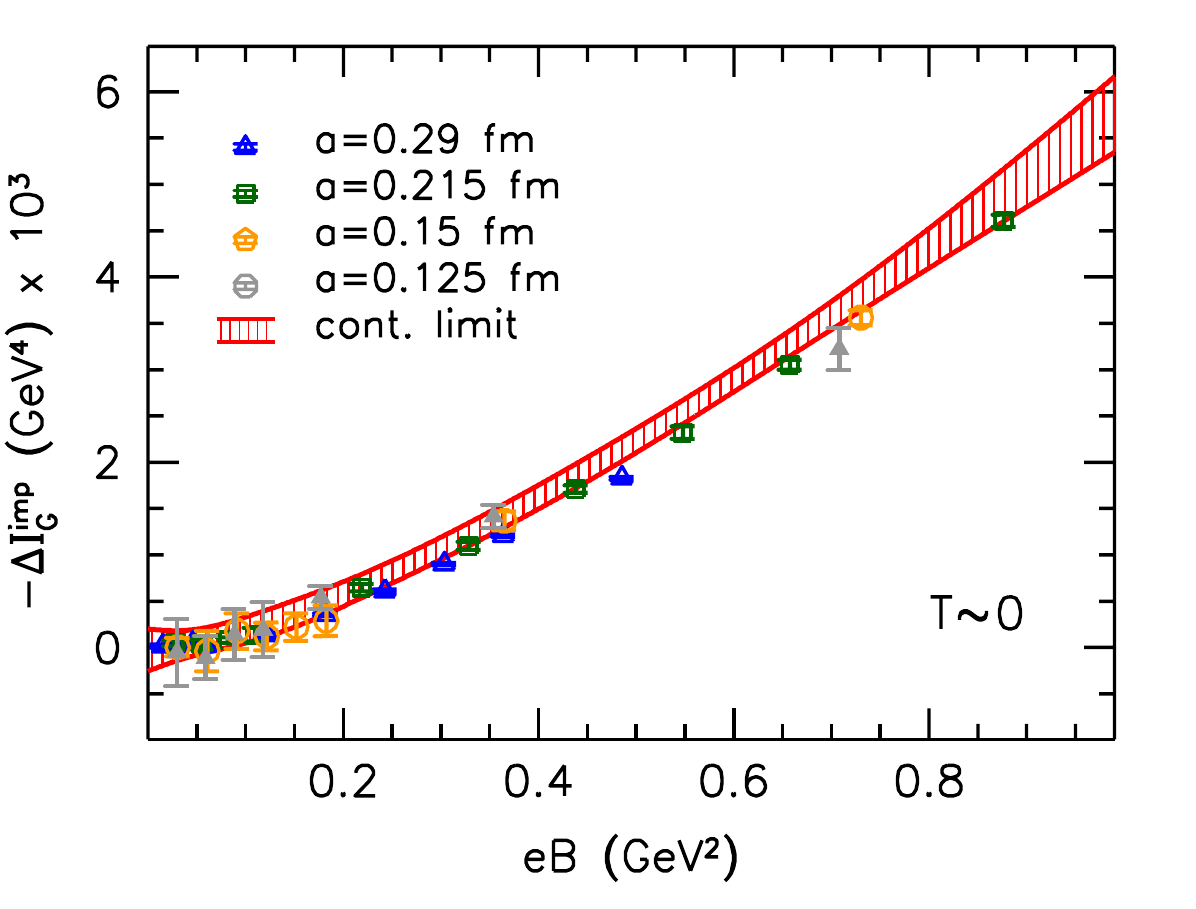}\qquad
 \includegraphics[width=8cm]{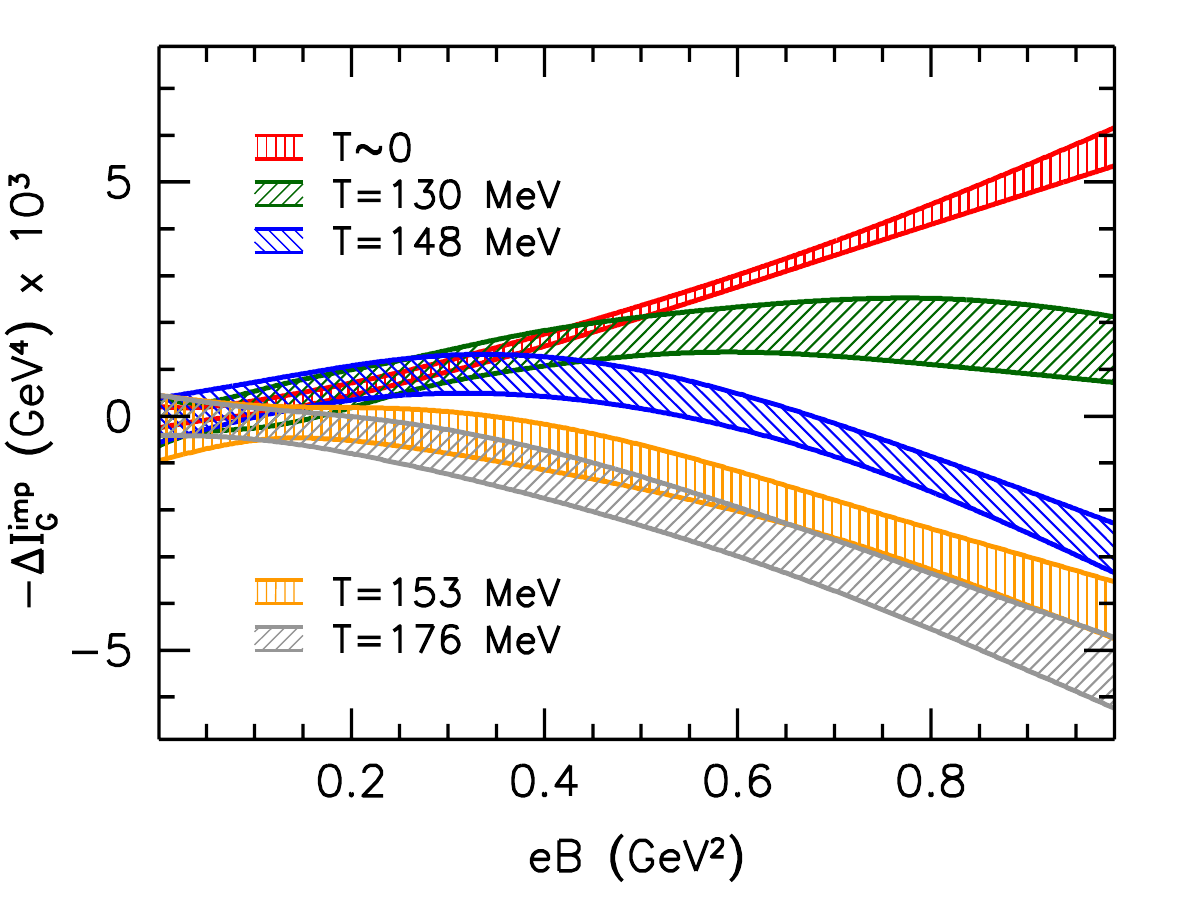}\caption{
 \label{fig:had_intmeas}
 The negative of the gluonic contribution to the interaction measure as a function of the magnetic field at low (left panel) and high temperatures (right panel), compared to its $B=0$ value~\cite{Bali:2013esa}. The data points in the left panel represent results for different lattice spacings, while the dashed bands the continuum extrapolations. Both figures are adapted from~\cite{Bali:2013esa}.
 }
\end{figure}

According to the results of~\cite{Bali:2013esa} and~\cite{Bali:2012zg}, the magnetic field has a surprisingly similar effect on $-I_G$ and on $I_F$. At low temperature, both these observables increase with $B$ and behave approximately linear for strong magnetic fields. The gluonic observable is shown\footnote{We note that the study~\cite{Bali:2013esa} developed an improvement scheme for reducing lattice artefacts in $I_G$ in order to facilitate taking the continuum limit. The improved observable is denoted by $I_G^{\rm imp}$ in Fig.~\ref{fig:had_intmeas}.}, after subtracting its $B=0$ value, in the left panel of Fig.~\ref{fig:had_intmeas}. For completeness, we briefly discuss here the behavior of $I_G$ at nonzero temperatures, too. In this case, a non-monotonous dependence arises~\cite{Bali:2013esa}, turning the observable around for temperatures in the transition region $T\approx 150\MeV$. This is also reminiscent of the behavior of the quark condensate~\cite{Bali:2012zg}, which will be relevant for our discussion of the phase diagram in Chap.~\ref{chap:pd}. The dependence of $-I_G$ on $B$ and $T$ is shown in the right panel of Fig.~\ref{fig:had_intmeas}.

Since the gluon condensate (just as the interaction measure) is subject to additive renormalization, the above method is only suitable for calculating differences, e.g.\ $\Delta I_G=I_G-I_G(B=0)$. The renormalized gluon condensate of the vacuum can be calculated via a different method, by determining the amplitude of the exponential decay in field strength correlators. The latter has been calculated in~\cite{DElia:2015eey} in a fully gauge invariant manner for a range of magnetic fields. This study, using unimproved staggered quarks, explored the Lorentz covariant structures appearing in general correlators of $\F_{\nu\rho}$ and the corresponding anisotropies at nonzero magnetic fields.

\subsubsection{Wilson loops and string tension}
\label{sec:had_stringtension}

One of the most prominent attributes of the confining nature of the QCD vacuum is the linearly rising potential between a static quark-antiquark pair. This potential can be probed via the expectation value of rectangular Wilson loops\footnote{Interestingly, albeit purely gluonic observables, Wilson loops themselves can also be obtained through fermionic observables in the presence of homogeneous magnetic fields. More precisely, one needs to consider the quark condensate defined from the two-dimensional Dirac operator restricted to $x_1-x_2$ planes of the lattice in the presence of valence magnetic fields. The Fourier transform with respect to $B$ gives the so-called dressed Wilson loops, which reproduce, in the large quark mass limit, ordinary Wilson loops~\cite{Bruckmann:2011zx}.} $W(\bm n, n_4)$ defined in~\eqref{eq:lat_Wilsonloops},
\be
\expv{W(\bm n, n_4)} \propto \exp\left[ -aV(\bm n) \,n_4 \right]\,.
\label{eq:had_Wilsonloopdef}
\ee
and is often fitted by the Cornell parameterization,
\be
aV(\bm n) = -\frac{\alpha}{|\bm n|} + aV_0 + a^2\sigma |\bm n| \,,
\label{eq:had_cornell0}
\ee
involving the string tension $\sigma$ in the large distance region.
In turn, for nonzero temperatures, the potential can be determined from the negative logarithm of Polyakov loop correlators, 
to which we will get back to in Sec.~\ref{sec:pd_LLdag}.

For $B=0$, the potential is isotropic, i.e.\ $V(\bm n)=V(|\bm n|)$. 
In turn, the presence of the magnetic field may induce anisotropies in the Wilson loops, and through them, in the static potential. In that case, the string tension may also be anisotropic and depend on the direction $\hat{\bm n}=\bm n/|\bm n|$,
\be
\frac{V(\bm n)}{a|\bm n|} \xrightarrow{|\bm n|\to\infty} \sigma(\hat{\bm n}) = \sigma_{B=0} + C \cdot B^2 + D \cdot (\hat{\bm n} \cdot \bm B )^2+\O(B^4)\,.
\label{eq:had_cornell1}
\ee
Here, we considered an expansion of $\sigma$ in the magnetic field, used that it may only depend on the scalar products $\bm B^2$ and $\hat{\bm n}\cdot \bm B$, and that only even powers of $B$ may appear due to charge conjugation symmetry. Denoting the angle between $\hat{\bm n}$ and $\bm B$ by $\vartheta$,~\eqref{eq:had_cornell1} implies that  the string tension is $\sigma(\vartheta)=\sigma_{B=0} + B^2(C+D\cos^2\vartheta)$ for weak fields. In particular, in the direction parallel to the magnetic field one has $\sigma_\parallel\equiv \sigma(\vartheta=0)=\sigma_{B=0}+B^2 (C+D)$, while in the perpendicular direction $\sigma_\perp\equiv\sigma(\vartheta=\pi/2)=\sigma_{B=0}+B^2 C$. 

\begin{figure}
 \centering
 \includegraphics[width=8cm]{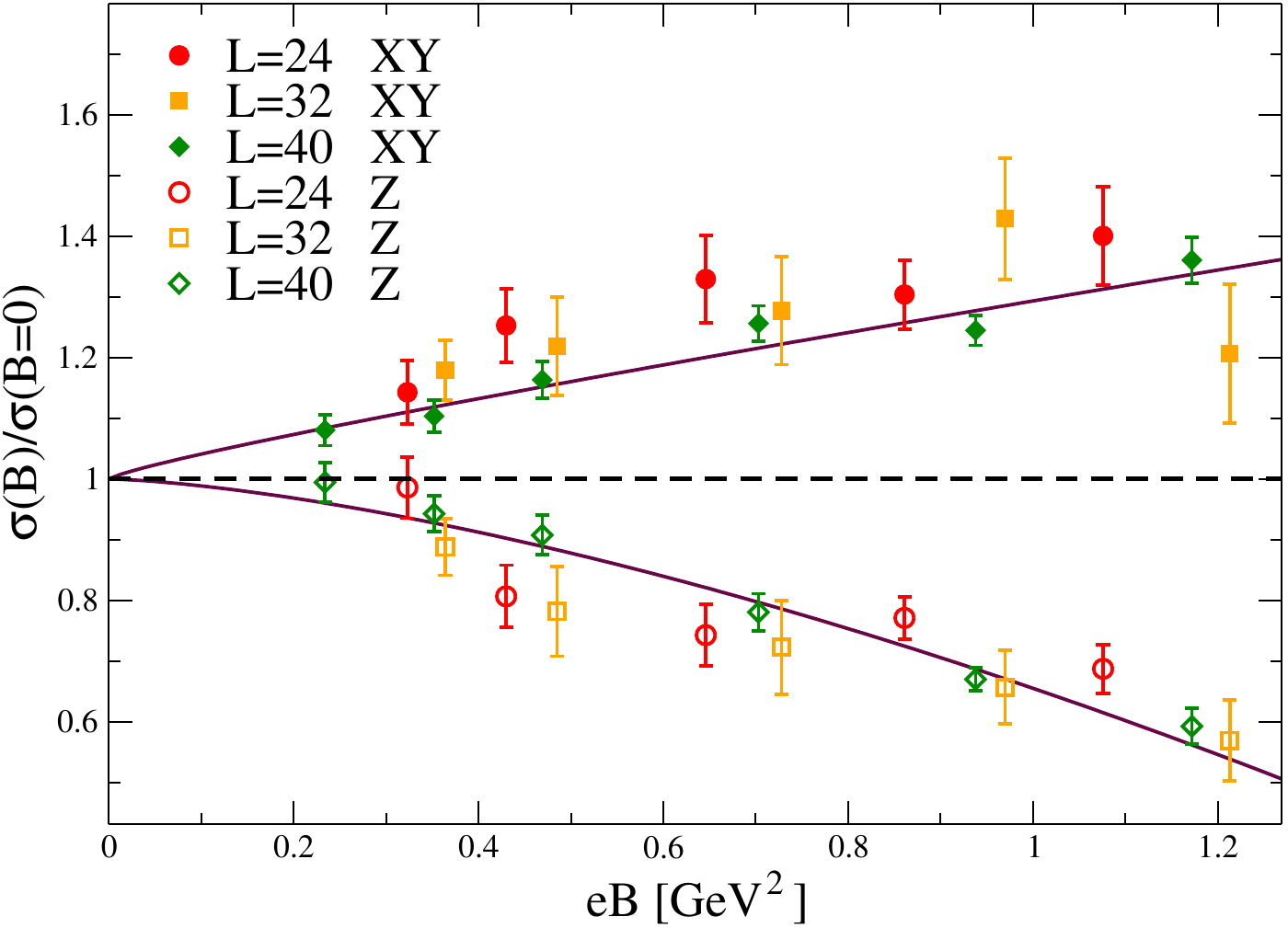}\qquad
 \includegraphics[width=8cm]{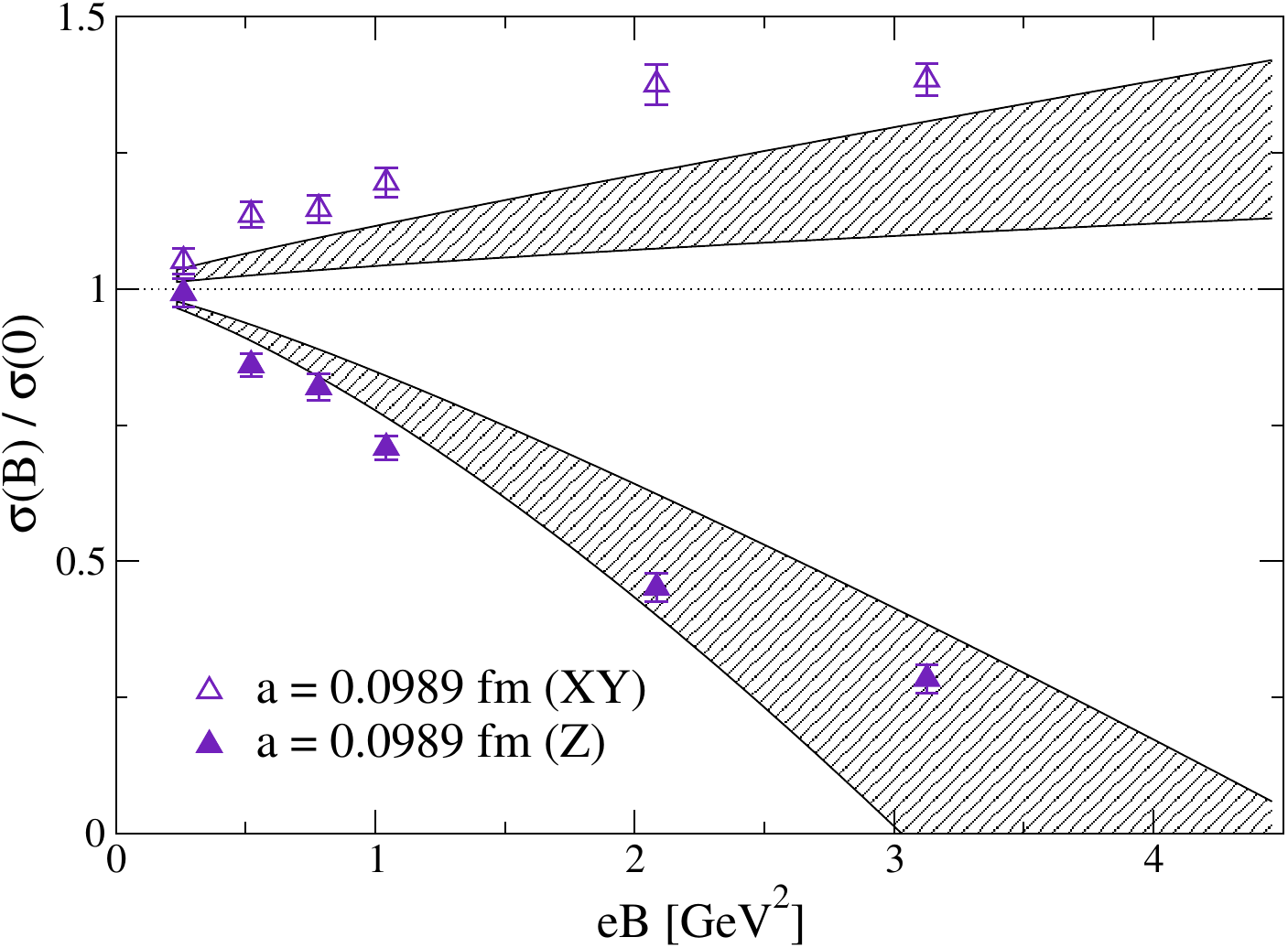}
 \caption{ 
 \label{fig:had_stringtension}
 Left panel: the string tensions in the direction parallel (labeled by $Z$) and perpendicular ($XY$) to the magnetic field, normalized by the $B=0$ value, as a function of $eB$~\cite{Bonati:2014ksa}. The different points represent data obtained for different lattice spacings and the solid line a power-law fit.
 Right panel: perpendicular and parallel string tensions (with the same notation as in the left panel) as a function of $eB$~\cite{Bonati:2016kxj}. The dashed bands represent the continuum limit based on four lattice spacings up to $eB\approx 1\GeVsq$ and extrapolated beyond that value, while the purple points are the results on the finest lattice spacing.
 }
\end{figure}

The first study of $V(\bm n)$ was performed in~\cite{Bali:2011qj} with stout-improved staggered quarks with physical masses. Here the directional dependence on $\vartheta$ was not resolved but an average over all spatial orientations of $\bm n$ was taken. The result showed no significant effects due to the magnetic field. A more detailed study in~\cite{Bonati:2014ksa}, with the same staggered action, discussed the spatial directions separately. The results for the string tension are shown in the left panel of Fig.~\ref{fig:had_stringtension}, revealing a significant anisotropy: the confining force becomes stronger in the direction perpendicular to the magnetic field, while it gets weaker parallel to it. In fact, the average string tension is approximately independent of $B$, in line with the previous findings~\cite{Bali:2011qj}.
The observed anisotropic behavior is in fact consistent with a similar hierarchy of anisotropies in the gluon action expectation values~\cite{Ilgenfritz:2012fw,Bali:2013esa}. The Coulomb part of the static potential was also observed to have anisotropic contributions~\cite{Bonati:2014ksa}.

In the $B\to\infty$ limit, QCD is described by an effective theory~\cite{Miransky:2002rp}, which we will discuss in more detail in Sec.~\ref{sec:pd_largeB}. This limiting theory predicts the vanishing of the parallel string tension~\cite{Endrodi:2015oba}.
An intriguing question, raised by~\cite{Bonati:2014ksa} was whether $\sigma_\parallel=0$ sets in already at a finite value of the magnetic field at zero temperature.
To answer this question, the study~\cite{Bonati:2016kxj} -- again using stout improved staggered quarks -- pushed the investigation of the anisotropic potential to stronger fields and also carried out the continuum extrapolation in the range $0\le eB\le1\GeVsq$. The continuum limit extrapolation was also extended to even stronger magnetic fields and further results at the finest lattice spacing were included, shown in the right panel of Fig.~\ref{fig:had_stringtension}. The results indicate a monotonous reduction of $\sigma_\parallel$, potentially vanishing for $eB\gtrsim 4\GeVsq$. We note that this work also discussed the complete angular dependence of the potential, with a parameterization similar to~\eqref{eq:had_cornell1}.

Recently, continuum extrapolations for $\sigma_\parallel$ and $\sigma_\perp$ were performed with the same lattice action, employing three lattice spacings at $eB=4\GeVsq$ and $eB=9\GeVsq$~\cite{DElia:2021tfb}. The parallel string tension still does not vanish at these extreme magnetic fields, but becomes as low as $\sim 7\%$ of its vacuum value at the strongest field. In view of the fact that for such magnetic fields, the deconfinement transition becomes first-order~\cite{DElia:2021yvk} -- as we will discuss in Sec.~\ref{sec:pd_largeB} -- this indicates that the QCD medium undergoes drastic changes in this domain.

In addition, the quark-antiquark potential was also investigated through Polyakov loop correlators at nonzero temperatures~\cite{Bonati:2016kxj}. The main effect of the magnetic field was observed to be the suppression of the potential in all directions and a corresponding reduction in the string tensions. This is consistent with the reduction of the transition temperature by the magnetic field, which we will discuss in Chap.~\ref{chap:pd}.

A realistic picture of the confining potential between the static quark and antiquark is provided by the color flux tube. In this picture, the gluon field energy density is concentrated in a relatively thin string between the sources, and the energy
density per unit length of the ﬂux tube is given by the string tension. The flux tube has been studied using lattice simulations with stout-improved staggered quarks in~\cite{Bonati:2018uwh}. The observable to calculate is the correlator of Wilson loops $W$ and the imaginary part of the plaquette $P_{\nu\rho}$, made gauge invariant by including a parallel transporter $\U$ between them,
\be
\rho^{\rm conn}_{\nu\rho} = \frac{\expv{\tr(W\,\U P_{\nu\rho}(\bm r)\,\U^\dagger)}}{\expv{\tr (W)}}- \frac{1}{3}\frac{\expv{\tr(P_{\nu\rho}(\bm r))\,\tr(W)}}{\expv{\tr(W)}}\,,
\label{eq:had_conncorr_wilson}
\ee
where the plaquette operator is inserted at a spatial distance $\bm r$ away from the Wilson loop, perpendicular to the plane in which the latter lies.
The expression~\eqref{eq:had_conncorr_wilson} is the so-called connected correlator~\cite{Bonati:2018uwh} that measures the gluon field strength (related to the imaginary part of the plaquette) due to the presence of the quark-antiquark pair. 

\begin{figure}
 \centering
 \includegraphics[width=8cm]{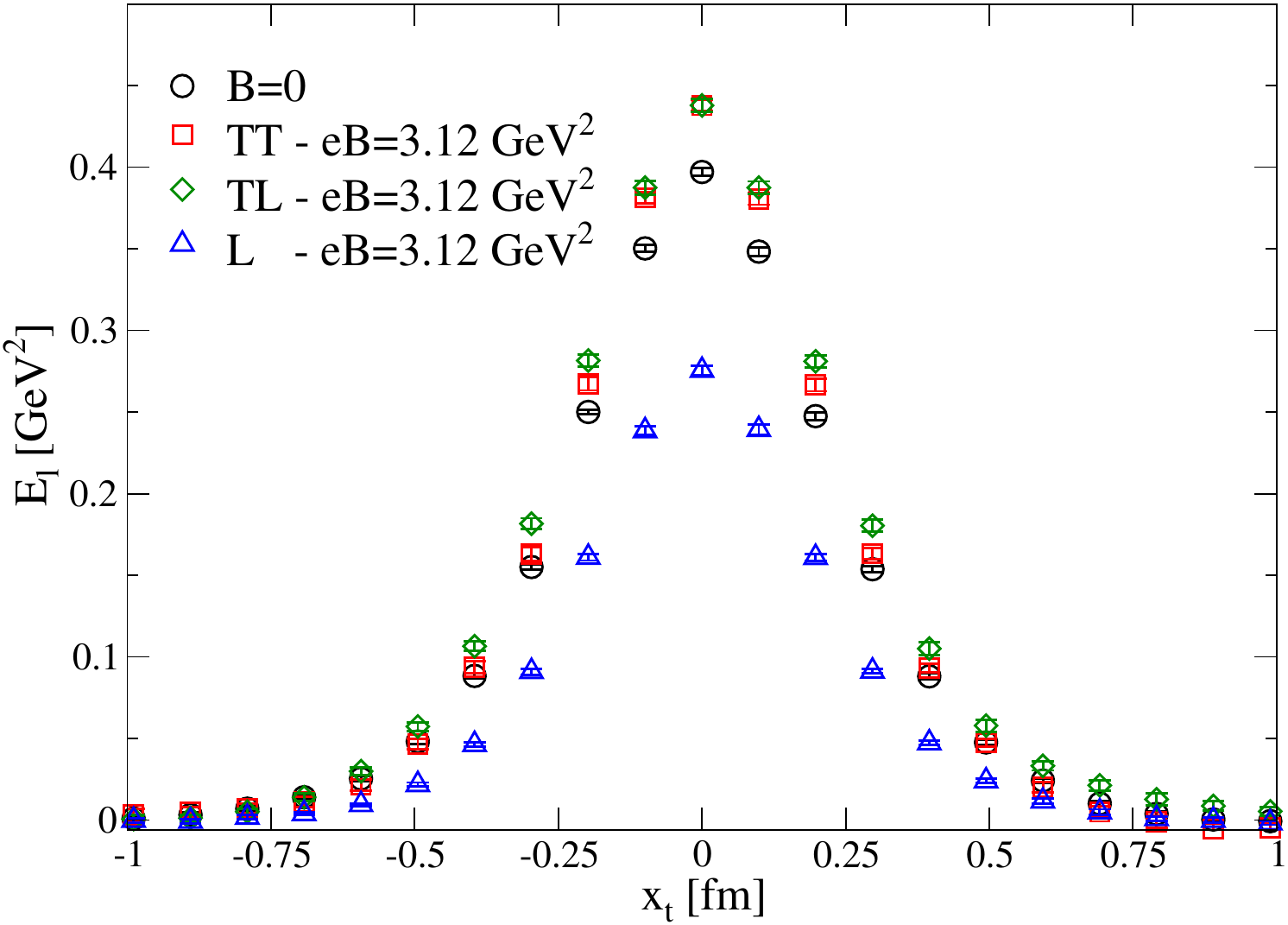}
 \caption{
 \label{fig:had_profil_fluxtube}
 Spatial profile of the flux tube between a static quark and antiquark source in the transversal direction (denoted by $x_t=|\bm d|$), for differently oriented magnetic fields, and in comparison to the $B=0$ case~\cite{Bonati:2018uwh}.
 }
\end{figure}

The most substantial field component was found to be the chromoelectric field $\E_l$ parallel to the quark-antiquark separation $\bm d$. The profile of this component, as measured in~\cite{Bonati:2018uwh}, is shown in Fig.~\ref{fig:had_profil_fluxtube} as a function of $|\bm r|$ for a quark-antiquark distance of $|\bm d|=0.7\fm$. There are three inequivalent orientations of the magnetic field: when $\bm d\parallel \bm B$ (denoted by $L$ in the figure), when $\bm d\perp \bm B  \parallel \bm r$ ($TL$) and when $\bm d\perp\bm B \perp \bm r$ ($TT$).
The results reinforce the findings about the string tension discussed above, i.e.\ a reduction of the energy density stored in the flux tube parallel to the magnetic field.
In fact, the magnetic field-dependence of the total integral of $\E_l^2$ inside the flux tube was shown to agree with that of $\sigma_\parallel$. In addition, the flux tube was found to become thinner as $B$ grows~\cite{Bonati:2018uwh}.

Finally, in light of the impact of the magnetic field on the static potential, the spectrum of heavy quarkonia is also expected to be modified substantially by $B$.
In fact, magnetic field-related effects on heavy flavors might be the most 
relevant in the context of heavy-ion collisions. This is because
heavy quarkonia are most sensitive to the conditions in the early stages of the collisions, where the magnetic field is the strongest.
To carry out this analysis, one needs to use the non-relativistic Hamiltonian for the two-body problem, based on the static potential discussed above and phenomenological parameterizations of spin-spin interactions~\cite{Alford:2013jva,Bonati:2015dka}. The anisotropy of the static potential was observed to significantly affect charmonia and bottomonia, by increasing their masses with respect to the case with isotropic potentials~\cite{Bonati:2015dka}.

\subsection{Lessons learned}

In this chapter we discussed the impact of background electromagnetic fields on the confining properties of the QCD vacuum and on the hadron spectrum. For weak fields, hadrons respond to linear order via their magnetic moments and electric dipole moments (if a CP-odd source is also considered). The quadratic order, in turn, is described by magnetic and electric polarizabilities. There are numerous technical and conceptual challenges to be tackled in calculations of these observables and significant computational efforts were recently devoted to this subject by many research groups world-wide.

The behavior of hadron masses -- in particular that of the $\rho^\pm$ meson -- for strong magnetic fields was identified as a key factor for the discussion of a possible superconducting phase in QCD. All existing results so far suggest that $\E_{\rho^\pm}(B)$ remains nonzero up to very strong magnetic fields, speaking against this scenario. 
While it was pointed out that inhomogeneous condensates might still be compatible with these findings~\cite{Chernodub:2012zx,Chernodub:2013uja}, no signal of a phase transition into a superconducting phase was found in dynamical QCD at low temperatures up to very high magnetic fields~\cite{DElia:2021tfb}.

Strong magnetic fields were found to lead to intriguing results  for charged and neutral pions as well. Charged pions exhibit new weak decay channels, characterized by different decay constants, which have been determined by now on the lattice. 
In turn, the lattice calculations of pion energies for strong magnetic fields lead to open questions.  
Currently, there is a tension between $\E_{\pi^\pm}(B)$ for strong magnetic fields determined from quenched Wilson quarks and HISQ fermions, which should be resolved.
Regarding the neutral pion, a better understanding of the flavor mixing in the light quark sector is required. This should be settled by a dedicated study of disconnected diagrams at $B>0$. Finally, the $B$-dependence of the energy of the $\Delta^{++}$ baryon, which was recently identified as an important ingredient for heavy-ion collision phenomenology~\cite{Vovchenko:2024wbg,Marczenko:2024kko}, should be determined on the lattice for strong magnetic fields, in order to check predictions by hadron resonance gas model approaches.

The confining nature of gluon fields in the QCD medium is affected indirectly by magnetic fields in non-trivial ways. The gluon condensate of the vacuum, related to the interaction measure, is enhanced as $B$ grows. Moreover, the static potential becomes anisotropic in the presence of the magnetic fields: color charges are more strongly confined when separated perpendicular to $B$. Specifically, the parallel string tension  decreases steadily with growing $B$. Whether $\sigma_\parallel=0$ is reached for a finite magnetic field, remains an open question for future research.

	\clearpage
	\section{Deconfinement and the phase diagram}
	\label{chap:pd}

It is well known that QCD exhibits (at least) two phases with qualitatively different properties. The low-energy regime features confinement and the spontaneous breaking of chiral symmetry. At high energies, the system becomes deconfined and chiral symmetry is restored. The two regimes are separated by a transition that can be best discussed via the respective symmetries of the theory and the corresponding order parameters. In fact, QCD with physical quark masses has no exact symmetries but only approximate ones -- correspondingly, one speaks about approximate order parameters.

The first relevant symmetry is the chiral symmetry of the massless QCD action, which is broken explicitly by the light quark masses. The associated order parameter is the average of the light quark condensates~\eqref{eq:lat_pbpdef},
\be
\expv{\bar\psi\psi}\equiv \frac{1}{2}\sum_{f=u,d}\expv{\bar\psi_f\psi_f}\,.
\label{eq:pd_pbpdef}
\ee
As an approximate order parameter, the light quark condensate is nonzero in both phases. 
Still, it exhibits a distinct behavior in the transition region that makes it a useful observable to describe the thermal behavior of QCD matter. In the massless limit, chiral symmetry is intact and $\expv{\bar\psi\psi}$ acts as an exact order parameter with vanishing value in the high-temperature phase.

Similarly, there is a symmetry associated to confinement as well: the $\mathrm{Z}(3)$ center symmetry of the gluon action. Center transformations can be written as large $\mathrm{SU}(3)$ gauge transformations that twist the $\A_4$ field in the imaginary time direction in a way compatible with the periodic boundary conditions. For details, we refer the reader to the review~\cite{Fukushima:2017csk}. Center symmetry is broken spontaneously at high temperature and is restored at low temperatures, and the corresponding order parameter is the expectation value $\expv{P}$ of the Polyakov loop, defined above in~\eqref{eq:lat_ploopavg}. 
As we used already in~\eqref{eq:lat_ploopfQ} and in the discussion in Sec.~\ref{sec:ev_ploop}, $\expv{P}$
is related to the negative exponential of the free energy of a static color charge~\cite{Fukushima:2017csk}. In a confined system this free energy is infinite, corresponding to $\expv{P}=0$. Similarly, deconfinement translates to $\expv{P}\neq0$.
Just like chiral symmetry, the center symmetry is also broken explicitly in full QCD and only becomes exact in a specific limit: in this case this is the infinite quark mass limit. For $m_f\to\infty$, quarks decouple from the theory and one recovers pure gauge theory.
In full QCD, the Polyakov loop therefore only acts as approximate order parameter. But just like $\expv{\bar\psi\psi}$, it provides useful information about the finite temperature QCD transition.

In the infinite quark mass limit, i.e.\ in pure gauge theory, the QCD transition is a first-order phase transition, for which the order parameter $\expv{P}$ exhibits a discontinuity at the critical temperature $T_c$. 
In QCD with physical quark masses, it is well known that the finite temperature transition is not a real phase transition anymore but merely an analytic crossover~\cite{Aoki:2006we,Bhattacharya:2014ara}. Here, both $\expv{P}$ and $\expv{\bar\psi\psi}$ are smooth functions of the temperature. Accordingly, the definition of a transition temperature is not unique. Different definitions, based on different observables and their different qualitative behaviors are known to deliver different results. For this reason, the transition temperature is called a pseudo-critical temperature. Still, it is often denoted by the same symbol, $T_c$, and this is the notation that we follow in this review, too. In some cases a superscript will indicate, whether the definition involves the quark condensate, the quark number susceptibility or the Polyakov loop ($T_c^{\bar\psi\psi}$, $T_c^{\chi_s}$, $T_c^P$ or similar).

In the opposite limit of massless quarks, the nature of the finite temperature transition is the subject of active ongoing research. Whether the transition is of first or of second order depends on the number of massless quarks and potentially on the masses of the remaining quarks. In this context often the up, down and strange quarks are discussed and the light quarks are considered degenerate, $m_u=m_d=m_\ell$. 
Considerable attention was devoted to the nature of the transition in the $m_\ell-m_s$ plane, the so-called Columbia plot, see e.g.~\cite{Aarts:2023vsf}.

Background electromagnetic fields impact the above characteristics of QCD thermodynamics in a non-trivial manner. This chapter serves to summarize our knowledge about the influence of the background fields on the approximate order parameters and the transition temperature. The latter dependence enables us to construct the QCD phase diagram in the background field - temperature plane. Specifically, we will consider homogeneous magnetic fields, spatially localized magnetic fields as well as homogeneous imaginary electric fields. In all cases we will concentrate on the above introduced approximate order parameters: the quark condensate and the Polyakov loop.

\subsection{Quark condensate}

After carrying out the fermionic path integral, the expectation value of the quark condensate takes the form~\eqref{eq:lat_pbpdef_0}.
Using the eigensystem~\eqref{eq:ev_Diracev} of the Dirac operator,
denoting the eigenvalues of $\Dsf$ by $\lambda_{fn}$, the trace can be spanned as,
\be
\expv{\bar\psi\psi} 
= \frac{1}{2}\sum_{f=u,d} \frac{T}{V}\sum_n \Expv{\frac{1}{m_f+i\lambda_{fn}}}
= \frac{1}{2}\sum_{f=u,d} \frac{T}{V}\sum_n \Expv{\frac{m_f}{\lambda_{fn}^{2}+m_f^2}}\,,
\label{eq:pd_BC1}
\ee
where in the second step, we used the chiral symmetry~\eqref{eq:ev_chiralsymm} of the Dirac operator, allowing us to add up the contributions of positive and negative eigenvalues. When the expression~\eqref{eq:pd_BC1} is evaluated with rooted staggered fermions on the lattice, an overall factor $1/4$ is to be included, as we already discussed under~\eqref{eq:lat_staggered_Dslash}.

Before we consider the impact of background electromagnetic fields on the quark condensate, it is useful to include a slight detour about the massless limit of $\expv{\bar\psi\psi}$, which will facilitate the interpretation of this observable.

\subsubsection{Banks-Casher relation and magnetic catalysis}
\label{sec:pd_magncat}

The spontaneous breaking of chiral symmetry in QCD with massless quarks is an essentially non-perturbative phenomenon that occurs in the infinite volume. For any finite volume, the massless limit of the condensate vanishes, just as the zero magnetic field limit of the expectation value of the magnetization vanishes in the Ising model for finite systems. The nonzero value of the condensate for massless quarks can be constructed as the double limit $V\to\infty$, followed by $m_\ell\to0$ (where we considered degenerate up and down quark masses $m_u=m_d=m_\ell$). In the thermodynamic limit, the spectral sum in~\eqref{eq:pd_BC1} can be turned into an integral, introducing the spectral density $\rho^f(\lambda)=\frac{T}{V}\sum_n \expv{\delta(\lambda-\lambda_n^f)}$ of the Dirac operator,
\be
\lim_{m_\ell\to0}\lim_{V\to\infty}\expv{\bar\psi\psi}
= \lim_{m_\ell\to0}\, \frac{1}{2}\sum_{f=u,d} \int \dd \lambda \,  \frac{m_\ell}{\lambda^2+m_\ell^2}\, \rho^f(\lambda) =
\frac{\pi}{2} \sum_{f=u,d}  \rho^f(0)\,,
\label{eq:pd_BCrel}
\ee
where we used that the massless limit of the kernel under the $\lambda$-integral is a $\delta$-distribution centered at zero.
This is the celebrated Banks-Casher relation~\cite{Banks:1979yr}, which relates the chiral condensate to the spectral density of the Dirac operator at the origin.

In Sec.~\ref{sec:lat_landaulevelQCD}, we learned that homogeneous background magnetic fields generally lead to an accumulation of the Dirac eigenvalues around zero, proportionally to $B$. In view of the Banks-Casher relation~\eqref{eq:pd_BCrel}, the proliferation of low eigenvalues directly amounts to an enhancement the condensate. In particular, one finds that the chiral limit of $\expv{\bar\psi\psi}$ grows linearly in $B$. This result readily generalizes to an overall enhancement of the quark condensate due to the magnetic field for any value of the quark mass. In that case, the behavior at small masses translates to the strong-field regime $B\to\infty$.

The general phenomenon of the enhancement of the condensate due to $B$ has been dubbed `magnetic catalysis'~\cite{Gusynin:1994re}. It has proven to be a very robust concept, valid for a broad class of theories, e.g.\ models and effective theories of QCD,
see the reviews~\cite{Shovkovy:2012zn,Andersen:2014xxa}.
While magnetic catalysis was originally viewed as a concept driven by strong magnetic fields, it has also been demonstrated to be universal in the weak-field region.
As we discussed in Sec.~\ref{sec:lat_magncat_paramag}, the enhancement of the condensate is quadratic in $B$ in the weak-field regime and the associated proportionality constant is related to the lowest-order coefficient of the QED $\beta$-function. Its positivity therefore fixes the leading enhancement of $\expv{\bar\psi\psi}$ by the magnetic field~\cite{Bali:2013txa,Endrodi:2014vza}.

\subsubsection{Quark condensate in the vacuum of QCD}
\label{sec:pbp_QCDT0}

\begin{figure}
 \centering
 \raisebox{5.5cm}{\includegraphics[angle=-90,width=9cm]{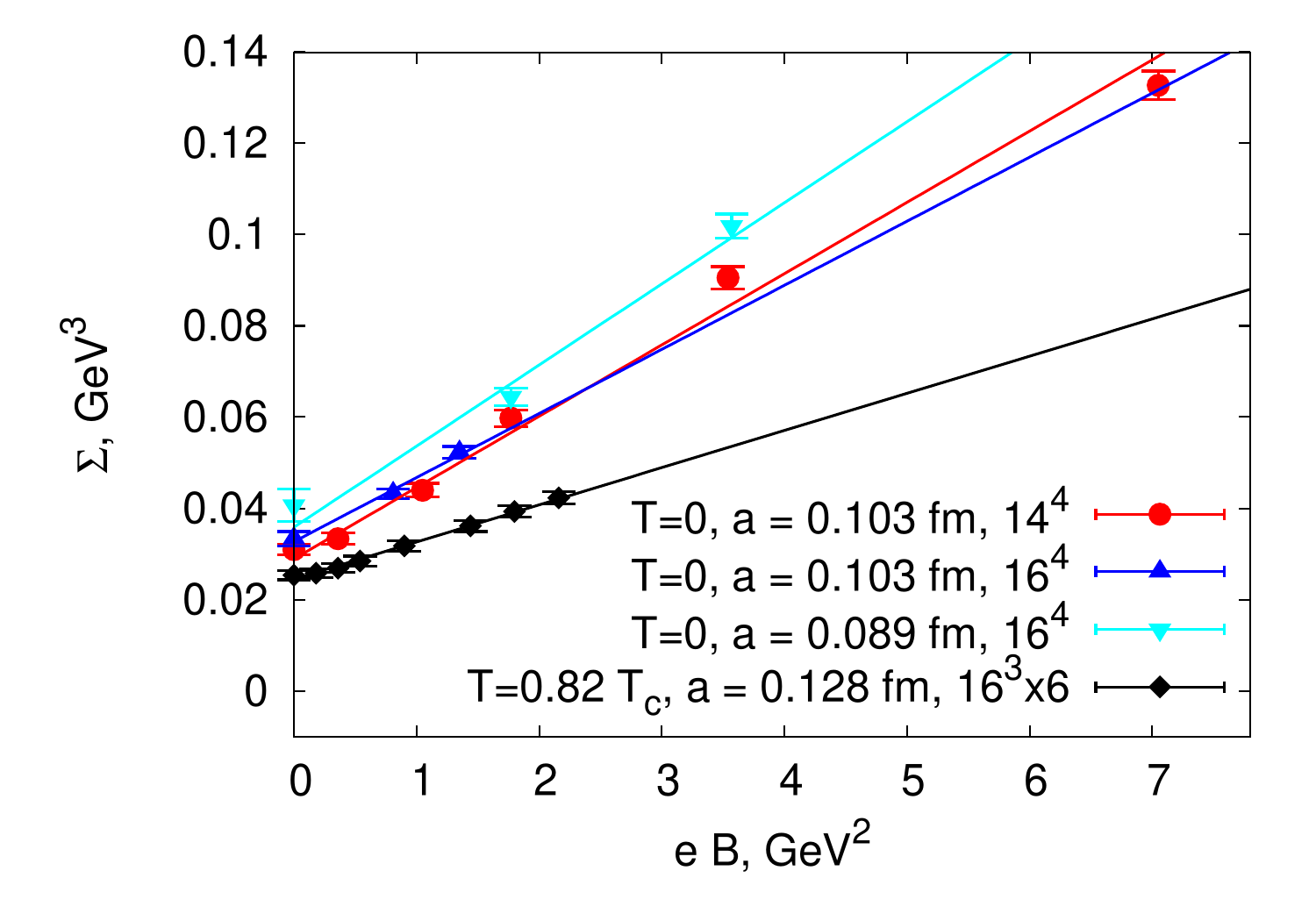}}
 \qquad
 \includegraphics[width=7.1cm]{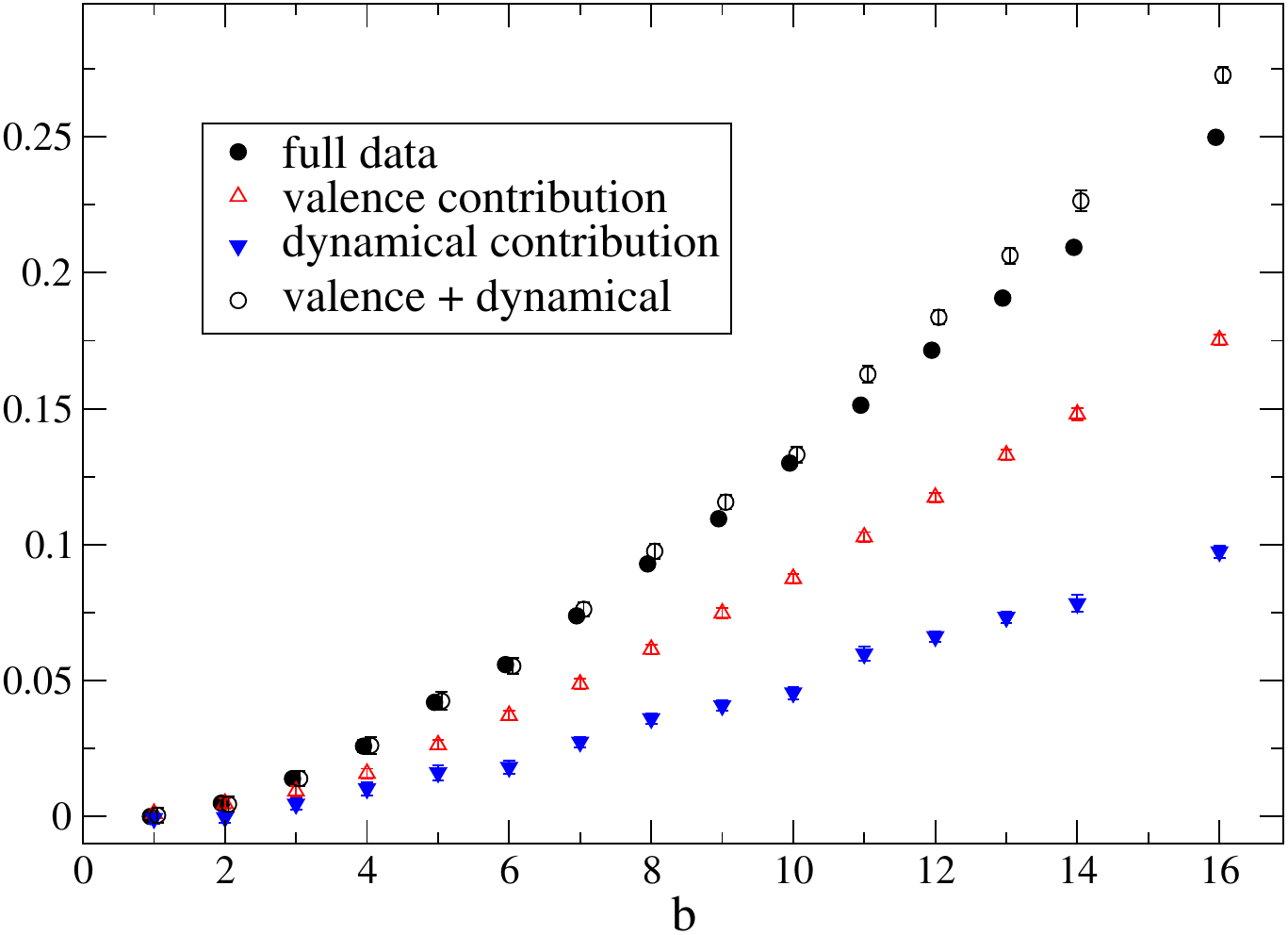}
 \caption{\label{fig:pd_QCDcondensateT0}
  Left panel: down quark condensate $\expv{\bar\psi_d\psi_d}$ (denoted here by $\Sigma$) in homogeneous magnetic fields on the lattice with quenched overlap fermions~\cite{Buividovich:2008wf}. Right panel: relative average light quark condensate (for details, see the text) using dynamical rooted staggered quarks with heavier-than-physical masses. Along with the full condensate, the sea and valence contributions are also plotted as a function of the magnetic flux quantum $N_b$ (denoted here by $b$)~\cite{DElia:2011koc}. For comparison, here $b=16$ corresponds to $eB\approx 0.5\GeVsq$.
 }
\end{figure}

The first lattice results about the magnetic catalysis of the quark condensate were obtained in~\cite{Buividovich:2008wf} using overlap quarks in the quenched approximation of two-color QCD. The down quark condensate was calculated here using the Banks-Casher relation~\eqref{eq:pd_BCrel}. The results revealed an accumulation of the Dirac eigenvalues around zero as the magnetic field grows and, accordingly, an increase in $\expv{\bar\psi_d\psi_d}$, see the left panel of Fig.~\ref{fig:pd_QCDcondensateT0}. Moreover, it was observed that the enhancement prevails at nonzero temperature, albeit the effect becomes milder there.

The same conclusion was reached in~\cite{DElia:2010abb,DElia:2011koc} using two flavors of dynamical rooted staggered quarks as well as in~\cite{Ilgenfritz:2012fw,Ilgenfritz:2013ara} in two-color QCD: a substantial enhancement of the condensate at low temperatures and a weaker, but still positive response at higher temperatures. We first focus on the low-temperature behavior and get back to high temperatures later.
The results are shown in the right panel of Fig.~\ref{fig:pd_QCDcondensateT0}, where the excess average condensate $\Delta \expv{\bar\psi\psi}=\expv{\bar\psi\psi}_B-\expv{\bar\psi\psi}_{B=0}$ is plotted (in units of the $B=0$ condensate).
The data reveal an enhancement, which is first quadratic, then turning into an approximately linear one, in line with our intuition from the free case in Fig.~\ref{fig:lat_freecondensate}. 

The study~\cite{DElia:2011koc} performed dynamical lattice simulations, where the magnetic field affects both the quark propagator as well as the distribution of gluon fields under the path integral through the quark determinant. In the language introduced in Sec.~\ref{sec:lat_valence_sea}, this means that both valence quarks and sea quarks are affected by the magnetic field. We will see that the individual contributions from sea and valence quarks play a crucial role for the physical understanding of the results at high temperature, so it is useful to discuss these features already at low $T$. Specifically, the valence and sea (or, in the notation of~\cite{DElia:2011koc}, dynamical) quark condensate is defined as\footnote{For rooted staggered quarks,~\eqref{eq:pd_valencedef} and~\eqref{eq:pd_seadef} should be modified to include the fourth root of the determinants as well as an overall factor of $1/4$ in front of the traces, as discussed above in Sec.~\ref{sec:lat_latDiracops}.}
\begin{align}
\expv{\bar\psi_f\psi_f}^{\rm val}_{B} &= \frac{1}{\Z(B=0)}
\int \D\U\, \exp\left[-\beta S_g\right] \, \prod_{f'} \det M_{f'}(B=0) \,\Tr \left[ M_f^{-1}(B)\right] \equiv
\Expv{\Tr
 \left[M_f^{-1}(B)\right] }_{B=0}\,, \label{eq:pd_valencedef}\\
\expv{\bar\psi_f\psi_f}^{\rm sea}_{B} &= \frac{1}{\Z(B)}
\int \D\U\, \exp\left[-\beta S_g\right] \, \prod_{f'} \det M_{f'}(B) \,\Tr \left[ M_f^{-1}(B=0)\right] \equiv
\Expv{\Tr
 \left[M_f^{-1}(B=0)\right] }_{B}\,. \label{eq:pd_seadef}
\end{align}
Similarly to 
the full observable, we can analogously introduce the excess (light quark average) valence and sea condensates due to the magnetic field, $\Delta \expv{\bar\psi\psi}^{\rm val}$ and $\Delta\expv{\bar\psi\psi}^{\rm sea}$. For weak magnetic fields, one can perform a leading Taylor-expansion in $B$ in order to show (see App.~\ref{app:valsea}),
\be
\Delta \expv{\bar\psi\psi}=\Delta \expv{\bar\psi\psi}^{\rm val} +\Delta \expv{\bar\psi\psi}^{\rm sea} + \epsilon\,\mathcal{O}(B^2)+ \mathcal{O}(B^4)\,.
\label{eq:pd_additivity}
\ee
The third term on the right hand side\footnote{This $\mathcal{O}(B^2)$ term was ignored in~\cite{DElia:2011koc}. We explain why it arises in App.~\ref{app:valsea}.} happens to come with a tiny coefficient $\epsilon$.
Thus, the valence-sea separation is reasonable for weak magnetic fields. The valence and sea contributions are also plotted in the right panel of Fig.~\ref{fig:pd_QCDcondensateT0}, demonstrating that the valence contribution is the dominant one and that the approximate leading-order additivity~\eqref{eq:pd_additivity} is indeed satisfied.
It is important to stress that this study used quite coarse lattices, with $a\approx 0.3\fm$ and an unimproved action, as well as heavier-than-physical quark masses. For the low-temperature results considered here, both of these aspects turned out to be unproblematic. However, for high temperatures -- in particular for the impact of the magnetic field on the transition temperature -- we will see that lattice artefacts and the specific value of the light quark mass play a substantial role.

Continuum extrapolated lattice results for the low-temperature magnetic catalysis were first presented in~\cite{Bali:2012zg}. This study used $2+1$ flavors of stout-improved rooted staggered quarks with physical masses, a tree-level Symanzik-improved gauge action and five different lattice spacings ranging down to $a=0.1\fm$.
For carrying out the continuum extrapolation, it is important that the observable is free of both multiplicative and additive divergences. As we discussed in Sec.~\ref{sec:lat_renormpbp}, this is the case for the combination $\Sigma_f$ defined in~\eqref{eq:lat_pbpren}. Similarly to the full observable, using~\eqref{eq:pd_valencedef} and~\eqref{eq:pd_seadef}, we can also define the normalized valence condensate $\Sigma^{\rm val}_f$ and sea condensate $\Sigma^{\rm sea}_f$. These are also renormalized quantities~\cite{Bruckmann:2013oba}.

\begin{figure}
 \centering
 \includegraphics[width=8.4cm]{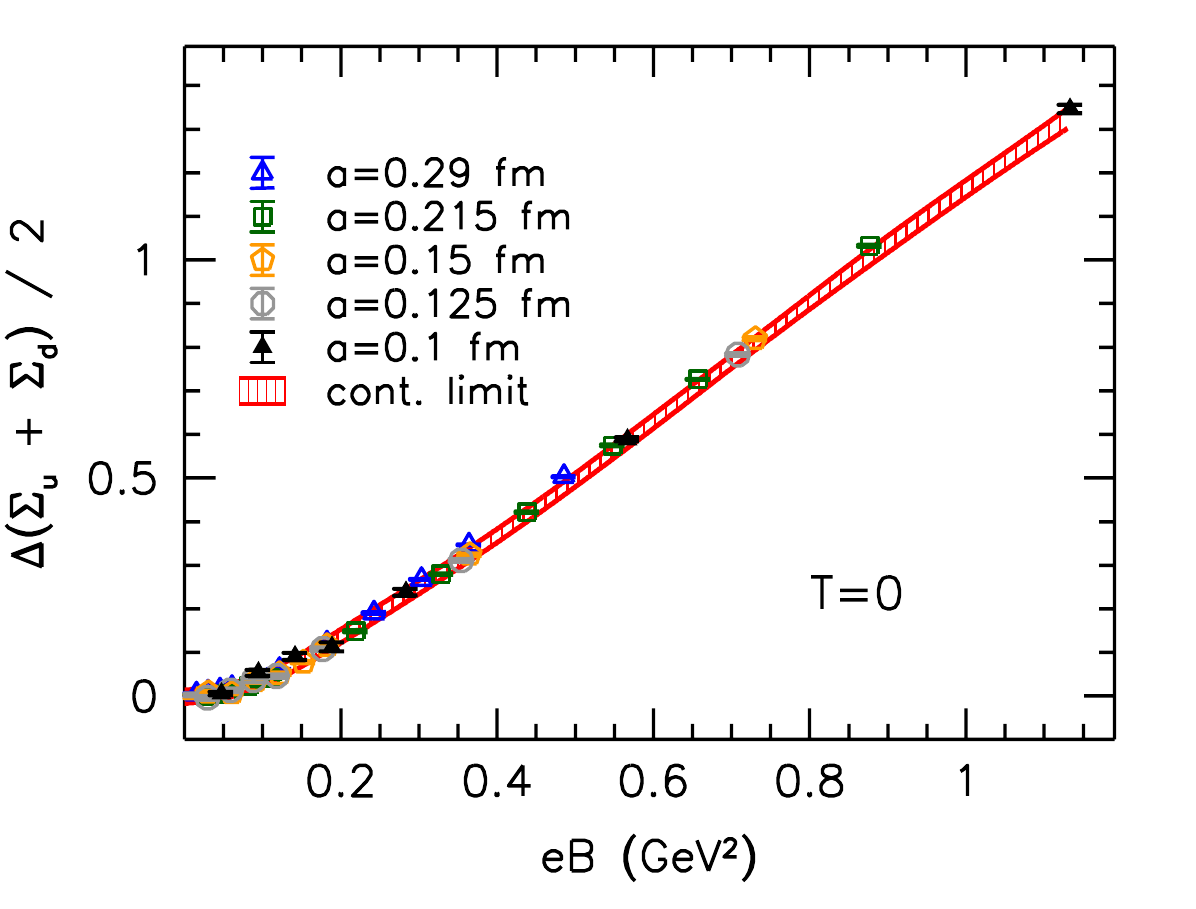} \quad
 \includegraphics[width=8.4cm]{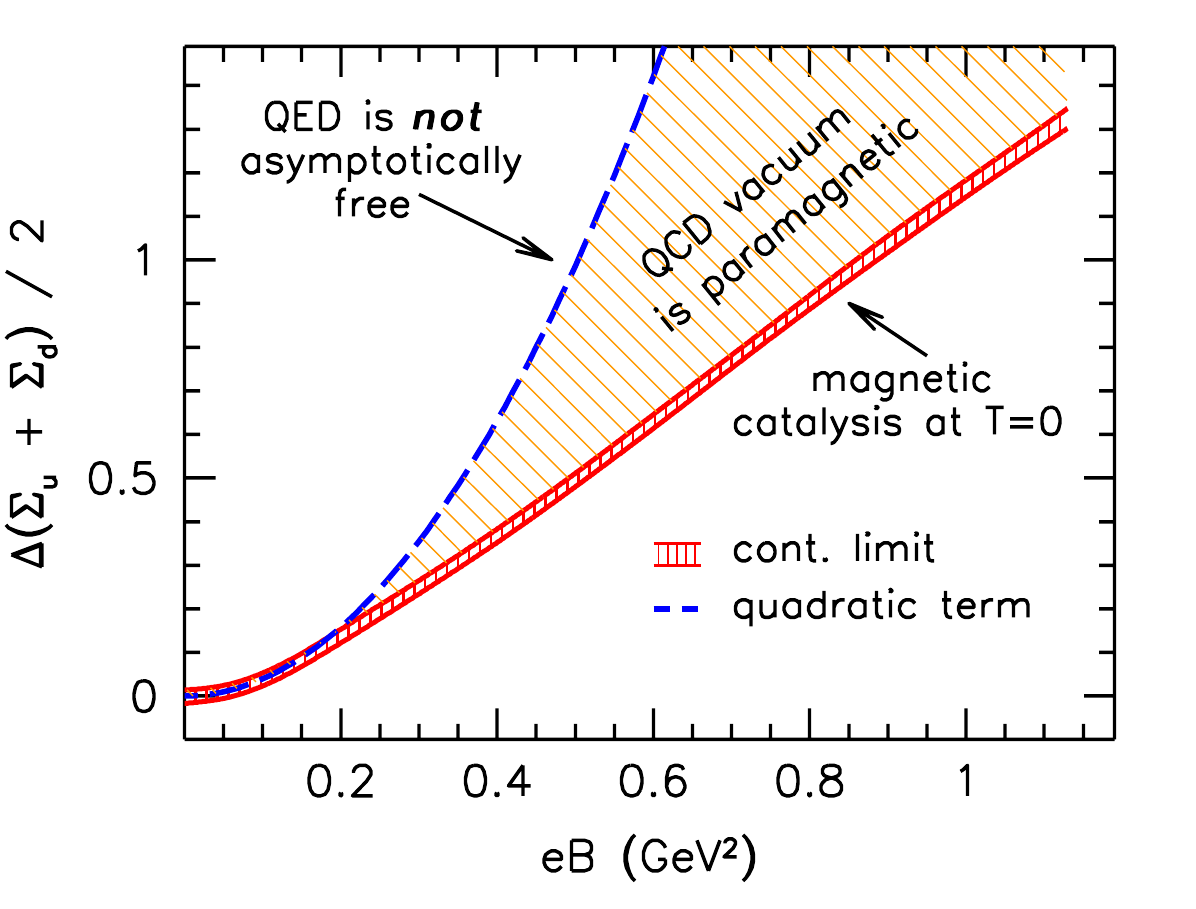}
 \caption{\label{fig:pd_QCDcondensateT0_2}
  Left panel: excess renormalized quark condensate due to the magnetic field. A continuum extrapolation is performed based on various different lattice spacings~\cite{Bali:2012zg}. Right panel: the interpretation of the continuum extrapolation from the left panel in terms of magnetic catalysis, the $\O(B^4)$ paramagnetism of the QCD vacuum and the absence of asymptotic freedom in (scalar) QED~\cite{Bali:2013txa}.
 }
\end{figure}

The continuum extrapolation for the average light quark condensate after $B=0$ subtraction, $(\Delta \Sigma_u + \Delta \Sigma_d)/2$ is shown in the left panel of Fig.~\ref{fig:pd_QCDcondensateT0_2}, clearly revealing the quadratic increase for weak fields and the quasi-linear behavior in the strong-field regime. It also shows that lattice artefacts for this observable (with this action and these lattice spacings) are completely under control.
This result constitutes a complete proof of the magnetic catalysis phenomenon in full QCD. Incidentally, this is a good time to compare the magnetic catalysis of the quark condensate~\cite{Bali:2012zg} to that of the gluon condensate~\cite{Bali:2013esa}, shown in the left panel of Fig.~\ref{fig:had_intmeas}, revealing consistent qualitative behaviors for these two observables.

Having obtained the $B$-dependence of the quark condensate in the vacuum, we can revisit the argument presented in Sec.~\ref{sec:lat_magncat_paramag} about relation to the $\O(B^4)$ paramagnetism of the QCD vacuum and the $\beta$-function coefficient appearing in the renormalization.
In this case, the effective degrees of freedom are charged pions, and the relevant $\beta$-function coefficient is the one for scalar QED~\cite{Bali:2013txa,Bali:2014kia}. It is positive just like $\beta_1$ for QED and matches the weak-field behavior of $\Delta\Sigma$, see Fig.~\ref{fig:pd_QCDcondensateT0_2}. 
Comparing to Fig.~\ref{fig:lat_freecondensate}, one again concludes that the higher-order terms in $B$ predestine the QCD vacuum to be paramagnetic to $\mathcal{O}(B^4)$~\cite{Bali:2013txa} -- a finding that we will get back to in Sec.~\ref{sec:eos_thermo_largeB}. This is also indicated in the right panel of Fig.~\ref{fig:pd_QCDcondensateT0_2}.

\begin{figure}
 \centering
 \includegraphics[width=8.4cm]{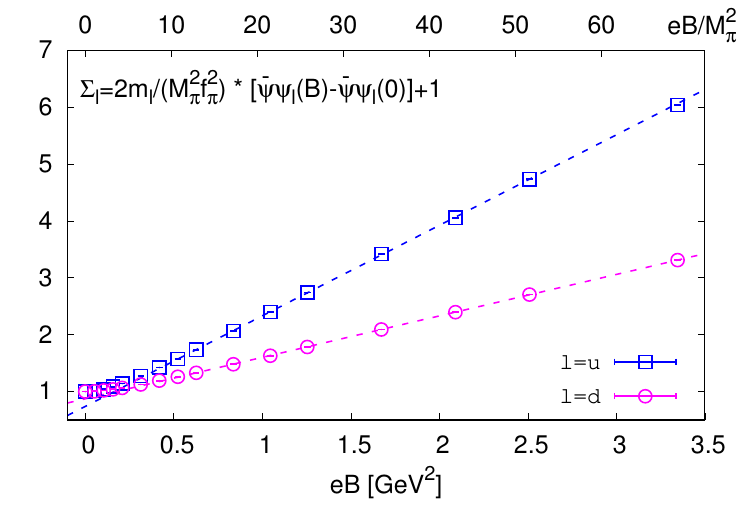}
 \caption{\label{fig:pd_condensate_u_d}
  Renormalized up and down quark condensates in strong magnetic fields~\cite{Ding:2020hxw} for a vacuum pion mass of $M_\pi=220\MeV$.
 }
\end{figure}

Finally, we note that the linear behavior of the condensate for strong fields was later confirmed by~\cite{Ding:2020hxw}  using HISQ fermions up to $eB=3.5\GeVsq$. The corresponding results, both for $\Sigma_u$ and for $\Sigma_d$, are shown in Fig.~\ref{fig:pd_condensate_u_d}. This study also demonstrated that the increase of the condensate is in a one-to-one correspondence with the reduction of the energy $\E_{\pi^0}$ of the (connected) neutral pion (see Fig.~\ref{fig:had_pi_masses2}) via a Ward-Takahashi identity~\cite{Ding:2020hxw}.
Concerning the strong-field regime; the study~\cite{DElia:2021tfb} used stout-improved quarks to push the limit for the strongest magnetic field up to $eB=9\GeVsq$. Below we will get back to the role of such strong magnetic fields when we discuss nonzero temperatures.

\subsubsection{Quark condensate in thermal QCD}
\label{sec:pd_condT}

The above results for the condensate were obtained at (approximately) zero temperature. In order to learn about the impact of the magnetic field on the QCD phase diagram, we need to extend the discussion to nonzero temperatures.
For $T>0$, the spectral representation~\eqref{eq:pd_BC1} of the quark condensate retains the same form as at $T=0$ -- we just need to evaluate the sum using the Dirac eigenvalues determined at finite temperature. As we argued in Sec.~\ref{sec:lat_landaulevelQCD}, the magnetic field always enhances the density of low eigenvalues, irrespectively of the underlying gluonic fields. Through the Banks-Casher relation~\eqref{eq:pd_BCrel}, this implies that for the value of the condensate operator evaluated on a given gluon field, is always increased when the magnetic field is switched on.

\begin{figure}
 \centering
 \mbox{
 \includegraphics[width=7.4cm]{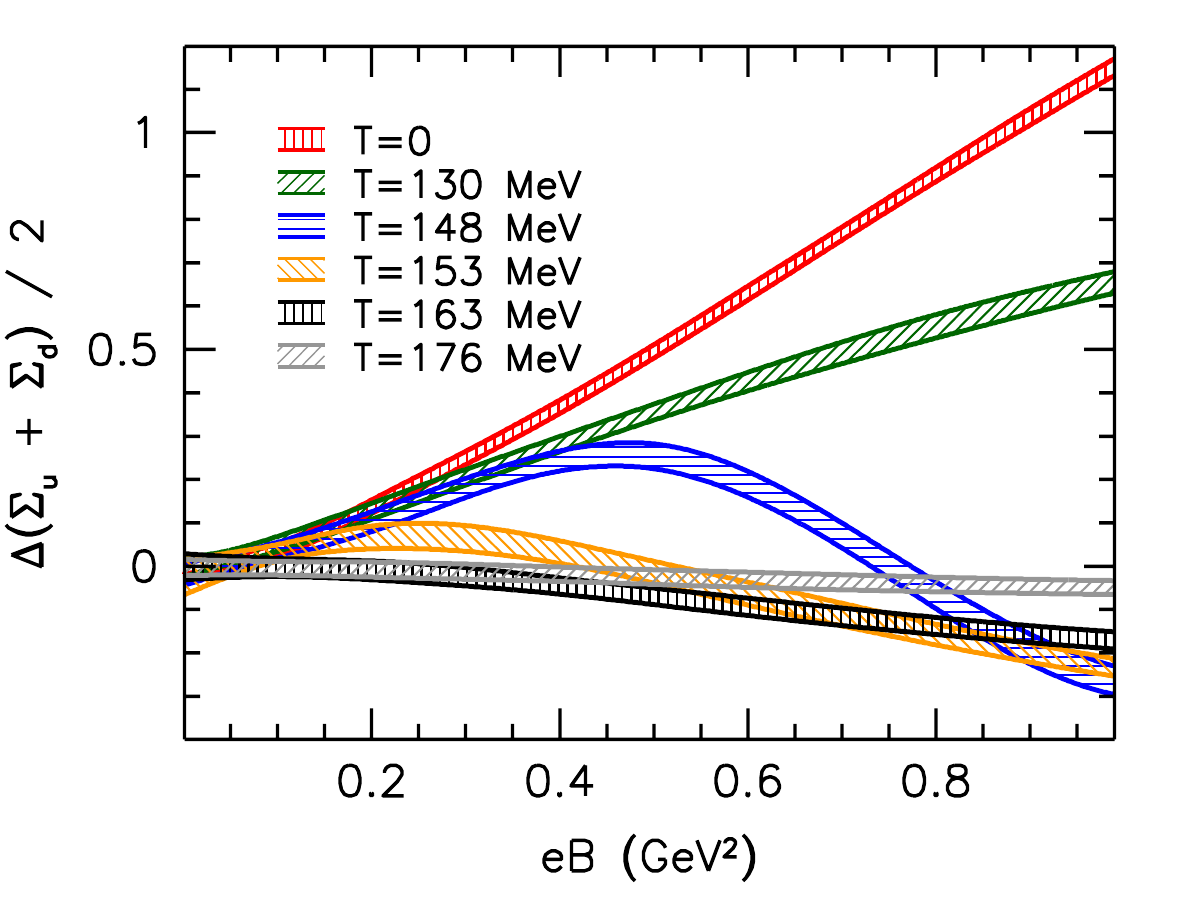}\hspace*{-.5cm}
 \raisebox{.4cm}{\includegraphics[width=5.9cm]{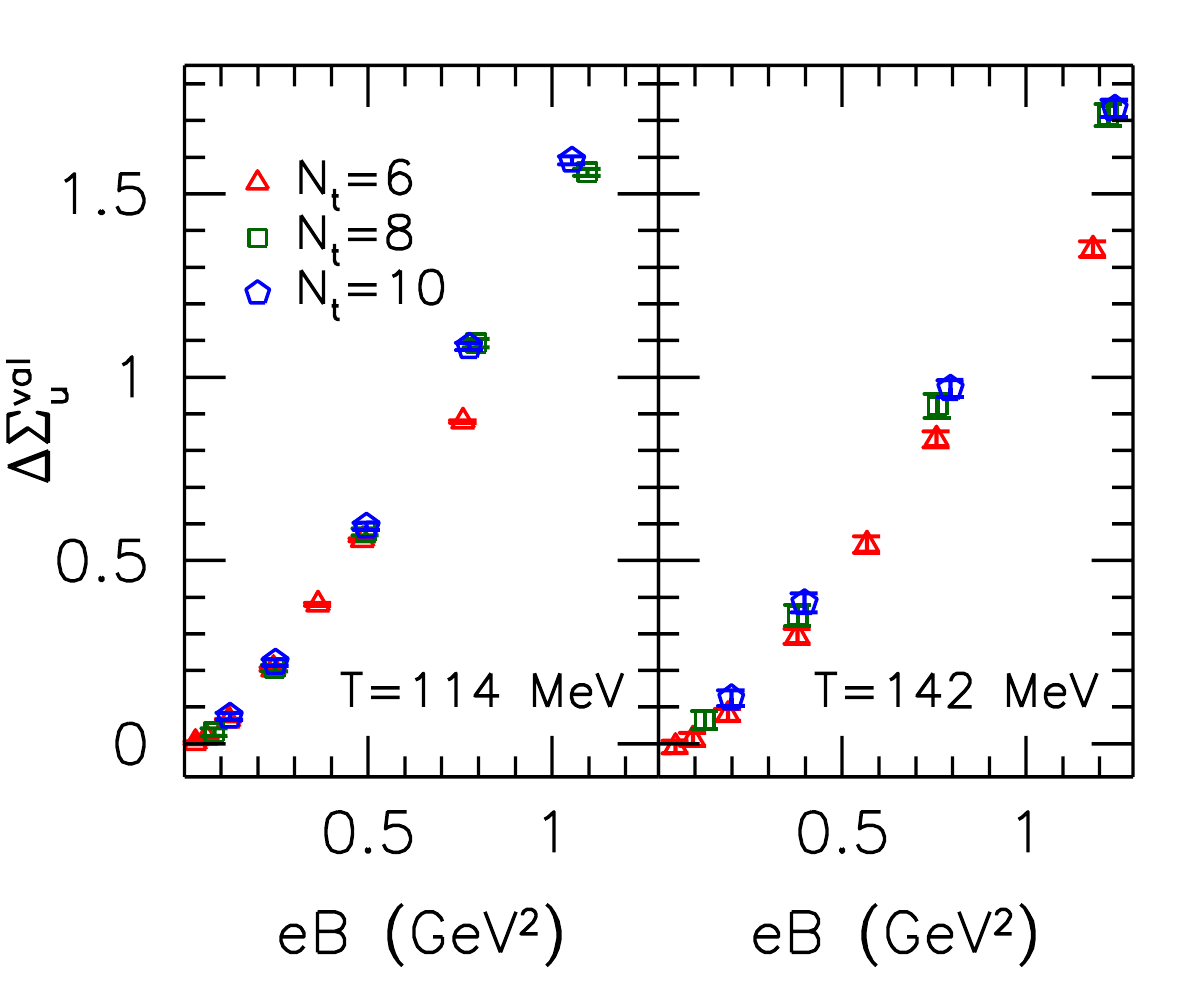}\hspace*{-.3cm}
 \includegraphics[width=5.9cm]{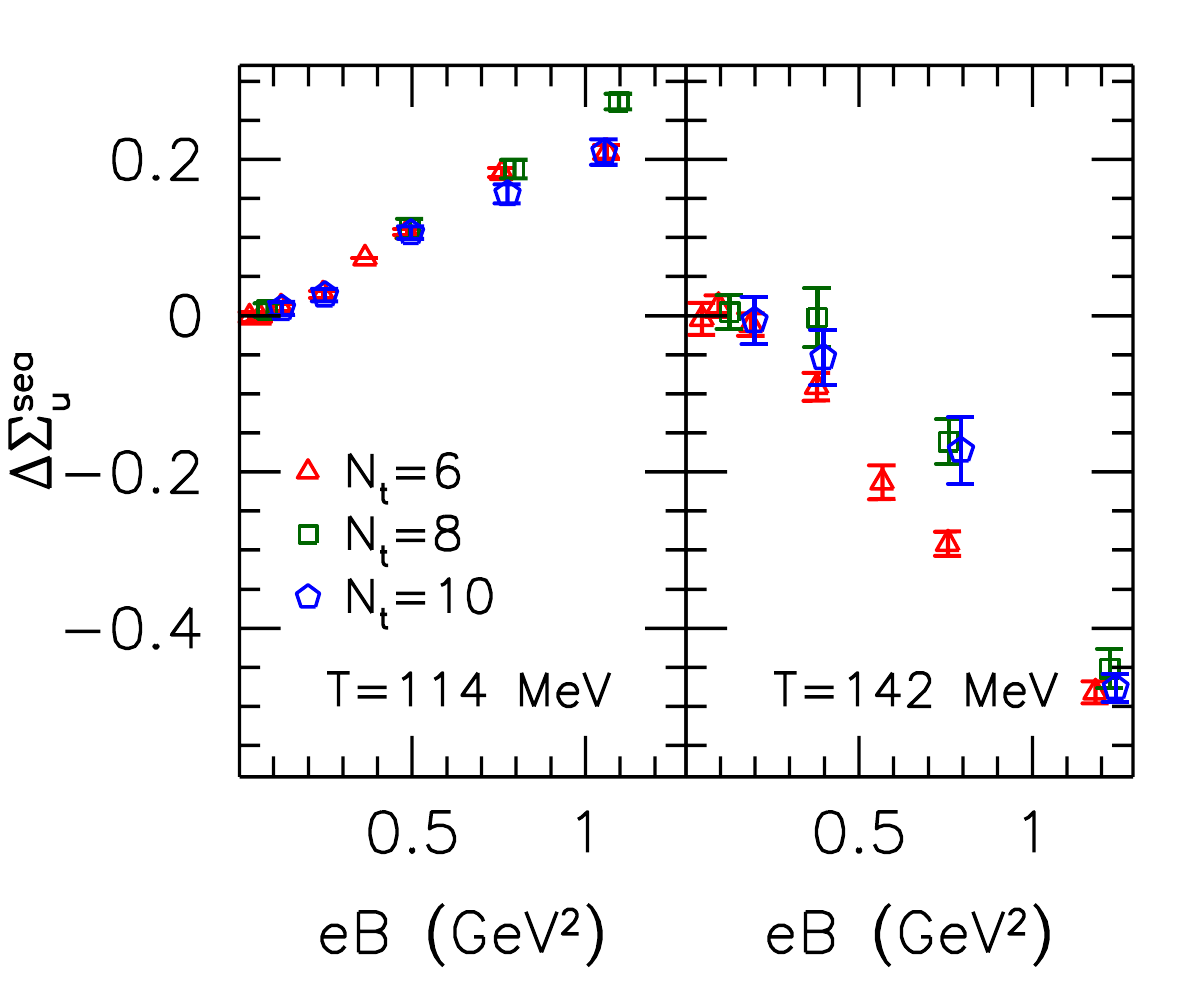}}}
 \caption{\label{fig:pd_QCDcondensateT}
  Left panel: dependence of the subtracted average light quark condensate on the magnetic field at different temperatures~\cite{Bali:2012zg}. Middle and right panels: Valence and sea condensates at two different temperatures~\cite{Bruckmann:2013oba}.
 }
\end{figure}

Notice that this statement concerns the value of the condensate operator on a given gluon configuration and does not say anything about how the gluon field configurations that dominate the path integral are themselves impacted by $B$.
In other words, the enhancement of the lowest Landau-level in QCD and the Banks-Casher relation imply that the {\it valence} quark condensate increases
\be
\Sigma^{\rm val}_f(B,T) >
\Sigma_f^{\rm val}(B=0,T) \qquad \forall \,T\,.
\label{eq:pd_valpbppos}
\ee
While there is no strict proof that~\eqref{eq:pd_valpbppos} holds for an arbitrary gluon field configuration, this inequality has been confirmed for a broad range of temperatures and quark masses~\cite{DElia:2011koc,Bruckmann:2013oba,DElia:2018xwo,Endrodi:2019zrl}. Two example temperatures from~\cite{Bruckmann:2013oba} are shown in the middle panel of Fig.~\ref{fig:pd_QCDcondensateT}. We note that the enhancement of the valence condensate is deeply related to the dependence of the so-called dressed Wilson loops~\cite{Bruckmann:2011zx} on the magnetic field and, thus, to confinement (see Sec.~\ref{sec:had_stringtension}).

However, the full expectation value is affected by the magnetic field not only in the condensate operator, but also in the quark determinants, i.e.\ not just via the valence but also via the sea effect.  To account for both effects, one needs to perform the full path integral, including dynamical, electrically charged quarks.
The first dynamical finite temperature study was carried out with heavier-than-physical rooted staggered quarks on coarse lattices in~\cite{DElia:2010abb}.
The results indicated that even after including the sea effect, the full condensate is increased by $B$ for all temperatures. By determining the inflection point of the quark condensate, the transition temperature $T_c(B)$ was shown to be a monotonously increasing function. This conclusion agreed with the prediction of most low-energy models and effective theories of QCD~\cite{Andersen:2014xxa}. However, later it was realized that the lattice results of~\cite{DElia:2010abb} were severely affected by lattice artefacts and missed the correct behavior in the continuum.

The first study that used physical quark masses and performed a continuum extrapolation for the condensate at nonzero temperatures and magnetic fields was carried out in~\cite{Bali:2011qj,Bali:2012zg}. A surprising result was observed: while the condensate undergoes magnetic catalysis at low temperature, it is reduced by the magnetic field in the transition region $T\approx T_c$.
This phenomenon was dubbed `inverse magnetic catalysis' in~\cite{Bruckmann:2013oba}. The results of~\cite{Bali:2012zg} for the average light quark condensate are shown as a function of the magnetic field for several temperatures in the left panel of Fig.~\ref{fig:pd_QCDcondensateT}. Just as at $T\approx0$, this behavior is again very similar to the dependence of the gluon condensate on $B$ and $T$~\cite{Bali:2013esa}, see the right panel of Fig.~\ref{fig:had_intmeas}.

For weak fields, the valence and sea effects have been shown to be approximately additive in the sense of~\eqref{eq:pd_additivity}. Thus the reduction of $\expv{\bar\psi\psi}$ around the crossover transition must be due to the {\it sea} contribution. This was explicitly demonstrated in~\cite{Bruckmann:2013oba}, showing that for the light quarks $f=u,d$ with physical quark masses,
\be
\Sigma_f^{\rm sea}(B,T) <
\Sigma_f^{\rm sea}(B=0,T) \qquad T\approx T_c\,.
\ee
This is visualized in the right panel of Fig.~\ref{fig:pd_QCDcondensateT}.
Inverse magnetic catalysis is therefore rooted in the sea contribution, i.e.\ in the impact of the magnetic field on the typical gluon field configurations that dominate the path integral. We will see below in Sec.~\ref{sec:pd_ploop} that the most important feature of the gluon fields that is affected by $B$ is the Polyakov loop. Following~\cite{Bruckmann:2013oba}, we will argue on general grounds that the Polyakov loop is enhanced by the magnetic field and relate this behavior to the sea quark condensate.

\begin{figure}
 \centering
 \includegraphics[width=8.cm]{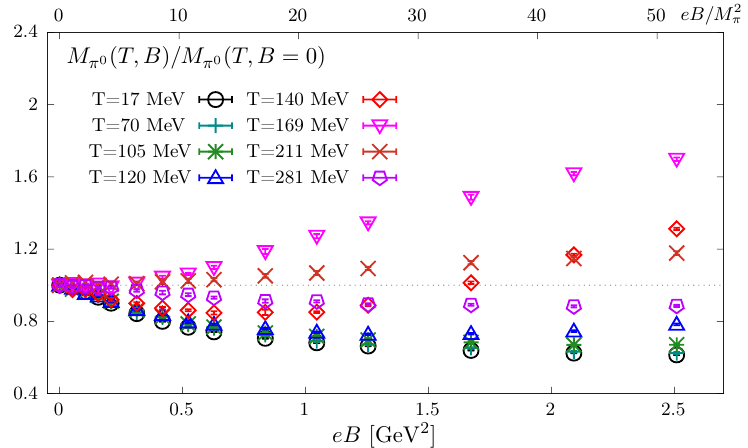}
 \mbox{
 \includegraphics[width=8.cm]{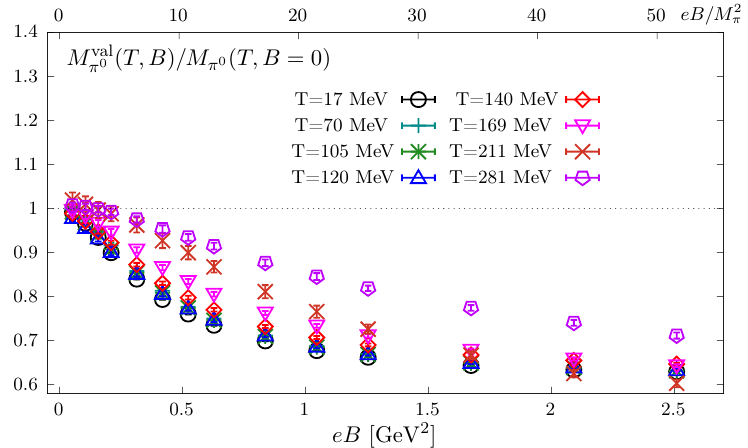}\qquad
 \includegraphics[width=8.cm]{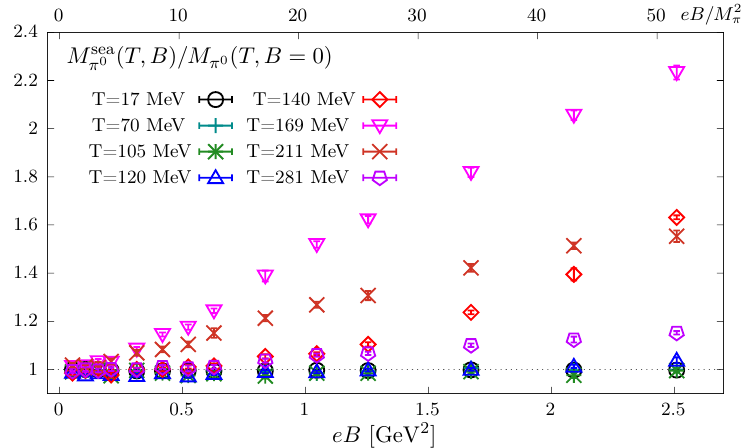}
 }
 \caption{
 \label{fig:pd_screeningmasses}
 Screening masses of the (connected) neutral pion as a function of the magnetic field for different temperatures, in units of the $B=0$ mass. The full magnetic field-dependence (upper panel) is decomposed into valence (lower left) and sea (lower right) contributions~\cite{Ding:2022tqn}.
 }
\end{figure}

In Sec.~\ref{sec:pbp_QCDT0} we noted that the zero-temperature magnetic catalysis of $\Sigma$ was found to be tied to the reduction of the (connected) neutral pion energy $\E_{\pi^0}$ in the vacuum via a Ward-Takahashi identity~\cite{Ding:2020hxw}. An analogous identity at nonzero temperature can be used to relate the intricate behavior of the condensate as a function of $T$ and $B$ to screening masses\footnote{Screening masses are determined from spatial correlation functions in a similar way as the vacuum energies from temporal correlators at $T=0$ discussed in Chap.~\ref{chap:hadron}~\cite{Aarts:2017rrl}. For $B>0$, the spatial correlation functions parallel and perpendicular to the magnetic field may be different. The study~\cite{Ding:2022tqn} discussed the parallel correlators.}, as shown in~\cite{Ding:2022tqn}. To quantify this, in Fig.~\ref{fig:pd_screeningmasses} we show the dependence of $\E_{\pi^0}$ on $T$ and $B$, revealing the already observed suppression by the magnetic field in the vacuum (see Fig.~\ref{fig:had_pi_masses2}), as well as an enhancement by $B$ in the transition region. Just as for the condensate, the impact of $B$ can be separated into a valence and a sea part~\cite{Ding:2022tqn}. The valence part always acts to reduce $\E_{\pi^0}$, while the sea contribution increases the energy. This is in complete correspondence with the behavior observed for the condensate in Fig.~\ref{fig:pd_QCDcondensateT}.

\subsubsection{Inhomogeneous magnetic fields}

Recently, the magnetic catalysis and inverse magnetic catalysis phenomena were generalized to the case of inhomogeneous magnetic fields~\cite{Brandt:2023dir}. In particular, the localized profile~\eqref{eq:lat_loc_Bfield} was employed in dynamical simulations involving stout-improved staggered quarks with physical masses. Both the quark condensate and the Polyakov loop were shown to develop non-trivial spatial profiles as a response to the inhomogeneous magnetic field. At low temperatures, the quark condensate was shown to undergo local magnetic catalysis by following approximately the $B(x_1)$ dependence in its spatial profile. In turn, in the transition region the condensate develops a more complicated profile due to the interplay of the direct effect and the indirect gluonic fluctuations~\cite{Brandt:2023dir}. 

In addition, the same type of inhomogeneous magnetic fields were found to lead to the appearance of local electric currents~\cite{Brandt:2024blb}. These are steady currents that flow in equilibrium in accordance with Amp{\'e}re's law and can be employed to determine the magnetic susceptibility of the QCD medium, as we discuss below in Sec.~\ref{sec:eos_ampere}.

\subsection{Phase diagram in the magnetic field -- temperature plane}
\label{sec:pd_phasediagBT}

\begin{figure}
 \centering
 \includegraphics[width=8.4cm]{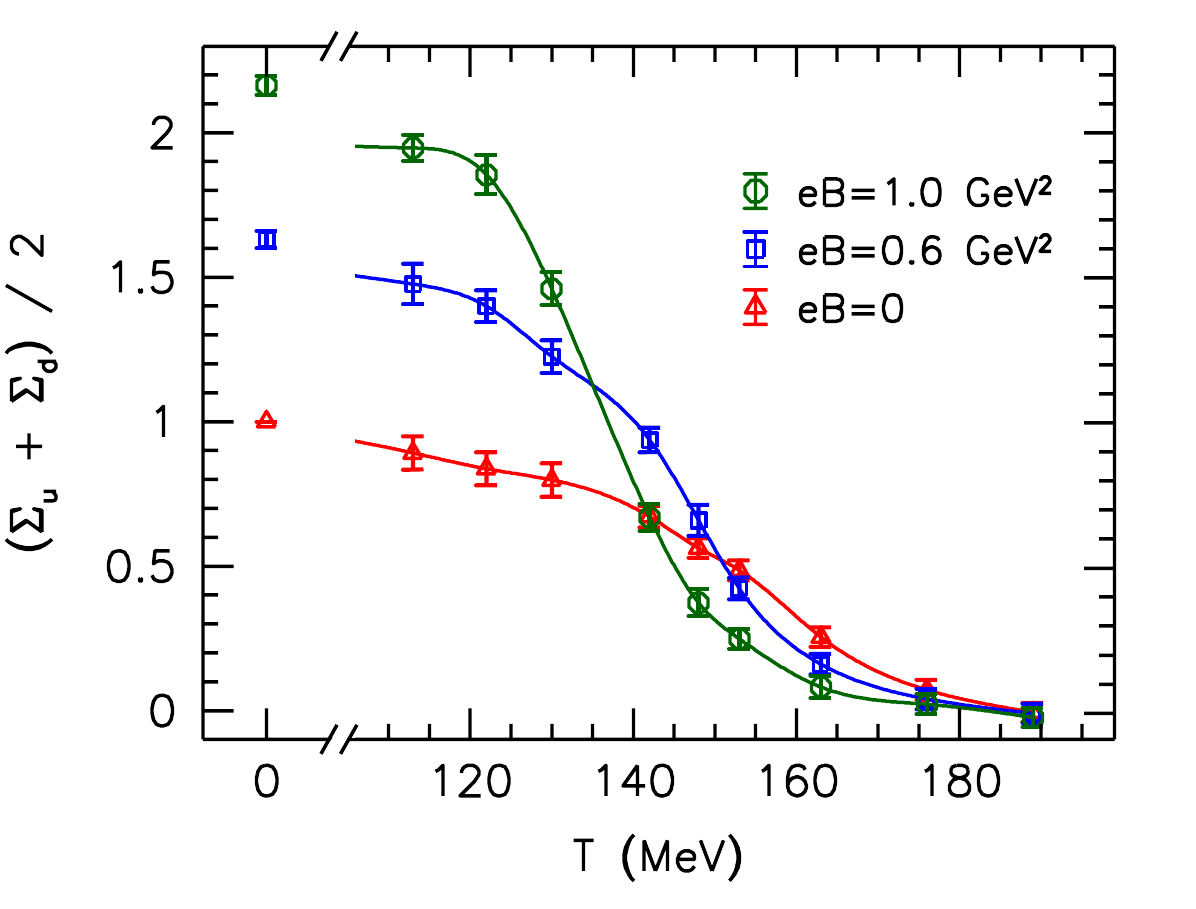} \quad
 \raisebox{.03cm}{\includegraphics[width=8.2cm]{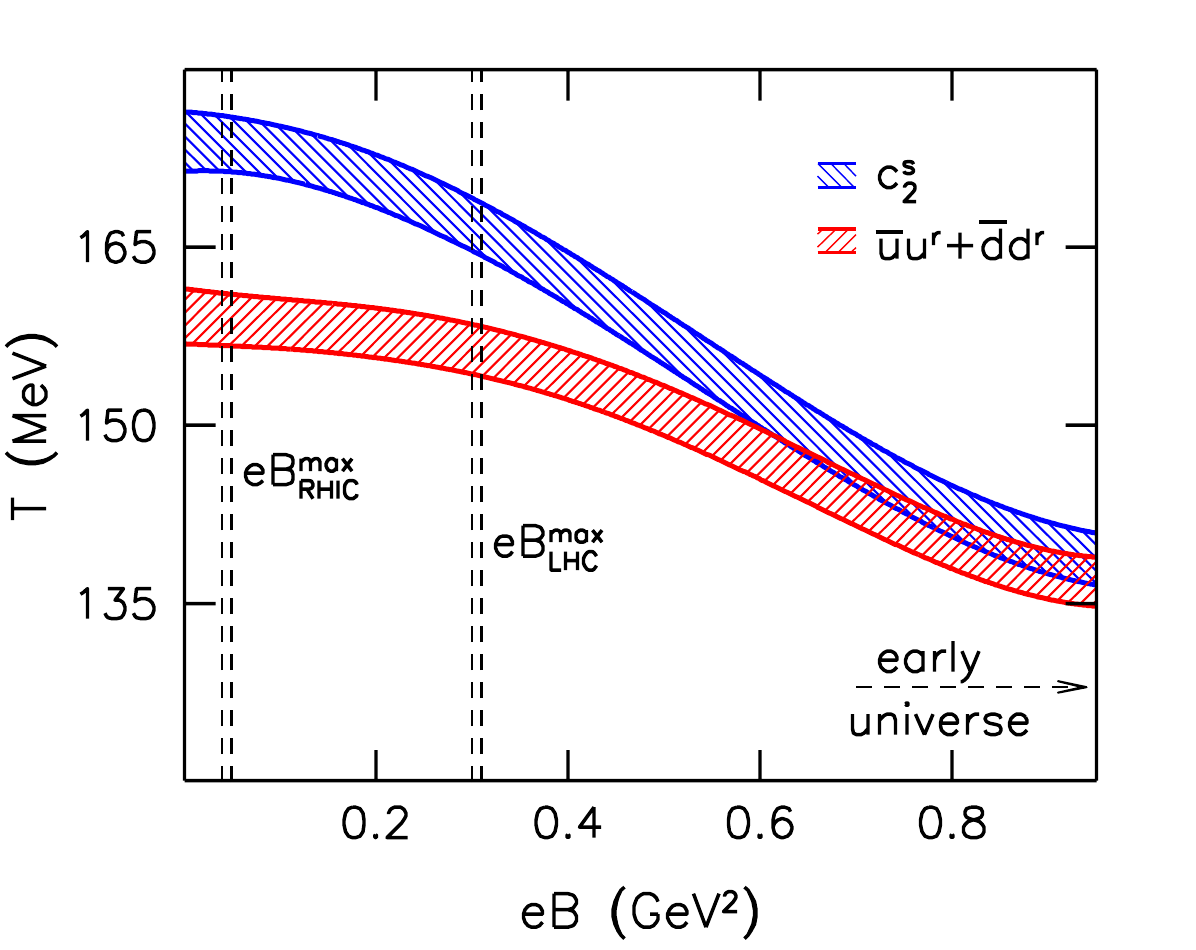}}
 \caption{\label{fig:pd_phasediag}
  Left panel: renormalized average light quark condensate as a function of the temperature for three different magnetic field values~\cite{Bali:2012zg}. The figure is taken from~\cite{Endrodi:2014vza}. Right panel: phase diagram in the temperature-magnetic field plane based on the inflection point of $\Sigma$ (red band, denoted as $\bar u u^r+\bar d d^r$ in the legend) and on that of the strange quark number susceptibility $\chi_2^{\S}$ (blue band, denoted as $c_2^s$)~\cite{Bali:2011qj}.
 }
\end{figure}

Before we turn to the Polyakov loop, we continue the discussion with the full quark condensate and the impact of $B$ on the QCD phase diagram. The continuum extrapolated results of~\cite{Bali:2012zg} for the average light quark condensate, in the normalization~\eqref{eq:lat_pbpren}, are shown in the left panel of Fig.~\ref{fig:pd_phasediag}. This is the same set of results as in the left panel of Fig.~\ref{fig:pd_QCDcondensateT}, just this time as a function of the temperature for various magnetic fields. Here, both the low-temperature magnetic catalysis as well as the inverse magnetic catalysis in the transition region is visible. In fact, for sufficiently high temperature, inverse magnetic catalysis again ceases to take place and $\Delta\Sigma$ becomes positive.

Presenting the temperature-dependence $\Sigma(T)$ is helpful to visualize how the transition temperature changes as the magnetic field is varied. Given the crossover nature of the transition, the pseudocritical temperature is not uniquely defined, as we discussed in the beginning of Chap.~\ref{chap:pd}. One possibility is to associate the inflection point of $\Sigma(T)$ with $T_c$. From the left panel of Fig.~\ref{fig:pd_phasediag}, it is clear that the so defined transition temperature decreases as $B$ grows. This curve was calculated using the continuum extrapolated staggered results in~\cite{Bali:2011qj} and is shown in the right panel of Fig.~\ref{fig:pd_phasediag}. Another alternative is the inflection point of the strange quark number susceptibility $\chi_2^{\S}$. We will discuss this observable in more detail in Sec.~\ref{sec:pd_fluc}; here it suffices to say that $\chi_2^{\S}$ measures the fluctuations of the strangeness quantum number in the QCD medium and is also sensitive to the crossover transition. Both definitions lead to a monotonically decreasing $T_c(B)$ function, as visible in the right panel of Fig.~\ref{fig:pd_phasediag}. This behavior, observed in~\cite{Bali:2011qj} for $eB\le 1\GeVsq$ was later shown to persist up to $eB=3.25\GeVsq$~\cite{Endrodi:2015oba}.

Historically, the inverse magnetic catalysis of the condensate in the transition region and the reduction of the transition temperature $T_c(B)$ by the magnetic field were often assumed to be equivalent phenomena. It is important to stress that due to the crossover nature of the transition, this is not the case. The inflection point of $\Sigma(T)$ may shift to lower temperatures as $B$ grows even if $\Delta\Sigma$ is positive for all temperatures. Vice versa, $\Delta\Sigma<0$ for some temperatures does not necessarily lead to a reduction of the inflection point.
It was the study~\cite{DElia:2018xwo} that first demonstrated explicitly the independence of these two aspects. Here, the analysis of the light quark condensates was repeated with the same improved action (using one lattice spacing at $N_t=6$) as in~\cite{Bali:2012zg}, but with heavier-than-physical quark masses. In particular, the results showed that for a pion mass of $M_\pi=664 \MeV$, inverse magnetic catalysis does not take place, but the transition temperature is still a decreasing function of $B$. This study also demonstrated that the sea contribution still reduces the condensate in the transition region -- but, unlike at the physical point, for such heavy quarks the valence effect is the dominant one. It was later shown (at one lattice spacing, $N_t=6$) that the limiting pion mass, where inverse magnetic catalysis disappears is at $M_\pi\approx 500 \MeV$~\cite{Endrodi:2019zrl}.

We note that unimproved lattice results with almost physical quark masses also showed signs of magnetic catalysis and an increasing $T_c(B)$~\cite{Ding:2020inp}, see also~\cite{Cea:2015ifa} for results in $N_f=1$ QCD and~\cite{Ilgenfritz:2013ara} for results in $N_f=4$ two-color QCD. Preliminary results using an improved (HISQ) staggered action observed inverse magnetic catalysis at different light pion masses~\cite{Tomiya:2019nym}.
The only study that used a different fermionic discretization is~\cite{Bornyakov:2013eya}, where two flavors of dynamical overlap quarks with a pion mass of $M_\pi\approx 500 \MeV$ were simulated on a single lattice spacing ($N_t=6$). The results indicated an enhancement of the Polyakov loop and a weak inverse magnetic catalysis at one temperature in the transition region.

In summary, based on the above discussed results for the phase diagram and the corresponding relevant observables, there is very strong evidence from several lattice groups for the following picture.
The magnetic field always enhances the valence quark condensate due to the proliferation of low eigenvalues on the Landau-levels via the Banks-Casher relation. Simultaneously, it also reduces the sea quark condensate (via a coupling to the Polyakov loop, see below in Sec.~\ref{sec:pd_ploop}). There is therefore a competition between the two effects: for heavy quarks, the valence effect is dominant and for light quarks the sea effect.
Irrespective of this behavior of the quark condensate, the magnetic field always appears to enhance the Polyakov loop and therefore reduce the transition temperature.
The reduction of $T_c(B)$ by the magnetic field is therefore less the manifestation of the inverse magnetic catalysis of the condensate but more a gluonic effect encoded in the behavior of the Polyakov loop at nonzero magnetic fields. The study~\cite{DElia:2018xwo} proposed to refer to this notion as `deconfinement catalysis'.

As mentioned above, most low-energy models and effective theories of QCD fail to describe these features of the quark condensate and the phase diagram~\cite{Andersen:2014xxa}. 
The lattice results~\cite{Bali:2011qj,Bali:2012zg,Bruckmann:2013oba} for $T_c(B)$ and $\Sigma(B,T)$ have been employed to improve such QCD models on many occasions, for example by considering magnetic field dependent model parameters, see e.g.~\cite{Fraga:2013ova,Farias:2014eca,Ferreira:2014kpa,Ayala:2014iba,Moreira:2022dwo}. For a summary of these approaches, we refer the reader to the overviews~\cite{Andersen:2014xxa,Cao:2021rwx,Andersen:2021lnk}.

We have already seen that the dependence of the quark condensate on $T$ and $B$ agrees very well with that of the energy of the (connected) neutral pion. From a slightly different viewpoint, the reduction of $T_c$ by the magnetic field may also be interpreted in terms of the impact of $B$ on  the lightest hadronic excitation. In fact, the dependences $\E_{\pi^0}(B)$ and $T_c(B)$ compare quantitatively well to each other~\cite{Bali:2017ian,Ding:2020hxw}. Finally, it is worth pointing out that the qualitative features of the phase diagram in the $T-B$ plane are quite different compared to the one defined in the case of background chromomagnetic fields, see the recent work~\cite{Cea:2024efe}.

\subsection{Very strong magnetic fields}
\label{sec:pd_largeB}

As discussed above, the magnetic field increases the number of low-lying eigenvalues of the Dirac operator and, via the spectral representation~\eqref{eq:pd_BC1}, the valence quark condensate. 
As the magnetic field grows and considerably exceeds all dimensionful scales in the system, these low eigenvalues are determined more and more by $B$ and less and less by the gluon field. Eventually, in the asymptotic magnetic field limit, the quark and gluon degrees of freedom decouple from each other. The system factorizes into a non-interacting quark theory -- which is irrelevant for the phase diagram -- and a pure gluon theory. This pure gluon theory inherits the spatial anisotropy induced by the magnetic field.

More specifically, the $B\to\infty$ limit is described by an effective field theory, in which the residual quark effects can be obtained by integrating out the magnetized fermion degrees of freedom. The result is an anisotropic gauge theory with the Euclidean continuum action (again assuming that $B$ points in the positive $x_3$ direction),
\be
S_G^{\rm aniso} = 
\frac{1}{2{g'}^2} \int \dd^3x \int_0^{1/T} \!\!\dd x_4 \;\left[ 
\tr \,\F_{\nu\rho}(x) \F_{\nu\rho}(x)
+g'^2\kappa(B) \,\tr \, \F_{43}\F_{43}\,,
\right]
\label{eq:pd_sganiso}
\ee
differing from the original gluon action $S_G$ in~\eqref{eq:lat_action_cont} by an overall renormalization of the coupling $g'$ as well as by the anisotropy, described by the parameter $\kappa$,
\be
\kappa(B) = \frac{1}{24\pi^2}\sum_f \left|\frac{q_f}{e}\right| \frac{|eB|}{m_f^2}\,.
\ee
This result can be derived by considering the gluon propagator in strong magnetic fields~\cite{Miransky:2002rp} as well as via the Euler-Heisenberg action for strong background magnetic fields and weak chromoelectric and chromomagnetic fields~\cite{Endrodi:2015oba}. Notice that the anisotropy $\kappa(B)>0$ enhances the coefficient of the chromoelectric field $\F_{43}=\E_3$ parallel to the background magnetic field, in effect suppressing its fluctuations. In the asymptotic magnetic field limit only the minimum, $\F_{43}=0$ contributes.

Being a pure gluon theory, this system is expected to exhibit a first-order deconfinement phase transition~\cite{Miransky:2002rp,Cohen:2013zja} with the Polyakov loop $P$ as the order parameter. The action~\eqref{eq:pd_sganiso} can be discretized just as the usual gauge action $S_G$ and simulated on the lattice. This was carried out using the anisotropic version of the tree-level Symanzik improved action in~\cite{Endrodi:2015oba} using one lattice spacing ($N_t=4$). Via a finite-size scaling analysis, the results indeed revealed a first-order phase transition, for which the Polyakov loop susceptibility becomes singular at the critical temperature in the infinite volume limit.
To corroborate this finding, a double peak structure was shown to appear in the histogram of the order parameter at the critical temperature -- a telltale sign of a first-order phase transition.

The first-order phase transition at $B\to\infty$ has a finite latent heat. As the magnetic field is reduced, this latent heat can change continuously and disappear at a critical magnetic field $B_{\rm CP}$, where the transition turns into a second-order phase transition, marking a critical point in the phase diagram. This criticality has been expected to be visible when approaching the critical point from lower magnetic fields. In fact, signs for the strengthening of the crossover transition as $B$ grows were already observed in~\cite{DElia:2010abb,Bali:2011qj,Endrodi:2015oba,Ding:2020inp}. Based on the behavior of the chiral susceptibility peak for increasing magnetic fields, an estimate $eB_{\rm CP}\approx 10(2)\GeVsq$ was given~\cite{Endrodi:2015oba}. This prediction was made more accurate by the study~\cite{DElia:2021yvk}, which explicitly simulated magnetic fields of comparable strengths using stout-improved staggered quarks. At $eB= 9\GeVsq$, a first-order phase transition was indeed finally observed.

\begin{figure}
 \centering
 \includegraphics[width=8.2cm]{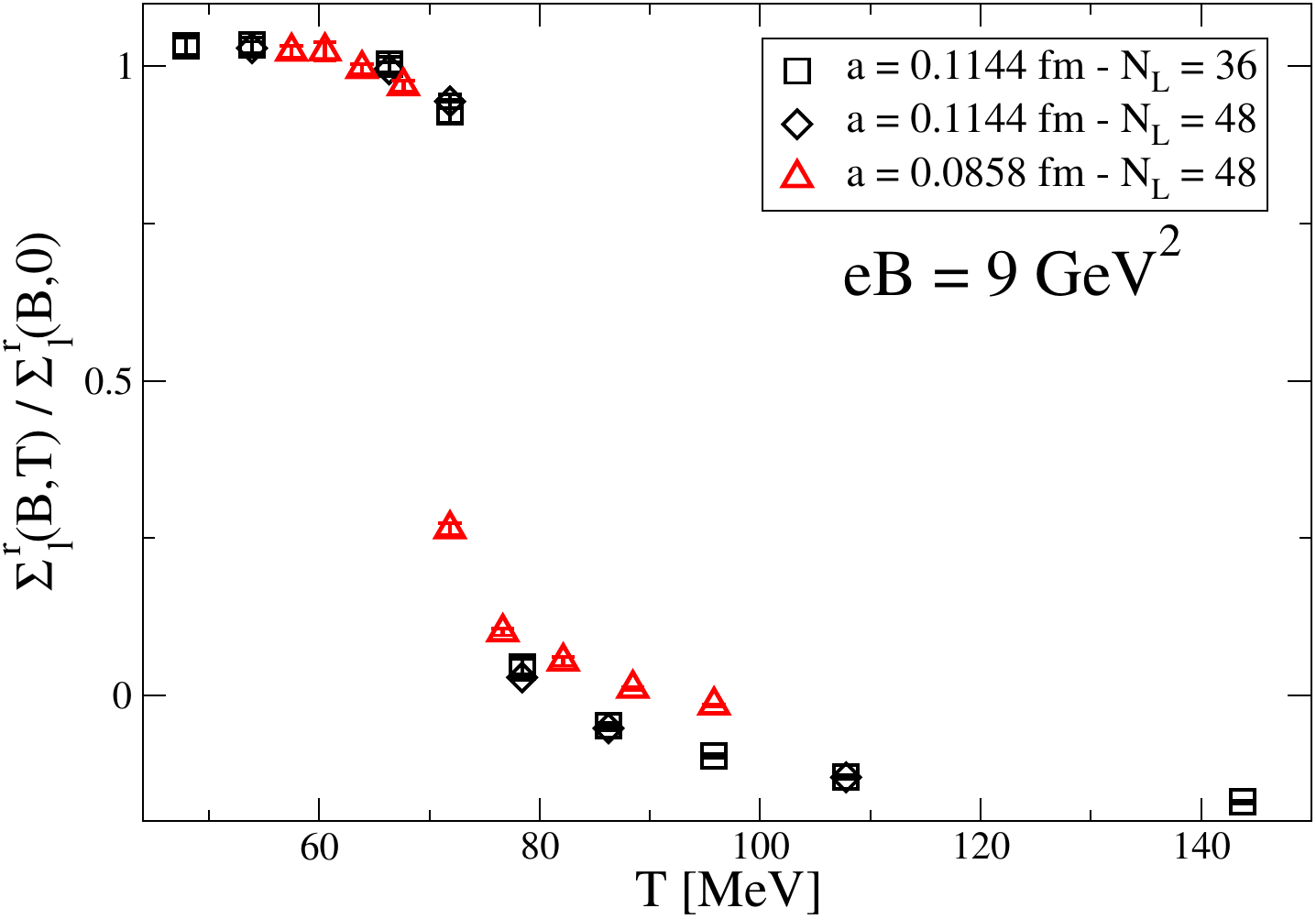}\qquad
 \raisebox{-.15cm}{\includegraphics[width=8.2cm]{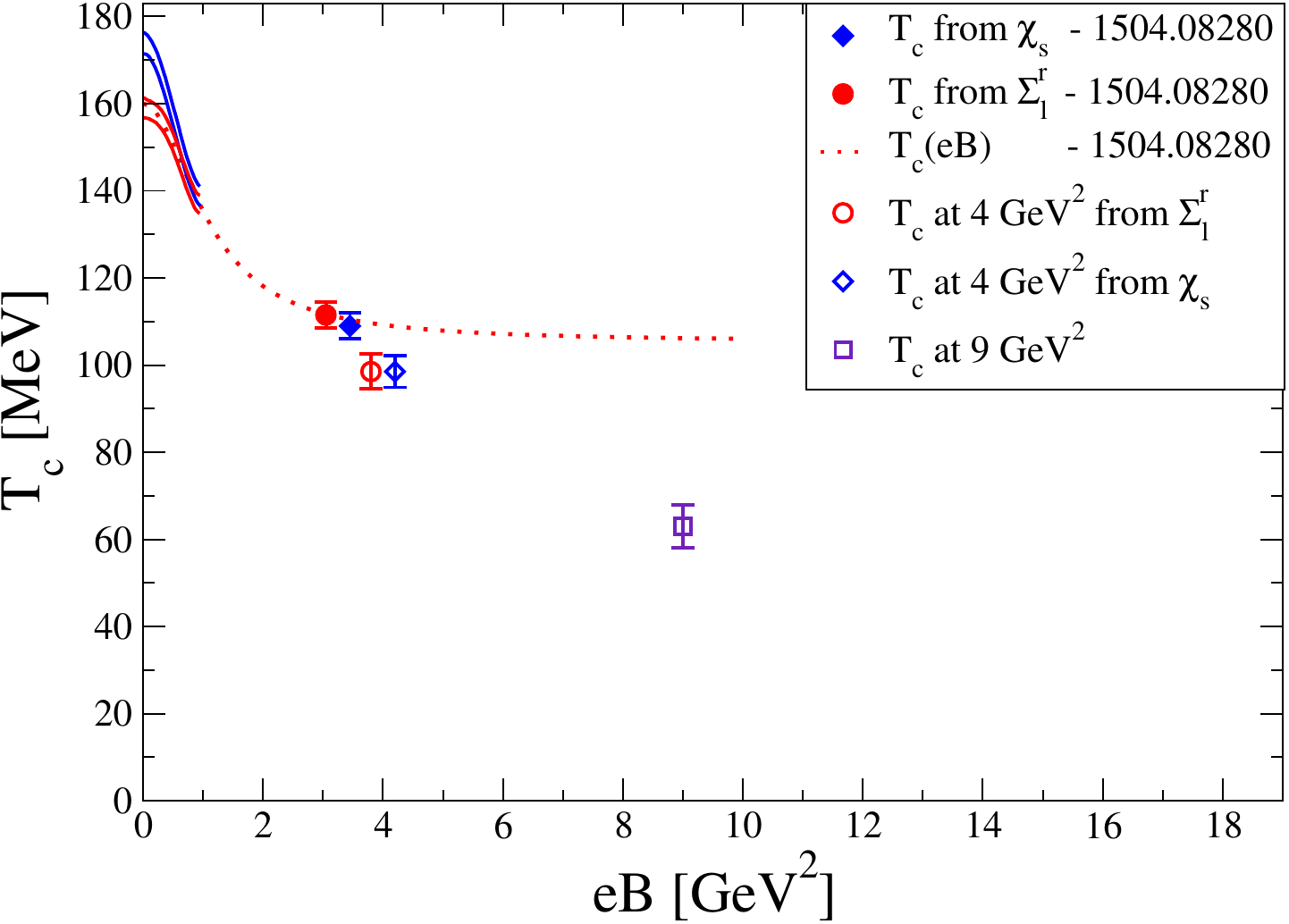}}
 \caption{\label{fig:pd_phasediag2}
  Left panel: renormalized average light quark condensate (here denoted as $\Sigma_l^r$) as a function of the temperature at $eB=9\GeVsq$~\cite{DElia:2021yvk}. The transition is of first order, as revealed by the distinct jump of the condensate at the critical temperature $T_c\approx 70\MeV$. Right panel: the phase diagram in the temperature-magnetic field plane~\cite{DElia:2021yvk}, collecting results on the condensate (red points) and on the strange quark number susceptibility (blue points) from~\cite{Bali:2011qj,Endrodi:2015oba,DElia:2021yvk}. The purple point is based on both observables and the dotted line the interpolation constructed in~\cite{Endrodi:2015oba}.
 }
\end{figure}

The biggest challenge in simulating such large values of $B$ is that the lattice must be sufficiently fine in order to resolve the magnetic field (cf.\ the discussion at the end of Sec.~\ref{sec:lat_contlimit}). In turn, this makes it very difficult to reach large physical volumes. In particular,~\cite{DElia:2021yvk} used rather small aspect ratios $N_s/N_t$ and a fixed-$\beta$ approach with two different lattice spacings. The results for $\Sigma$ (normalized by its value at $T\approx0$, in order to further reduce lattice artefacts) are plotted in the left panel of Fig.~\ref{fig:pd_phasediag2}. Notice the pronounced jump in the temperature-dependence of $\Sigma$, very different from the steep but clearly continuous behavior at weaker magnetic fields, as visible in the left panel of Fig.~\ref{fig:pd_phasediag}. This work also investigated a weaker magnetic field, $eB=4\GeVsq$ -- here the transition was still found to be a crossover~\cite{DElia:2021yvk}. Altogether, this lead to the conclusion that the critical point lies in the interval $4\GeVsq < eB_{\rm CP} < 9 \GeVsq$.
The histogram of the quark condensate was shown to exhibit a double-peak structure at the critical temperature for $eB=9\GeVsq$, confirming the first-order nature of the phase transition.

We close this section with a discussion about the relation of this phase transition to chiral symmetry restoration.
The lattice findings for $eB\le 4\GeVsq$ reveal the strengthening of the crossover transition, manifested in the temperature-dependence of both the quark condensate and the Polyakov loop (see below). The results at $eB=9\GeVsq$ demonstrate a first-order transition in the condensate\footnote{While the Polyakov loop expectation value was not calculated in~\cite{DElia:2021yvk}, a jump in $\expv{P}$ at the same temperature where the condensate exhibits a jump, is strongly expected~\cite{Massimo_privcomm}.}.
In turn, quarks decouple completely in the asymptotic magnetic field limit and in this region the transition manifests itself in the discontinuous behavior of the Polyakov loop. The condensate is determined exclusively by the magnetic field and is insensitive to gluons.
This tendency is supposed to be approached as one increases the magnetic field in lattice simulations.
Here we sketch how this might occur.
Notice that while the condensate exhibits a substantial jump, its overall scale is given by $\Sigma(B,T=0)$. This scale is linearly proportional to the magnetic field~\cite{Bali:2012zg,DElia:2021tfb} and assumes the value $\Sigma(B=9\GeVsq,T=0)\approx 13$~\cite{DElia:2021tfb} i.e.\ thirteen times larger than the vacuum value. Thus, while the condensate is discontinuous, its jump is getting relatively smaller. In order for this scenario to hold, one should show that the high-temperature value of the magnetic field still scales linearly with $B$.

Finally, we note that the findings of~\cite{DElia:2021tfb} constitute the first ever determination of a first-order phase transition in full QCD with physical parameters using an improved lattice action. The critical point in the $T-B$ plane might be discussed analogously to the critical point that is expected to emerge at large baryon chemical potentials~\cite{Aarts:2023vsf}. One might even conjecture that, if the latter indeed exists, the two critical points are continuously connected by a critical line~\cite{MarquesValois:2023ehu}. 
Another possibility is that the magnetic critical point is connected to the endpoint of the Roberge-Weiss transition~\cite{Roberge:1986mm} at imaginary chemical potentials~\cite{KZ_latticetalk}.
Such connections might provide 
alternative routes for lattice practitioners to study the critical behavior of hot and dense QCD.

\subsection{Polyakov loop}
\label{sec:pd_ploop}

In Secs.~\ref{sec:pd_condT} and~\ref{sec:pd_phasediagBT} we argued that both the reduction of the transition temperature by $B$ and the inverse magnetic catalysis of the sea quark condensate $\Sigma^{\rm sea}$ in the transition region are tied to gluonic effects: the typical gluon fields that dominate the path integral change as the magnetic field varies. The physical components of the gluon field may be expressed in terms of gauge invariant observables -- traces of products of gluon links $\U_\nu(n)$ along closed loops. The most important such observable, relevant for the high-temperature behavior of the system is the Polyakov loop $P[\U]$, defined above in~\eqref{eq:lat_ploopavg}. A measure for how the important gluon fields $\{\U\}$ are affected by the magnetic field may therefore be obtained by the expectation value $\expv{P}_{B,T}$. More information is contained in the Polyakov loop effective potential\footnote{For rooted staggered quarks, the determinant under the path integral is raised to the power $1/4$.},
\be
\Omega^P_{B,T}(p) = -\frac{T}{V}\log\int \D\U \, \exp\left[-\beta S_g\right] \, \prod_{f} \det \left[ \Dsf(B)+m_f\right] \, \delta\!\left(P[\U] - p\,\right)\,,
\label{eq:pd_OmegaP}
\ee
which has its minimum at $p=\expv{P}_{B,T}$ and its shape is related to the nature of the transition. For a first-order phase transition, $\Omega^P_{B,T=T_c}$ has several degenerate minima, corresponding to the different phases and a flat region in between.\footnote{In a system with finite volume, the degenerate minima are separated by a local maximum. In the infinite volume limit, this maximum disappears and the effective potential becomes flat, see e.g.~\cite{ORaifeartaigh:1986axd,Endrodi:2021kur}.} For a crossover transition, it is a smooth function whose minimum moves as $T$ changes and $\Omega^P$ is nearly flat in the transition region. More details about center symmetry and the Polyakov loop effective potential
can be found in~\cite{Fukushima:2017csk}, which we also build on below.

To explore the effective potential, we need to understand how a change in $P$ affects the weight of a configuration. 
Let us first consider pure gauge theory, i.e.\ exclude the determinant from~\eqref{eq:pd_OmegaP}. On the one hand, the Haar measure $\D\U$ has three degenerate minima, which forces the local Polyakov loop~\eqref{eq:lat_ploopdef1} to the three center sectors,
\be
-\pi/3\le \arg P(\bm n) <\pi/3, \qquad
\pi/3\le \arg P(\bm n) <\pi, \qquad
-\pi\le \arg P(\bm n) <-\pi/3\,.
\label{eq:pd_centersectors}
\ee
On the other hand, the gauge action controls the spatial fluctuations of $P(\bm n)$: it allows large fluctuations at low temperature, where the theory is strongly coupled, but suppresses them in the weakly coupled, high-temperature regime. This mechanism of spontaneous center symmetry breaking leads to a disordered average Polyakov loop (confinement) at low $T$ and an ordered one (deconfinement) at high $T$. The critical temperature of this first-order phase transition is at around $T_c=300\MeV$.

The pure gauge theory can be thought of as the $m_f\to\infty$ limit in~\eqref{eq:pd_OmegaP}. We can include dynamical fermions by including a finite quark mass in the determinant. This changes the above sketched picture in two major ways. First, the determinant breaks center symmetry and prefers the real Polyakov loop sector (the first item in~\eqref{eq:pd_centersectors}). Losing the exact symmetry weakens the transition and renders it an analytic crossover for sufficiently large values of $1/m_f$. Second, the determinant generically favors deconfinement over confinement, as we demonstrated in Sec.~\ref{sec:ev_ploop} based on the behavior of the lowest eigenvalues of $\Dsf$. In effect, this lowers the transition temperature as the quark masses are reduced. For physical quark masses, the pseudocritical temperature is around $T_c\approx 150\MeV$. 

Turning on the magnetic field leads to the proliferation of low eigenvalues, see Sec.~\ref{sec:lat_landaulevelQCD}. Thus, a larger portion of the spectrum responds to the Polyakov loop backgrounds, amplifying the effect of the fermion determinant. This further enhances the preference of deconfinement over confinement and lowers the transition temperature even more, as mentioned already in Sec.~\ref{sec:ev_ploop}. This mechanism has been discussed in more detail in~\cite{Bruckmann:2013oba}.

\begin{figure}
 \centering
 \includegraphics[width=8.2cm]{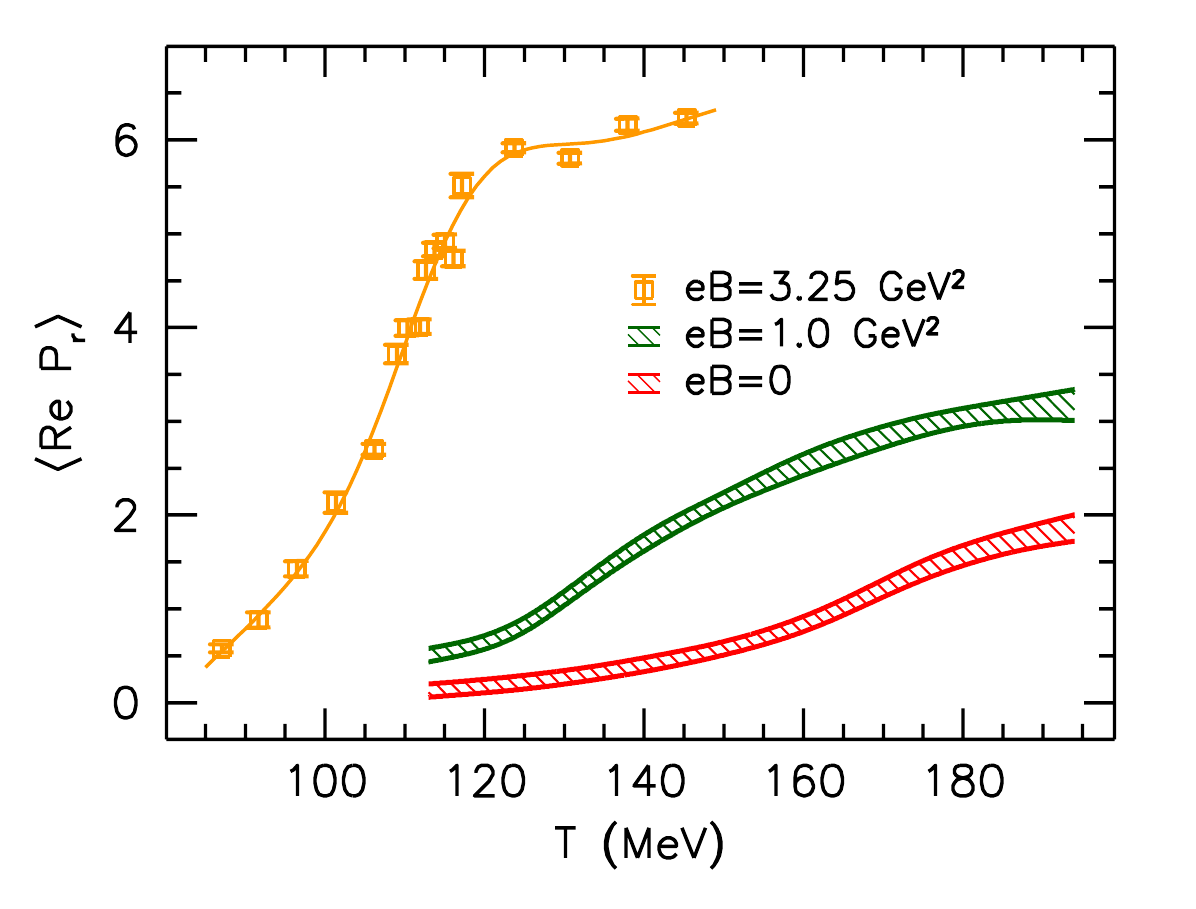}\qquad
 \caption{\label{fig:pd_ploop1}
  Renormalized Polyakov loop for different magnetic fields as a function of the temperature~\cite{Bruckmann:2013oba,Endrodi:2015oba}. Figure from~\cite{Endrodi:2015oba}.
 }
\end{figure}

The first continuum extrapolated results for the Polyakov loop expectation value in the presence of magnetic fields were obtained in~\cite{Bruckmann:2013oba}. Here, the renormalization of $P$ was carried out via~\eqref{eq:lat_ploop_renorm_Tstar}. These results showed that -- in accordance with the above expectations -- $\expv{P^r}$ is increased by the magnetic field for all temperatures. This effect was observed to be most prominent in the transition region, in effect shifting the characteristic rise of the Polyakov loop to lower temperatures.
This investigation was extended to stronger magnetic fields up to $eB=3.25\GeVsq$ in~\cite{Endrodi:2015oba}. The results for the renormalized observable are shown in Fig.~\ref{fig:pd_ploop1}. For the strongest field, the inflection point of the Polyakov loop was shown to coincide with that of the quark condensate $\Sigma$, corroborating the picture described above in Sec.~\ref{sec:pd_largeB}, involving coincident critical behaviors of these two observables for very strong fields. We note that more recently, the above described dependence of the renormalized Polyakov loop
on the temperature and the magnetic field was confirmed in~\cite{Braguta:2019yci}, using a renormalization via the gradient flow.

We mention that using the expectation value of the renormalized Polyakov loop at nonzero temperatures, one can gain further information about the single quark free energy density, which we already encountered in Sec.~\ref{sec:lat_renormploop},
\be
f^r_Q -f^r= -\frac{T}{V}\log \expv{P^r}\,.
\ee
Similarly to the formulation of the equation of state of the QCD medium, to be discussed in Sec.~\ref{sec:eos_thermo_rel}, the derivatives of $f^r_Q-f^r$ with respect to external parameters like the temperature and the background magnetic field give the single quark entropy and single quark magnetization~\cite{Braguta:2019yci},
\be
s_Q=-\frac{\partial (f^r_Q-f^r)}{\partial T}, \qquad
\M_Q=-\frac{\partial (f^r_Q-f^r)}{\partial (eB)}\,.
\ee
The enhancement of $\expv{P^r}$ implies that the single quark magnetization is positive i.e.\ the static quark represented by the Polyakov loop contributes paramagnetically to the total magnetization~\cite{Braguta:2019yci}. As we discussed already, the Polyakov loop is a purely gluonic quantity. In other words, the static quark associated with it is electrically neutral and this contribution arises indirectly, via the modification of dynamical sea quarks.

\subsubsection{Local Polyakov loops and their correlators}
\label{sec:pd_LLdag}

An important ingredient for the description of Polyakov loop fluctuations is the spatial correlator of local Polyakov loop operators. In particular, using the local untraced Polyakov loop $L(\bm n)$ from~\eqref{eq:lat_ploopdef1}, the magnetic and electric correlators can be written as~\cite{Bonati:2017uvz},
\be
C_M(\bm n)=\frac{1}{2}\textmd{Re}\left[ C_{LL}(\bm n)+C_{LL^\dagger}(\bm n)\right]-|\expv{\tr L(\bm n)}|^2, \qquad
C_E(\bm n)=\frac{-1}{2}\textmd{Re}\left[ C_{LL}(\bm n)-C_{LL^\dagger}(\bm n)\right]\,,
\ee
where
\be
C_{LL^\dagger}(\bm n-\bm k) = \expv{\tr \, L(\bm n) \, \tr L^\dagger(\bm k)}, \qquad
C_{LL}(\bm n-\bm k) = \expv{\tr \, L(\bm n) \, \tr L(\bm k)}\,.
\label{eq:pd_LLdag_corr}
\ee
The last term in $C_M(\bm n)$ is equivalent to $|\expv{P}|^2$. Such a disconnected term is not present in $C_E(\bm x)$ due to charge conjugation symmetry~\cite{Bonati:2017uvz}.

Using the correlators $C_M$ and $C_E$, the magnetic and electric screening masses can be determined via fits of the type
\be
C_M(\bm n)\xrightarrow{n_i\to\infty} \exp\left(-am_{Mi} \,n_i\right)/n_i, \qquad
C_E(\bm n)\xrightarrow{n_i\to\infty} \exp\left(-am_{Ei} \,n_i\right)/n_i\,,
\ee
similarly to the relation~\eqref{eq:had_Wilsonloopdef} of Wilson loops to the static quark-antiquark potential.
These magnetic and electric screening masses represent the characteristic inverse lengths over which gluonic interactions are screened in the QCD medium. These have been determined using stout-improved staggered quarks at physical masses for a range of magnetic fields and temperatures in~\cite{Bonati:2017uvz}. The results indicate that both masses are in general reduced by $T$ and enhanced by $B$, i.e.\ that the magnetic field appears to catalyze deconfinement. Moreover, $m_E$ always exceeds $m_B$, that is to say, electric components of gluons are screened more strongly. Finally, while there is an anisotropy in the magnetic sector, $m_{M3}<m_{M1}=m_{M2}$, within statistical errors no anisotropy was observed among the electric masses~\cite{Bonati:2017uvz}.

Turning to the local traced Polyakov loop $P(\bm n)$, the spatial distribution of this observable can be used to define so-called center clusters: connected domains of neighboring spatial sites $\bm n$ that belong to the same center sector~\eqref{eq:pd_centersectors}.
In fact, the confinement-deconfinement transition may be understood as a percolation phenomenon of these center clusters~\cite{Gattringer:2010ms,Gattringer:2010ug}. In this picture, the transition temperature is identified with the temperature where the largest center cluster starts to percolate through the full volume. In pure gauge theory, this percolation sets in abruptly, in accordance with the first-order nature of the phase transition~\cite{Gattringer:2010ms,Gattringer:2010ug,Endrodi:2014yaa}. 
In the presence of dynamical quarks, the transition can still be described via percolation~\cite{Borsanyi:2010cw,Stokes:2013oaa}, which now sets in continuously with $T$.
The cluster percolation probability and the associated transition temperature $T_c$ has been determined in the presence of background magnetic fields in~\cite{Schafer:2015wja}. The results of this study, obtained with stout-smeared staggered quarks at physical masses, showed that the so obtained $T_c$ decreases with $B$, in line with findings from the global observables we discussed in this chapter.

\subsubsection{Phase diagram in the electric field -- temperature plane}

\begin{figure}
 \centering
 \includegraphics[width=8.2cm]{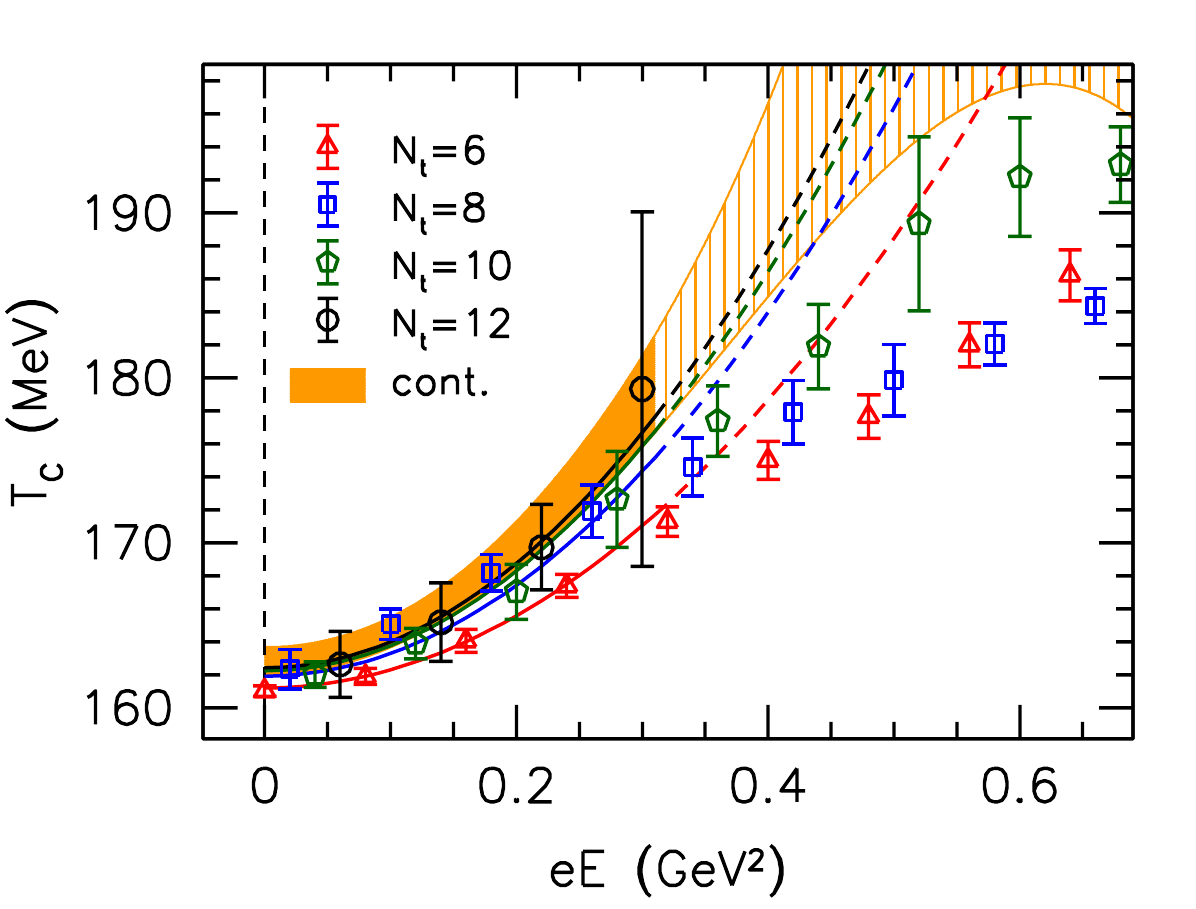}
 \caption{\label{fig:pd_electric}
  Phase diagram in the temperature-electric field plane defined through the behavior of the renormalized Polyakov loop, using a leading-order Taylor-expansion~\cite{Endrodi:2023wwf}. The filled orange band indicates the continuum extrapolation in the range of electric fields, where the leading-order expansion is expected to be reliable.
 }
\end{figure}

The only lattice result concerning the impact of background electric fields on the QCD phase diagram so far was provided by~\cite{Endrodi:2023wwf} using stout-smeared staggered quarks with physical masses. Here, the transition temperature $T_c(E)$ was defined using the characteristic behavior of the renormalized Polyakov loop $\expv{P^r}$, in particular via the implicit equation $\expv{P^r}_{E,T_c(E)} = \expv{P^r}_{E=0,T_c(E=0)}$. The electric field-dependence was determined through a leading-order Taylor expansion in $E$, which we will describe in detail in Sec.~\ref{sec:eos_elsusc}. The results for the transition temperature, obtained using four lattice spacings and extrapolated to the continuum limit, reveal a significant enhancement of $T_c(E)$, see Fig.~\ref{fig:pd_electric}.

We note that the introduction of the electric field at nonzero temperatures entails an ambiguity related to how the thermodynamic limit is taken~\cite{Endrodi:2022wym} -- related to the mismatch of relevant ensembles discussed in Sec.~\ref{sec:ev_electricfields}. In particular in a perturbative approach, using the exact infinite-volume Schwinger propagator at $E>0$~\cite{Elmfors:1998ee} results in a different $\O(E^2)$ effect in the free energy as the weak-field expansion, formulated via the photon vacuum polarization diagram~\cite{Endrodi:2022wym,Ferreira:2023cqw}. The lattice result~\cite{Endrodi:2023wwf} corresponds to the latter approach, while existing model calculations, used the former method~\cite{Ozaki:2015yja,Tavares:2019mvq,Tavares:2023ybt}. Therefore, a meaningful comparison of these findings is, at this point, not yet possible.

\subsection{Dense and magnetized QCD}
\label{sec:pd_densemagnetized}

Finally, we turn to dense and magnetized QCD. Nonzero density can be conveniently described on the lattice within the grand canonical ensemble, using chemical potentials $\mu_f$ coupled to each quark flavor $f$. In general, nonzero chemical potentials render the QCD action complex, hindering standard Monte-Carlo simulations, as we discussed in Sec.~\ref{sec:lat_signproblem}. Most lattice studies circumvent this problem by means of Taylor expansions in $\mu_f$ or via analytical continuation from imaginary chemical potentials. 
We will get back to the specific definition of chemical potentials and further details regarding nonzero density in Sec.~\ref{sec:pd_fluc}.

To investigate QCD with background magnetic fields and nonzero baryon density (more precisely, in the setup $\mu_u=\mu_d=\mu_\L$ and $\mu_s=0$), the work~\cite{Braguta:2019yci} followed the imaginary chemical potential approach. In particular, this study performed simulations at $i\mu_\L\neq0$ at a series of background magnetic fields using stout-improved staggered quarks. The results were used to carry out the analytic continuation to real chemical potentials, constructing the phase diagram for low $\mu_\L$.

In particular, fitting the temperature-dependence of the light quark condensate, its inflection point $T_c$ and the width $\delta T_c$ of the transition region (referred to as the thermal width) were determined as functions of $i\mu_\L$ and $eB$. For the analytic continuation, polynomials in $eB$ and in $\mu_\L$ were used, defining the curvature $\kappa_2$ of the transition temperature as well as that of the width, $\delta\kappa_2$. 
Specifically, these were defined as~\cite{Braguta:2019yci},
\begin{equation}
 \frac{T_c(i\mu_\L,B)}{T_c(0,B)} = 1 + \kappa_2(B) \left(\frac{3\mu_\L}{T_c(0,B)}\right)^2 + \O(\mu_L^4), \qquad
 \frac{\delta T_c(i\mu_\L,B)}{\delta T_c(0,B)} = 1 + \delta \kappa_2(B) \left(\frac{3\mu_\L}{\delta T_c(0,B)}\right)^2 + \O(\mu_L^4)\,.
\end{equation}
Besides confirming the reduction of $T_c(B)$ by the magnetic field and the strengthening of the transition by $B$ -- i.e., a reduction in $\delta T_c(B)$ -- this study delivered several further interesting findings. The curvature $\kappa_2$ was found to be rather insensitive to the magnetic field. In contrast, the curvature of the thermal width was observed to depend non-trivially on $B$, in a way that $\delta\kappa_2$ is negative for weak fields but turns sign at around $eB\approx 0.6\GeVsq$ as the magnetic field grows. This implies a mild strengthening of the transition due to real chemical potentials at low $B$, and a weakening due to $\mu_\L$ for strong magnetic fields.
The analogous analysis was also carried out in~\cite{Braguta:2019yci} for the $T$-dependence of the Polyakov loop with similar findings, except that in this case the transition was found to become weaker due to $\mu_\L$ for all magnetic fields.

\subsection{Lessons learned}

In this chapter we summarized the current status of studies devoted to the determination of the QCD phase diagram at nonzero background magnetic fields, based on the behavior of the quark condensate and the Polyakov loop.
At low temperature, the quark condensate undergoes magnetic catalysis -- a finding supported by all existing lattice studies. In turn, around $T_c$ the condensate is reduced by the magnetic field -- this inverse magnetic catalysis phenomenon is only found in works that employ quark masses near the physical point and sufficiently fine lattices.

The behavior of the quark condensate can be explained by separating the contributions to it due to valence and sea quarks. The valence contribution is always found to be positive, while the sea contribution becomes negative in the transition region for light quarks near the continuum limit.
Both types of behavior can be understood in terms of the low Dirac eigenvalues: their proliferation due to the magnetic field on arbitrary gluon configurations and the impact of Polyakov loop backgrounds and magnetic fields on them, respectively.
Based on the dependence of the quark condensate on the temperature and the magnetic field, the phase diagram exhibits a decreasing $T_c(B)$ transition curve.

By now, the impact of lattice artefacts and quark masses on the inverse magnetic catalysis phenomenon and on the reduction of $T_c(B)$ are well understood. Moreover, it has also been recognized that the two notions -- albeit strongly related -- are strictly speaking independent phenomena. In particular, the inverse magnetic catalysis of $\expv{\bar\psi\psi}$ ceases  to take place for vacuum pion masses $M_\pi\gtrsim500\MeV$, but $T_c$ was found to be reduced by $B$ for all pion masses investigated so far. In turn, lattice artefacts appear to impact both types of behavior simultaneously.

It has been long known that magnetic fields lead to a strengthening of the transition. 
Since recently, there is evidence that the QCD phase diagram exhibits a critical point for very strong magnetic fields $4\GeVsq\le eB_{\rm CP}\le 9\GeVsq$. For magnetic fields beyond $B_{\rm CP}$, the transition turns first order. Aspects for future research include precise determinations of the rich physics associated with the critical point as well as a better understanding of the deconfinement/chiral symmetry restoration nature of the transition and of the decoupling of quarks for large $B$. 
Whether the magnetic critical point is related to the conjectured QCD critical point at nonzero baryon chemical potentials or to the Roberge-Weiss endpoint at imaginary chemical potentials, is an open question as well. Lattice investigations of the dense and magnetized QCD medium can address this question and have been started recently.

Due to the fact that the inverse magnetic catalysis phenomenon and the reduction of $T_c(B)$ are contrary to the initial predictions of most low-energy models and effective theories of QCD, these surprising findings have lead to major developments in our understanding of many of these models.
The analogous treatment of QCD thermodynamics in the presence of background electric fields is more complicated due to the inhomogeneous nature of the equilibrium state and the need of an infrared regularization, but continuum extrapolated lattice results for the phase diagram are now available also in this setting.

	\clearpage
	\section{Equation of state}
	\label{chap:eos}

The equilibrium description of QCD matter is provided by the equation of state, i.e.\ the dependence $\epsilon(p)$ of the energy density on the pressure. This relationship plays a central role in a series of physical settings. In neutron star physics, it is required for a consistent solution of the Tolman-Oppenheimer-Volkoff equations to find the mass and radius of gravitationally stable compact objects~\cite{Lattimer:2000nx}. In cosmology, it enters the Friedmann equation required for the determination of the isentropic expansion of the early Universe~\cite{Boyanovsky:2006bf}. In heavy-ion collision phenomenology, the equation of state and its dependence on the chemical potentials coupled to conserved charges is necessary to find the hydrodynamic evolution of the quark-gluon plasma and the subsequent hadronic freeze-out~\cite{Teaney:2001av,Kolb:2003dz}.

In this chapter, we will discuss the QCD equation of state in the presence of background electromagnetic fields. 
A special emphasis is put on the weak-field behavior encoded by the magnetic and electric susceptibilities of QCD matter. There is a multitude of lattice techniques that have been developed in the last decade to calculate these susceptibilities, which we will review. Moreover, we will also discuss some of the characteristic concepts of the magnetized equation of state, like the anisotropy of the pressures or the decomposition of the magnetization into spin and angular momentum contributions. A summary of recent lattice studies on dense and magnetized, as well as electrically polarized systems is also included here.

\subsection{Thermodynamic relations}
\label{sec:eos_thermo_rel}

All observables related to the equation of state can be derived from the matter free energy $F$ defined in terms of the partition function $\Z$ in~\eqref{eq:lat_Fdef}. First we consider the case without background electromagnetic fields. We work in the grand canonical ensemble, so that $F=F(T,\mu_f, V)$ is a function of the state variables: the temperature, the chemical potentials $\mu_f$ coupled to conserved particle numbers $N_f$, as well as the volume. The conserved charges are given by the expectation values $\expv{V_4^f}$ of the temporal components of the vector currents defined in~\eqref{eq:lat_VV5S5T}. Each conjugate quantity can be found by differentiating $F$ with respect to the corresponding state variable (with all other parameters kept constant). In particular, the entropy density and the quark number density read,
\be
s = -\frac{1}{V}\frac{\partial F}{\partial T}\,, \qquad
n_f = -\frac{1}{V}\frac{\partial F}{\partial \mu_f}\,.
\label{eq:eos_s_n_def}
\ee
The energy density follows from the thermodynamic relation,
\be
\epsilon= f+Ts + \sum_f \mu_f n_f\,,
\ee
and can equivalently be obtained via the derivative
\be
\epsilon=-\frac{1}{V}\frac{\log\Z}{\partial (1/T)}\,.
\ee

Finally, to complete the standard relations of thermodynamics, we need the definition of the pressure $p$ as the derivative of $F$ with respect to the volume. More specifically, the pressure components $p_i$ are written as derivatives with respect to the corresponding spatial sizes $L_i$,
\be
p_i=-\frac{L_i}{V}\frac{\partial F}{\partial L_i}\,.
\label{eq:eos_p_def}
\ee
In an isotropic system, $L_1=L_2=L_3$ holds and the derivative above can be traded for a derivative with respect to the volume $V=L_1L_2L_3$. Thus, the pressure is isotropic,
\be
\forall i\qquad p_i=- \frac{L_i}{V}\frac{\partial F}{\partial V} \frac{\partial V}{\partial L_i} = -\frac{\partial F}{\partial V} \equiv p\,.
\ee
Moreover,
in the thermodynamic limit, the differentiation with respect to $V$ can be exchanged through a normalization by $V$ under the assumption of homogeneity,
\be
p\xrightarrow{V\to\infty} -\frac{F}{V} = -f\,.
\label{eq:eos_p_thermolim}
\ee

The energy density and the pressures constitute the diagonal components of the stress-energy tensor $\T_{\nu\rho}$. Its trace gives the so-called trace anomaly\,,
\be
I= \T^\nu_{\nu} =  \epsilon-p_1-p_2-p_3 \,,
\label{eq:eos_intmeasure}
\ee
simplifying to $I=\epsilon-3p$ for isotropic systems. An equivalent definition of $I$ is given in terms of the response of the system to an overall change of the scale, as in~\eqref{eq:had_intmeasdef1}. We note that $I$ vanishes for an ideal gas and therefore reflects the extent of interactions in the system. We will refer to it as interaction measure.

\subsubsection{Magnetic and electric susceptibility}

Next, we switch on background electromagnetic fields. Without loss of generalization, the fields are assumed to point in the $x_3$ direction.
The observables conjugate to the magnetic and electric fields are the local magnetization and polarization densities~\cite{landau1995electrodynamics} obtained via functional differentiation\footnote{The derivatives are normalized so that the renormalization group invariant combinations $eB$ and $eE$ appear in order to avoid the need to introduce multiplicative renormalization factors, see Sec.~\ref{sec:lat_renormfreeenergy}.
},
\be
\M(\bm x)=-\frac{\delta F}{\delta [eB(\bm x)]}, \qquad
\P(\bm x)=-\frac{\delta F}{\delta [eE(\bm x)]}\,.
\ee
The linear response is encoded by the susceptibilities,
\be
\chi(\bm x-\bm y)=\frac{\delta \M(\bm x)}{\delta [eB(\bm y)]}, \qquad
\xi(\bm x-\bm y)=\frac{\delta \P(\bm x)}{\delta [eE(\bm y)]}\,,
\ee
where we used the translational invariance of the medium (as well as its isotropy at $E=B=0$, which dictates that the induced response is parallel to the background field). 
Thus, for weak background fields,
\be
\M(\bm x) = \frac{1}{V}\int \dd^3 \bm y \,\chi(\bm x-\bm y) \,eB(\bm y), \qquad
\P(\bm x) = \frac{1}{V}\int \dd^3 \bm y \,\xi(x-\bm y) \,eE(\bm y)\,,
\label{eq:eos_susc_coord}
\ee
or -- equivalently -- in momentum space,
\be
\widetilde{\M}(\bm p) = \widetilde{\chi}(\bm p)\,e\widetilde{B}(\bm p), \qquad
\widetilde{\P}(\bm p) = \widetilde{\xi}(\bm p)\,e\widetilde{E}(\bm p)\,.
\label{eq:eos_susc_Fourier}
\ee

For homogeneous background fields, the conjugate observables become simple derivatives,
\be
\M=-\frac{1}{V}\frac{\partial F}{\partial (eB)}, \qquad
\P=-\frac{1}{V}\frac{\partial F}{\partial (eE)}\,,
\label{eq:eos_magnetization_def}
\ee
and the leading, quadratic effect is encoded by the corresponding susceptibilities that were already introduced in~\eqref{eq:lat_susc_def},
\be
\chi \equiv \widetilde{\chi}(0)= \left.\frac{\partial \M}{\partial (eB)}\right|_{B=0}, \qquad \left.\xi\equiv\widetilde{\xi}(0)=\frac{\partial \P}{\partial (eE)}\right|_{E=0}\,.
\label{eq:eos_magnsusc_elsusc}
\ee
as indicated here, the susceptibilities with respect to the homogeneous background fields are given by the zero momentum limits of the Fourier transformed susceptibilities.
Below we will often simplify the notation by indicating only the component of the coordinate or of the momentum on which the observable depends, i.e.\ $\widetilde{\chi}(p_1)$ stands for $\widetilde{\chi}(\bm p = (p_1,0,0))$ and $\M(x_1)$ for $\M(\bm x=(x_1,0,0))$.

\subsubsection{Pressure anisotropy in magnetic fields}

\begin{figure}[t]
 \centering
 \vspace*{-.3cm}
 \includegraphics[width=3.5cm]{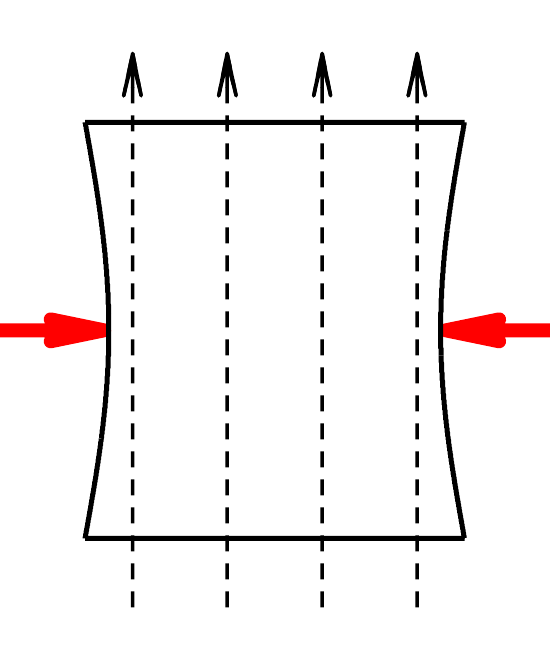} \quad\quad\quad\quad
 \includegraphics[width=3.5cm]{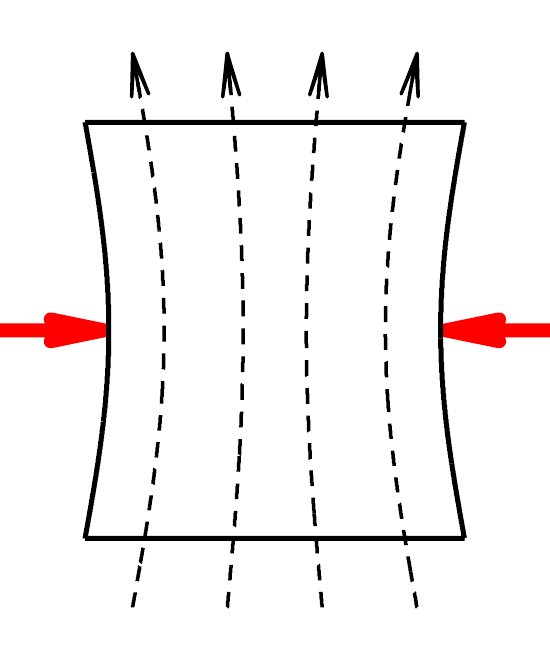}
 \vspace*{-.3cm}
 \caption{ \label{fig:eos_illustr}
  Illustration of the definition of the pressure perpendicular to the magnetic field in the $B$-scheme (left panel) and the $\Phi$-scheme (right panel)~\cite{Bali:2013txa}.
 }
\end{figure}

We proceed by pointing out a peculiarity of the definition of the pressure in the presence of homogeneous background magnetic fields~\cite{Bali:2013esa}. 
The orientation of the magnetic field breaks the isotropy of the system implying that the derivatives~\eqref{eq:eos_p_def} with respect to the spatial sizes have to be taken with care. In particular, one needs to specify the thermodynamic variables that are kept fixed for the partial derivatives. 
There are two obvious possibilities: either keeping the magnetic field $B$ fixed or keeping the flux $\Phi\equiv\Phi_{12}=BL_1L_2$ of the magnetic field fixed, as illustrated in Fig.~\ref{fig:eos_illustr}. In the language of~\cite{Bali:2013esa}, we will refer to these choices as the $B$-scheme and the $\Phi$-scheme and indicate them by superscripts $(B)$ and $(\Phi)$, respectively. Certain observables will depend on this choice of the scheme, while others will not -- for the latter the superscript will be suppressed.

For a large homogeneous system, the free energy is expected to be a linear function of the volume, $F(B,L_i)=L_1L_2L_3\cdot f(B)$, just as in~\eqref{eq:eos_p_thermolim}. 
For fixed $B$, the only dependence of the free energy on the spatial extents is via the explicit prefactor, resulting in isotropic pressures. In turn, in the $\Phi$-scheme there is an implicit dependence on $L_1$ and $L_2$ through the magnetic field-dependence, $F(\Phi,L_i)=L_1L_2L_3\cdot f(\Phi/(L_1L_2))$, rendering the pressures anisotropic~\cite{Bali:2013esa}. This implies that the parallel pressure component $p_3$ is the same in both schemes, while the perpendicular components $p_1$ and $p_2$ differ by a term proportional to the magnetization~\eqref{eq:eos_magnetization_def},
\be
p_1^{(B)}=p_2^{(B)}=p_3\,, \qquad p_1^{(\Phi)}=p_2^{(\Phi)}=p_3-eB\cdot \M\,.
\label{eq:eos_p_aniso}
\ee
The energy density, the entropy density and the particle number densities are the same in both schemes. In turn, the interaction measure satisfies $I^{(\Phi)} = I^{(B)}+2M\cdot eB$. More details on the scheme-dependence of various observables can be found in~\cite{Bali:2014kia}.
We also note that one finds the same result~\eqref{eq:eos_p_aniso}, if the pressures are defined as diagonal components of the stress-energy tensor~\cite{Ferrer:2010wz} -- in this case the choice of the scheme appears in how the components $\T_{\nu\rho}$ are defined via variations of the action with respect to the components of the metric~\cite{Bali:2014kia}.

Intuitively, the perpendicular pressures depend on how the compression of the system proceeds: either pushing the magnetic field lines together with the medium ($\Phi$-scheme) or leaving the field lines unaffected ($B$-scheme), see Fig.~\ref{fig:eos_illustr}. 
Which scheme is applicable to a certain system depends on the physical situation. For a perfectly conducting plasma for example, magnetic flux is conserved~\cite{Vachaspati:1991nm,Kandus:2010nw}, making the $\Phi$-scheme the relevant choice. Nevertheless, one should keep in mind that the above discussion refers to the pressure of the thermal medium, which does not include the contribution of the external devices that generate the background field.

\subsection{Susceptibilities on the torus}
\label{sec:eos_susctorus}

The quantization condition~\eqref{eq:lat_flux_quant_hom} of the magnetic flux
on the torus implies that homogeneous magnetic fields cannot be changed continuously. Thus, the partial derivative in the definition of the magnetic susceptibility~\eqref{eq:eos_magnsusc_elsusc} cannot be evaluated directly. This calls for alternative approaches in order to determine the response of the QCD medium to weak background fields.
During the last several years, various methods have been developed in order to circumvent this issue and we discuss each of them in chronological order next.

\subsubsection{Anisotropy method}
\label{sec:eos_aniso}

Due to the quantization of the magnetic flux in a finite periodic volume, the lattice setup inherently realizes the $\Phi$-scheme, where~\eqref{eq:eos_p_aniso} holds. Thus, at any nonzero $B$ the magnetization can be expressed as the difference of parallel and perpendicular pressures. The latter can be derived from the expectation values of the diagonal components of the stress-energy tensor $\T_{\nu\rho}$,
\be
eB\cdot \M = \expv{\T_{33}}-\expv{\T_{11}}\,.
\label{eq:eos_aniso_magnet_1}
\ee
To calculate the individual components, one needs to work with anisotropic lattice spacings~\cite{Engels:1981qx,Karsch:1982ve,Karsch:1989fu,Levkova:2006gn}. This entails the introduction of anisotropy coefficients that control how the bare lattice parameters need to be tuned to achieve a certain physical anisotropy. There are two such parameters ($\zeta_g$ and $\hat\zeta_g$) needed in the gluon action and one parameter ($\zeta_f$) in the fermion action. Within perturbation theory, all of them are of the form $1+\O(g^2)$. 

In terms of these coefficients and the gluon and fermion actions, the difference~\eqref{eq:eos_aniso_magnet_1} of the stress-energy tensor components reads~\cite{Bali:2013esa},
\be
\expv{T_{33}}-\expv{T_{11}} = (\zeta_g+\hat\zeta_g) \cdot \frac{T}{V}\frac{\beta}{6}\Expv{\Tr\left[ \F_{23}^2-\F_{12}^2 + \F_{43}^2-\F_{41}^2\right]} + \zeta_f \cdot \frac{T}{V}\sum_f \Expv{\Tr \left[ M_f^{-1}\left(\gamma_1 D_{f1} - \gamma_3 D_{f3}\right) \right]}
\label{eq:eos_aniso_magnet_2}
\ee
where the lattice discretizations of the components of the fermionic and gluonic actions~\eqref{eq:lat_action_cont} enter. Above, $\Tr$ denotes a trace in coordinate space in the gluonic term and a trace over coordinate and internal spaces in the fermion term.

Once one measured $\M$ using~\eqref{eq:eos_aniso_magnet_1} and~\eqref{eq:eos_aniso_magnet_2} at different values of $B$, the susceptibility can be obtained via a numerical differentation.
This formalism was put into practice in~\cite{Bali:2013esa} at zero temperature and in~\cite{Bali:2013owa} at high temperatures. In both works, the tree-level values $\zeta_g=\hat\zeta_g=\zeta_f=1$ were used in~\eqref{eq:eos_aniso_magnet_2}.

\subsubsection{Half-half method}
\label{sec:eos_half}

In Sec.~\ref{sec:lat_oscillatory_fields} we learned that there are magnetic field profiles that have continuous amplitudes even on the torus. The oscillatory magnetic field $B(x_1)=B\cos(p_1x_1)$ corresponds to a nonzero momentum component $\widetilde{B}(p_1)$ of the Fourier-transformed magnetic field and thus probes the component $\widetilde{\chi}(p_1)$ according to~\eqref{eq:eos_susc_Fourier}. The half-half profile $B(x_1)=B\,\sgn(x_1-L_1/2)$, as introduced in~\eqref{eq:lat_half_Bfield}, also has vanishing flux and is, in some sense, more similar to the homogeneous field. While the gradient of the field vanishes in most of the volume, it also involves two singularities required to ensure the periodicity of the vector potential. The response of the medium to a weak magnetic field with this profile can in general be written as a linear combination of $\widetilde{\chi}(p_1)$ with all different $p_1$.

The Taylor-expansion in the amplitude of the half-half-type magnetic field can be constructed very similarly to a standard Taylor-expansion in chemical potentials, for example. This approach was developed in~\cite{Levkova:2013qda}, where it was also employed to determine the magnetic susceptibility at nonzero temperatures using an improved staggered action and slightly heavier-than-physical quarks. The same method was also used at zero temperature in~\cite{Bali:2014kia,Bali:2015msa} and generalized to the case of nonzero isospin chemical potentials in~\cite{Endrodi:2014lja} using an unimproved action.

We also note that a half-half-type real electric field was used in~\cite{Yamamoto:2012bd} to calculate electric polarization effects at zero and nonzero temperatures.

\subsubsection{Finite difference method}
\label{sec:eos_finitediffmethod}

Let us come back to the prescription~\eqref{eq:lat_hom_links} of the $\mathrm{U}(1)$ links generating a homogeneous magnetic field on the lattice and recall that the down quark sets the flux unit, i.e.\ $N_b=N_b^d=N_b^s$ and $-2N_b=N_b^u$, leading to the field value $eB=6\pi N_b/L^2$, see~\eqref{eq:lat_flux_quant_hom}.
The flux quantum $N_b$ is only allowed to assume integer values, otherwise the periodic boundary conditions on the torus are violated. In fact, one may view non-integer values of $N_b$ as a superposition of a homogeneous magnetic field plus a Dirac string that pierces the torus through the plaquette at $n_1=n_2=N_s-1$. For any $N_b^f\not\in\mathds{Z}$, the quark with charge $q_f$ experiences this Dirac string.

\begin{figure}
 \centering
 \includegraphics[width=7cm]{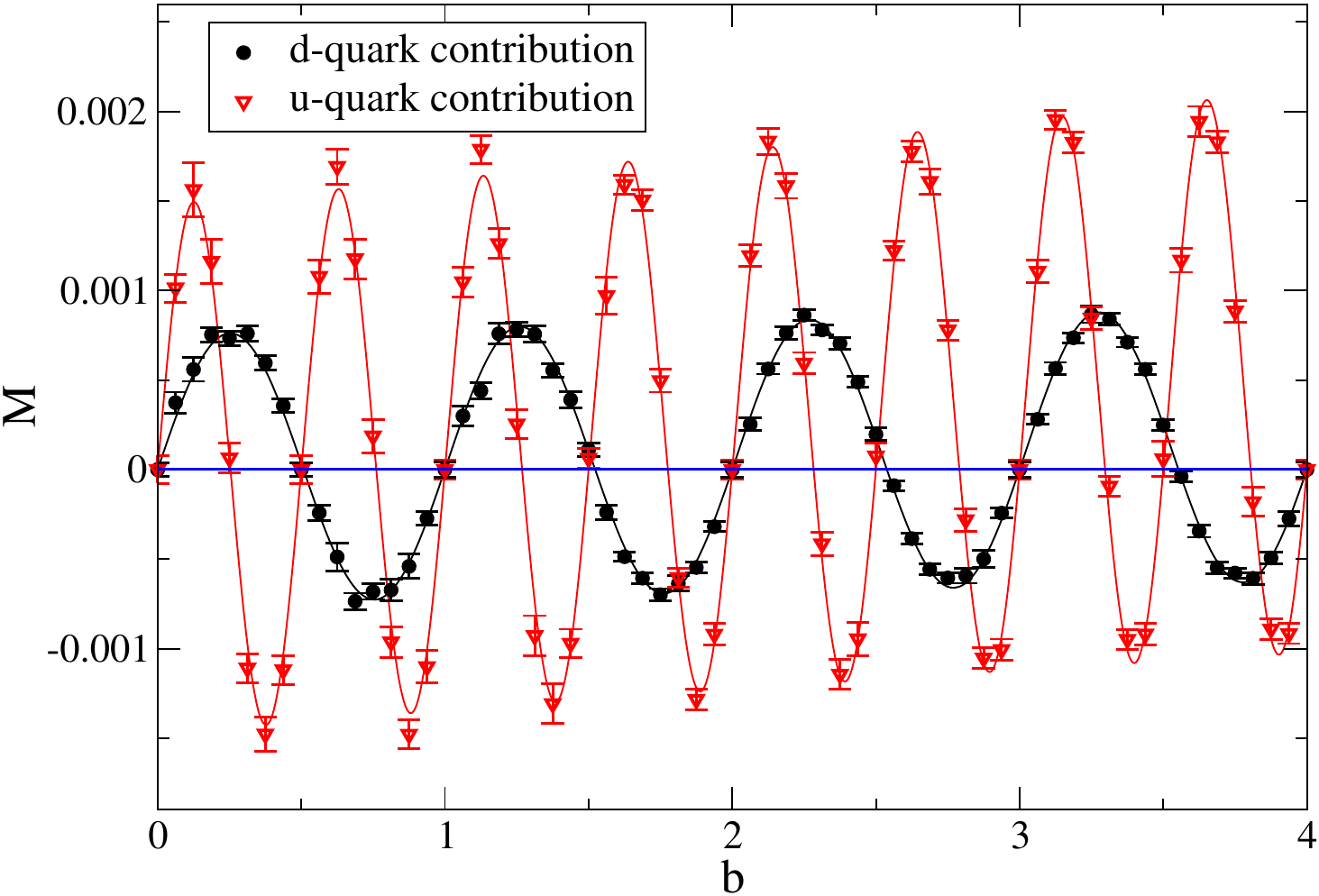}
 \caption{ \label{fig:eos_finitediffmethod}
  Derivative of the matter free energy with respect to the magnetic field variable $N_b$ (denoted here by $b$) on the torus. Between integer values of the flux, this unphysical quantity integrates to physical free energy differences~\cite{Bonati:2013lca}.
 }
\end{figure}

Irrespective of this undesired contribution, one may analytically continue the Dirac operator to real values of $N_b$ using the photon links, and calculate the derivative of the matter free energy $F$ with respect to the magnetic field. While this is an unphysical quantity, its integral in $N_b$ will reproduce physical free energy differences, if we integrate between integer values, for example,
\be
F(B)-F(0) = \int_0^{eBL^2/(6\pi)} \!\dd N_b \,\frac{\partial F}{\partial N_b}\,.
\label{eq:eos_finitediffmethod}
\ee
This approach was developed in~\cite{Bonati:2013lca} and dubbed finite difference method. The dependence of $-\partial F/\partial N_b$ on the flux, continued in this manner to real values, is shown in Fig.~\ref{fig:eos_finitediffmethod} for an example case of $16^3\times 4$ lattices. The unphysical oscillations correspond to the impact of the Dirac string. Here the contributions to the integrand are separately shown for different flavors, revealing that the oscillation frequency is set by the quark charge ratio.

Having obtained the free energy differences, the magnetic susceptibility is again determined via numerical differentiation.
The first results using this approach were obtained in~\cite{Bonati:2013lca} for two flavors of unimproved staggered quarks with heavier-than-physical masses. The method was also employed with an improved staggered action and physical quark masses in~\cite{Bonati:2013vba}.

\subsubsection{Generalized integral method}
\label{sec:eos_genintmethod}

Standard simulation algorithms allow for the direct determination of derivatives of the free energy, but not the free energy itself. This poses a problem already for the calculation of the equation of state at zero magnetic fields. One solution is the so-called integral method: similarly to~\eqref{eq:eos_finitediffmethod}, one integrates the partial derivatives in order to recover the difference of free energies at different temperatures~\cite{Engels:1990vr}. Changing the temperature is realized by changing the inverse gauge coupling $\beta$. In order to remain on the line of constant physics, the lattice quark masses $m_fa$ need to be tuned as a function of the former. Thus, one needs to integrate the derivative with respect to $\beta$ -- the lattice gauge action $\expv{S_g}$ of~\eqref{eq:lat_sgdef} -- as well as the derivatives with respect to the quark masses $m_fa$ -- the quark condensates $\expv{\bar\psi_f\psi_f}$ of~\eqref{eq:lat_pbpdef_0}. This approach can be trivially generalized to nonzero magnetic fields and gives the difference of pressures at different magnetic fields at a fixed magnetic flux.

\begin{figure}
 \centering
 \includegraphics[width=7cm]{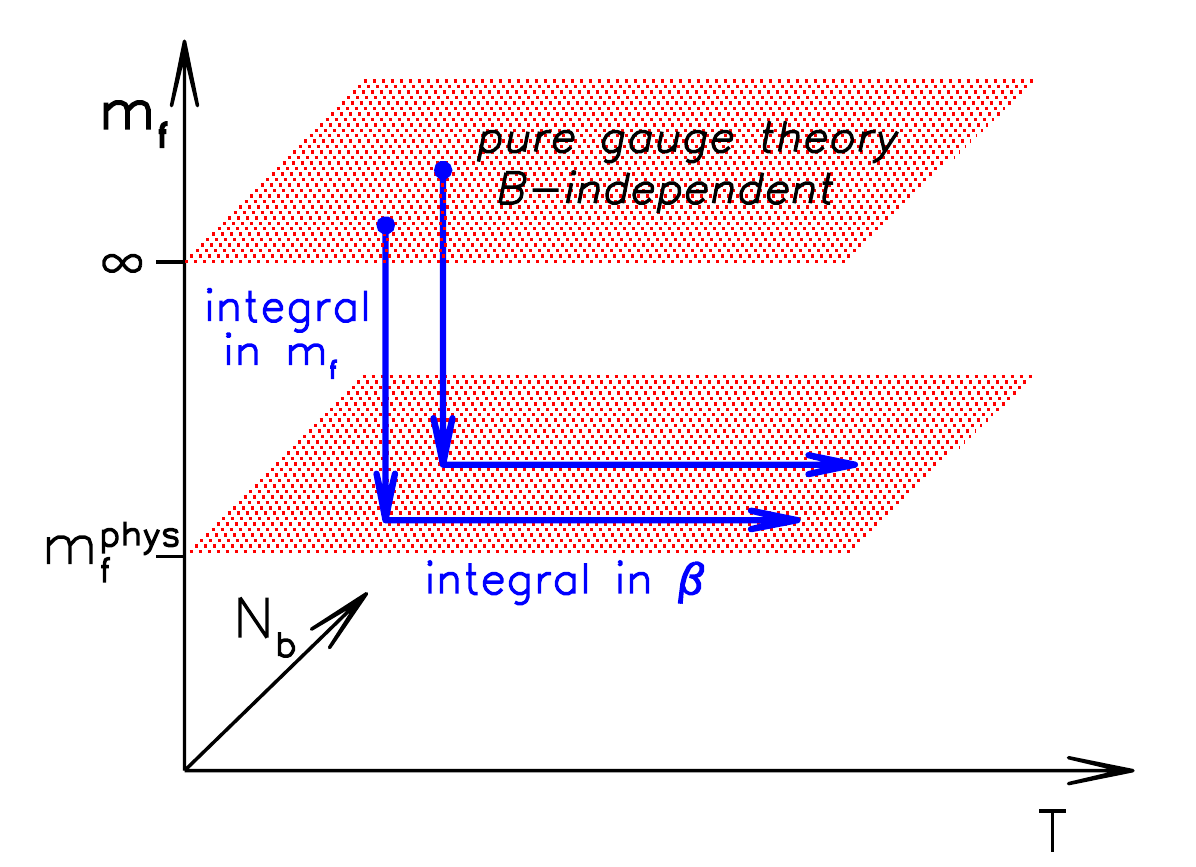}\qquad
 \includegraphics[width=7cm]{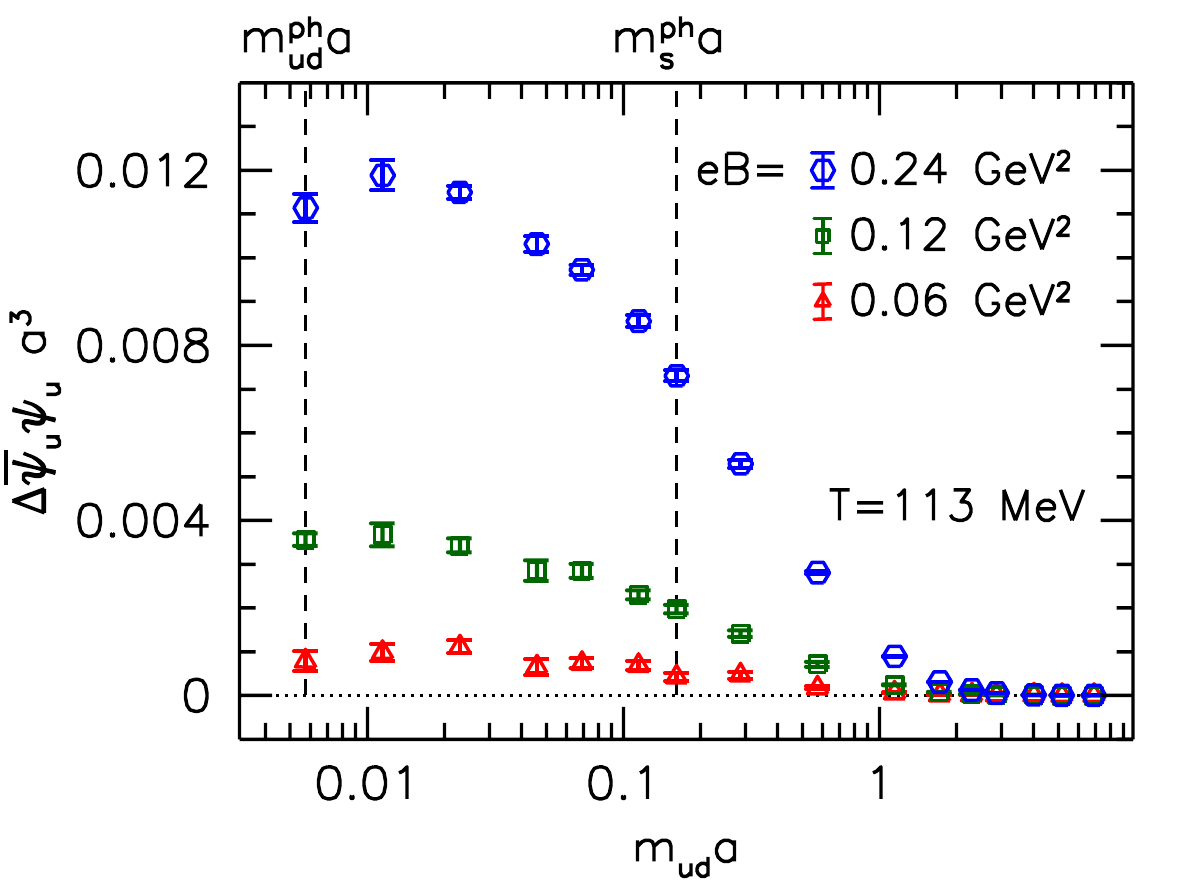}
 \caption{ \label{fig:eos_genintmethod}
  Left panel: illustration of the generalized integral method to obtain the dependence of thermodynamic observables on the temperature and the magnetic field~\cite{Bali:2013txa}. Right panel: the integrand along the vertical integration contour for several magnetic fields~\cite{Bali:2014kia}.
 }
\end{figure}

The remaining difficulty is in obtaining the integration constant. For $B=0$ this is calculated by starting the integration at a sufficiently low temperature, where the free energy (i.e.\ the pressure, cf.~\eqref{eq:eos_p_thermolim}) is negligibly small. For $B>0$, the magnetic field affects the free energy already in the vacuum, therefore one needs an alternative way to set the integration constant. One possibility is to consider as starting point for the integration the infinite quark mass limit of QCD, where the magnetic field has no effect. This leads to the integration contours depicted in the left panel of Fig.~\ref{fig:eos_genintmethod}. Along the vertical integration contour, we need to integrate the quark condensate differences
\be
\Delta f(B)=\sum_f \int_{m_f^{\rm ph}}^\infty \dd m_f \,\Delta \expv{\bar\psi_f\psi_f}\,,
\label{eq:eos_massintegral}
\ee
a relation that we already encountered in its renormalized form in~\eqref{eq:lat_magncat_paramag}. The subtracted up quark condensate is shown in the right panel of Fig.~\ref{fig:eos_genintmethod} along the integration contour, revealing the expected suppression as the quarks become heavier, which ensures the convergence of the integral.\footnote{In fact, towards the continuum limit this integral develops an ultraviolet divergent term proportional to $\log a$, in accordance with the renormalization prescription given in~\eqref{eq:lat_fr_muQED_def}.}
Finally, we only need to supply $f(B=0)$ -- following the same argument as above, this vanishes to a good approximation if the temperature is sufficiently low.

Using the $B$-dependence of the free energy, the magnetic susceptibility can be extracted via numerical differentiation, just like for the finite difference method.
This method was developed in~\cite{Bali:2013txa,Bali:2014kia} and was used to determine the equation of state with an improved staggered action at physical quark masses for a broad range of temperatures and magnetic fields in~\cite{Bali:2014kia}. The standard integral method was also used in~\cite{Cea:2015ifa} for $N_f=1$ QCD. This study focused on the doubly subtracted free energy density, $\Delta f(B,T)-\Delta f(B,T=0)$, for which the determination of the integration constant via the mass integral~\eqref{eq:eos_massintegral} is not necessary.

\subsubsection{Current-current correlator method}
\label{sec:eos_currentcurrent}

There is yet another alternative to circumvent the flux quantization condition, by constructing the homogeneous magnetic field as a limit of harmonic magnetic fields, as introduced in~\eqref{eq:lat_osc_Bfield}.
In the action, the corresponding vector potential couples to the electromagnetic current $j_\nu$ of~\eqref{eq:lat_jnucurrentdef}.
The second derivative of the matter free energy with respect to such external fields can therefore be related to the vacuum polarization tensor,
\be
\Pi_{\nu\rho}(p)=\int \dd^4x\, e^{ipx}\,\expv{j_\nu(x)j_\rho(0)}, \qquad j_\nu(x)=\sum_f \frac{q_f}{e}\,\bar\psi_f(x)\gamma_\nu\psi_f(x)\,.
\label{eq:eos_vacpol1}
\ee
For the specific gauge choice~\eqref{eq:lat_osc_Bfield}, where the magnetic field is represented by $A_2(x_1)$, we thus need to focus on the $p_1$-dependence of the $\Pi_{22}$ component. One finds that the oscillatory susceptibility can be written as~\cite{Bali:2015msa,Bali:2020bcn},
\be
\widetilde{\chi}(p_1) = -\frac{\Pi_{22}(p_1)}{p_1^2}\,,
\label{eq:eos_chip1}
\ee
which, in the $p_1\to0$ limit, becomes the second derivative of the vacuum polarization~\eqref{eq:eos_vacpol1} with respect to $p_1$.
The same result, which in fact dates back to the seminal paper~\cite{Weldon:1982aq}, can be obtained using standard linear response theory~\cite{Buividovich:2021fsa}.

The susceptibility~\eqref{eq:eos_chip1} can be evaluated conveniently in terms of the mixed-representation current-current correlator,
\be
G(x_1)=\int\dd x_2\,\dd x_2\, \dd x_3 \, \expv{j_2(x)j_2(0)}\,.
\ee
In fact this object, at zero temperature, plays a crucial role in the determination of the hadronic contribution to the anomalous magnetic moment of the muon, see the review~\cite{Meyer:2018til}.
For the zero momentum limit, care has to be taken so that the integration kernel respects the periodic boundary conditions, leading to~\cite{Bali:2015msa,Bali:2020bcn},
\be
\chi=\lim_{p_1\to0}\widetilde{\chi}(p_1) = \frac{1}{2}\int_0^{L_1} \dd x_1\, G(x_1) \cdot \,
\begin{cases}
 x_1^2,  & x_1\le L_1/2 \\
 (L_1-x_1)^2, & x_1>L_1/2 \\
\end{cases}\,.
\label{eq:eos_magnsusc_repres}
\ee
The cusp of the kernel at $x_1=L_1/2$ is multiplied by the exponentially decaying current-current correlator and was found to be unproblematic in practice~\cite{Bali:2015msa,Bali:2020bcn}. The magnetic susceptibility was determined using an improved action and physical quark masses at zero temperature in~\cite{Bali:2015msa} and finite temperature in~\cite{Bali:2020bcn}. The same method was also used in~\cite{Buividovich:2021fsa} for two-color QCD.

One advantage of this approach is that it can be generalized for background electric fields, although the interpretation of equilibrium is more complicated in this case~\cite{Endrodi:2022wym}. For a harmonic electric field of the type~\eqref{eq:lat_osc_Efield}, this time one needs the $p_3$ dependence of the $\Pi_{44}$ component of the vacuum polarization tensor~\cite{Weldon:1982aq}. The equivalent of~\eqref{eq:eos_magnsusc_repres} was derived in~\cite{Endrodi:2022wym} and reads,
\be
\xi= \frac{1}{2}\int_0^{L_3} \dd x_3\, K(x_3) \cdot \,
\begin{cases}
 x_3^2,  & x_3\le L_3/2 \\
 (L_3-x_3)^2, & x_3>L_3/2 \\
\end{cases}\,, \qquad
K(x_3)=\int\dd x_1\,\dd x_2\, \dd x_4 \, \expv{j_4(x)j_4(0)}\,.
\label{eq:eos_elsusc_currcurr}
\ee
This formulation was used in~\cite{Endrodi:2023wwf} to calculate the electric susceptibility in high-temperature QCD.

\subsubsection{Amp\`{e}re's law}
\label{sec:eos_ampere}

For the last approach, we consider the coordinate space representation~\eqref{eq:eos_susc_coord} of the magnetization for a general magnetic field profile. The magnetization can be related to the induced current flowing in the medium via Amp\`{e}re's law, $\expv{\bm j} = \bm \nabla \times \bm \M$. 
Note that such currents are only present for inhomogeneous magnetizations (i.e.\ inhomogeneous background fields), cf.\ the discussion around~\eqref{eq:lat_inducedcurrent_totalcurrent}.

For a magnetic field (and, thus, magnetization) that points in the $x_3$ direction and varies in the $x_1$ direction, the relation~\eqref{eq:eos_susc_coord} implies~\cite{Brandt:2024blb},
\be
\expv{j_2(x_1)} = \frac{\partial \M(x)}{\partial x_1} = \frac{1}{L_1}\int \dd y_1 \, \chi(x_1-y_1)\, \frac{\partial eB(y_1)}{\partial y_1} + \O(B^3)\,,
\label{eq:eos_ampere}
\ee
where we used the parity symmetry of the susceptibility and integrated by parts. 

In Fourier space, the relation~\eqref{eq:eos_ampere} can be easily inverted to find the momentum-space susceptibility $\widetilde{\chi}(p_1)$ based on the induced current in an arbitrary inhomogeneous magnetic background,
\be
\widetilde{\chi}(p_1)=\frac{-i}{p_1}\frac{\expv{\widetilde{j_2}(p_1)}}{e\widetilde{B}(p_1)} + \O(B^2)\,.
\ee
The susceptibility is obtained in the $p_1\to0$ limit. To that end one expresses $\widetilde{\chi}(p_1)$ via a coordinate space integral of the current.
Just like in Sec.~\ref{sec:eos_currentcurrent}, care has to be taken so that the integration kernel satisfies the periodic boundary conditions. For simplicity, a magnetic profile $B(x_1)=B(L_1-x_1)$ is used, implying $\expv{j_2(x_1)}=-\expv{j_2(L_1-x_1)}$. The result for the susceptibility then reads~\cite{Brandt:2024blb},
\be
\chi= - \lim_{B\to0}\int_0^{L_1} \!\dd x_1 \,W(x_1) \, \expv{j_2(x_1)}\Bigg/
\int_0^{L_1} \!\dd x_1 \,W(x_1) \, \frac{\partial eB}{\partial x_1}, \qquad
W(x_1)=
\begin{cases}
 -x_1, & x\le L_1/4\\
 x_1-L_1/2, & L_1/4<x_1\le 3L_1/4 \\
 L_1-x_1, & 3L_1/4<x_1\\
\end{cases}\,.
\label{eq:eos_ampere_final}
\ee
The susceptibility is obtained via a numerical extrapolation of this expression to $B\to0$, as indicated in~\eqref{eq:eos_ampere_final}.

This approach was developed in~\cite{Brandt:2024blb} and employed for the localized magnetic profile~\eqref{eq:lat_loc_Bfield}. The susceptibility was calculated throughout the transition region using an improved action and physical quark masses.

\subsection{The magnetic susceptibility in thermal QCD}
\label{sec:eos_magsusc_QCD}

\begin{figure}
 \centering
 \includegraphics[width=8cm]{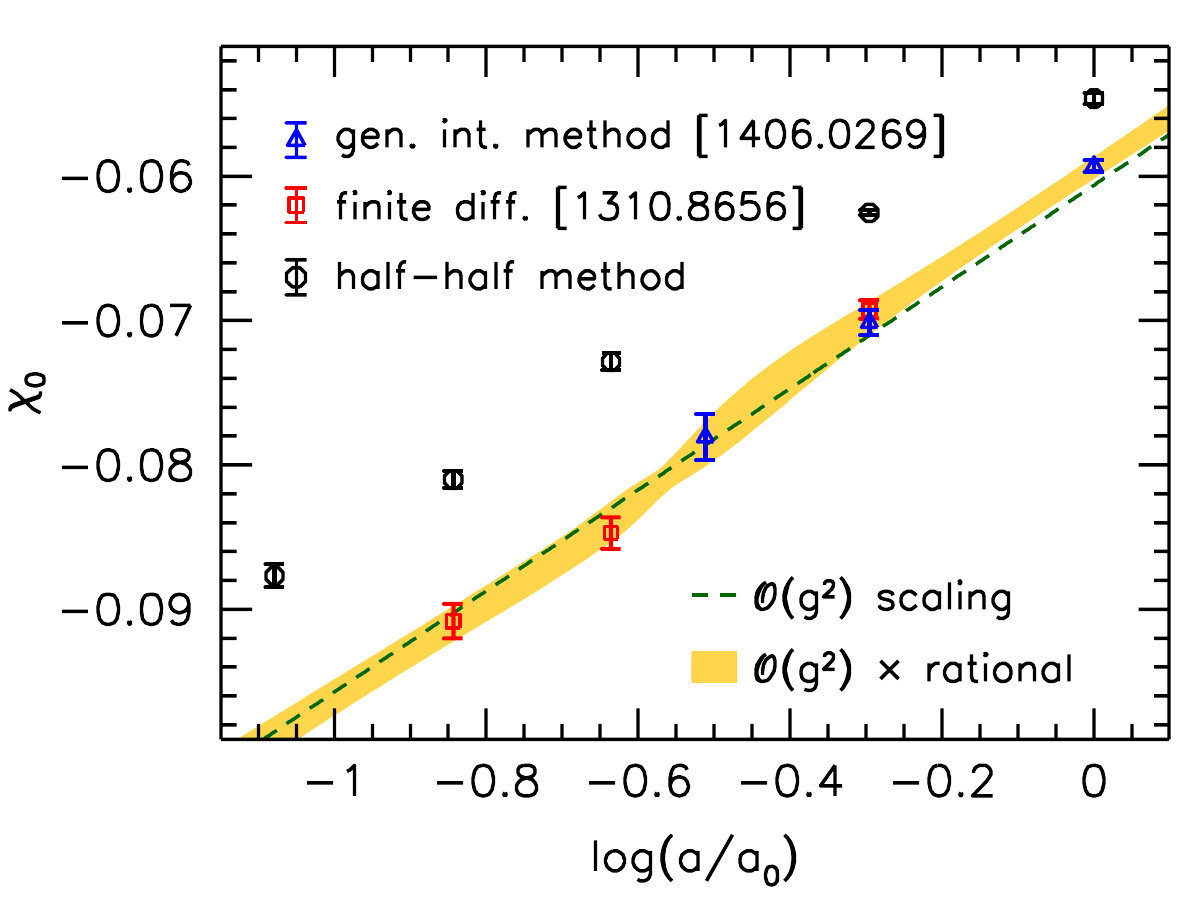}\qquad
 \includegraphics[width=8.2cm]{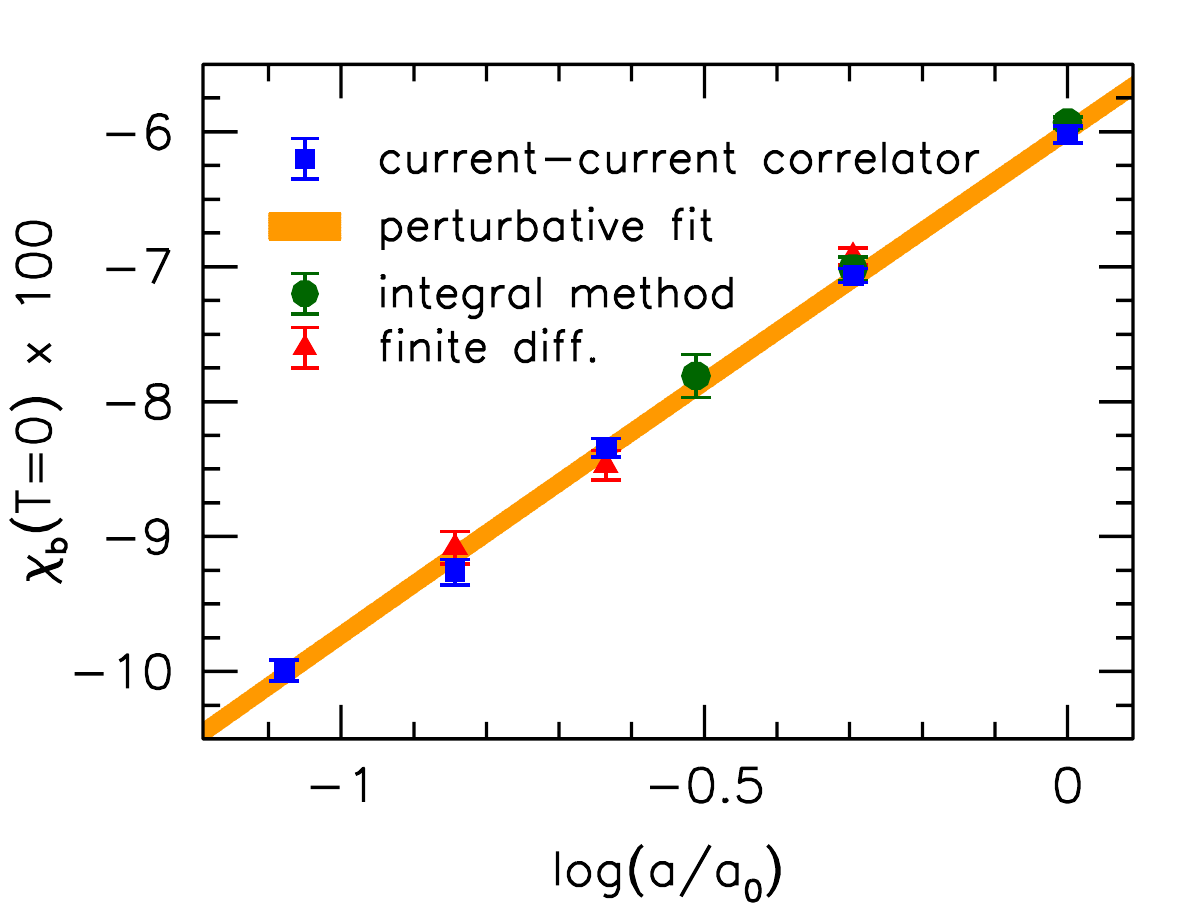}
 \caption{ \label{fig:eos_chi_T0}
  A collection of results for the bare magnetic susceptibility $\chi$ (denoted here by $\chi_0$ in the left and $\chi_b$ in the right panel) at zero temperature using various approaches, together with fits revealing the logarithmic scaling. Here, $a_0=1.46\GeVin$ is a fixed constant. The results were obtained using the half-half method~\cite{Bali:2015msa} (left panel), the finite difference method~\cite{Bonati:2013vba}, the generalized integral method~\cite{Bali:2014kia}, as well as the current-current correlator method~\cite{Bali:2020bcn} (right panel) and all correspond to the same lattice action.
  The perturbative scaling~\eqref{eq:lat_fr_muQED_def} is linear in $\log a$ with the proportionality constant $\beta_1$ being the lowest-order QED $\beta$-function coefficient. The figures are taken from~\cite{Bali:2015msa} and~\cite{Bali:2020bcn}.
 }
\end{figure}

All of the methods discussed in Sec.~\ref{sec:eos_susctorus} were employed in the literature to determine the magnetic susceptibility. A subset of these results is shown in Fig.~\ref{fig:eos_chi_T0} for the bare susceptibility at zero temperature (notice that at $T=0$, $\chi=\xi$ due to Lorentz invariance). Note that all data points shown here were obtained using the same lattice action -- stout improved staggered quarks at physical masses. One sees that all approaches agree and follow the perturbative expectation that follows from~\eqref{eq:lat_fr_muQED_def}, namely a logarithmic scaling with the lattice spacing. One exception is the half-half method, for which the results of~\cite{Bali:2015msa} are shown. The mismatch visible in the left panel of Fig.~\ref{fig:eos_chi_T0} is due to enhanced finite volume effects, resulting from the discontinuities of the magnetic field in this approach. In fact, these finite volume effects were observed to cancel to a large extent in the renormalized magnetic susceptibility~\cite{Bali:2015msa,Endrodi:2014lja}.

\begin{figure}
 \centering
 \raisebox{-.15cm}{\includegraphics[width=8.4cm]{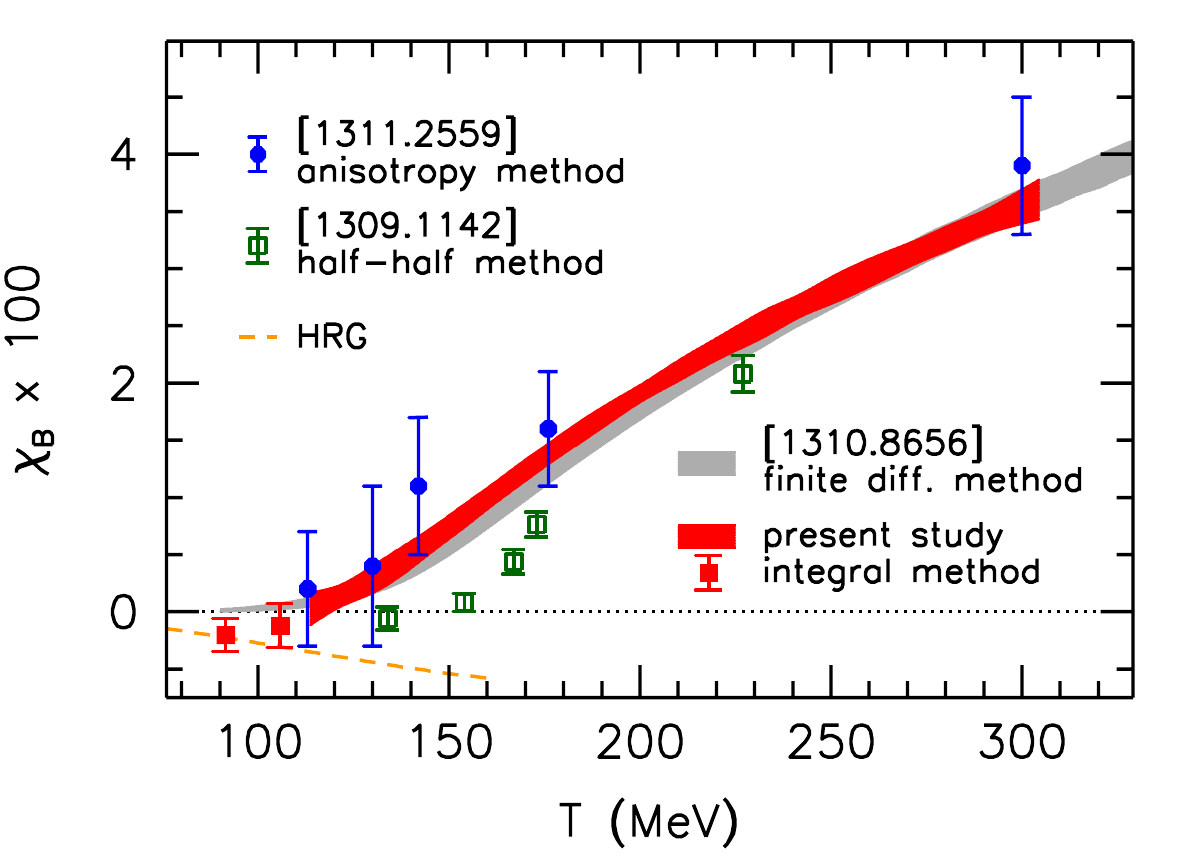}}\qquad
 \includegraphics[width=8cm]{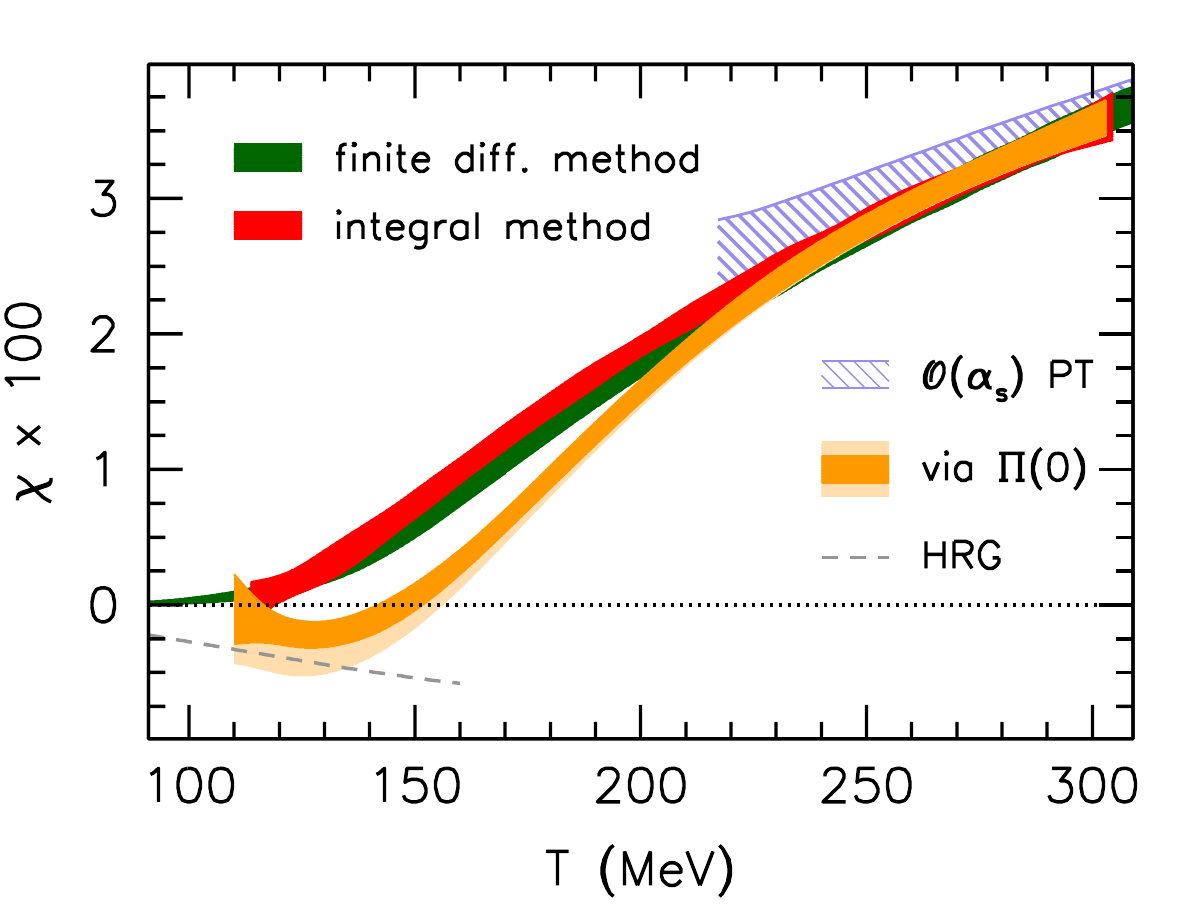}
 \caption{ \label{fig:eos_chi_T}
  A collection of results for the temperature-dependence of the renormalized magnetic susceptibility $\chi^r$ (denoted here by $\chi_B$ in the left and $\chi$ in the right panel) using various approaches. Comparisons to the hadron resonance gas model and to perturbation theory (in the right panel) are also shown. The results were obtained using the half-half method~\cite{Levkova:2013qda}, the anisotropy method~\cite{Bali:2013owa} (left panel), the finite difference method~\cite{Bonati:2013vba}, the generalized integral method~\cite{Bali:2014kia}, as well as the current-current correlator method~\cite{Bali:2020bcn} (right panel). Note that the half-half method was employed using a lattice action different from the other approaches. The figures are taken from~\cite{Bali:2014kia} and~\cite{Bali:2020bcn}.
 }
\end{figure}

The renormalized susceptibility $\chi^r$ is obtained by zero-temperature subtraction according to~\eqref{eq:lat_magsusc_renorm}. The results are shown in Fig.~\ref{fig:eos_chi_T} for a variety of methods, revealing qualitative agreement about the gradual enhancement of $\chi^r$ as the temperature grows. Let us discuss this qualitative behavior first, before we tend to the differences between the various approaches. Clearly, there are two different regimes: at high temperature $\chi^r$ is large and positive, whereas at low temperature $\chi^r$ is suppressed and slightly negative. In fact, these two types of behavior follow from simple considerations, and reflect the different effective nature of QCD matter at high and low $T$, respectively. At high temperatures, the system consists of quasi-free quarks that respond to the magnetic field via their spins, giving rise to paramagnetism, i.e.\ $\chi^r>0$. This can be quantified by perturbation theory, which predicts $\chi^r=2\beta_1\log (T/\mu_{\rm QED})$ to leading order~\cite{Bali:2014kia,Bali:2020bcn}. 

In turn, the effective degrees of freedom that respond to the magnetic field at low temperatures are charged hadrons, most importantly charged pions. We have seen in Sec.~\ref{sec:had_hadelmagfields} that charged meson energies are increased to leading order by the magnetic field, leading to an overall increase in the free energy of the system i.e.\ $\chi^r<0$. This may also be understood from the fact that the linear coupling in the meson energy~\eqref{eq:had_energy_hadron_spin} stems from orbital angular momentum. The response of the latter to the magnetic field is always diamagnetic owing to Lenz's law. A more quantitative statement for $\chi^r$ may be obtained using the Hadron Resonance Gas (HRG) model, which indeed predicts a negative susceptibility at low $T$~\cite{Bali:2020bcn}. The transition between the two regimes occurs at $T\approx T_c$, where the susceptibility is then expected to change its sign.

The strong paramagnetic behavior at high $T$ is confirmed by all lattice results~\cite{Levkova:2013qda,Bali:2013owa,Bonati:2013vba,Bali:2014kia,Bali:2020bcn} shown in Fig.~\ref{fig:eos_chi_T}. 
Here, the results obtained with different methods used the same, stout-improved staggered action, except for the study~\cite{Levkova:2013qda} employing the half-half method, which used the HISQ staggered action and slightly heavier-than-physical quark masses. The deviation between these results and the rest in the left panel of Fig.~\ref{fig:eos_chi_T} is most probably due to the finite size effects related to the half-half-profile, as well as the quark mass difference, which is most relevant in the transition region and pushes $T_c$ to somewhat higher values.

The diamagnetic behavior at low $T$ is expected to be much weaker and is only visible in the results of~\cite{Bali:2020bcn}. The reason behind the deviation between the results shown in the right panel of Fig.~\ref{fig:eos_chi_T} appears to lie in the continuum extrapolations. In fact, the taste-splitting lattice artefacts of the staggered formulation are expected to imply a strong suppression of $\chi^r$ and a slow approach to the negative continuum limit~\cite{Bali:2020bcn}. The mismatch between the finite difference method~\cite{Bonati:2013vba} and the current-current correlator method~\cite{Bali:2020bcn} is yet to be reconciled by dedicated simulations in the low-temperature regime.

\begin{figure}[t]
 \centering
 \includegraphics[width=6.7cm]{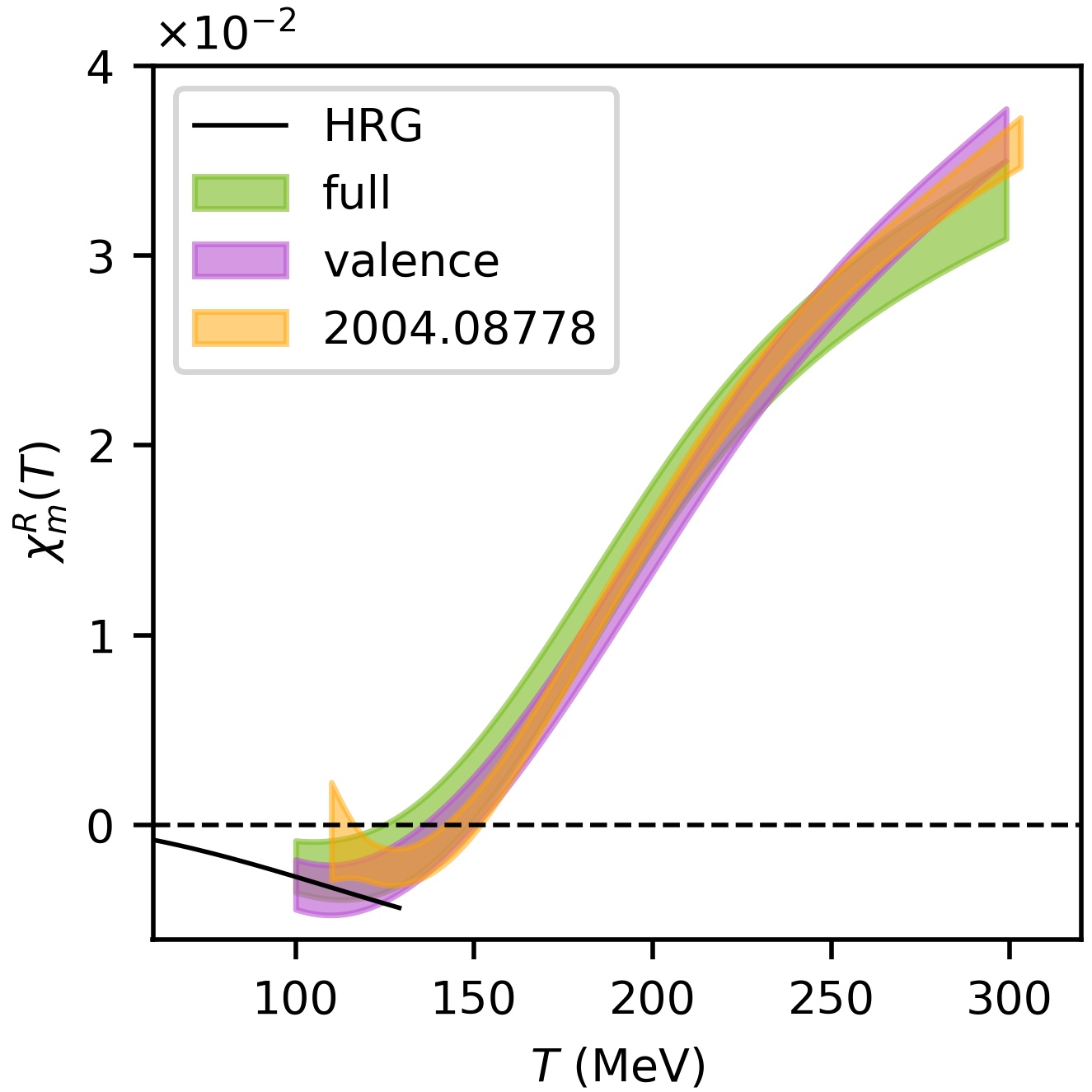}\qquad\qquad
 \includegraphics[width=6.7cm]{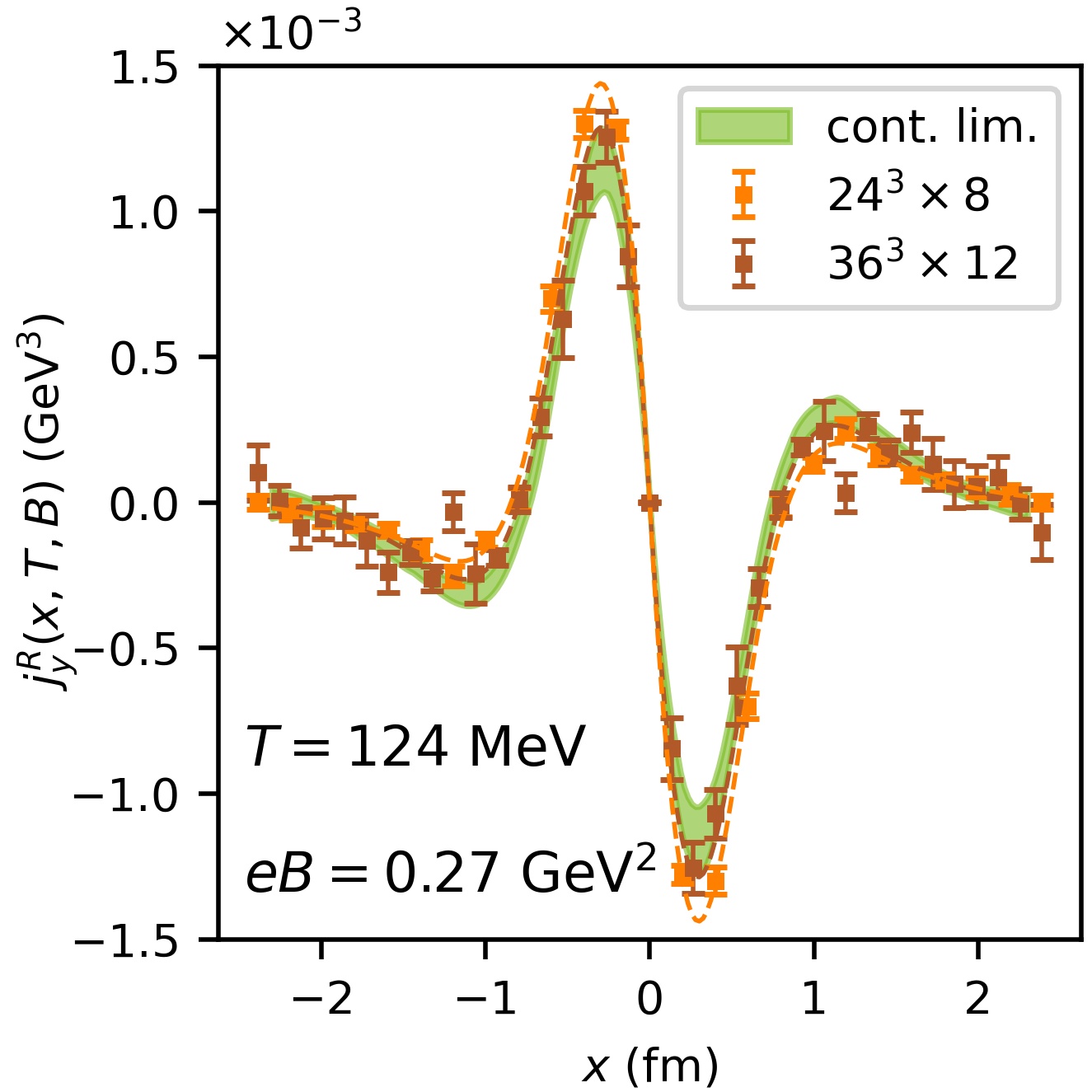}
 \caption{ \label{fig:eos_chi_ampere}
  Left panel: the renormalized magnetic susceptibility $\chi^r$ (denoted here as $\chi^R_m$) as a function of the temperature, determined using Amp\'{e}re's law~\cite{Brandt:2024blb}. The valence approximation to the susceptibility is also included, along with the results of the current-current correlator method~\cite{Bali:2020bcn} and the HRG model~\cite{Bali:2020bcn}.
  Right panel: renormalized electric current (denoted by $j^R$) induced by an inhomogeneous background magnetic field~\eqref{eq:lat_loc_Bfield} with a profile width $\epsilon\approx0.6\fm$~\cite{Brandt:2024blb}.
 }
\end{figure}

We note that a very recent study also calculated $\chi^r$ for the same temperature range using Amp\'{e}re's law~\cite{Brandt:2024blb} and the localized magnetic field profile~\eqref{eq:lat_loc_Bfield}, and found results consistent with the current-current correlator method~\cite{Bali:2020bcn}, see the left panel of Fig.~\ref{fig:eos_chi_ampere}. Besides the full result for $\chi^r$, this study also considered the valence approximation of the induced current and, thus, for the magnetic susceptibility. For this observable, sea and valence quark effects satisfy an additivity relation similar to~\eqref{eq:pd_additivity}, but this time it is exact to order $B$~\cite{Brandt:2024blb},
\be
\expv{j_\nu}_B=\expv{j_\nu}_B^{\rm val}+\expv{j_\nu}_B^{\rm sea}+\O(B^3)\,.
\ee
Interestingly, sea quark effects for $\chi^r$ were found to be negligibly small, so that the valence approximation and the full result for the magnetic susceptibility agree within statistical errors. This comparison is also included in the left panel of Fig.~\ref{fig:eos_chi_ampere}. For completeness, in the right panel of Fig.~\ref{fig:eos_chi_ampere} we show the results of~\cite{Brandt:2024blb} for the induced current at nonzero magnetic field, after having performed its renormalization according to~\eqref{eq:lat_renorm_induced_current}.

Finally, we mention that a simple parameterization for $\chi^r(T)$ that connects the HRG model, the continuum extrapolated lattice results of~\cite{Bali:2020bcn} and the perturbative behavior, was constructed in~\cite{Bali:2020bcn}, to be used in phenomenological models or comparisons to analytical approaches.

\subsubsection{Spin contribution to the susceptibility and the photon distribution amplitude}
\label{sec:eos_tensor}

The response of the QCD medium to the background field may be decomposed into a contribution from quark spins and one from quark and gluon orbital angular momenta. While such decompositions are, in a gauge theory, often ambiguous, for the case at hand it is possible to construct a quark spin contribution that is well defined up to multiplicative renormalization~\cite{Bali:2020bcn}. The spin term is related to the dependence of the tensor fermion bilinear $\expv{\bar\psi_f\sigma_{\nu\rho}\psi_f}$ on the homogeneous background magnetic field\footnote{We note that for general fermion bilinears $\bar\psi_f\Gamma\psi_f$ at nonzero magnetic field $B=F_{12}$, there are two operators with nonzero expectation values: $\Gamma=\mathds{1}$ for the quark condensate and $\Gamma=\sigma_{12}$ for the tensor bilinear. For CP-odd environments, like a specific instanton configuration or in correlators with the topological charge density, the electric dipole component $\Gamma=\sigma_{34}$ may also develop nonzero values. We get back to this point in Sec.~\ref{sec:anom_lorentz_cov}.
}. Therefore we discuss this observable next.

The explicit breaking of Lorentz-symmetry by the background field picks a preferred plane (the $x_\nu-x_\rho$ plane)
and induces nonzero expectation value for the tensor operator, as was pointed out already in~\eqref{eq:lat_def_tensorcoeff}. For a weak magnetic field in the $x_3$ direction, this expectation value is parameterized by the leading coefficient,
\be
\tau_f = \left.\frac{\partial \expv{\bar\psi_f\sigma_{12}\psi_f}}{\partial (q_fB)}\right|_{B=0}\,,
\label{eq:eos_def_tensorcoeff}
\ee
the so-called tensor coefficient. 
It undergoes both multiplicative and additive renormalization, as discussed in~\eqref{eq:lat_renorm_tensorcoeff}.
Occasionally, in the literature the ratio $\tau_f/\expv{\bar\psi_f\psi_f}$ is considered and called the magnetic susceptibility of the condensate, but we will not use this nomenclature here. 
Furthermore, the observable~\eqref{eq:eos_def_tensorcoeff} gives 
the normalization of the leading-twist photon distribution amplitude~\cite{Balitsky:1989ry}, relevant for photon to quark-antiquark dissociation and radiative heavy meson decays~\cite{Rohrwild:2007yt}.

\begin{figure}
 \centering
 \includegraphics[width=8cm]{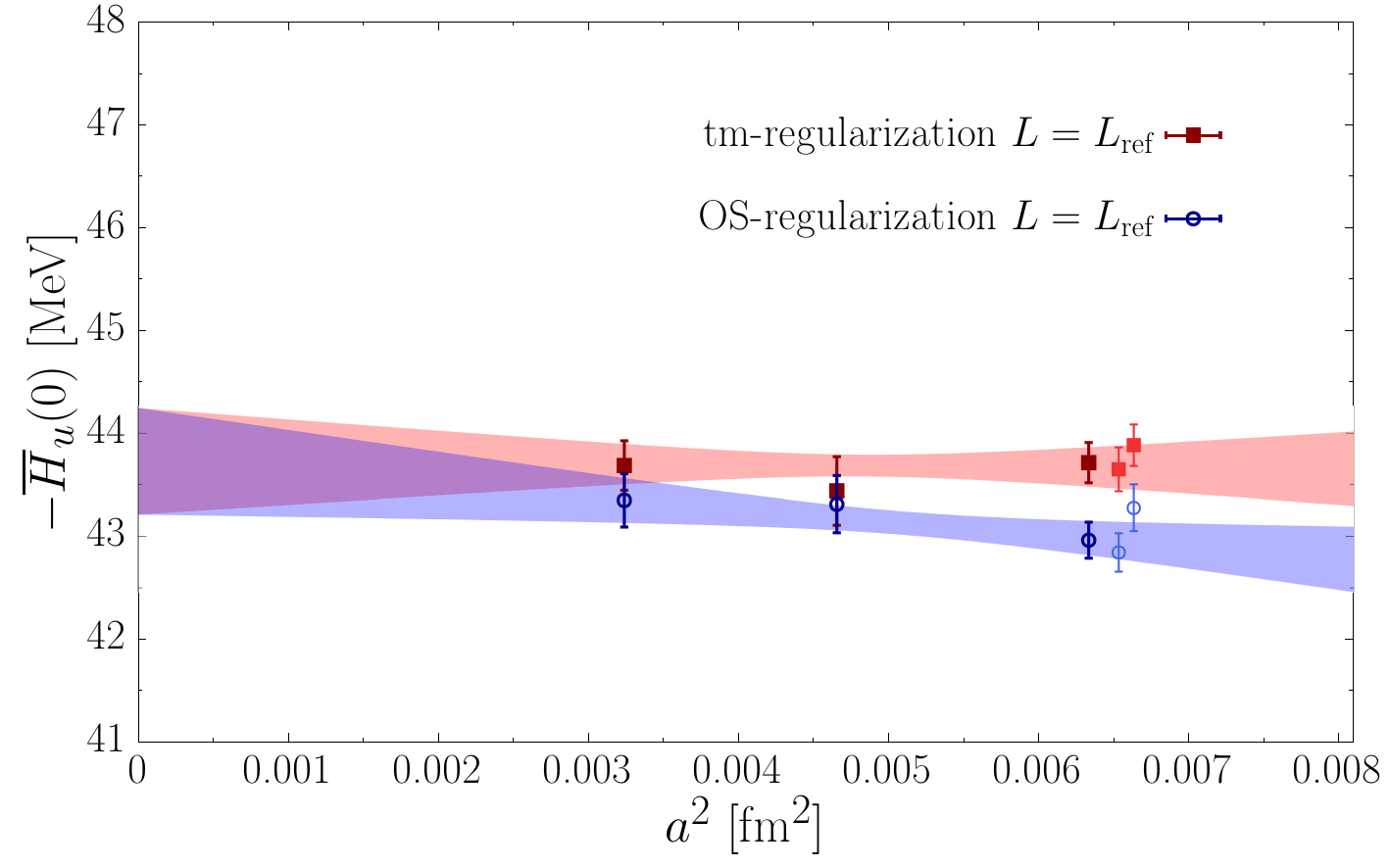}\qquad
 \raisebox{-.6cm}{\includegraphics[width=8cm]{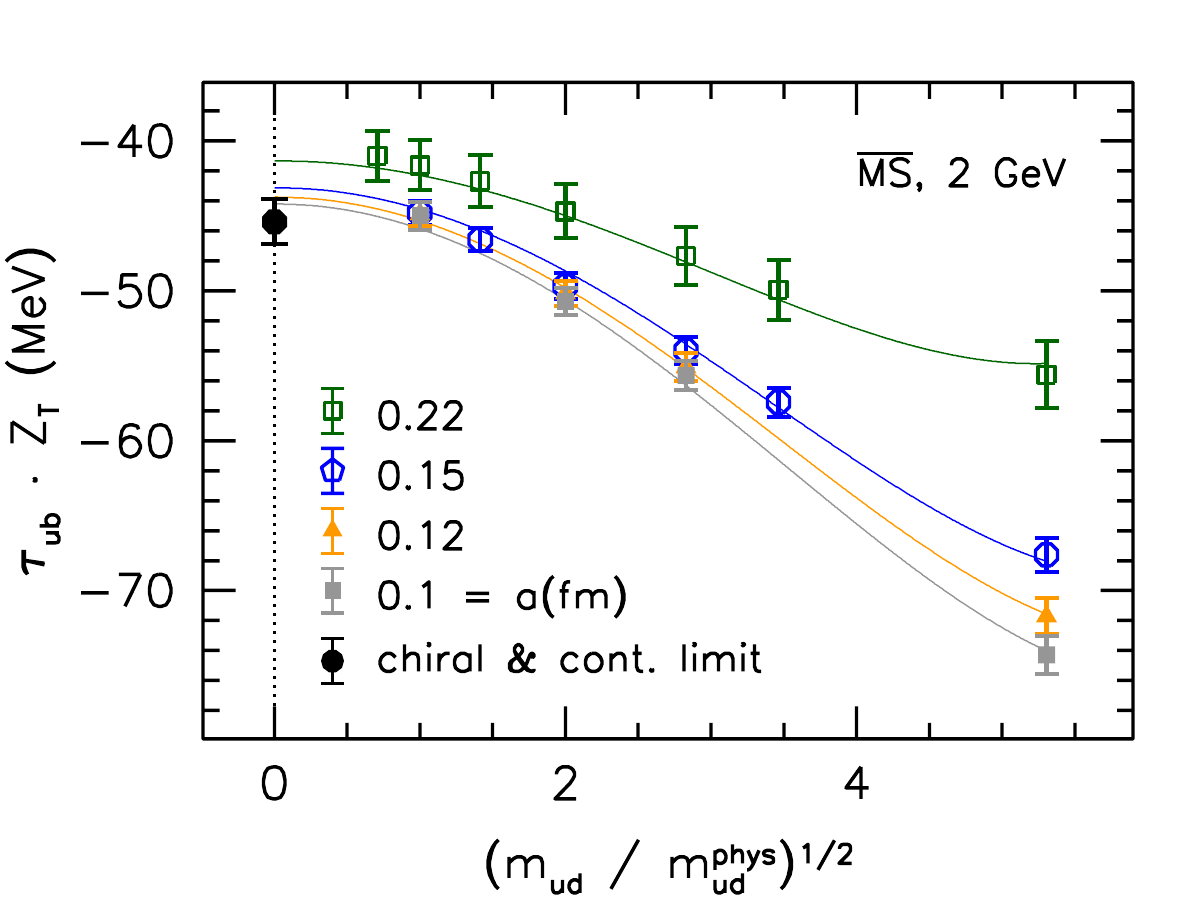}}
 \caption{ \label{fig:eos_res_tensorcoeff}
  Left panel: continuum extrapolation of the negative of the tensor coefficient of the up quark  -- more precisely, renormalized as $Z_T(1-m_u\partial_{m_u})\tau_u$, denoted here by $\bar H_u(0)$ -- using Wilson-clover twisted-mass sea quarks~\cite{Bacchio:2024dce}. Right panel: combined chiral and continuum extrapolation of the tensor coefficient using stout-improved staggered quarks~\cite{Bali:2020bcn}.
 }
\end{figure}

As a one-point function, the tensor bilinear can be straightforwardly calculated in lattice simulations, just like the quark condensate. The tensor coefficient~\eqref{eq:eos_def_tensorcoeff} can then be determined by numerical differentiation with respect to the magnetic field.
The bare tensor bilinear was first calculated on the lattice in~\cite{Buividovich:2009ih} using the quenched approximation of two-color QCD with overlap valence quarks, and later updated with a similar quenched setup in three-color QCD~\cite{Braguta:2010ej}. Later, dynamical simulations using $2+1$ flavors of improved staggered quarks at physical masses were used to calculate the same observable, extrapolated to the continuum limit~\cite{Bali:2012jv,Bali:2020bcn}. Here, the complete renormalization of $\tau_f$ was also performed, first perturbatively~\cite{Bali:2012jv} and then non-perturbatively~\cite{Bali:2020bcn}. 

The tensor coefficient was recently also calculated in~\cite{Bacchio:2024dce} using $2+1+1$ flavors of Wilson-clover twisted-mass fermions in the sea and different valence quark regularizations. In this study, $\tau_f$ was directly determined via two-point functions, an alternative approach also discussed in~\cite{Bali:2020bcn}, in spirit similar to the current-current correlator method of Sec.~\ref{sec:eos_currentcurrent} for the calculation of $\chi$. The continuum extrapolations of the renormalized tensor coefficients, using this direct method and the approach of~\cite{Bali:2020bcn} discussed above, agree within errors, as visible in Fig.~\ref{fig:eos_res_tensorcoeff}.
Regarding the phenomenological value of the photon distribution amplitude, detailed summaries on the different approaches and numerical comparisons were given in~\cite{Bali:2020bcn,Bacchio:2024dce}.

\begin{figure}[t]
 \centering
 \includegraphics[width=8cm]{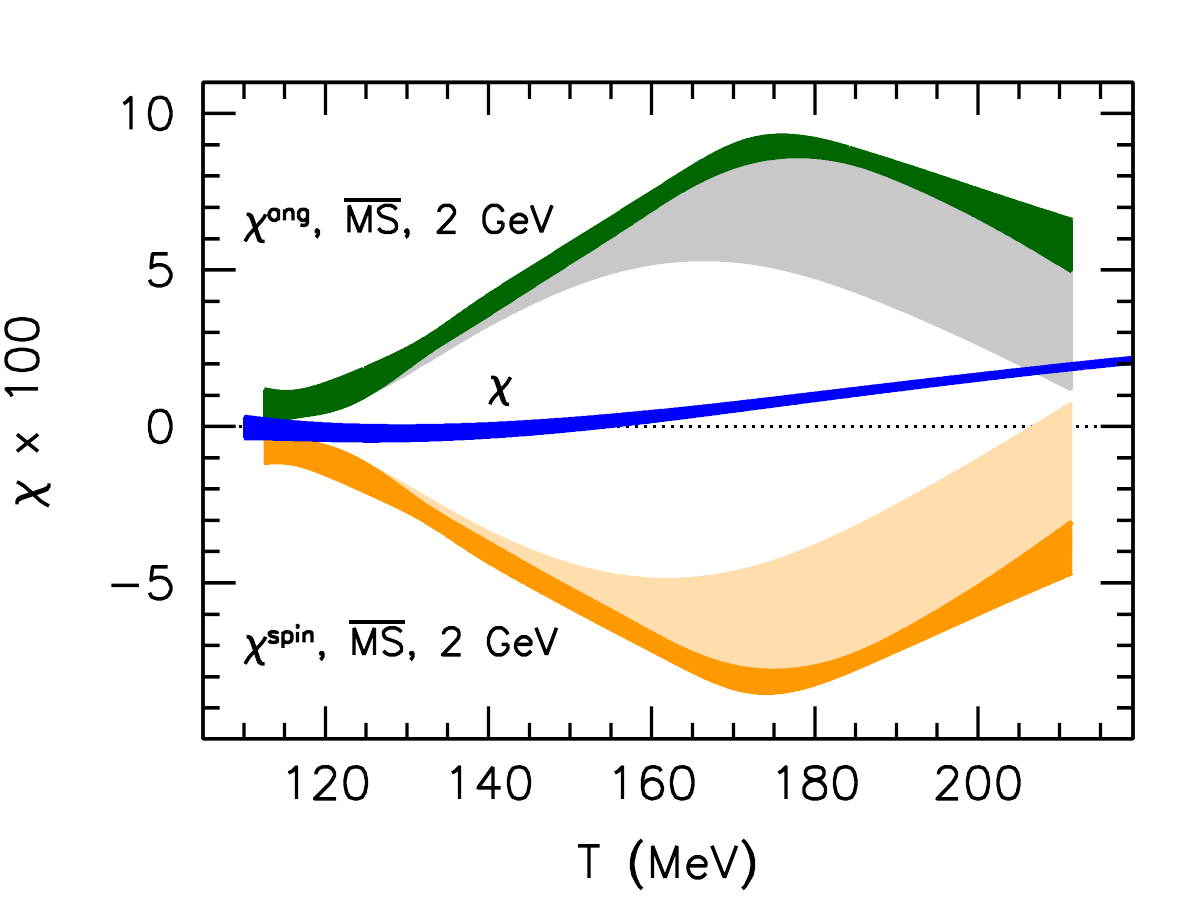}
 \caption{ \label{fig:eos_chi_decomp}
  Continuum estimate of the decomposition of the renormalized magnetic susceptibility into contributions from spin and angular momentum (denoted here by $\chi$, $\chi^{\rm spin}$ and $\chi^{\rm ang}$, respectively)~\cite{Bali:2020bcn}. The renormalization constants, required for the separate contributions, were calculated in the $\overline{\rm MS}$ scheme at a QCD renormalization scale $\mu_{\rm QCD}=2\GeV$.
 }
\end{figure}

Finally, we turn to the temperature dependence of the renormalized tensor coefficient $\tau_f^r$. As mentioned above, it plays a role in the decomposition of the magnetic susceptibility into contributions from spin and orbital angular momenta, $\chi^r=\chi^r_{\rm spin} + \chi^r_{\rm ang}$. This decomposition may be derived by identifying the tensor bilinear in the magnetization~\cite{Bali:2020bcn},
\be
\M =\sum_f \frac{q_f/e}{2m_f} \left[ 1 -\lim_{m_f^{\rm val}\to0} \right] \langle \bar\psi_f \sigma_{12}\psi_f + \bar\psi_f L_{12}\psi_f\rangle\,,
\label{eq:eos_magn_separation}
\ee
which emerges from a simple manipulation of the Dirac trace involved in $\M$.
Above, $L_{12}$ is an orbital angular momentum operator (for its specific definition, see~\cite{Bali:2020bcn}) and $m_f^{\rm val}$ denotes the value of the valence quark mass, cf.\ Sec.~\ref{sec:lat_valence_sea}. Now we differentiate~\eqref{eq:eos_magn_separation} with respect to $eB$: on the left hand side we recover $\chi$, while in the first term on the right hand side we can discover the tensor coefficient. Performing the additive and multiplicative renormalization, the spin contribution is obtained as~\cite{Bali:2020bcn},
\be
\chi^r_{\rm spin}=\sum_f\frac{(q_f/e)^2}{m_f} \left[ \tau^r_f(m_f^{\rm val}=m_f)-\tau_f^r(m_f^{\rm val}=0)\right] \cdot Z_TZ_S\,.
\ee
This term, together with the total susceptibility $\chi^r$ and the angular momentum contribution, $\chi^r_{\rm ang}=\chi^r-\chi^r_{\rm spin}$ is shown in Fig.~\ref{fig:eos_chi_decomp} from~\cite{Bali:2020bcn}. The zero valence quark mass limit is computationally expensive and was estimated in this study from a fit of the mass-dependence of $\tau_f^r$. This entails large systematic errors, which are included in the figure as shaded bands. Interestingly, the total susceptibility arises as a cancellation of spin and orbital angular momentum contributions. An interpretation of these terms and a comparison to the free-quark picture involving Pauli paramagnetism and Landau diamagnetism is discussed in~\cite{Bali:2020bcn}.

\subsection{Thermodynamic observables in strong magnetic fields}
\label{sec:eos_thermo_largeB}

The observables relevant for the equation of state can be obtained using the thermodynamic relations listed in Sec.~\ref{sec:eos_thermo_rel}.
For weak homogeneous background magnetic fields, all of these observables can be expanded in $eB$ to leading order and, in turn, expressed in terms of their $B=0$ values and the renormalized magnetic susceptibility $\chi^r$.
For example, the parallel pressure, this expansion takes the form $p_3(B)=p(0)+\chi^r (eB)^2/2$. This dependence was constructed up to $eB=0.3\GeVsq$ using the susceptibility results from the finite difference method in~\cite{Bonati:2013vba}. The expressions for all other observables can be found analogously. These formulae were collected in~\cite{Bali:2020bcn} and constructed from the parameterization of the $B=0$ equation of state and the susceptibility for weak magnetic fields and arbitrary temperatures~\cite{Bali:2020bcn}.

\begin{figure}
 \centering
 \includegraphics[width=8cm]{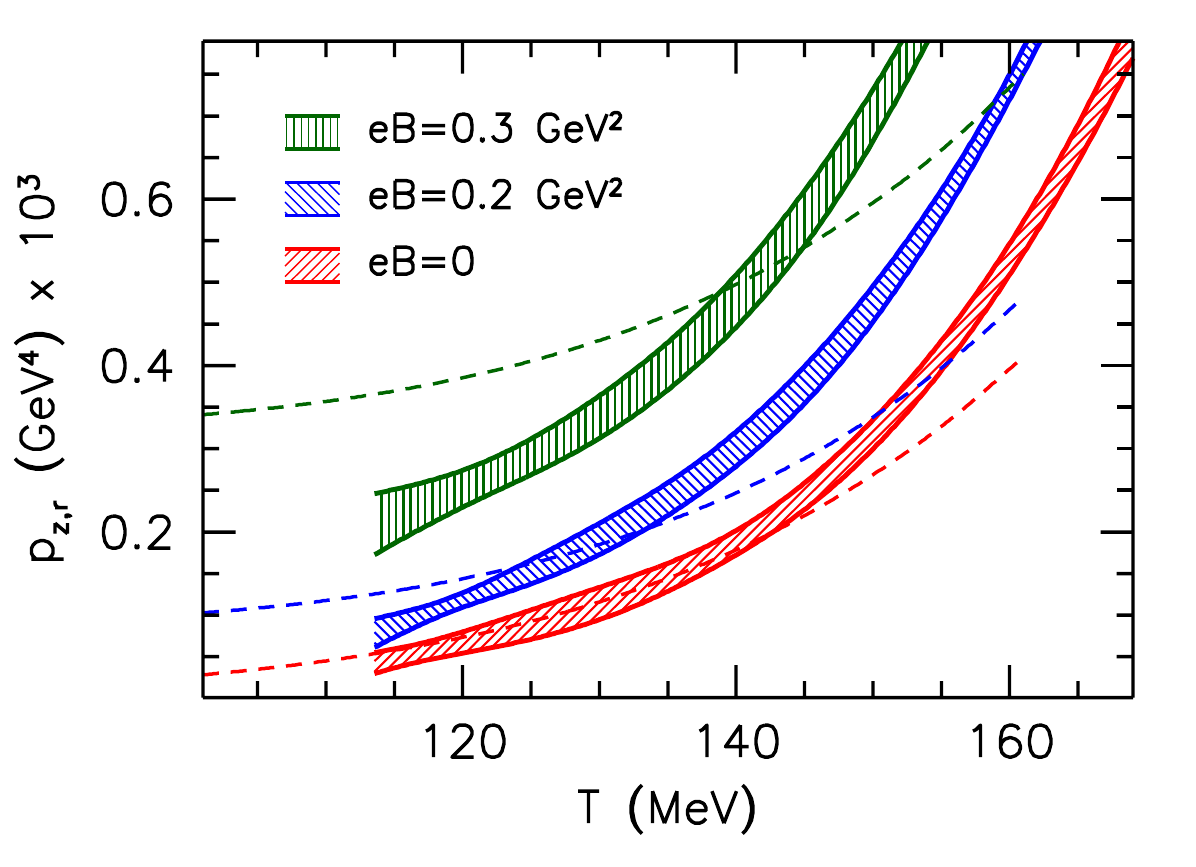}\qquad
 \includegraphics[width=8cm]{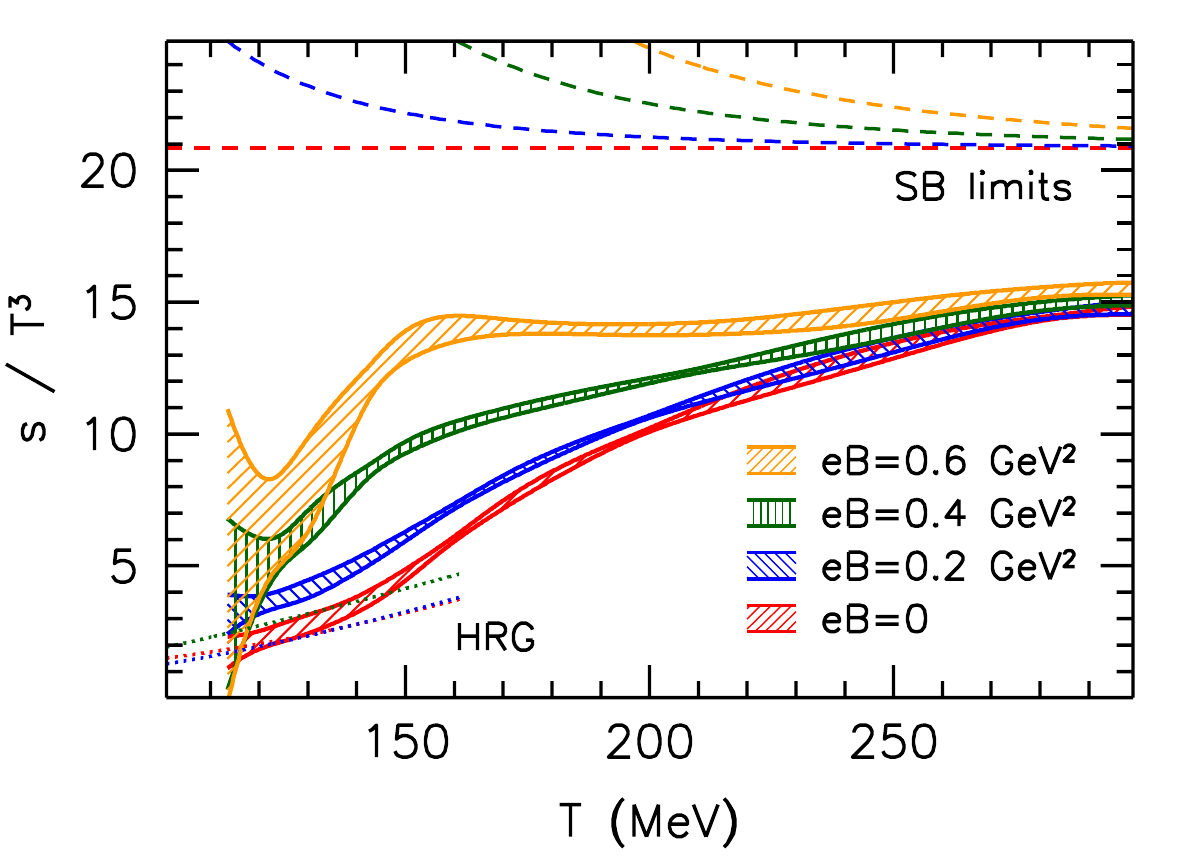}
 \caption{ \label{fig:eos_thermoobsB}
  The parallel pressure $p_3$ (denoted here as $p_{z,r}$, left panel) and the entropy density (right panel) as functions of the temperature for various values of the magnetic field, obtained using the generalized integral method~\cite{Bali:2014kia}. Also included is a comparison to the HRG model~\cite{Endrodi:2013cs} at low temperature, as well as to the prediction of leading-order perturbation theory at high $T$ in the right panel.
 }
\end{figure}

Alternatively, one can calculate the thermodynamic observables immediately at $B>0$ using the approaches that work directly at nonzero homogeneous magnetic field values, i.e.\ the anisotropy method (Sec.~\ref{sec:eos_aniso}), the finite difference method (Sec.~\ref{sec:eos_finitediffmethod}) and the generalized integral method (Sec.~\ref{sec:eos_genintmethod}).
The anisotropy method was employed to determine the magnetization as a function of $B$ at zero~\cite{Bali:2013esa} and nonzero temperatures~\cite{Bali:2013owa}, while in~\cite{Bali:2014kia} all observables of Sec.~\ref{sec:eos_thermo_rel} were determined up to $eB=0.7\GeVsq$ and tabulated in ancillary files. We show the results of~\cite{Bali:2014kia} for the parallel pressure and the entropy density in Fig.~\ref{fig:eos_thermoobsB}.
We remark that these results represent a continuum estimate based on three lattice spacings ($N_t=6,8$ and $10$). Later it was shown~\cite{Bali:2020bcn} that these lattice spacings are not sufficiently close to the continuum limit at low temperatures -- they tend to overestimate the magnetic susceptibility, see the right panel of Fig.~\ref{fig:eos_chi_T}.

At this point, we stress that at $T=0$, the renormalized magnetic susceptibility vanishes (see Fig.~\ref{fig:eos_chi_T}), but the QCD medium still responds to the background field starting at $\O(B^4)$, see~\eqref{eq:lat_fr_def}. As visible from the dependence of the pressure $p_3(B)$ at low temperature (see the left panel of Fig.~\ref{fig:eos_thermoobsB}), this contribution is positive, consistent with the results of~\cite{Bali:2013esa} for the $T=0$ magnetization. Thus, we can conclude that at $T=0$ the QCD vacuum is paramagnetic to $\O(B^4)$, but becomes diamagnetic to $\O(B^2)$ at low temperatures.
This tendency is also predicted by the HRG model~\cite{Endrodi:2013cs,Bali:2020bcn}. Here, charged pions result in $\chi^r<0$, building up the low-temperature $\O(B^2)$ response, but other hadrons reduce the matter free energy at $\O(B^4)$, resulting in a positive magnetization at strictly $T=0$. This picture is consistent with our findings about hadron magnetic moments in Sec.~\ref{sec:had_hadelmagfields}.

Finally, we note that the usual normalization of the above discussed observables, for example $p/T^4$ for the pressure, is disadvantageous for $B>0$ due to the just discussed magnetic field-dependent terms that are present already at $T=0$. For this reason, the pressure is best plotted without the normalization by $T^4$ in the left panel of Fig.~\ref{fig:eos_thermoobsB}. The entropy density, shown in the right panel, is an exception to this rule, since it vanishes at $T=0$ even for $B>0$. This is because the vacuum term stems from quantum fluctuations related to the interaction of virtual particles with the magnetic field and not from physical particles that could create entropy~\cite{Bali:2014kia}.

All results presented in this section so far were obtained using the stout-improved staggered action with physical quark masses in $N_f=2+1$ QCD.
In addition to the above studies, the equation of state at nonzero $B$ was also studied in $N_f=1$ QCD via the Schr\"odinger functional approach to introduce the magnetic field~\cite{Cea:2015ifa}. In this study, doubly subtracted observables like $\Delta p_3(B,T)-\Delta p_3(B,T=0)$ were calculated.

\subsection{Dense and magnetized QCD}
\label{sec:pd_fluc}

The equation of state discussed up to now corresponds to a system with zero net particle number for each quark flavor $f$. We proceed now to discuss dense systems, addressed in the context of the phase diagram already in Sec.~\ref{sec:pd_densemagnetized}. In the grand canonical ensemble, we use chemical potentials $\mu_f$ to control the densities $n_f$ according to~\eqref{eq:eos_s_n_def}. The equation of state at $\mu_f\neq0$ can be rewritten in terms of fluctuations and higher moments of conserved charges. In turn, these moments are themselves important for heavy-ion phenomenology due to their sensitivity to critical behavior and for a better understanding of the nature of effective degrees of freedom in the hot QCD medium.

For three-flavor QCD, the conserved charges\footnote{In QCD, the number of each quark flavor is conserved and so are arbitrary linear combinations of them. In the Standard Model, only $\B$ and $\Q$ correspond to conserved numbers.} are the baryon number $\B$, the electric charge $\Q$ and the strangeness $\S$. In terms of the individual flavors, the chemical potentials in this basis are
\be
\mu_u = \frac{1}{3} \mu_\B + \frac{2}{3}\mu_\Q, \qquad
\mu_d = \frac{1}{3} \mu_\B - \frac{1}{3}\mu_\Q, \qquad
\mu_s = \frac{1}{3} \mu_\B - \frac{1}{3}\mu_\Q - \mu_\S\,.
\label{eq:eos_baryonmu}
\ee
For lattice simulations, an alternative basis is often more convenient, where we consider the isospin charge $\I$, the so-called light baryon number $\L$ and the strangeness,
\be
\mu_u = \frac{1}{3} \mu_\L + \frac{1}{2}\mu_\I, \qquad
\mu_d = \frac{1}{3} \mu_\L - \frac{1}{2}\mu_\I, \qquad
\mu_s = - \mu_\S\,.
\ee

For most choices of these chemical potentials, the fermion determinants under the path integral become complex and standard Monte-Carlo algorithms fail, as we discussed in Sec.~\ref{sec:lat_signproblem}.
In this context, the isospin axis is special, since the fermion action remains real and positive for degenerate $u$ and $d$ quarks for any $\mu_\I\neq0$, enabling direct simulations along this axis. However, the positivity is lost as soon as the light quarks are not exactly degenerate. Thus, the complex action problem re-emerges for $m_u\neq m_d$ or when background electromagnetic fields, coupled to $q_u\neq q_d$, are present.

In order to explore the impact of nonzero chemical potentials on the equation of state, the most popular approaches are the Taylor-expansion method and simulations at $i\mu_f\neq0$. Moreover, simulations at $\mu_\I>0$ for weak background magnetic fields have also been carried out. In this section we summarize these studies.

An essential detail in the Taylor- and imaginary $\mu_f$-approaches is the direction, in which one probes the three-dimensional space of chemical potentials spanned by the bases $\mu_{u,d,s}$, $\mu_{\B,\Q,\S}$ or $\mu_{\L,\I,\S}$. With the heavy-ion collisions as application in mind, it is desirable to consider strangeness neutral systems, $n_\S=0$. The electric charge over baryon number in these collisions is also fixed by the initial conditions, so that $n_\Q/n_\B$ is around $0.4-0.45$ depending on the colliding ions.
An alternative choice that is simple to implement and that has been considered often in the literature is to expand along the $\mu_\L$ axis in the $\L-\I-\S$ basis, while keeping $\mu_\S=\mu_\I=0$. However, this does not correspond to a strangeness neutral system nor to the above specified charge to baryon ratio. The correct $\mu_\S(\mu_\B)$ and $\mu_\Q(\mu_\B)$
trajectories need to be calculated self-consistently using the Taylor-expansion of the above constrains~\cite{Bazavov:2012vg}. In the presence of a magnetic field, the determination of these trajectories is even more challenging due to the absence of the $u-d$ symmetry. This was considered in the studies~\cite{Ding:2023bft,MarquesValois:2023ehu}.

\begin{figure}
 \centering
 \includegraphics[width=7.8cm]{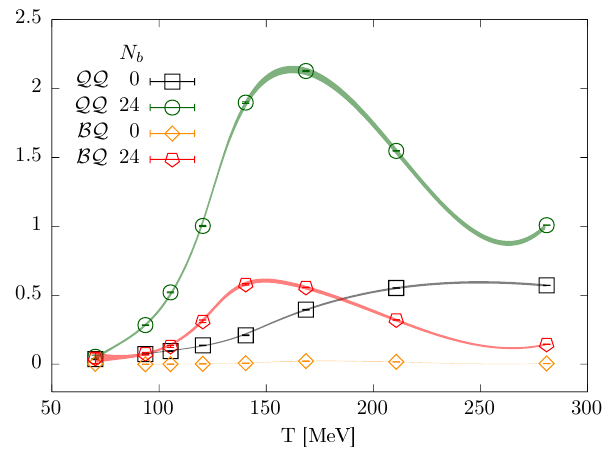}\qquad
 \raisebox{-.2cm}{\includegraphics[width=8.3cm]{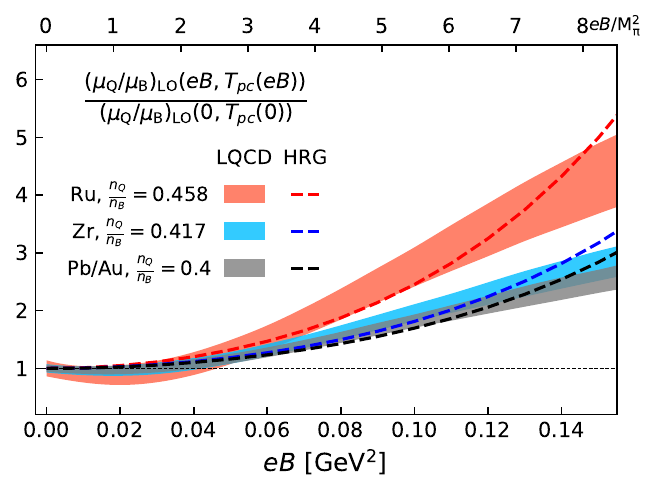}}
 \caption{
 \label{fig:eos_c2Q}
 Left panel: fluctuations of conserved charges -- $\chi_{2}^{\Q}$, denoted as $\Q\Q$ and $\chi_{11}^{\B\Q}$, denoted as $\B\Q$ -- as functions of the temperature for different magnetic fields, labeled by the flux quantum $N_b$. The flux value $N_b=24$ corresponds to a magnetic field of $eB\approx 1.25\GeVsq$. Figure adapted from~\cite{Ding:2021cwv}.
 Right panel: leading-order value of $\mu_\Q/\mu_\B$ at the transition temperature, corresponding to the conditions of strangeness neutrality and charge to baryon ratio specified in the legend~\cite{Ding:2023bft}.
 }
\end{figure}

The Taylor-expansion method was employed in~\cite{Ding:2021cwv} using HISQ fermions with a vacuum pion mass of $M_\pi\approx 220\MeV$. Here, the leading-order susceptibilities were measured\footnote{
In the following, indices with zero derivatives are suppressed and the same superscript is only written out once, e.g.\ $\chi_2^{\B}$ is the second derivative with respect to $\mu_\B$.},
\be
\chi_{ijk}^{\B\Q\S}=-\left.\frac{\partial^{i+j+k} f/T^4}{\partial (\mu_\B/T)^i\,\partial (\mu_\Q/T)^j\,\partial (\mu_\S/T)^k}\right|_{\mu_\B=\mu_\Q=\mu_\S=0}\,.
\ee
As an example, the results of~\cite{Ding:2021cwv} for the fluctuations $\chi_2^\Q$ of the electric charge and for the baryon-charge correlation $\chi_{11}^{\B\Q}$, are shown in the left panel of Fig.~\ref{fig:eos_c2Q} for $B=0$ and $B\approx 1.25\GeVsq$. For $\chi_2^\Q$, one can observe the development of a pronounced peak around the transition temperature for strong fields, contrasting the well-known monotonous rise at $B=0$. Incidentally, this peak was also observed to shift to lower temperatures as $B$ grows, signaling the decreasing behavior in $T_c(B)$, known from other observables discussed in Sec.~\ref{sec:pd_phasediagBT}.
Other diagonal as well as off-diagonal second-order susceptibilities were shown to exhibit very similar dependences on $T$ and $B$~\cite{Ding:2021cwv}. For example, $\chi_{11}^{\B\Q}$ shows a pronounced response to the magnetic field, see the left panel of Fig.~\ref{fig:eos_c2Q}. We mention that this study also investigated the effects of flavor symmetry-breaking in the light quark sector due to the magnetic field in more detail.

Already~\cite{Ding:2021cwv} suggested to use such second-order susceptibilities as proxies to determine the magnetic field strength in off-central heavy-ion collisions. This idea of using observables as `magnetometers' was developed further in~\cite{Ding:2023bft}, focusing on the second-order correlation $\chi_{11}^{\B\Q}$ of baryon number and the electric charge, which was found to be enhanced most prominently by $B$ around $T_c$. This study also
constructed the ratio $\mu_\Q/\mu_\B$ along the conditions $n_\S=0$ and $n_\Q/n_\B\approx 0.4$, using the leading-order behavior in the chemical potentials. This ratio was also found to be strongly enhanced by $B$, see the right panel of Fig.~\ref{fig:eos_c2Q}. We note that a comparison of these lattice observables to the experimental hadron yields is typically performed via a HRG-type model. The interpretation of such comparisons involving the HRG model at nonzero $B$ were scrutinized recently in~\cite{Vovchenko:2024wbg,Marczenko:2024kko}.
Finally we mention that the line of strangeness neutrality and $n_\Q/n_\B=0.4$ was also considered for various second-order fluctuations in~\cite{MarquesValois:2023ehu} in the presence of strong magnetic fields using stout-improved staggered quarks.

\begin{figure}
 \centering
 \includegraphics[width=7cm]{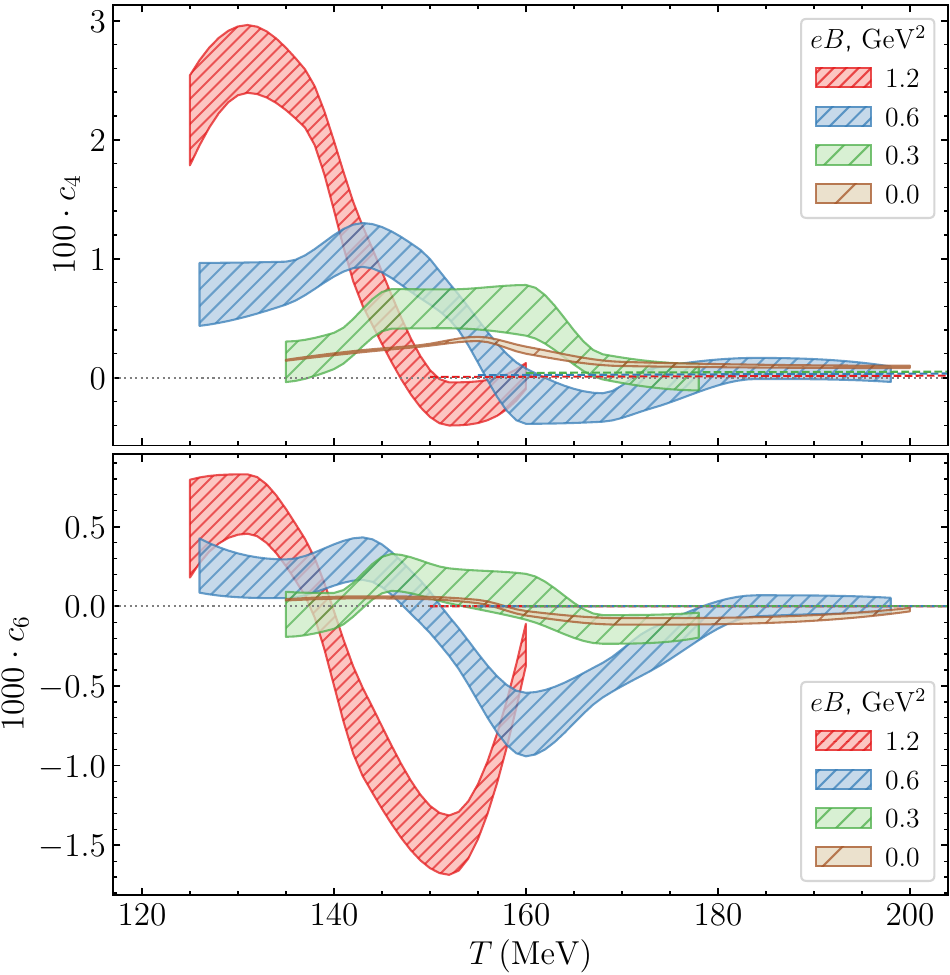} \qquad
 \raisebox{.9cm}{\includegraphics[width=8cm]{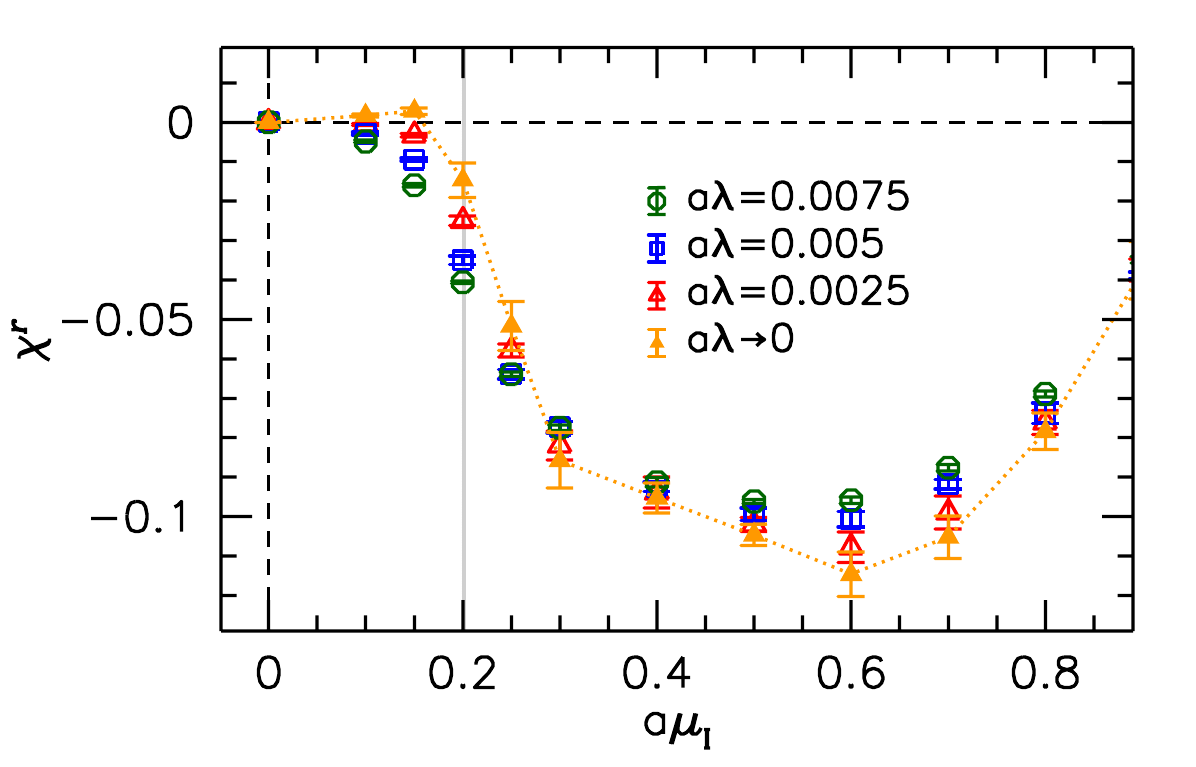}}
 \caption{
 \label{fig:eos_c46}
 Left panel: temperature-dependence of the coefficients of $\mu_\L^4$ (upper panel) and $\mu_\L^6$ (lower panel) in the Taylor-expansion of the pressure, constructed through simulations at imaginary chemical potentials and background magnetic fields. The bands represent continuum extrapolations using three lattice spacings and the colors correspond to the different values of $eB$~\cite{Astrakhantsev:2024mat}.
 Right panel: renormalized magnetic susceptibility as a function of the isospin chemical potential for different pion source parameters and in the physical, $\lambda\to0$ limit~\cite{Endrodi:2014lja}. The onset of pion condensation is at $a\mu_I\approx 0.2$, indicated by the gray vertical line.
}
\end{figure}

Recently, the imaginary chemical potential approach was followed in~\cite{Astrakhantsev:2024mat}, where simulations at $i\mu_\L\neq0$ were carried out using stout-improved staggered quarks with physical masses. The analytic continuation to real $\mu_\L$ was performed via polynomial fits up to $\O(\mu_\L^6)$ for different values of the magnetic field and the pressure $p(T,\mu_\L,eB)$ was constructed. The magnetic field was observed to enhance all Taylor-coefficients and to shift their characteristic temperature-dependence to lower values of $T$, as shown in the left panel of  Fig.~\ref{fig:eos_c46} for the quartic and sextic coefficients. Again, these findings are consistent with the reduction of $T_c(B)$ and the strengthening of the transition as $B$ grows~\cite{Astrakhantsev:2024mat}.

Finally, there is one lattice study that investigated the impact of weak magnetic fields on the isospin-dense QCD medium, using simulations at real isospin chemical potentials $\mu_\I>0$ and $\mu_\L=\mu_\S=0$~\cite{Endrodi:2014lja}. It is well known that the ground state of QCD changes from the usual vacuum state to a pion condensed phase at $\mu_\I=M_\pi$ through a second-order phase transition at low temperatures\footnote{The work~\cite{Endrodi:2014lja} used a different normalization convention for the isospin chemical potential, for which the transition occurs at $M_\pi/2$. We note moreover that in the absence of background fields, the phase diagram~\cite{Brandt:2017oyy,Brandt:2018omg} and the equation of state~\cite{Brandt:2022hwy,Abbott:2023coj,Abbott:2024vhj} have been determined on the lattice for nonzero isospin chemical potentials.}. The condensed phase involves charged pions that respond strongly to the background magnetic field. To simulate this setup, an auxiliary pion source parameter $\lambda$ needs to be introduced in the fermion matrix $M_f$ in order to trigger pion condensation~\cite{Kogut:2002zg}, which is to be extrapolated to zero at the end of the analysis.

Due to $\mu_u\neq\mu_d$, in this setup the magnetic field introduces a sign problem, which was avoided in~\cite{Endrodi:2014lja} by using a half-half-type magnetic field profile -- as discussed in Sec.~\ref{sec:eos_half} -- and performing a Taylor-expansion in its amplitude. This exploratory study simulated two flavors of unimproved staggered quarks on small lattices at low temperature. The results for the renormalized magnetic susceptibility $\chi^r$ are shown in the right panel of Fig.~\ref{fig:eos_c46}. In the limit of vanishing pion source parameters, the susceptibility exhibits a Silver-Blaze-type behavior up to the pion condensation onset and a strong diamagnetic behavior inside the condensed phase~\cite{Endrodi:2014lja}. This is expected due to the fact that charged pions contribute to the magnetization of the medium only via their orbital angular momentum, which is inherently diamagnetic, as we discussed already in Sec.~\ref{sec:eos_magsusc_QCD}.

Finally, in the context of magnetized and dense systems we mention the studies~\cite{Lenz:2023gsq,Lenz:2023wvk}, which investigated the 2+1-dimensional Gross-Neveu model at nonzero chemical potentials and magnetic fields. This model features a rich phase diagram and exhibits many QCD-like properties, in particular the magnetic catalysis of the quark condensate.

\subsection{Background electric fields}
\label{sec:eos_elsusc}

In the last section of this chapter, we turn to the response of the QCD medium to background electric fields. This can be considered on the same footing as the response to background magnetic fields.
However, there are two major complications that render the corresponding calculations considerably more challenging.

First, in the presence of nonzero electric fields, the fermion determinant is complex as we discussed in Sec.~\ref{sec:lat_signproblem}. This is very similar to the situation at a nonzero chemical potential $\mu$. More specifically, as the electric field couples to the quark charges $q_f$, it is instructive to compare its effect to that of a charge chemical potential $\mu_\Q$. The most established approaches to circumvent the complex action (sign) problem include simulations at imaginary electric fields and Taylor-expansions in the electric field variable. In fact, we already encountered these two methods for the lattice calculations of the electric polarizability $\alpha$ of hadrons at zero temperature -- the direct method of Sec.~\ref{sec:had_directmethod}, involving simulations at $iE\neq0$, and the weak-field expansion of Sec.~\ref{sec:had_weakfield}, which amounts to a Taylor-expansion in $E$. In the context of the electric susceptibility $\xi$, we are not interested in the impact of the electric field on hadronic two-point functions, but on the free energy $F$ of the medium itself, at nonzero $T$.

In addition, there is a second complication that arises only at nonzero temperature and renders the approach with imaginary electric fields cumbersome.
As we have seen in Sec.~\ref{sec:ev_electricfields}, the system at any $iE\neq0$ automatically corresponds to the canonical thermodynamic ensemble with zero total electric charge, instead of the usual grand canonical ensemble relevant for $E=0$. Already for free fermions, this leads to a singular behavior of spatially averaged observables, as was demonstrated in~\cite{Endrodi:2022wym,Endrodi:2023wwf}. Local observables, like the local electric charge density $\expv{j_4(x_3)}$ or the local chiral condensate $\expv{\bar\psi(x_3)\psi(x_3)}$ exhibit interesting inhomogeneous profiles, as was observed in~\cite{Yang:2022zob,Yang:2023zzx}.

\begin{figure}
 \centering
 \includegraphics[width=8cm]{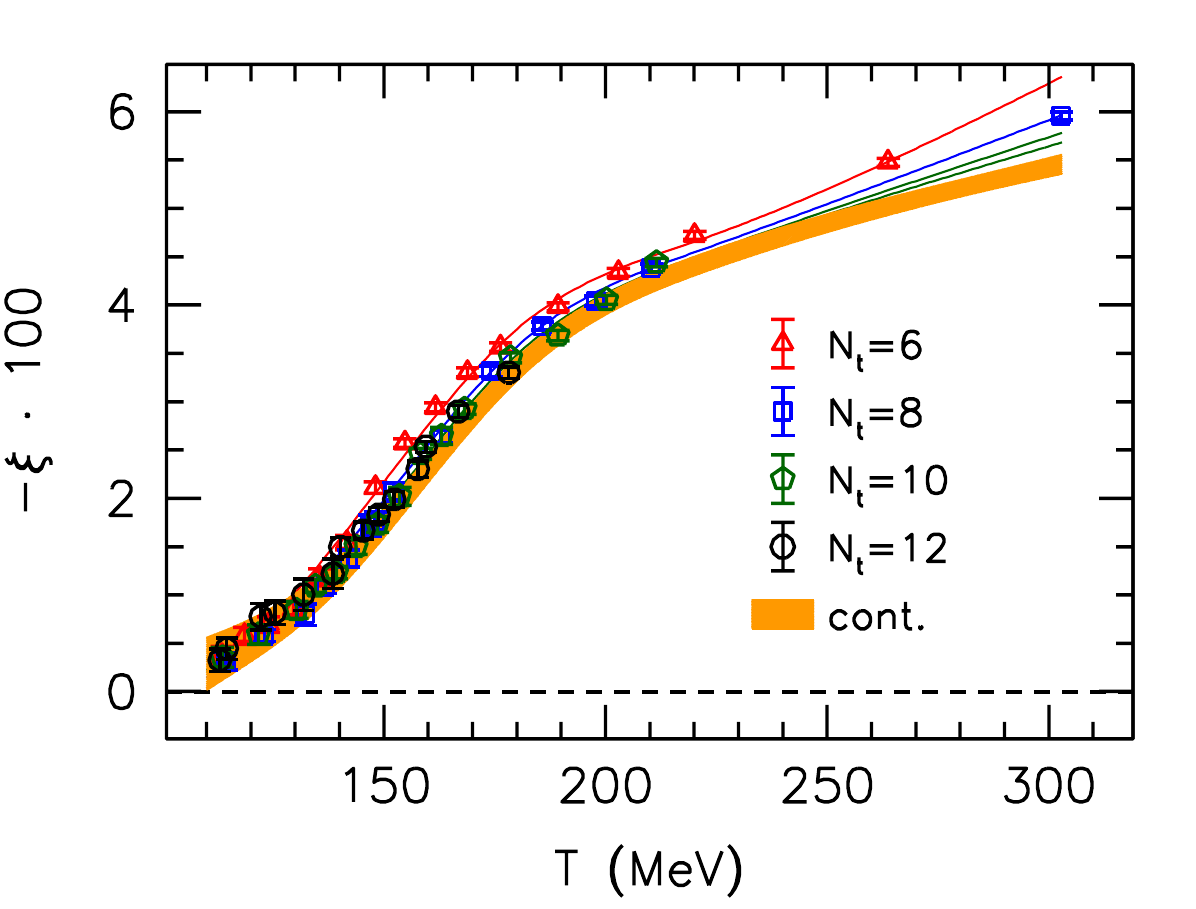}\qquad
 \raisebox{.4cm}{\includegraphics[width=7.8cm]{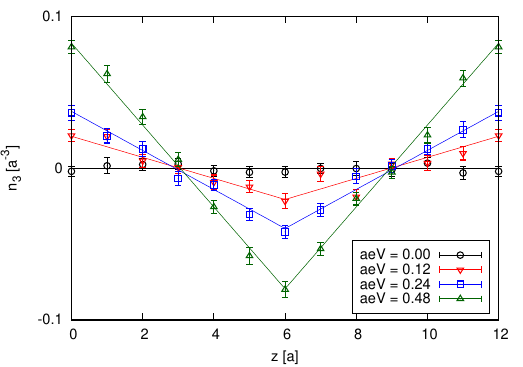}}
 \caption{ \label{fig:eos_el_susc_cont}
  Left panel: the negative of the renormalized electric susceptibility $\xi^r$ (denoted here as $\xi$) as a function of the temperature. The data points for different lattice spacings are extrapolated to the continuum limit (orange band). Figure from~\cite{Endrodi:2023wwf}. Right panel: the isospin charge (denoted here as $n_3(z)$) separated due to the real isospin electric field at high temperature~\cite{Yamamoto:2012bd}. The amplitude of the field is related to the potential $eV$ appearing in the legend as $eE=2V/ L_3$.
 }
\end{figure}

To avoid this second issue, one is therefore forced to consider the Taylor expansion of $F$ in the electric field. Due to the quantization condition~\eqref{eq:lat_elquant}, this requires using oscillatory profiles~\eqref{eq:lat_osc_Efield}.
The calculation of $\xi$ with oscillatory fields was discussed in Sec.~\ref{sec:eos_currentcurrent} and lead to the representation~\eqref{eq:eos_elsusc_currcurr} of $\xi$. Using this approach, the electric susceptibility was calculated using stout-improved staggered quarks in~\cite{Endrodi:2023wwf}. The results for the renormalized susceptibility are shown in the left panel of Fig.~\ref{fig:eos_el_susc_cont}, revealing a negative $\xi^r$ for all temperatures, a characteristic feature of plasmas. The high-temperature behavior was observed~\cite{Endrodi:2023wwf} to be consistent with leading-order perturbation theory~\cite{Endrodi:2022wym}, similarly to the case of the magnetic susceptibility shown in the right panel of Fig.~\ref{fig:eos_chi_T}.

Besides using imaginary fields or a Taylor-expansion, there is a third approach to circumvent the sign problem in QCD at nonzero real electric fields. As we discussed in Sec.~\ref{sec:lat_signproblem}, for the special (unphysical) choice of electric charges $q_u=-q_d$ and $q_s=0$, the path integral weight remains real and positive. This choice is analogous to considering an isospin chemical potential, $\mu_u=\mu_\I/2$, $\mu_d=-\mu_\I/2$ and $\mu_s=0$ in QCD and thus may be referred to as an isospin electric field.
To comply with the periodic boundary conditions, one cannot use homogeneous real fields, but instead the half-half setup of Sec.~\ref{sec:eos_half}. This approach was followed in~\cite{Yamamoto:2012bd} to calculate the real isospin charge profile $n_\I(x_3)$. The results, shown in the right panel of Fig.~\ref{fig:eos_el_susc_cont}, demonstrate that the isospin charge is effectively separated by the isospin electric field at high temperatures. In contrast, at low temperatures the analogous response was found to be suppressed for weak fields. This is consistent with the results found for the susceptibility $\xi^r$ above, whose magnitude was observed to increase monotonously with $T$.

Finally, we note that homogeneous imaginary electric fields also affect the local Polyakov loop $\expv{P(x_3)}$ in a non-trivial manner. This may be understood in terms of the effect that imaginary chemical potentials have on the Polyakov loop -- these tend to force the gluon field to non-trivial Polyakov loop sectors through so-called Roberge-Weiss transitions~\cite{Roberge:1986mm}. Given the similarity between the electric field and an inhomogeneous imaginary chemical potential (noted below~\eqref{eq:lat_osc_Efield}), imaginary electric fields in general prefer inhomogeneous Polyakov loops that interpolate between different center sectors.
Depending on the temperature, the electric field strength and the size $L_3$ of the system parallel to the electric field, different patterns were found for the variation of the Polyakov loop angle in the complex plane~\cite{Yang:2022zob,Yang:2023zzx}. In fact, this behavior is completely analogous to the impact of homogeneous magnetic fields in a system, where one spatial extent perpendicular to the magnetic field is small~\cite{DElia:2016kpz}.

\subsection{Lessons learned}

In this chapter we discussed the lattice results revolving around the QCD equation of state in the presence of background electromagnetic fields. First of all, we pointed out that the definition of the pressure becomes more involved in a homogeneous magnetic field -- in particular, the components $p_i$  may become anisotropic if they are defined at fixed magnetic flux, as done usually when one works with the stress-energy tensor.

We mainly focused on results concerning the weak magnetic field behavior of the free energy density, encoded by the magnetic susceptibility. Currently, there are six completely different approaches on the market that can be used to calculate $\chi$. Of these, four methods operate with measurements carried out at nonzero background fields (anisotropy method, finite difference method, generalized integral method and Amp\'{e}re's law), while two are formulated using zero-field expectation values (half-half method and current-current correlator method). The results of the different methods for the renormalized magnetic susceptibility compare well with each other, although currently there is a tension between two continuum extrapolations of $\chi^r$ at low temperatures.

The second type of methods can be generalized for the case of the electric response as well, and the renormalized electric susceptibility $\xi^r$ has been measured using the current-current correlator method.
While the magnetic susceptibility is observed to flip sign at the QCD transition (signaling diamagnetism in the confined and paramagnetism in the deconfined regimes, respectively), the electric susceptibility shows a plasma-type behavior for all temperatures ($\xi^r<0$).
Moreover, the magnetic susceptibility was also decomposed into spin- and orbital angular momentum-related contributions and the spin term was demonstrated to enter the description of the photon distribution amplitude, relevant for radiative decays.

Besides the weak-field response, the equation of state --  including the pressures, energy density, entropy density and further observables -- have been measured on the lattice for strong magnetic fields as well. These investigations were also extended to nonzero quark densities. In this context,  the identification of observables that exhibit a strong response to the magnetic field has been recently recognized as an important task. Such observables may be used as magnetometers in the benchmarking of the initial magnetic field in heavy-ion collision experiments -- a crucial ingredient for searches for the chiral magnetic effect, which will be in the focus of our interest in Chap.~\ref{chap:anom}.

	\clearpage
	\section{Transport, chirality and topology}
	\label{chap:anom} 

In the preceding parts of this review we were discussing aspects of QCD in background electromagnetic fields in equilibrium. Transport phenomena do not fall into this category, as they typically describe time-dependent and thus out-of-equilibrium processes. The transport of electric charge, for example, is characterized by the electric conductivity $\sigma$ of the QCD medium. In an effective, hydrodynamic description of QCD, this gives the electric current that flows due to an electric field in accordance with Ohm's law,
\be
\expv{\bm j} = \sigma \bm E\,.
\label{eq:anom_sigmae}
\ee
More precisely, in the quantum field theoretical context, the conductivity is defined via the zero-frequency behavior of the retarded correlator of electromagnetic currents.
This may be contrasted with the information that is encoded in the (time-independent) Euclidean correlator of electromagnetic currents. As we have seen in Sec.~\ref{sec:eos_currentcurrent}, the latter describe the equilibrium properties of the medium, i.e.\ the magnetic and electric susceptibilities.

In the presence of background magnetic fields and non-trivial chirality, novel transport phenomena arise. As these are intimately related to the axial anomaly that can convert topology and chirality into one another, these phenomena are referred to as anomalous transport phenomena~\cite{Kharzeev:2013ffa,Huang:2015oca,Landsteiner:2016led}. The prime examples for such effects include the chiral magnetic effect (CME) and the chiral separation effect (CSE), describing the generation of electric currents and axial currents, respectively, in the presence of a background magnetic field,
\be
\expv{\bm j} = \sigma_{\rm CME} \bm B, \qquad
\expv{\bm j_5} = \sigma_{\rm CSE} \bm B\,.
\label{eq:anom_sigmacsme}
\ee
For the CME, the magnetized system needs to feature non-trivial chirality~\cite{Kharzeev:2007jp,Fukushima:2008xe}, whereas the CSE arises in a magnetized and dense system~\cite{Son:2004tq,Metlitski:2005pr}.
The prime focus for the CME is in heavy-ion experiments, where its signatures have been sought for since long~\cite{STAR:2013ksd,STAR:2014uiw,STAR:2021mii,Kharzeev:2024zzm}, but it has also been observed in condensed matter systems~\cite{Li:2014bha} and argued to have substantial impact in astrophysics and cosmology~\cite{Kamada:2022nyt}. A combination of the CME and CSE effects may give rise to a collective excitation called the chiral magnetic wave~\cite{Kharzeev:2010gd}, which is also the subject of intense experimental searches~\cite{Huang:2015oca}.

\begin{figure}
 \centering
 \includegraphics[width=8cm]{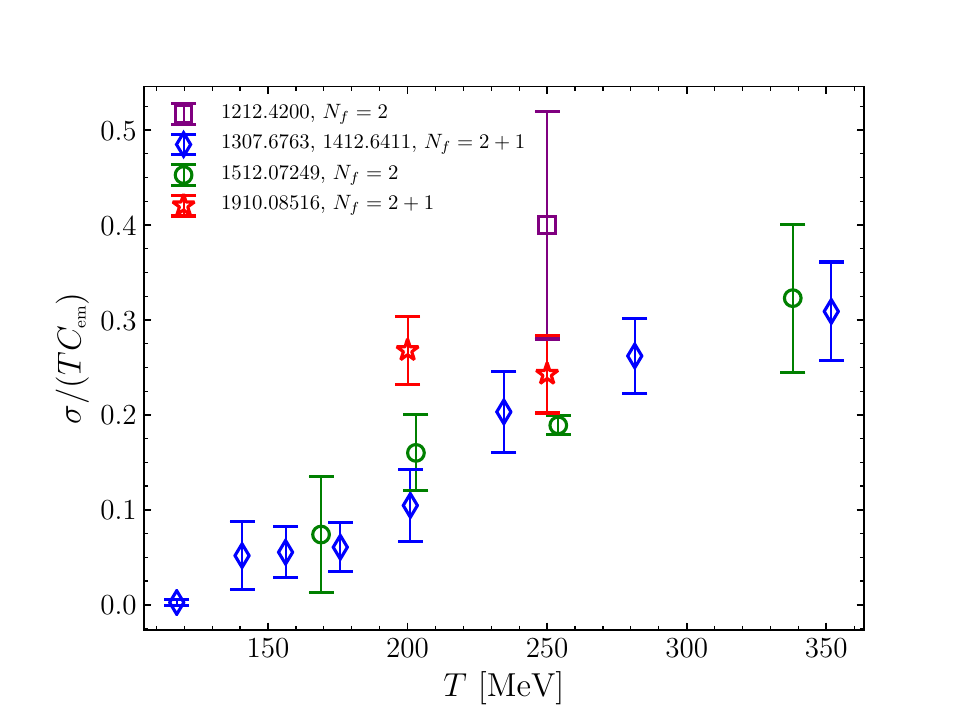}
 \caption{
 \label{fig:anom_conductivity_summary}
 Summary of dynamical QCD results for the electric conductivity in the absence of background fields~\cite{Aarts:2020dda}. The data points correspond to~\cite{Brandt:2012jc} (purple),~\cite{Amato:2013naa} (blue),~\cite{Brandt:2015aqk} (green) and~\cite{Astrakhantsev:2019zkr} (red). 
 The conductivity $\sigma$ is normalized by the temperature and the charge factor $C_{\rm em}=\sum_f (q_f/e)^2$. Figure taken from~\cite{Aarts:2020dda}.
 }
\end{figure}

Just like the electric conductivity in~\eqref{eq:anom_sigmae}, the determination of the out-of-equilibrium CME and CSE conductivities~\eqref{eq:anom_sigmacsme} requires one to go beyond the standard Euclidean path integral. The traditional method to access $\sigma$ in lattice simulations involves a spectral reconstruction that relates the Euclidean correlator to the retarded correlator, or, directly to the spectral function. The spectral reconstruction, being an inherently ill-posed procedure, has been considered using various numerical strategies. 
Reviews of these approaches and of the general formulation can be found in~\cite{Meyer:2011gj,Aarts:2020dda,Kaczmarek:2022ffn}.
Here, we merely show a summary of recent results for the temperature-dependence of the conductivity at vanishing background fields. In particular, the collection of~\cite{Aarts:2020dda} in Fig.~\ref{fig:anom_conductivity_summary} contains recent results obtained using dynamical QCD simulations~\cite{Brandt:2012jc,Amato:2013naa,Brandt:2015aqk,Astrakhantsev:2019zkr}.
We note that the electric conductivity has also been calculated in two-color QCD for a broad range of temperatures and densities~\cite{Buividovich:2020dks}.

In addition to the discussion of transport phenomena, in this chapter we will summarize the in-equilibrium lattice calculations that discuss CME-related observables.
The latter will involve fermionic observables, including the vector, axial vector, pseudoscalar and tensor currents~\eqref{eq:lat_VV5S5T}, as well as gluonic observables like the topological charge~\eqref{eq:lat_topcharge_def}. Some of the topology-related observables will also turn out to be relevant for the physics of the axion, a potential dark matter candidate particle.

\subsection{Lorentz-covariant expectation values}
\label{sec:anom_lorentz_cov}

Lorentz-covariance and parity symmetry enables one to describe the possible fermion bilinears that can develop an expectation value in the presence of weak external parameters -- a guiding principle that will turn out to be useful throughout this chapter. To leading order, a background electromagnetic field $F_{\nu\rho}$ induces a nonzero expectation value for the tensor bilinear of~\eqref{eq:lat_VV5S5T},
\be
\expv{T_{\nu\rho}^f} \propto q_fF_{\nu\rho}\,,
\ee
implying the polarization of spins in a magnetic field when both indices $\nu$, $\rho$ are spatial, and an electric dipole moment in electric fields when one of the indices is temporal. The former case was discussed already in Sec.~\ref{sec:eos_tensor}, where we studied the tensor polarization $\tau_f$ as well as its relation to the photon distribution amplitude and the spin contribution to the magnetic susceptibility.

Another important type of background is a topological gluon field, either in a local form, $q_{\rm top}(n)$ or globally, $Q_{\rm top}$, as defined in~\eqref{eq:lat_topcharge_def}. The pseudoscalar nature of this field enables the global or local expectation values of the correlators to leading order,
\be
\expv{Q_{\rm top} T_{\nu\rho}^f } \propto \epsilon_{\nu\rho\alpha\beta}\, q_fF_{\alpha\beta}, \qquad
\expv{q_{\rm top}(n)\,T_{\nu\rho}^f(n) } \propto \epsilon_{\nu\rho\alpha\beta}\, q_fF_{\alpha\beta}\,.
\label{eq:anom_LC_qtensor}
\ee
The same type of backgrounds also correlate with gradients of the vector currents locally,
\be
\expv{q_{\rm top}(n) \,\partial_\nu V_{\rho}^f(n) } \propto \epsilon_{\nu\rho\alpha\beta}\, q_fF_{\alpha\beta}\,.
\label{eq:anom_LC_qgradvector}
\ee
For a magnetic field in the $x_3$ direction, $F_{\alpha\beta}=F_{12}=B$, both of the local equations above are related to the electric dipole moment of quarks, but in a slightly different manner. While the second equation in~\eqref{eq:anom_LC_qtensor} involves the point-like dipole polarization $T_{34}^f(n)$,~\eqref{eq:anom_LC_qgradvector} describes the separation of electric charge, $\partial_3 V_4^f(n)$ over extended spatial distances.

Based on the above, we see that nonzero expectation values for $V_\nu^f$ or $A_\nu^f$ are not possible in equilibrium, if only the field strength tensor and topological fluctuations are present. 
Vector and axial vector currents become nonzero only when a corresponding vector-type source is switched on in the action. The fourth components of the latter are the chemical potential $\mu$ and the chiral chemical potential $\mu_5$, respectively.  To leading order, the expectation values are\footnote{More precisely, these chemical potentials are understood to couple to all fermion flavors, i.e.\ they are baryon chemical potentials $\mu_\B$ and $\mu_{\B5}$ in the sense of~\eqref{eq:eos_baryonmu}. However, we suppress the subscript $\B$ in this chapter for brevity.},
\be
\expv{V_4^f} \propto \mu, \qquad
\expv{V_{45}^f} \propto \mu_5\,.
\label{eq:anom_LC_vec_mu}
\ee

The combination of electromagnetic fields and chemical potentials leads to the anomalous transport phenomena in the focus of this chapter. Requiring parity symmetry, we obtain for the chiral magnetic effect and the chiral separation effect to leading order,
\be
\expv{V_{\nu}^f} \propto \epsilon_{\nu\rho\alpha4}\,q_fF_{\rho\alpha}\, \mu_5, \qquad
\expv{V_{\nu 5}^f} \propto \epsilon_{\nu\rho\alpha4}\,q_fF_{\rho\alpha}\, \mu\,.
\label{eq:anom_LC_vector}
\ee
While both of these equations are allowed by Lorentz-covariance and parity symmetry, we will see later in Sec.~\ref{sec:anom_CME} that $\expv{V_\nu^f}=0$, i.e.\ the chiral magnetic effect for homogeneous magnetic fields and chiral chemical potentials vanishes in equilibrium QCD. In turn, $\expv{V_{\nu5}^f}\neq0$ does arise at nonzero magnetic fields and chemical potentials, implying a nonzero chiral separation effect, see Sec.~\ref{sec:anom_CSE}.

Finally, let us note that the proportionality factors in all of the above equations are in general independent and may be determined by the lattice calculation of the corresponding one- or two-point functions in the presence of the required (weak) background fields and chemical potentials.

\subsection{Observables related to chirality and spin}
\label{sec:anom_chir_spin}

Historically, the first attempt to gain insight into the CME relation in~\eqref{eq:anom_sigmacsme} was an indirect one, where the fluctuations of the electric current $j_\nu$ were determined as functions of $B$~\cite{Buividovich:2009wi} using the quenched approximation of two-color QCD and overlap valence quarks. 
Using a magnetic field in the $x_3$ direction, this study observed an enhancement of $\expv{j_4^2}$ and of $\expv{j_3^2}$ due to the magnetic field, and  interpreted this as a signature of the CME. The transverse components $\expv{j_{1,2}^2}$ were also found to be enhanced, but less than the longitudinal ones. Moreover, the fluctuations of the chirality $S_5$ (defined above in~\eqref{eq:lat_VV5S5T}) were also found to increase as $B$ grows~\cite{Buividovich:2009wi}. The same behavior was later found also in the physical case of three colors (and again within the quenched approximation)~\cite{Braguta:2010ej}.

Similar tendencies for the current fluctuations were also shown to hold on an instanton configuration with nonzero $Q_{\rm top}$~\cite{Buividovich:2009wi}. Such topological gluonic configurations were also employed by the study~\cite{Abramczyk:2009gb} in order to investigate indirect CME signatures. In particular, 
this study calculated the electric charge density $j_4(x_3)$ on an instanton configuration in the direction of the magnetic field, probing the relation~\eqref{eq:anom_LC_qgradvector}. In accordance with the latter, an excess of positive (negative) charge was observed above (below) the center of the instanton. However, systematic effects on QCD ensembles could not be identified~\cite{Abramczyk:2009gb}.

The relation~\eqref{eq:anom_LC_qgradvector} captures the physical effect of the electric charge being separated in the presence of magnetic fields and CP-odd gluonic defects. The physical interpretation of~\eqref{eq:anom_LC_qtensor} is quite similar and involves the local electric dipole moments of quarks, as we mentioned above. The latter relation was first studied on the lattice in~\cite{Buividovich:2009my}. On an instanton configuration, the local form of~\eqref{eq:anom_LC_qtensor} was demonstrated to hold. Regarding dynamical gluons, this study calculated the fluctuations of the electric $T_{34}^f$ and magnetic dipole moments $T_{12}^f$ using the quenched approximation of two-color QCD. Both of these components were shown to fluctuate more as the magnetic field grows. Moreover, the correlator $\expv{S_{5}^f\,T_{34}^f}$ of the chirality and the electric dipole moment was shown to be proportional to $B$ for weak fields~\cite{Buividovich:2009my}.

\begin{figure}
 \centering
 \includegraphics[width=8cm]{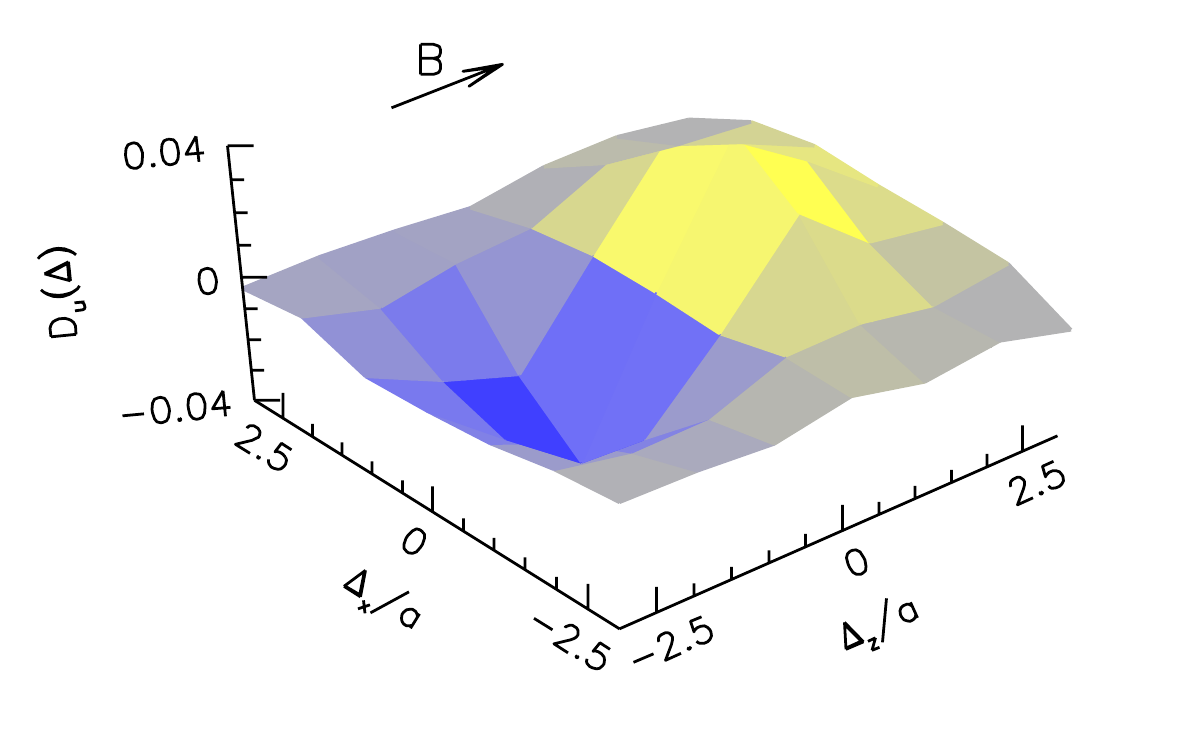}
 \caption{
 \label{fig:anom_2dcorrelator}
 Separation of the electric charge by the magnetic field (oriented along the $x_3$ direction) near a topological fluctuation in the $x_1-x_3$ plane~\cite{Bali:2014vja}. For the definition of $D_u(\Delta)$, see the text.
 }
\end{figure}

The local correlator of~\eqref{eq:anom_LC_qtensor} and the correlator~\eqref{eq:anom_LC_qgradvector} were first determined directly in~\cite{Bali:2014vja} using stout-smeared staggered quarks with physical masses. Specifically, the dimensionless combinations considered by this study were
\be
C_f=
\frac
{\displaystyle{}
\expv{  q_{\rm top}(n) \cdot T^f_{34}(n)}}
{\displaystyle{}
{\sqrt{\expv{ q_{\rm top}^2(n)}}}
\expv{T^f_{12}}},\qquad
D_f(\Delta) = \frac{\displaystyle{}\expv{q_{\rm top}(n)\cdot V^f_4(n+\Delta/a)}}{\displaystyle{}\sqrt{\expv{q_{\rm top}^2(n)}} \expv{T^f_{12}}},
\ee
where the expectation values are understood to involve a global average over the sites $n$ and $\Delta$ is a spatial vector measuring the distance between the topological charge and the electric charge density operator insertions. Given these definitions, $C_f$ was shown to take values around $0.1$, which was found to be an order of magnitude smaller than a model prediction based on nearly massless quarks in domains of constant
topological gluon backgrounds.
Concerning the observable $D_f(\Delta)$, a clear signal for a spatially extended electric dipole structure was observed~\cite{Bali:2014vja}, visualized in Fig.~\ref{fig:anom_2dcorrelator}.

The above discussed effects related to charge separation depend both on the associated coefficients as well as on the abundance of topological defects in equilibrium QCD configurations. The background magnetic field may have an indirect effect on the latter, both for global fluctuations as well for local ones. 
The global effect is captured by the topological susceptibility and its dependence on $B$, to which we get back to below in Sec.~\ref{sec:anom_axion}.
Regarding the local effect, the topological charge density correlator $\expv{q_{\rm top}(0)q_{\rm top}(n)}$ was determined at zero temperature using stout-smeared staggered quarks in~\cite{Bali:2013esa}. The results implied that magnetic field-related effects in this correlator are very mild, and in particular, no anisotropy could be observed~\cite{Bali:2013esa}.

\subsection{Chiral magnetic effect in equilibrium}
\label{sec:anom_CME}

The first direct investigations of the CME relation of~\eqref{eq:anom_LC_vector} with nonzero magnetic fields and a chiral chemical potential $\mu_5$ were performed using dynamical Wilson quarks~\cite{Yamamoto:2011gk} and with quenched Wilson quarks~\cite{Yamamoto:2011ks}. 

Before we discuss the results, it is important to consider the specific definition of the electromagnetic current on the lattice. Using the Wilson fermion discretization~\eqref{eq:lat_Wilson_Dslash} of the Dirac operator, it is possible to construct a conserved vector current $V_\nu^f$ that satisfies the lattice vector Ward identity $\nabla_\nu V^f_\nu=0$, where $\nabla_\nu$ is a discretized derivative. This current is a point-split operator~\cite{Karsten:1980wd},
\be
V^f_\nu(n) = \frac{1}{2a} \left[ \bar\psi_f(n)(\gamma_\nu-\mathds{1}) \,\U_\nu(n) u_{\nu f}(n) \psi_f(n+\hat\nu) +\bar\psi_f(n)(\gamma_\nu+\mathds{1})
 \,\U^\dagger_\nu(n-\hat\nu)\,u_{\nu f}^*(n-\hat\nu)\, \psi_f(n-\hat\nu) \right]\,.
\ee
In turn, it is also usual to consider the local vector current $V_\nu^{f,\rm loc}(n)$,
\be
V_\nu^{f,\rm loc}(n)=\bar\psi_f(n)\gamma_\nu \psi_f(n)\,.
\ee
The latter has the same quantum numbers as $V_\nu^f$ and is often employed in hadron spectroscopy calculations. However, it does not satisfy a lattice Ward identity i.e.\ it is not conserved, $\nabla_\nu V_\nu^{f,\rm loc}\neq0$. For the study of the CME, this turns out to be a critical shortcoming.

\begin{figure}
 \centering
 \includegraphics[width=8cm]{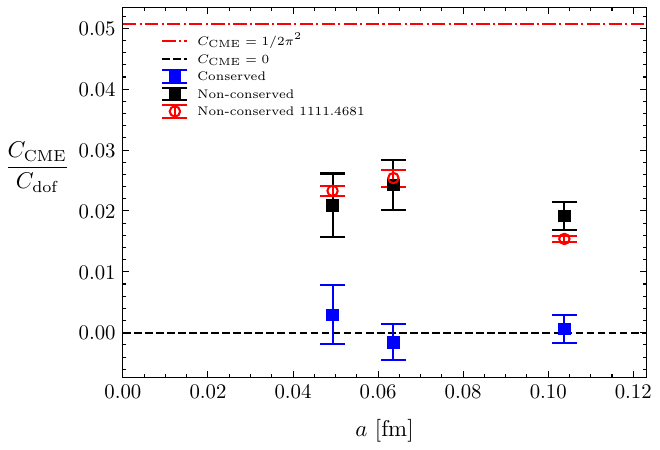}\qquad
 \includegraphics[width=8cm]{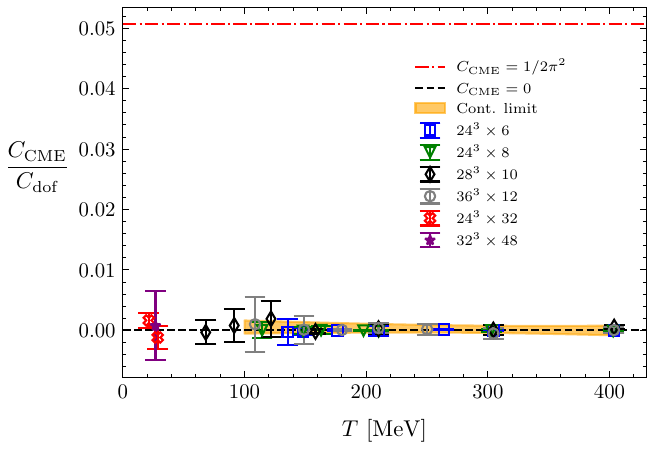}
 \caption{
 \label{fig:anom_CMEdyn}
 Left panel: lattice spacing-dependence of the CME coefficient defined using the conserved and the non-conserved vector currents in the quenched Wilson formulation~\cite{Brandt:2024wlw}. The results of~\cite{Yamamoto:2011ks} for the non-conserved vector current are also included.
 Right panel: demonstration of the vanishing of the CME coefficient defined using dynamical staggered quarks and a conserved vector current for a broad range of temperatures~\cite{Brandt:2024wlw}. In both panels, $C_{\rm dof}$ is the proportionality constant mentioned in footnote~\ref{fn:anom_1}.
 }
\end{figure}

Analogously to the conserved electromagnetic current $j_\nu$ in~\eqref{eq:lat_jnucurrentdef}, one can define a local electromagnetic current $j_\nu^{\rm loc}$ from $V_\nu^{f,\rm loc}$.
The studies~\cite{Yamamoto:2011gk,Yamamoto:2011ks} employed the non-conserved, local electric current for the calculation of the CME coefficient. The expectation value of the current was determined for weak magnetic fields and small chiral chemical potentials, 
\be
\expv{j_3^{\rm loc}} = C_{\rm CME}^{\rm loc} \cdot eB \cdot \mu_5\,.
\ee
The corresponding proportionality factor\footnote{\label{fn:anom_1}There is an overall constant of proportionality in the CME and CSE coefficients including the number $N_c$ of colors as well as electric charge factors. Throughout this chapter, we suppress such factors for brevity, so that the coefficient values can be directly compared to the free-case massless prediction of $1/(2\pi^2)\approx0.05$. In fact, the proportionality constant depends on what kind of axial current is considered -- that coupled to baryon number or that to electric charge. This aspect has been discussed in detail in~\cite{Brandt:2023wgf,Brandt:2024wlw}.}
was calculated using different lattice spacings and a continuum estimate of $C_{\rm CME}^{\rm loc}\approx 0.02-0.03$ was given~\cite{Yamamoto:2011ks}. The quenched results are included in the left panel of Fig.~\ref{fig:anom_CMEdyn}.

For a long period of time, this was the only lattice calculation of the CME coefficient in interacting QCD. Recently, the quenched Wilson lattice setup of~\cite{Yamamoto:2011ks} was revisited in~\cite{Brandt:2024wlw}. Instead of calculating the current expectation value at nonzero $B$ and $\mu_5$, the latter study evaluated the derivative of the current with respect to $\mu_5$ at $\mu_5=0$,
\be
C_{\rm CME} = \lim_{B\to0} \frac{1}{eB} \,\left.\frac{\partial \expv{j_3}}{ \partial \mu_5}\right|_{\mu_5=0}\,,
\label{eq:anom_CMEstrategy1}
\ee
at small magnetic fields. 
The analysis was carried out both for the conserved $j_\nu$ and the non-conserved current $j_\nu^{\rm loc}$. The approach of the two so defined coefficients $C_{\rm CME}$ and $C_{\rm CME}^{\rm loc}$ towards the continuum limit is shown in the left panel of Fig.~\ref{fig:anom_CMEdyn}. The results demonstrate that $C_{\rm CME}=0$ for the correct setup with the conserved current. In contrast, $C_{\rm CME}^{\rm loc}$ is consistent with~\cite{Yamamoto:2011ks} and gives an unphysical, non-vanishing result.

The study~\cite{Brandt:2024wlw} also discussed the impact of regularization on the CME coefficient for non-interacting fermions and showed that in regularization schemes compatible with vector current conservation, the (spatially averaged) CME current vanishes. In fact, this finding is consistent with Bloch's theorem, which -- when generalized to the quantum field theoretical context -- prohibits global conserved currents to flow in equilibrium~\cite{Yamamoto:2015fxa}. We note that the importance of the ultraviolet regularization for the CME was also discussed in~\cite{Hou:2011ze,Buividovich:2013hza,Zubkov:2016tcp,Horvath:2019dvl}.\footnote{
For a different viewpoint on the regularization issue, it is instructive to compare the in-equilibrium CME coefficient to the axial anomaly. Both quantities can be shown to originate from the triangle diagram (a quark loop with two vector and an axial vector leg) and involve an improper integral that depends on the regularization despite naively being ultraviolet finite. The lack of a gauge invariant regularization can incorrectly render the CME coefficient nonzero in the same way as it can lead to an incorrect result for the axial anomaly. For details, see~\cite{Brandt:2024wlw} and compare to the textbook~\cite{Itzykson:1980rh}.}

A vanishing result for the equilibrium CME conductivity was also found in the free case using overlap and Wilson fermions in~\cite{Buividovich:2013hza}. Contrary to the strategy of~\cite{Brandt:2024wlw}, where the conductivity was obtained from the $\mu_5$-derivative~\eqref{eq:anom_CMEstrategy1} at nonzero $B$, here the opposite route was taken, i.e.\ the $B$-derivative of $\expv{j_3}$ was evaluated at $\mu_5>0$.
Similarly to the treatment of the magnetic susceptibility in Sec.~\ref{sec:eos_currentcurrent}, the $B$-derivative leads to a current-current correlator of the form~\eqref{eq:eos_vacpol1}, namely the zero-momentum limit of the off-diagonal component $\Pi_{23}(p_1)$~\cite{Buividovich:2013hza},
\be
C_{\rm CME} = \lim_{\mu_5\to0} \frac{i}{\mu_5} \left.\frac{\partial\, \Pi_{23}(p_1)}{\partial p_1}\right|_{p_1=0}\,.
\label{eq:anom_CMEstrategy2}
\ee
This study also demonstrated that a nonzero result for $C_{\rm CME}^{\rm loc}\neq0$ may arise via the use of non-conserved vector currents. Moreover, relations between the asymptotic momentum limit of the correlator and Ward identities were derived~\cite{Buividovich:2013hza}.

Finally, the first lattice simulations with dynamical quarks were performed recently in~\cite{Brandt:2024wlw}, using the relation~\eqref{eq:anom_CMEstrategy1}, including connected and disconnected diagrams. Using stout-improved staggered quarks with physical masses, this study found $C_{\rm CME}=0$ for all considered temperatures. This result is shown in the right panel of Fig.~\ref{fig:anom_CMEdyn}. Just like for Wilson quarks, the specific choice of the conserved vector current was also found to be crucial for staggered fermions~\cite{Brandt:2024wlw}. In particular, the introduction of a nonzero $\mu_5$ in the staggered Dirac operator requires special care.

Before we conclude with the in-equilibrium CME, a discussion on the physical meaning of the chiral chemical potential $\mu_5$ is in order. This parameter is not a usual chemical potential, as the axial current it couples to is not conserved. Thus, $\mu_5$ should rather be viewed as an external source similar to the quark mass. 
In the context of the CME, it is important to check that $\mu_5>0$ indeed induces a nonzero chiral density $\expv{V_{45}^f}$, as indicated in the second relation of~\eqref{eq:anom_LC_vec_mu}. The strength of this response is characterized by the so-called axial susceptibility,
\be
\chi_5=-\left.\frac{\partial f}{\partial \mu_5^2}\right|_{\mu_5=0} = \sum_f\left.\frac{\partial \expv{V_{45}^f}}{\partial \mu_5}\right|_{\mu_5=0}\,.
\ee
This observable is subject to additive renormalization, similarly to the quark condensate.
Recently, the renormalized $\chi_5$ has been determined on the lattice using stout-improved staggered quarks with physical quark masses~\cite{Brandt:2024wlw}. The results indicated that $\chi_5$ is suppressed at low temperatures, exhibits a slow rise around the crossover temperature and gradually approaches its massless non-interacting value of $N_c/(2\pi^2)$ in the high temperature limit. Thus, for $T\gtrsim T_c$, the chiral chemical potential can indeed be used to parameterize the chiral imbalance in the equilibrium QCD medium. We note that a nonzero $\expv{V_{45}^f}$ was also found at $\mu_5>0$ in~\cite{Astrakhantsev:2019wnp}.

In summary, we can conclude that for homogeneous background magnetic fields and chiral chemical potentials, a chiral imbalance can be created but nevertheless the chiral magnetic effect is absent in equilibrium QCD. Previous lattice results in the literature that predicted nonzero CME coefficients were plagued by the effects of using a non-conserved electric current.
We stress that the above findings strictly correspond to physical QCD, i.e.\ a gauge theory with massive fermions and a vector gauge field (and no axial gauge field).
We also note that in the context of holographic theories, where vector and axial vector gauge fields and massless fermions are often included, see e.g.~\cite{Gynther:2010ed,Landsteiner:2013aba,Ammon:2020rvg,Matthias_privcomm}, the conclusions about CME are more delicate and depend on whether one uses so-called consistent or covariant currents~\cite{Rebhan:2009vc,Landsteiner:2012kd,Landsteiner:2013aba}. In QCD, such ambiguities are absent since the vector and axial vector currents are fixed by Ward identities, in particular the conservation of the electric current.
Finally, we highlight that the generalized Bloch's theorem~\cite{Yamamoto:2015fxa} only prohibits global currents, and an inhomogeneous {\it local} CME current, which has vanishing global average, can still flow in equilibrium QCD. Such local currents have been observed in the presence of inhomogeneous magnetic fields of the type~\eqref{eq:lat_loc_Bfield} in full QCD very recently~\cite{Brandt:2024fpc}.

\subsection{Chiral separation effect}
\label{sec:anom_CSE}

The CSE can be discussed quite analogously to the CME. Here, the axial current~\eqref{eq:lat_jnu5currentdef} needs to be calculated at nonzero chemical potentials and magnetic fields. 
However, in contrast to the CME, the chiral separation effect turns out to be less sensitive to regularization effects. Moreover, in this case there is a non-trivial effect already in equilibrium.

The CSE conductivity was first discussed on the lattice in the free case using massless overlap fermions in~\cite{Buividovich:2013hza}.
Here, the magnetic field-derivative of the axial current was evaluated using the off-diagonal components of the vector-axial vector correlator, similarly to the strategy of~\cite{Buividovich:2013hza} discussed above in~\eqref{eq:anom_CMEstrategy2} for the CME,
\be
C_{\rm CSE} = \lim_{\mu\to0} \frac{i}{\mu} \left.\frac{\partial\, \Pi^{(5)}_{23}(p_1)}{\partial p_1}\right|_{p_1=0}\,, \qquad
\Pi^{(5)}_{\nu\rho}(p)=\int \dd^4x\, e^{ipx}\,\expv{j_\nu(x)j_{\rho5}(0)}\,.
\label{eq:anom_CSEstrategy2}
\ee
This study confirmed the result $C_{\rm CSE}=1/(2\pi^2)$ of analytical continuum methods for massless quarks~\cite{Son:2004tq,Metlitski:2005pr}.
The impact of color interactions were investigated later in~\cite{Puhr:2016kzp} using massless overlap quarks in the quenched approximation. Here, simulations at $B>0$ and $\mu>0$ were performed both in the confined and in the deconfined phase of quenched QCD, with a decomposition of the gauge ensembles into topological sectors. The results showed an agreement with the massless free-case value of $C_{\rm CSE}$ for all temperatures and topological sectors~\cite{Puhr:2016kzp}. This finding was argued to signal the absence of non-perturbative corrections to the CSE conductivity. Nevertheless, it was pointed out that the quenched approximation might lead to uncontrolled systematic uncertainties for $C_{\rm CSE}$~\cite{Puhr:2016kzp}. 

Abandoning the quenched approximation, the first dynamical lattice simulations were carried out in two-color QCD using staggered quarks in the sea and Wilson and domain wall fermions in the valence sector~\cite{Buividovich:2020gnl}. Following the strategy described above in~\eqref{eq:anom_CSEstrategy2}, the vector-axial vector correlator was measured at nonzero chemical potentials, taking into account both connected and disconnected diagrams. The results indicated a value for $C_{\rm CSE}$ close to the massless free-case value at high temperatures and a gradual suppression towards the low-temperature confined phase.

\begin{figure}
 \centering
 \includegraphics[width=8cm]{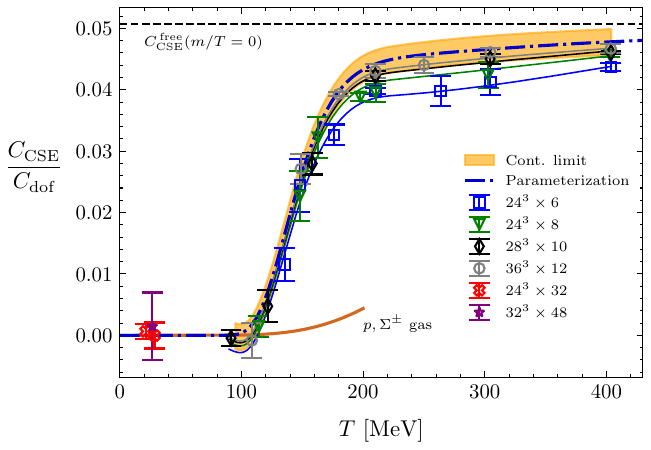}\qquad
 \caption{
 \label{fig:anom_CSEdyn}
 The temperature-dependence of the CSE coefficient using dynamical staggered quarks~\cite{Brandt:2023wgf}. The yellow band indicates the continuum extrapolation based on four lattice spacings, the solid brown line the prediction of a baryon gas model, while the dashed line the massless perturbative value.
 The normalization $C_{\rm dof}$ is the proportionality constant mentioned in footnote~\ref{fn:anom_1}.
 }
\end{figure}

Finally, the CSE coefficient has been determined for a range of temperatures in~\cite{Brandt:2023wgf} using dynamical stout-improved staggered quarks with physical masses. Here, the equivalent of the strategy for~\eqref{eq:anom_CMEstrategy2} was used, i.e.\ the $\mu$-derivative was carried out explicitly,
\be
C_{\rm CSE} = \lim_{B\to0} \frac{1}{eB} \,\left.\frac{\partial \expv{j_{35}}}{ \partial \mu}\right|_{\mu=0}\,.
\label{eq:anom_CSEstrategy1}
\ee
The continuum limit for $C_{\rm CSE}(T)$ was taken using four different lattice spacings. The results indicated a suppression of the coefficient at low temperatures, a distinct rise in the crossover region and an approach towards the massless non-interacting value for high $T$, see Fig.~\ref{fig:anom_CSEdyn}.
We note that the deviation of $C_{\rm CSE}$ from the massless free-case value $1/(2\pi^2)$ is an interesting observable on its own, as it is related to the $\pi^0\to\gamma\gamma$ decay amplitude, see~\cite{Newman:2005as}.

\subsection{Out-of-equilibrium effects}

In~\eqref{eq:anom_sigmae} we introduced the electric conductivity $\sigma$ and summarized the existing lattice determinations of it in the absence of background fields in Fig.~\ref{fig:anom_conductivity_summary}. The capability of the QCD medium to conduct electric currents can be modified by background magnetic fields. In particular, due to the breaking of isotropy, the medium may resist currents differently depending on the angle $\vartheta$ between the current and the magnetic field. The conductivity $\sigma_\perp=\sigma(\vartheta=\pi/2)$ describing currents perpendicular to the magnetic field is in general expected to be reduced by the magnetic field. This, so-called magnetoresistance behavior~\cite{pippard1989magnetoresistance} may be understood in terms of the Landau-levels that charge carriers can occupy. Indeed, on Landau-levels, the magnetic field tends to force charges on periodic orbits and constrain currents in the plane perpendicular to $\bm B$. We have already made a few observations in this review that support this picture: that Landau-levels may be defined in full QCD (see Sec.~\ref{sec:lat_landaulevelQCD}) and that the magnetic field tends to shrink typical confining structures in the perpendicular directions (see Sec.~\ref{sec:had_stringtension}).

The first lattice determination of the electric conductivity of the QCD medium in the presence of background magnetic fields was performed in~\cite{Buividovich:2010tn} using the quenched approximation of two-color QCD and overlap valence quarks. The correlator of the spatial components of the electric current (no sum over $i$),
\be
C_{ii}(x_4) = \int \dd^3 \bm x\,\expv{j_i(0) j_i(x)}\,,
\label{eq:anom_Euclcorr}
\ee
was determined, neglecting the disconnected contributions to it. We note that these correlators also enter the calculation of the magnetic field-dependence of the energy $\E_{\rho^0}$ of neutral mesons with different spin orientations, see Sec.~\ref{sec:had_massesB}. With the magnetic field pointing in the $x_3$ direction, the parallel  correlator $C_{33}(x_4)$ was observed to decay slower with $x_4$, while the perpendicular components $C_{11}(x_4)=C_{22}(x_4)$ exhibited a faster exponential decay~\cite{Buividovich:2010tn}. 

\begin{figure}
 \centering
 \includegraphics[width=8.3cm]{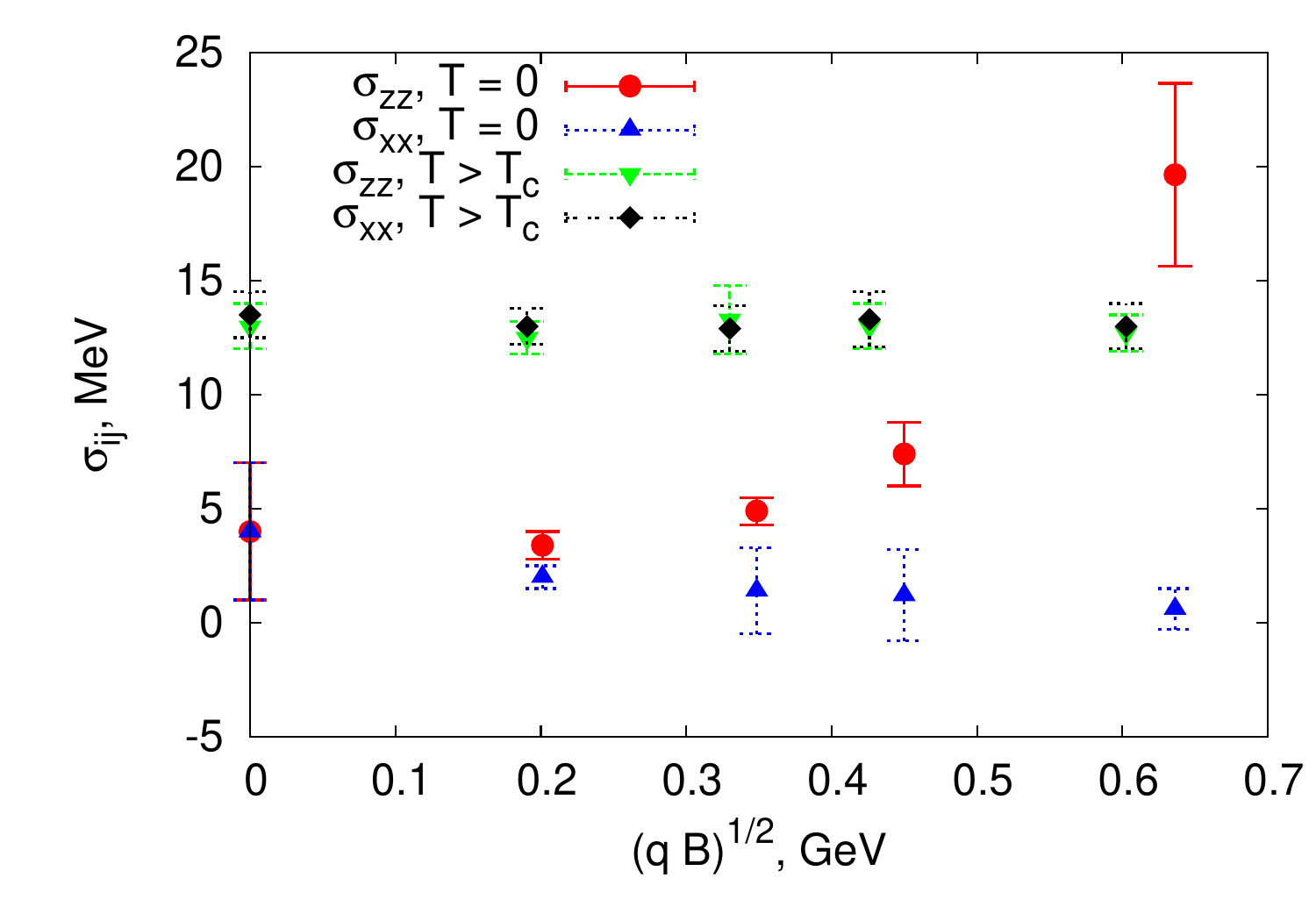}\qquad
 \raisebox{.2cm}{\includegraphics[width=8cm]{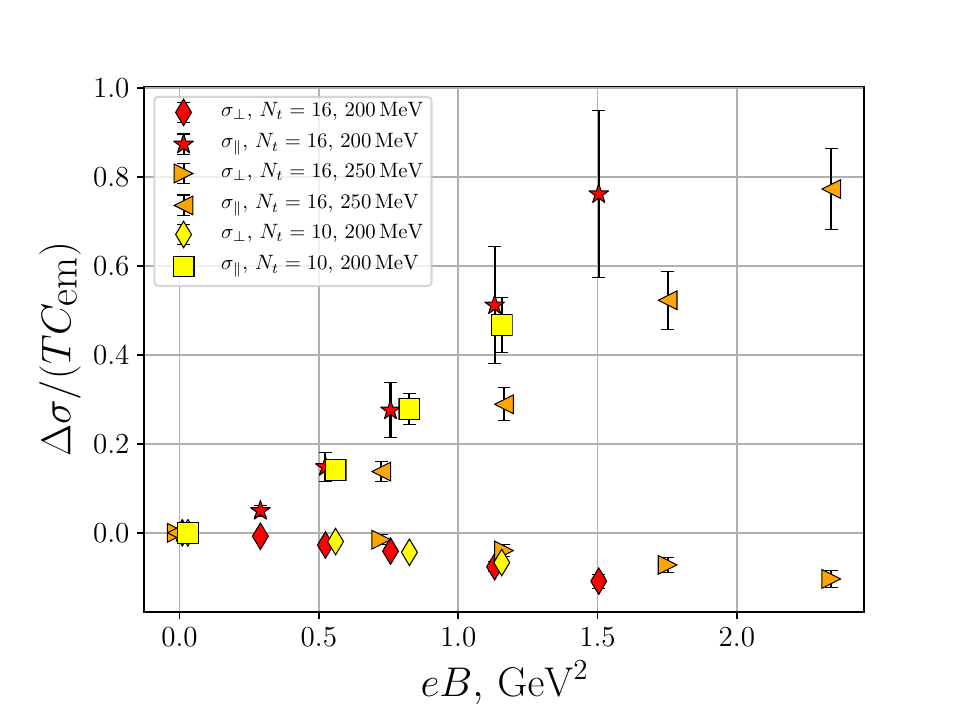}}
 \caption{
 \label{fig:anom_conductivity}
 Left panel: electric conductivities parallel and perpendicular to the background magnetic field (denoted here as $\sigma_{zz}$ and $\sigma_{xx}=\sigma_{yy}$) as functions of the square root of $q_dB$ for two different temperatures~\cite{Buividovich:2010tn}.
 Right panel: the excess conductivity $\Delta\sigma=\sigma(B)-\sigma(B=0)$ parallel and perpendicular to the magnetic field as a function of $eB$ at two high temperatures~\cite{Astrakhantsev:2019zkr}. Just as in Fig.~\ref{fig:anom_conductivity_summary}, the conductivities are normalized by the temperature and the charge factor $C_{\rm em}=\sum_f (q_f/e)^2$.
 }
\end{figure}

To calculate the electric conductivity, one needs the representation of the Euclidean correlator~\eqref{eq:anom_Euclcorr} in terms of the spectral function $\rho_{ii}(\omega)$,
\be
C_{ii}(x_4) = \int_0^\infty \frac{\dd \omega}{2\pi} \,K(\omega,x_4)\, \rho_{ii}(\omega), \qquad K(\omega,x_4) = \frac{\cosh\left[\omega(x_4-1/(2T))\right]}{\sinh\left[\omega/(2T)\right]}\,.
\label{eq:anom_spectreconstr}
\ee
and its zero-frequency limit~\cite{Buividovich:2010tn},
\be
\sigma_\perp = \lim_{\omega\to0}\frac{\rho_{11}(\omega)}{2\omega}
= \lim_{\omega\to0}\frac{\rho_{22}(\omega)}{2\omega}, \qquad
\sigma_\parallel = \lim_{\omega\to0}\frac{\rho_{33}(\omega)}{2\omega}\,.
\ee
The inversion of the first relation of~\eqref{eq:anom_spectreconstr} is, in general, an ill-posed numerical problem, as we mentioned already in the beginning of this chapter. The study~\cite{Buividovich:2010tn} employed the maximal entropy method to tackle it, and obtained estimates for the conductivities in the different directions. The results are shown in the left panel of Fig.~\ref{fig:anom_conductivity}. At low temperature, the perpendicular conductivity $\sigma_\perp$ was observed to remain zero, just as at $B=0$, while the parallel component $\sigma_\parallel$ was found to grow with the magnetic field. In turn, at high temperature, the results indicated that the conductivity remains isotropic and approximately independent of $B$~\cite{Buividovich:2010tn}.

The magnetic field-dependence of the electric conductivities was investigated in more detail for high temperatures in~\cite{Astrakhantsev:2019zkr} using stout-improved staggered quarks with physical masses. Here, the correlator was calculated neglecting disconnected diagrams and the spectral reconstruction was carried out for the correlator difference $C_{ii}(x_4;B)-C_{ii}(x_4;B=0)$ using the Tikhonov regularization method. The parallel conductivity was found to increase as the magnetic field grows, in disagreement with the high-temperature results of~\cite{Buividovich:2010tn}. The increasing behavior was interpreted as a possible manifestation of the CME~\cite{Astrakhantsev:2019zkr}. In turn, the perpendicular conductivities were found to be reduced by $B$ -- in line with the magnetoresistance phenomenon mentioned above. The results are shown in the right panel of Fig.~\ref{fig:anom_conductivity}. We note that the above observed tendencies, in particular the enhancement of $\sigma_\parallel(B)$, were found very recently to persist to magnetic fields as strong as $eB\approx 9\GeVsq$~\cite{Almirante:2024lqn}.

The spectral reconstruction for the conductivity pertaining to the out-of-equilibrium CME current (for a time-dependent chiral imbalance) involves the vector-axial vector correlator at nonzero magnetic field,
\be
C^{(5)}_{34}(x_4)=\expv{j_3(0)j_{45}(x_4)}\,.
\label{eq:anom_corr_c5ii}
\ee
This has only been discussed on the lattice so far in~\cite{Buividovich:2024bmu}. This study considered an electroquenched setup with staggered sea quarks and Wilson valence fermions and only connected contributions to~\eqref{eq:anom_corr_c5ii}. In place of performing a spectral reconstruction, the CME conductivity was argued to be related to the midpoint value of the correlator, $C^{(5)}_{34}(1/(2T))$. This was found to lie close to the results obtained with free Wilson fermions~\cite{Buividovich:2024bmu}.

Finally, we mention that real-time classical statistical lattice simulations of two-color QCD were also carried out to address anomalous transport phenomena in~\cite{Muller:2016jod,Mace:2016shq}. These studies considered the behavior of vector and axial vector currents during a sphaleron transition in a background magnetic field and observed characteristics of the chiral magnetic wave.

\subsection{Topology in background electromagnetic fields}
\label{sec:anom_axion}

Let us finally return to the realm of in-equilibrium QCD and focus on observables related to topology and their behavior in background electromagnetic fields. These turn out to be related to the properties of the axion -- a potential dark matter candidate that couples to the QCD topological charge in beyond-Standard-Model theories~\cite{Kim:1979if,Shifman:1979if,Dine:1981rt}. 

The axion field $a$ multiplies $q_{\rm top}/f_a$ in the Lagrangian, where $f_a$ is a scale parameter. Assuming homogeneous axion fields, an effective potential for the axion can be derived. The mass $m_a$ of the axion is proportional to the topological susceptibility,
\be
m_a^2 f_a^2 = \chi_{\rm top} = \frac{\expv{Q_{\rm top}^2}}{V}\,.
\ee
In the presence of background electromagnetic fields, the axion field also couples to the CP-odd combination $\bm E\cdot \bm B$. In particular, the Lagrangian contains the term $g_{a\gamma\gamma}^0 a \bm E\cdot \bm B$, where $g_{a\gamma\gamma}^0$ is the direct axion-photon coupling, which depends on the specific axion model. 
Since the axion can now also couple to photons through QCD loops, an indirect, model-independent coupling also arises in the axion effective potential,
\be
g_{a\gamma\gamma}^{\rm QCD} f_a = -i\,\frac{T}{V} \left.\frac{\partial \expv{Q_{\rm top}}}{\partial (\bm E\cdot \bm B)}\right|_{E=B=0}.
\ee
For a review on axions and a calculation of the axion mass and the axion-photon coupling within chiral perturbation theory, see~\cite{GrillidiCortona:2015jxo}.

The temperature-dependence of the topological susceptibility has been determined on the lattice by several groups with high precision, see e.g.~\cite{Bonati:2015vqz,Petreczky:2016vrs,Borsanyi:2016ksw,Jahn:2018dke}. However, there are only a few lattice results on the impact of background magnetic fields on $\chi_{\rm top}$~\cite{Brandt:2022jfk,Brandt:2023awt,Brandt:2024gso}. These studies suggest an enhancement of the susceptibility at low temperatures (in accordance with chiral perturbation theory predictions~\cite{Adhikari:2021lbl}) and a reduction in the transition region, following the general trend of magnetic catalysis and inverse magnetic catalysis observed for the quark condensate, see Sec.~\ref{sec:pd_condT}.

The impact of parallel magnetic and (imaginary) electric fields on the average topological charge was first discussed on the lattice in~\cite{DElia:2012ifm} using unimproved staggered quarks with larger-than-physical masses. This study calculated $\expv{Q_{\rm top}}$ in the presence of background fields with various orientations and demonstrated that it is only sensitive to the CP-odd combination $i\bm E\cdot \bm B$. The so calculated topology shift is related to the effective pseudoscalar QED-QCD interactions, potentially relevant for electric charge separation in heavy-ion collisions~\cite{Asakawa:2010bu}.
Recent preliminary results for the QCD contribution $g_{a\gamma\gamma}^{\rm QCD} f_a$ to the axion-photon coupling were also obtained using stout-improved staggered quarks with physical masses~\cite{Brandt:2022jfk,Brandt:2023awt}. A preliminary continuum estimate appears to lie close to the chiral perturbation theory prediction.

\subsection{Lessons learned}

In this chapter we focused on observables related to chirality and topology in equilibrium magnetized QCD, as well as out-of-equilibrium effects related to standard and anomalous transport phenomena.

The separation of electric charge in the presence of magnetic fields and topological defects has been demonstrated on the lattice by studying various different observables: local and extended electric dipole moments as well as their correlators with the QCD topological charge. Conceptually, these observables are related to the chiral magnetic effect, but they do not concern the electric currents themselves directly.
Gluonic observables describing the topological nature of the QCD vacuum were also found to respond nontrivially to background electromagnetic fields in equilibrium. These features also turn out to be relevant for the physics of the axion, a possible dark matter candidate.

The equilibrium formulation of the chiral separation effect and the chiral magnetic effect can be checked directly with Euclidean lattice simulations involving background magnetic fields and ordinary or chiral chemical potentials. For the CME, the details of the regularization turn out to be absolutely essential. In particular, for regularizations satisfying the conservation of the electric current, the CME coefficient in equilibrium QCD vanishes for homogeneous magnetic fields. Some of the existing lattice results in the literature that obtained nonzero results for the CME were misguided by the use of non-conserved electric currents. In turn, the CSE coefficient is nonzero in equilibrium, and recently it was found to exhibit a pronounced temperature-dependence, being suppressed in the confined and large in the deconfined phase.

Concerning out-of-equilibrium effects, the magnetic field has a distinct impact on the electric conductivity, rendering it anisotropic in a way that the QCD medium conducts currents more effectively parallel to the magnetic field as perpendicular to it. First lattice investigations of the out-of-equilibrium CME have been carried out, laying the ground for full-fledged investigations of the time-dependent anomalous transport effect.

	\clearpage
	\section{Summary and outlook}
	\label{chap:sum}

The goal of this review has been to provide a complete summary of the studies of strongly interacting matter in the presence of background electromagnetic fields, obtained via lattice simulations of the underlying theory, QCD.
This is a vast subject that bears relevance for a range of physical systems including magnetized neutron stars, the cosmological evolution of the early Universe and the phenomenology of off-central heavy-ion collisions. The theoretical understanding of these setups requires observables that are accessible in lattice QCD simulations and which have been discussed in the literature in the last several years. We divided these features up into four main chapters, related to confinement and the hadron spectrum; deconfinement and the phase diagram; the equation of state; and transport phenomena and topology.

To set the stage, we began the discussion with a formulation of lattice QCD in the presence of background electromagnetic fields in Chap.~\ref{chap:lat}, focusing on the most important aspects of the discretized theory and the renormalization of relevant observables. Many of the physical effects involving these observables can be understood qualitatively in terms of the spectrum of the Dirac operator. Therefore, Chap.~\ref{chap:ev} was devoted to the lattice studies of the Dirac eigenvalues in the presence of background electromagnetic fields. Armed with this knowledge, in Chap.~\ref{chap:hadron} we discussed magnetic and electric effects on the hadron spectrum and other properties of the confining QCD vacuum. This was followed by the analysis of the QCD phase diagram and the impact of background fields on observables related to deconfinement in Chap.~\ref{chap:pd}. Finally, Chap.~\ref{chap:eos} was devoted to the investigation of the equation of state of magnetized and electrically polarized QCD matter and Chap.~\ref{chap:anom} to the lattice determinations of standard and anomalous transport phenomena and the topological features of the QCD vacuum in the presence of magnetic and electric fields.

Each chapter was concluded with a brief summary, containing the most important lattice findings and the identification of the most pressing open problems and tasks for future research.
Here we reiterate some of these points:
\begin{enumerate}
 \item[a)]
 {\bf Hadron spectrum.} Currently, there are several open questions related to the impact of magnetic and electric fields on hadrons, which should be settled. Notably, these include the calculation of the neutral pion electric polarizability, the flavor content of the neutral pion, the charged pion energy for strong magnetic fields as well as that of the doubly charged $\Delta$ baryon. In general, novel techniques are needed that can go beyond the electroquenched approximation but still maintain a reasonable signal-to-noise ratio, enabling direct comparisons and predictions for experimentally relevant quantities.
 \item[b)]
 {\bf Phase diagram.} The critical point in the magnetic field-temperature phase diagram should be localized more precisely and its impact for the fluctuations of conserved charges should be determined on the lattice. This might provide useful information about potential signatures of the conjectured critical point at nonzero baryon density as well. Learning about the impact of nonzero chemical potentials on the magnetic critical point itself might help in bounding the location of the baryonic critical point, too.
 \item[c)]
 {\bf Equation of state.} Future lattice studies should make a closer contact to the heavy-ion collision setup by finding observables that carry strong imprints of the magnetic fields generated in the initial stage of the collisions and can be compared to experimental data. Such comparisons will necessarily involve a hadron resonance gas-type description of hadron abundances.
 \item[d)]
 {\bf Transport.} Anomalous transport coefficients should be determined on the lattice via spectral reconstruction methods. These can be included in (magneto-)hydrodynamic descriptions of the strongly interacting medium, to be used in the modeling of both magnetized neutron star matter and the expanding fireball in heavy-ion collisions. Once both in-equilibrium and out-of-equilibrium aspects of the chiral magnetic effect have been determined precisely on the lattice, these and hydrodynamic approaches should provide a systematic guidance to experimental searches of the CME.
\end{enumerate}

Finally, lattice results on the thermodynamic properties of QCD in the presence of background magnetic and electric fields should be used to improve low-energy models and effective theories via systematic comparisons. The incorporation of the inverse magnetic catalysis phenomenon, for example, into such models proved to be a major improvement of these approaches. Improving models of this type in regions, where lattice QCD results are available will also help us better understand the nonzero baryon density realm of QCD, where current lattice simulations break down, but where models with predictive power are still of use.

In summary, there are many exciting open questions and challenging tasks for the future. QCD with background electromagnetic fields remains to be a fascinating research subject for a broad community, and for a long time to come, it will surely keep us lattice practitioners occupied -- as well as entertained.

	
	\newpage
	\section*{Acknowledgements}
	It is a great pleasure to thank all of my colleagues with whom I collaborated on projects revolving around background electromagnetic fields: Gunnar Bali, Szabolcs Bors\'anyi, Bastian Brandt, Falk Bruckmann, Martha Constantinou, Marios Costa, Francesca Cuteri, Hannah Elfner, Zolt\'an Fodor, Eduardo Garnacho, Matteo Giordano, Benjamin Gl{\"a}{\ss}le, Jana G{\"u}nther, Javier Hern\'andez, Matthias Kaminski, Ruben Kara, S\'andor Katz, Tam\'as Kov\'acs, Stefan Krieg, Gergely Mark\'o, Haris Panagopoulos, Long-Gang Pang, Ferenc Pittler, Stefano Piemonte, Leon Sandbote, Andreas Sch\"afer, K\'alm\'an \mbox{Szab\'o}, Laurence Yaffe, Dean Valois, Jacob Wellnhofer and Jackson Wu.
	Moreover, I am very grateful for the numerous enlightening magnetized discussions that I had with Gert Aarts, Prabal Adhikari, Andrei Alexandru, Alejandro Ayala, Jens Oluf Andersen, Claudio Bonati, Vladimir Braun, Tom\'a\v{s} Brauner, Pavel Buividovich, Maxim Chernodub, Massimo D'Elia, Heng-Tong Ding, David Dudal, Gerald Dunne, Ricardo Sonego Farias, Christian Fischer, Eduardo Fraga, Kenji Fukushima, Christof Gattringer, Holger Gies, Antal Jakov\'ac, Dmitri Kharzeev, Guy Moore, Laurin Pannullo, Jan Pawlowski, Claudia Ratti, Urko Reinosa, Dirk Rischke, Marco Ruggieri, Hans-Peter Schadler, Andreas Schmitt, Norberto Scoccola, Igor Shovkovy, B\'alint T\'oth and Fuqiang Wang.
	I also thank Heng-Tong Ding for providing the lattice data for one of the figures in this review.
	Furthermore, my special thanks goes to Irene Kehler for her help with preparing this manuscript.
	This work is partly based on projects supported by the German Research Foundation (Emmy Noether Programme EN 1064/2-1 and Collaborative Research Centers SFB/TRR 55 and SFB/TRR 211) and received support from the Hungarian National Research, Development and Innovation Office (Research Grant Hungary 150241) and the European Research Council (Consolidator Grant 101125637 CoStaMM).
	Moreover, this work has been supported by STRONG-2020 ``The strong interaction at the frontier of knowledge: fundamental research and applications'' which received funding from the European Union's Horizon 2020 research and innovation programme under grant agreement No 824093.
	\bibliography{PPNP_review}
	
	\newpage
	\appendix
	\renewcommand*{\thesection}{\Alph{section}}
	\section{Valence and sea contributions to fermionic expectation values}
\label{app:valsea}

In this appendix, we consider the valence and sea contributions to the expectation value of the quark condensate $\expv{\bar\psi_f\psi_f}$, as defined in~\eqref{eq:lat_expvO_fermion},~\eqref{eq:pd_valencedef} and~\eqref{eq:pd_seadef} and discuss the approximate leading-order additivity relation~\eqref{eq:pd_additivity}, valid for weak background magnetic fields. The discussion follows~\cite{DElia:2011koc,Brandt:2024blb} and corrects one important point in the argument of~\cite{DElia:2011koc}.

Let us start by introducing a shorthand notation for the probability measure and the operator as
\be
W(\U,B)=\frac{e^{-\beta S_g}\,\prod_{f'} \det M_{f'}(B)}{\Z(B)}, \qquad
A(\U,B)=\Tr \left[ M_f^{-1}(B)\right]\,,
\ee
where we explicitly indicated the dependence on the gluon field $\U$. With this notation, the full, valence and sea expectation values respectively read
\be
\expv{A}_B=\int \D \U \, W(\U,B)\, A(\U,B), \qquad
\expv{A}^{\rm val}_B=\int \D \U \, W(\U,0)\, A(\U,B), \qquad
\expv{A}^{\rm sea}_B=\int \D \U \, W(\U,B)\, A(\U,0)\,.
\ee
We would now like to expand these observables for weak magnetic fields and exploit their symmetry properties.
All expectation values are even functions of $B$ due to the parity symmetry of the system. However, a single gauge configuration is not parity symmetric, i.e.\ $W(\U,B)\neq W(\U,-B)$ and $A(\U,B)\neq A(\U,-B)$, contrary to the statement in~\cite{DElia:2011koc}.
Therefore, the leading-order term in the weak magnetic field-expansion of $\expv{A}_B$ reads
\be
\left.\frac{\partial^2 \expv{A}_B}{\partial B^2}\right|_{B=0}
=
\int \D\U \left[ \left.\frac{\partial^2 W(\U,B)}{\partial B^2}\right|_{B=0} A(\U,0) + 
W(\U,0) \left.\frac{\partial^2 A(\U,B)}{\partial B^2}\right|_{B=0}
+2
\left.\frac{\partial W(\U,B)}{\partial B}\right|_{B=0}
\left.\frac{\partial A(\U,B)}{\partial B}\right|_{B=0}\,
\right]\,,
\label{eq:app_weakBexp}
\ee
where the three terms respectively correspond to the leading-order term in $\expv{A}^{\rm sea}_B$, that in $\expv{A}^{\rm val}_B$ as well as a mixed term.

\begin{figure}[ht!]
 \centering
 \tikzset{cross/.style={cross out, draw=black, minimum size=10*(#1-\pgflinewidth), inner sep=0pt, outer sep=0pt}, cross/.default={1pt}}
 \vspace*{.2cm}
\begin{tikzpicture}[scale=1,line width=0.18mm]
  \draw (-.5,0) circle (.8cm);
  \draw (2,0) circle (.8cm);
  \draw[decorate,decoration={gluon, amplitude=3pt,
    segment length=8pt, aspect=0}] (-1.3,0) -- (-2.5,0);
  \draw[decorate,decoration={gluon, amplitude=3pt,
    segment length=8pt, aspect=0}] (2.69,.4) -- (4,.4);
  \draw (2.69,-.4) node[cross,rotate=-20] {};
  \draw[decorate,decoration={gluon, amplitude=4pt,
    segment length=4.9pt}] (.3,0)   -- (1.2,0);
  \draw[decorate,decoration={gluon, amplitude=4pt,
    segment length=4.2pt}] (.2,.4)   -- (1.3,0.4);
  \draw[decorate,decoration={gluon, amplitude=4pt,
    segment length=4.2pt}] (.2,-.4)   -- (1.3,-.4);
\end{tikzpicture}
\caption{\label{fig:app_diagram1}
 The first nonzero diagram in perturbation theory contributing to the third term in~\protect\eqref{eq:app_weakBexp}. The solid lines denote quark propagators (at zero magnetic field), the wavy lines external photon legs, the curly lines gluon propagators, while the cross the scalar insertion.
}
\end{figure}
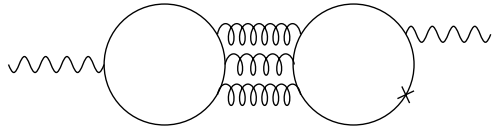

To explicitly show that the last contribution is nonzero, we consider it in perturbation theory. The factor $\partial W/\partial B$ leads to a trace over a quark propagator $M_{f'}^{-1}$ and an external photon leg (the magnetic field derivative). The factor $\partial A/\partial B$ involves two quark propagators $M_f^{-1}$, a scalar insertion and an external photon leg. Thus, the third term in~\eqref{eq:app_weakBexp} altogether consists of two quark loops, two external photon legs and a scalar insertion. To form a connected diagram, the two loops must be connected by internal gluon or quark propagators. The lowest order diagram that does not vanish is of $\mathcal{O}(g_s^6)$ and is depicted in Fig.~\ref{fig:app_diagram1}. For mass-degenerate quarks, it is proportional to $\sum_{f'} q_{f'}$, which vanishes in the three-flavor theory. Altogether we conclude that the third term in~\eqref{eq:app_weakBexp}, is highly suppressed and the full expectation value is almost entirely given by the sum of the valence and sea terms to $\mathcal{O}(B^2)$, proving~\eqref{eq:pd_additivity}.

We note that the diagram in question is the mass derivative of the same diagram without the scalar insertion. The latter has been calculated in the massless limit in~\cite{Baikov:2012zm}.

\end{document}